\documentclass{aa}  

\usepackage{graphicx}
\usepackage{txfonts}
\usepackage{enumerate}
\usepackage{paralist}
\usepackage[section]{placeins}
\bibpunct{(}{)}{;}{a}{}{,} 

\usepackage{wrapfig}

\usepackage[belowskip=-10pt,aboveskip=0pt]{caption}
\usepackage{enumitem}
\usepackage{mwe}

\RequirePackage[colorlinks,citecolor=blue,urlcolor=blue,pdfencoding=auto, psdextra]{hyperref}

\begin{document} 

\let\include\input

   \title{A closer look at the deep radio sky:  \\    Multi-component radio sources at 3-GHz VLA-COSMOS   }
   \titlerunning{Multi-component radio sources at 3-GHz COSMOS}
   \subtitle{}
\let\cleardoublepage\clearpage
  \author{E. Vardoulaki\inst{1}\fnmsep\thanks{email: elenivard@gmail.com},    
       		E. F. Jim\'{e}nez Andrade\inst{1}\fnmsep\inst{2},
                A. Karim\inst{1},
                M. Novak\inst{3}\fnmsep\inst{4},
                S. K. Leslie\inst{3},   
                K. Tisani\'{c} \inst{4},
                V. Smol\v{c}i\'{c}\inst{4},
                E. Schinnerer\inst{3}, 
                M. T. Sargent\inst{5},
                M. Bondi\inst{6}, 
                G. Zamorani\inst{7},
                B. Magnelli\inst{1},
                F. Bertoldi\inst{1},
                N. Herrera Ruiz\inst{8},
                K. P. Mooley\inst{9}\fnmsep\inst{10},                
                J. Delhaize\inst{4},
                S. T. Myers\inst{10},
                S. Marchesi\inst{11},
                A. M. Koekemoer\inst{12},    
                G. Gozaliasl\inst{13}\fnmsep\inst{14}\fnmsep\inst{15},   
                A. Finoguenov\inst{14}\fnmsep\inst{16},      
                E. Middleberg\inst{8},
                P. Ciliegi\inst{7} 
          }

\authorrunning{Vardoulaki et al.}



   \institute{Argelander-Institut f\"{u}r Astronomie, Auf dem H\"{u}gel 71, D-53121 Bonn, Germany
\and
International Max Planck Research School of Astronomy and Astrophysics at the Universities of Bonn and Cologne
	\and
	Max-Planck-Institut f\"{u}r Astronomie, K\"{o}nigstuhl 17, 69117, Heidelberg, Germany
	\and
	Department of Physics, Faculty of Science, University of Zagreb, Bijeni\v{c}ka cesta 32, 10000  Zagreb, Croatia
        \and
        Astronomy Centre, Department of Physics and Astronomy, University of Sussex, Brighton, BN1 9QH, UK
      \and
         INAF - Istituto di Radioastronomia, Via Gobetti 101, 40129 Bologna, Italy 
         \and 
         INAF-Osservatorio di Astrofisica e Scienza dello Spazio di Bologna, Via Piero Gobetti 93/3, I - 40129 Bologna, Italy
          \and     
                Astronomisches Institut, Ruhr-Universit\"{a}t Bochum, Universit\"{a}tsstrasse 150, 44801 Bochum, Germany
         \and
         Caltech, 1200 E. California Blvd. MC 249-17, Pasadena, CA 91125, USA 
         \and
         National Radio Astronomy Observatory, P.O. Box 0, Socorro, NM 87801, USA
         \and
         Department of Physics and Astronomy, Clemson University, Clemson, SC 29634, USA
         \and 
         Space Telescope Science Institute, 3700 San Martin Drive, Baltimore MD 21218, USA
         \and
         Finnish centre for Astronomy with ESO (FINCA), Quantum, Vesilinnantie 5, University of Turku, FI-20014, Turku, Finland 
         \and
         Department of Physics, University of Helsinki, P. O. Box 64, FI-00014 , Helsinki, Finland
         \and
         Helsinki Institute of Physics, University of Helsinki, P.O. Box 64, FI-00014, Helsinki, Finland
         \and
         Max-Planck Institut f\"{u}r extraterrestrische Physik, Postfach 1312, 85741 Garching bei M\"{u}nchen, Germany
          }         

   \date{Received ; accepted }
 
  \abstract
   {Given the unprecedented depth achieved in current large radio surveys, we are starting to probe populations of radio sources that have not been studied in the past. However, identifying and categorising these objects,  differing in size, shape and physical properties,  is  becoming a more difficult task.}
   {In this data paper we present and characterise the multi-component radio sources identified in the VLA-COSMOS Large Project at 3 GHz (0.75 arcsec resolution, 2.3 $\mu$Jy/beam $rms$), i.e. the radio sources which are composed of two or more radio blobs. }
   {The classification of objects into multi-components was done by visual inspection of 351 of the brightest and most extended blobs from a sample of 10,899 blobs identified by the automatic code \textsc{blobcat}. For that purpose we used multi-wavelength information of the field, such as the 1.4-GHz VLA-COSMOS data and the UltraVISTA stacked mosaic available for COSMOS.}
   {We have identified 67 multi-component radio sources at 3 GHz: 58 sources with AGN powered radio emission and 9 star-forming galaxies. We report 8 new detections that were not observed by the VLA-COSMOS Large Project at 1.4 GHz, due to the slightly larger area coverage at 3 GHz. The increased spatial resolution of 0.75 arcsec has allowed us to resolve (and isolate) multiple emission peaks of 28 extended radio sources not identified in the 1.4-GHz VLA-COSMOS map. We report the multi-frequency flux densities (324 MHz, 325 MHz, 1.4 GHz \& 3 GHz), star-formation-rates, and stellar masses of these objects. We find that multi-component objects at 3-GHz VLA-COSMOS inhabit mainly massive galaxies ($>10^{10.5}~\rm M_{\odot}$). The majority of the multi-component AGN lie below the main-sequence of star-forming galaxies (SFGs), in the green valley and the quiescent region. Furthermore, we provide detailed description of the objects and find that amongst the AGN there are 2 head-tail, 10 core-lobe, 9 wide-angle-tail (WAT), 8 double-double or Z-/X-shaped, 3 bent-tail radio sources, and 26 symmetric sources, while amongst the SFGs we find the only star-forming ring seen in radio emission in COSMOS. Additionally, we report a large number (32/58) of disturbed/bent multi-component AGN, 18 of which do not lie within X-ray groups in COSMOS (redshift range 0.08 $\leq z <$ 1.53). }
   {The high angular resolution and sensitivity of the 3-GHz VLA-COSMOS data-set give us the opportunity to identify peculiar radio structures and sub-structures of multi-component objects, and relate them to physical phenomena such as AGN or star-forming galaxies. This study illustrates the complexity of the $\mu$Jy radio-source population; at the sensitivity and resolution of 3-GHz VLA-COSMOS, the radio structures of AGN and SFG both emitting radio continuum emission, become comparable in the absence of clear, symmetrical jets. Thus, disentangling the AGN and SFG contributions using solely radio observations can be misleading in a number of cases. This has implications for future surveys, such as done by SKA and precursors, which will identify hundreds of thousand multi-component objects. }


   \maketitle
%

\section{Introduction}

For several decades astronomers have been exploring the radio part of the electromagnetic spectrum, probing the physical phenomena that are responsible for emitting at radio wavelengths \cite[see][for a review]{simpson17}. Thanks to the greater sensitivity of modern radio observatories such as the Karl G. Jansky Very Large Array (VLA), we now have the opportunity to study in greater detail and depth populations of objects.  Radio observations, being dust-free probes of star-formation (SF), pinpoint to the birthplace of stars within galaxies and can also highlight the complexity of radio structures emitted by active galactic nuclei (AGN). They further aid the study of feedback mechanisms from AGN \citep[e.g.][]{deyoung10,best14,williams15}. Radio observations, in combination with multi-wavelength observations are a powerful tool to study the physical processes behind the formation of the stars in galaxies, as well as the energy released by AGN in their environment in the form of kinetic energy, probed by synchrotron radiation \cite[e.g.][]{willott99,smolcic17c}.

For galaxies, the radio-frequency range below 10 GHz is dominated by non-thermal synchrotron radiation \citep[e.g.][]{condon92}. As a result, radio studies in this range pick up populations of radio sources that are either powered by AGN or SF, as both physical phenomena emit non-thermal synchrotron radiation. Additionally, in some cases both phenomena simultaneously occur (hybrid or composite objects, for example \cite{symeonidis13, seymour09}, but see \cite{padovani16} for a different point of view on the nomenclature of calling these hybrid objects). In the case of star formation, what we observe in the radio is synchrotron radiation from cosmic-ray electrons accelerated by supernova remnants \citep[e.g.][]{condon92, murphy09}, such that non-thermal radio emission traces the most recent episodes of massive star-formation. In the case of AGN, synchrotron radiation that is observed at radio wavelengths originates from relativistic electrons spiralling around the magnetic field associated with the central black hole region \citep[see][for a review of unified AGN models]{antonucci93, netzer15}. The so-called radio-mode feedback \citep[e.g.][]{fabian12} from AGN, or kinetic/jet feedback, which is thought to preventing massive galaxies in the present-day Universe from new star-formation activities \cite[e.g.][]{ishibashi14}, can have different observational signatures in the radio, giving rise to complex radio structures. 

Traditionally, radio AGN with extended radio structure are classified as edge-brightened or FRII (lobed-like radio sources) and as edge-darkened or FRI (jet-like radio sources) depending on the distribution of power along their radio structure \citep{fr74}. Thus in large radio surveys one can find radio sources of different types, shapes and sizes, some single-component and some composed of several components (or multi-component; e.g. lobes, jets, star-forming regions). These systems can exhibit peculiar and complex structures, which gives rise to questions regarding their formation and their interaction with the surrounding intergalactic (IGM) and/or intra-cluster medium (ICM). The complexity of the structures observed in radio surveys, as well as the different sizes and shapes, in combination with the two physical phenomena in place (AGN or SF), hold a challenge when it comes to current automatic identification and classification algorithms \cite[e.g. \textsc{blobcat}, \textsc{pyBDSF};][respectively]{hales12,mohan15}. These algorithms cannot fully undertake the task of identifying multiple components, and disentangling AGN and SF objects through their radio signatures. 

Here we present the multi-component radio sources in the VLA-COSMOS Large Project at 3 GHz \citep{smolcic17a}, and compare them to their 1.4-GHz analogues presented by \cite{schinnerer07, schinnerer10}. We discuss the difficulties of identifying multi-component radio sources in radio mosaics given the complexity and plethora of the radio structures found at such low flux densities, ranging from 100 $\mu$Jy down to $\sim$ 10 $\mu$Jy.  We present a value added catalogue for these objects with corrected radio positions and core flux-densities, as well as radio properties and properties of the host, such as star-formation-rate (SFR) and stellar mass (M$_{*}$), in Section~\ref{sec:sample}. We also create a tool (presented in Appendix~\ref{app:matching}) to cross-match multi-component sources to their analogues at other radio frequencies in COSMOS, by taking into account the different angular resolution of those surveys and the size of our objects. We present an analysis on the SFR and stellar masses of their host galaxies in Sec.~\ref{sec:hosts}. Furthermore we provide some general notes on the objects (Sec.~\ref{sec:notes}), while a more detailed description based on their radio structure can be found in Appendix~\ref{app:notes}. We close our paper with a discussion on the need for automatic identification methods which can identify multi-component sources and disentangle AGN and SF objects (Sec.~\ref{sec:discuss}). The objects presented here are shown in Fig.~\ref{fig:maps2} in the Appendix.

Throughout this paper we use the following convention for all spectral indices, $\alpha$: flux density is $S_{\nu} \propto \nu^{-\alpha}$,
where $\nu$ is the observing frequency. Also, a low-density, $\Lambda$-dominated Universe in which $H_{0}=70~ {\rm km~s^{-1}Mpc^{-1}}$, $\Omega_{\rm M}=0.3$ and $\Omega_{\Lambda}=0.7$ is assumed throughout. For the estimation of star-formation-rates and stellar masses, a \cite{chabrier03} initial mass function (IMF) was used.

\section{Analysis of multi-component sample}
\label{sec:sample}

In this paper we present the multi-component radio sources from the VLA-COSMOS Large Project at 3 GHz \citep[][3-GHz VLA-COSMOS henceforth]{smolcic17a}, i.e. objects that are composed of two or more radio blobs, or islands of pixels representing sources. The survey is complete down to 40 $\mu$Jy \citep[94\%; see Fig. 16 of][]{smolcic17a} at a resolution of 0.75 arcsec and $rms$ of 2.3 $\mu$Jy/beam. Details on the data reduction can be found in \cite{smolcic17a}. 

In the radio mosaic, radio emission of some morphologically complex galaxies can be split into multiple blobs if the surface brightness drops below the detection threshold. For example, this might happen when the radio emission is composed of a core, faint jets and bright lobes, or several disjoint star-forming regions. Given the sub-arcsec resolution of the 3-GHz data we expect that individual components of such sources are extended with non-Gaussian structures.

We note that the definition of multi-component sources in this paper differs from other definitions that deem multi-component sources those made up by multiple Gaussian components \citep[e.g.][]{schinnerer07}. Our definition of multi-component, as in \cite{smolcic17a}, has to do with a parent source being composed of two or more islands/blobs.

For the identification and classification of the objects we make use of the 1.4-GHz VLA-COSMOS Large Project \citep{schinnerer07, schinnerer10}, which surveyed the 2 deg$^{2}$ of the COSMOS field at a resolution of $\sim$ 1.5 arcsec. We also make use of the optical/near-infrared stacked $YJHK_{S}$ image from the Ultra Deep Survey with the VISTA telescope \citep[UltraVISTA; see][and references therein]{laigle16, smolcic17b}, including regions observed at $z^{++}$ with an upgrade of the Subaru Suprime-Cam \cite[see][]{taniguchi07, taniguchi15, smolcic17b}. \cite{laigle16} combined the near-IR images of UltraVISTA (YJHKs) with the optical $z^{++}$-band data from Subaru in order to probe the high-redshift Universe and provide a catalogue containing UV-luminous sources at $z>$ 2. The photometry was done using SExtractor in dual image mode and the stacked image created using SWARP (Bertin et al. 2002). The final product is a stacked image at $z^{++}YJHK_{S}$, which we will refer to from now on as UltraVISTA stacked or $\chi^{2}$ map, and has arbitrary units. This map helped significantly in identifying which blobs belong to which source, as is described further down. Without the use of multi-wavelength data, a solely radio-based classification is not conclusive (see Sec.~\ref{sec:disentangling} for an example).

   \begin{figure}[!ht]
    \resizebox{\hsize}{!}
            {
            \includegraphics{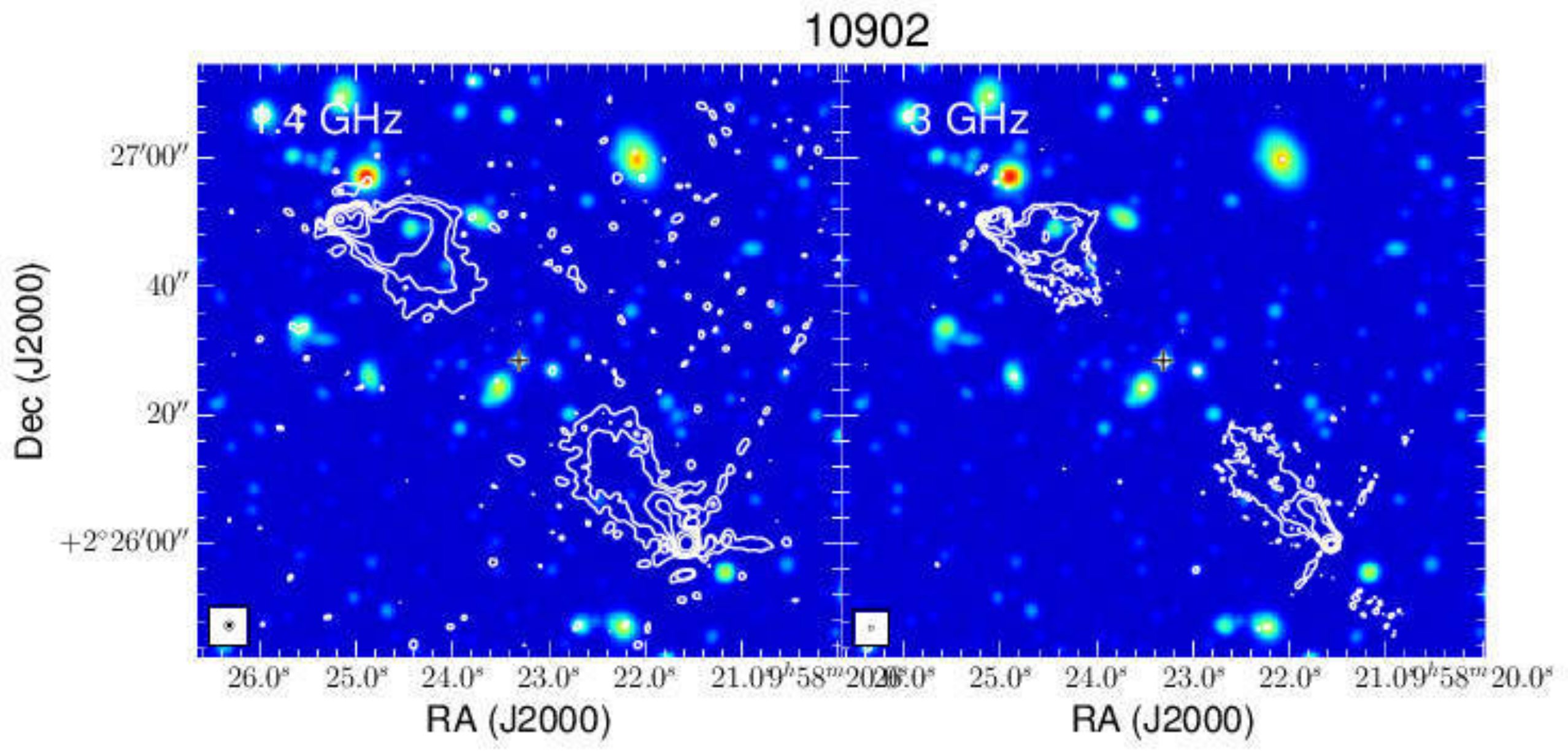}
           }
            \resizebox{\hsize}{!}
            {
            \includegraphics{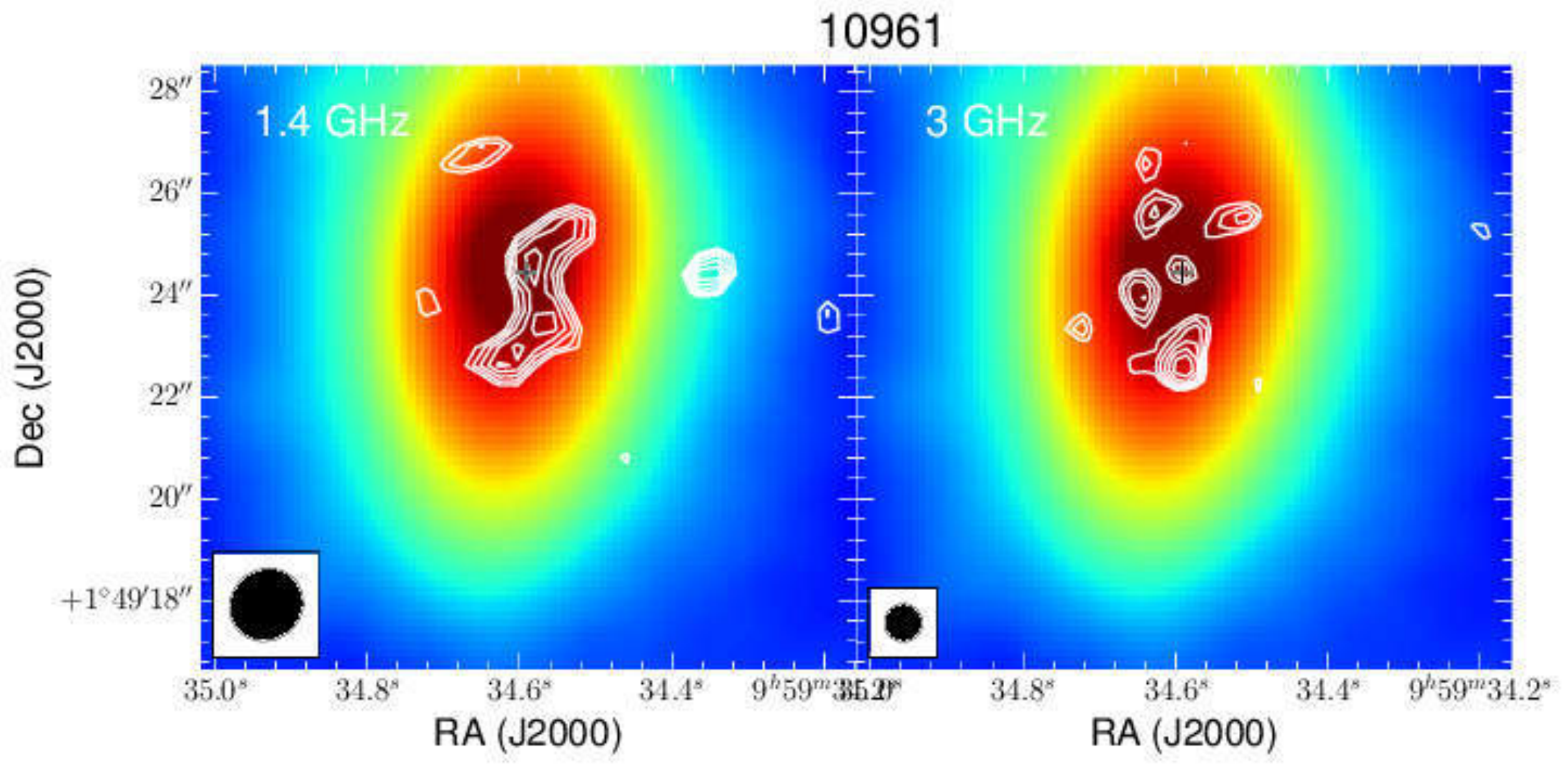}
            }
            
            \caption{Set of 1.4-GHz VLA ($left$) and 3-GHz VLA ($right$) stamps for two of the radio multi-component objects in the COSMOS field shown as white contours, overlaid on the UltraVISTA stacked image shown in colour-scale in arbitrary units. The beam size for the radio observations is at the bottom-left corner of the stamp: 1.4$\times$1.5 arcsec$^{2}$ for the 1.4-GHz and 0.75 arcsec FWHM for the 3-GHz maps. {The black cross denotes the 3-GHz radio position as in Table~\ref{table:data}.} The contour levels are equally spaced on a log-scale, where the lowest is set at 3 $\sigma$ and the highest at the maximum peak flux-density of the radio structure. The top set of stamps shows an example of radio AGN, while the bottom set shows a SFG. The remainder of the maps can be found throughout the text, grouped in categories depending on the type of the object and in the Appendix. \\
   }
              \label{fig:radmaps}%
    \end{figure}
%

\subsection{Source identification and matching of radio blobs}
\label{sec:sourceid}

To identify sources in the 3-GHz mosaic, \cite{smolcic17a} used the automatic algorithm \textsc{blobcat} \citep{hales12}, which locates islands, or radio blobs, assigns a catalogue number or ID, and performs flux density, $rms$, position and size measurements amongst others.  This and similar codes \citep[e.g. \textsc{pyBDSF};][]{mohan15} perform well for single source identification. Nevertheless, to date, even the very advanced codes available \citep[see][for a bayesian approach]{fan15} cannot completely substitute the time-consuming step of visual inspection. For that reason visual inspection of the sample is required, and not only as a validation test. \textsc{blobcat} produced a catalogue from the 3-GHz mosaic with 10,899 entries, which makes it time-consuming to visually inspect all radio blobs. Furthermore, as the visual inspection technique is quite subjective depending on the inspector and their knowledge of the physics of these objects, the exercise will need to be repeated independently by several inspectors to get to a reliable match.

\begin{figure}[!ht]
    \resizebox{\hsize}{!}
            {\includegraphics[trim={0cm 0cm 0cm 0.2cm},clip]{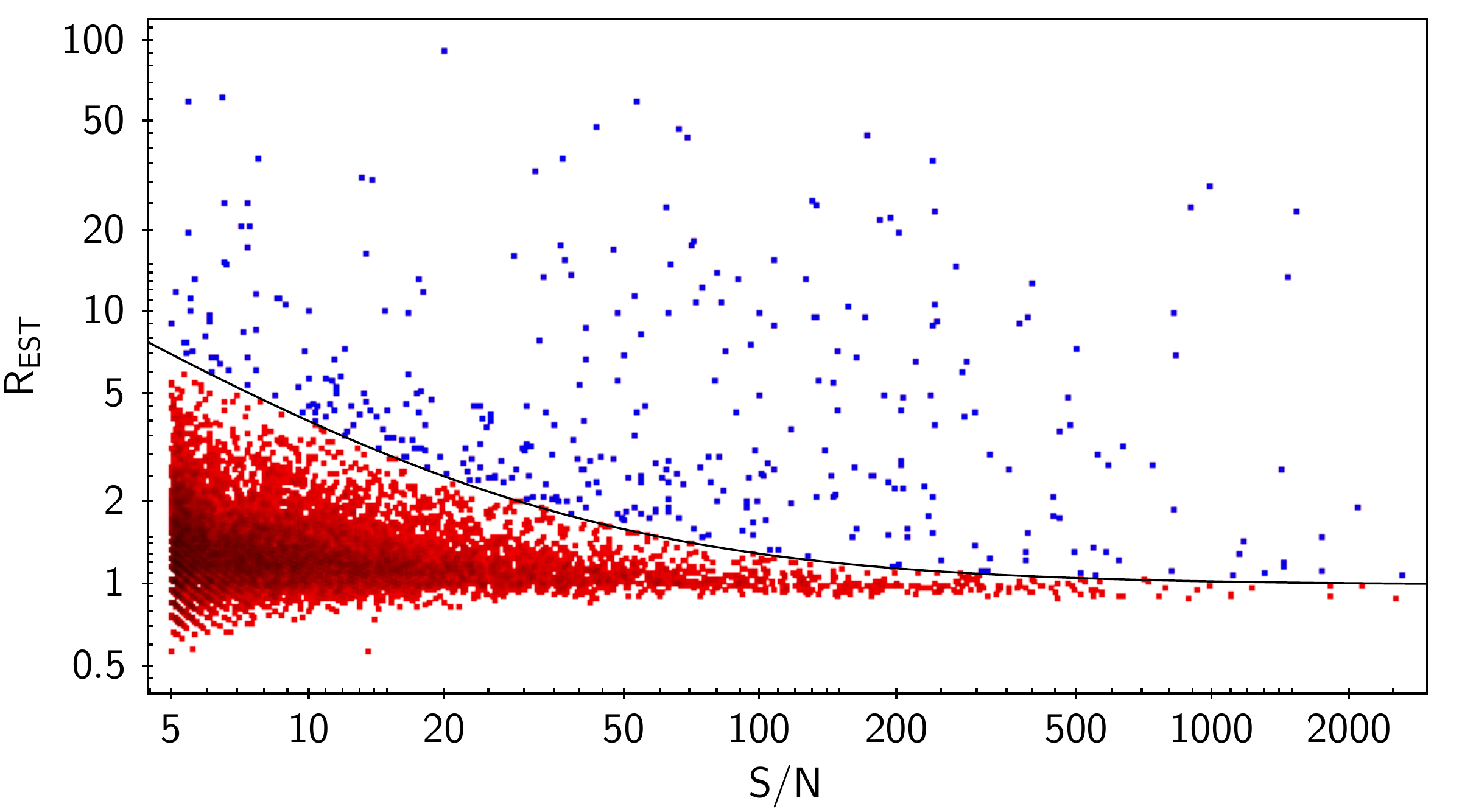}
 }
 \\
               \caption{ R$_{\rm EST}$ parameter \citep[a size estimate from \textsc{blobcat} not intended for quantitative analysis;][]{hales12} versus signal-to-noise ratio (S/N) for the 10,899 entries in the \textsc{blobcat} catalogue identified from the 3-GHz mosaic of the COSMOS field. The black solid line separates the sources that obey the relation R$_{\rm EST} >$ 1+30/(S/N) and were visually inspected to create the multi-component sample from the rest of the 3-GHz sources, as described in Sec.~\ref{sec:sourceid}.
   }
              \label{fig:rest_snr}%
    \end{figure}

To limit the number of components that require visual inspection, and to pick up objects composed of several radio blobs (e.g. lobes from an AGN that extend beyond the host galaxy), we utilise the $R_\text{EST}$ parameter reported by \textsc{blobcat}. This parameter provides a size estimate in the units of the sky area covered by an unresolved Gaussian blob with the same peak surface brightness (see Eq.~34 in \citealt{hales12}). The larger the value ($R_\text{EST} >$ 1.4; suited for automatically flagging complex blobs), the more likely the source has a complex radio structure, while the value of one corresponds to a point source. Random noise variations can also contribute to the size estimate, especially at low signal-to-noise ratios (S/N). For this reason, we have chosen by eye an envelope defined as $R_\text{EST} > 1+30\times(\text{S/N})^{-1}$ (shown in Fig.~\ref{fig:rest_snr} as black line) to select likely candidates for complex sources. Such a selection yields 351 components in total, shown in blue, which is a fairly small number of blobs to perform visual inspection on. These components were then combined into multi-component sources where necessary (see also Sect.~3.3 in \citealt{smolcic17a}). To make sure we are not missing FR-type multi-component objects due to our selection, the entire 2.6 deg$^{2}$ mosaic was visually inspected, despite this being highly impractical and time consuming.

Within these sources there are 127 known and previously observed extended radio AGN from the VLA-COSMOS survey at 1.4 GHz \citep{schinnerer07, schinnerer10}, which aided in the matching process\footnote{At the resolution of the 1.4-GHz observations (1.5 arcsec), their  analogues are not necessarily multi-component objects (see Fig~\ref{fig:maps2}), as their radio structure might not be divided in several radio blobs. The reason for this is partly the different and coarser resolution than at 3 GHz and partly due to diffuse emission being missed at 3 GHz.}. The R$_{\rm EST}$ envelope was selected to pick up the brightest and most extended radio components. Less bright components that might be part of the parent source and fall below the envelope in Fig.~\ref{fig:rest_snr} are picked up as components of the parent radio source after visual inspection.

In this paper we go beyond a radio-only classification, in order to avoid mismatches and mis-classifications. Thus we created overlays of the radio 3 GHz and the stacked UltraVISTA map, similar to what is presented in Figs~\ref{fig:radmaps}~\&~\ref{fig:maps2}. The maps were inspected visually by seven researchers of the group to match the radio blobs to the corresponding optical/near-IR counterpart. At a later stage, \cite{smolcic17b} performed counterpart association, i.e. they combined the 3-GHz radio data with optical, near-infrared  \citep[UltraVISTA; see][and references therein]{laigle16}, and mid-infrared Spitzer/IRAC data \citep{sanders07}, as well as X-ray data \citep[Chandra-COSMOS \& COSMOS-Legacy;][]{elvis09, civano12, civano16, marchesi16}, to match the radio sources to their corresponding hosts out to $z <$ 6. In this procedure, they made use of the latest photometric catalogue available for COSMOS (henceforth COSMOS2015; \citealt{laigle16}). For the multi-component objects, the locations of the hosts were re-examined by eye. There is only one mis-match with the counterpart catalogue, that of source ID 10904, where the radio position was given between two neighbouring galaxies. Thus we re-measured the radio position and we present it in Table~\ref{table:data}. 

After carefully inspecting and matching the blobs to single radio sources, we re-measured the radio positions and flux densities, as well as local $rms$ and added the matched sources as new entries in the VLA catalogue produced by \textsc{blobcat} after removing multiple entries. The matched components, deemed multi-component radio sources at 3-GHz VLA-COSMOS, were assigned a new ID starting from 10900-10966. The full sample of multi-component radio sources is presented in Table~\ref{table:data}. 

Furthermore, objects were classified in AGN or SFGs based on their characteristic radio structure with the aid of the 1.4-GHz maps and the UltraVISTA $\chi^{2}$ map which highlights the location of the host galaxy, in the following way:

\begin{enumerate}
      \item AGN: Multi-component sources with two or more blobs that belong to the same parent radio source and resemble jets/lobes produced by an AGN.
       \item SFGs: Multi-component sources composed of several blobs which are associated with the disk of the galaxy in the optical/near-IR, and they do not resemble jets/lobes produced by an AGN.
\end{enumerate}

In Fig.~\ref{fig:radmaps} we give examples of the two main categories of objects, AGN (top; 10902) and star-forming galaxies (SFGs; bottom; 10961), presented in this paper. The rest of the objects can be found in the Appendix in Fig.~\ref{fig:maps2}. For comparison, we also present the 1.4-GHz maps for each multi-component source, overlaid as contours on the UltraVISTA stacked image.

The total flux of the multi-component sources was derived in the following way. Firstly, a more reliable determination of the noise close to bright and extended sources was derived measuring the $rms$ of the total intensity image in a region at least 200$\times$200 pixels$^{2}$ nearby the radio source but free of radio emission. Then the image was blanked down to 2 times the $rms$ and the total flux density was measured using the AIPS task TVSTAT that allowed for the integration of the multi-component source flux density over irregular areas. Secondly, we measured the core flux-densities of AGN-type multi-component objects by fitting a Gaussian component around the radio core using the task IMFIT in CASA \citep{McMullin07}. When needed, radio positions were corrected to match that of the core, as the \textsc{blobcat} algorithm provides the position of the weighted mean, which in this case does not correspond to the core position. All values and radio positions are presented in Table~\ref{table:data}, along with redshift information and the 1.4-GHz analogue. Finally, we provide information about the corresponding VLBA source from \cite{noelia17}.

\begin{figure}[!ht]
    \resizebox{\hsize}{!}
            {\includegraphics[trim={0cm 0cm 0cm 0.2cm},clip]{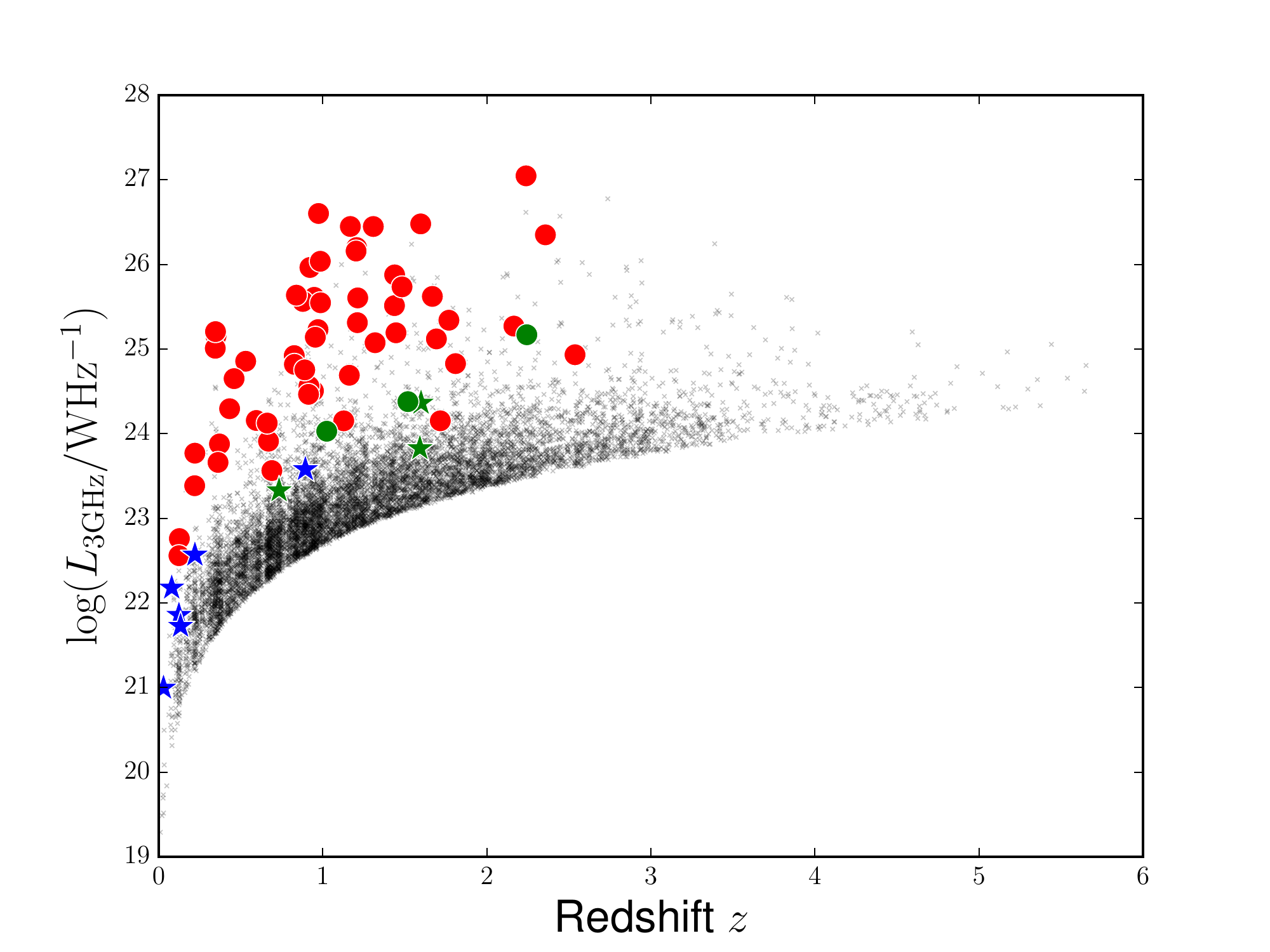}
            
 }
   \resizebox{\hsize}{!}
            {
            \includegraphics[trim={0cm 0cm 0cm 0.2cm},clip]{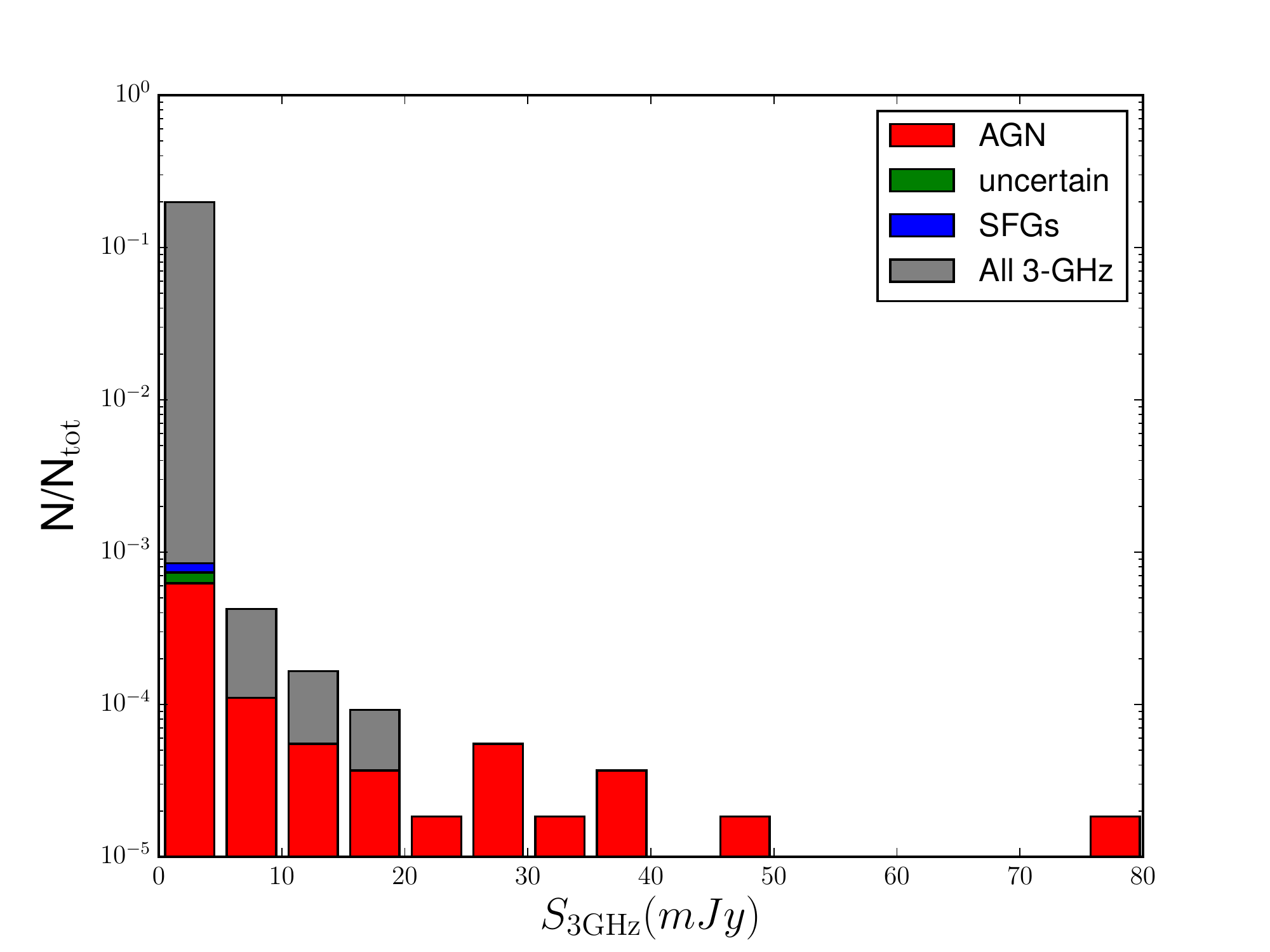}
            
 }
 \\
               \caption{{\bf Top}: Radio luminosity at 3 GHz versus redshift showing the multi-component radio sources compared to the single-component radio sources at 3 GHz. Symbols: red circles multi-component AGN; blue stars multi-component SFGs; green circles/stars for uncertain class, where circle is for AGN and star for SFG based on the radio excess (see Sec.~\ref{sec:hosts}); small black crosses show the single-component radio sources at 3 GHz. {\bf Bottom}: Flux-density histogram showing the multi-component objects (AGN in red, uncertain in green and SFGs in blue) relative to the full 3 GHz sample (grey).
   }
              \label{fig:l3z}%
    \end{figure}

In Fig.~\ref{fig:l3z}-$Top$ we compare the radio luminosities of the multi-component radio sources at 3 GHz to the rest of the objects at 3-GHz COSMOS which are single-component radio sources. We see that multi-component objects occupy the region in the $L-z$ diagram of higher luminosities than the single-component objects at the corresponding redshifts. This can be seen more clearly in the flux-density histogram in Fig.~\ref{fig:l3z}-$Bottom$, where at flux densities above 20 mJy we only find AGN multi-component objects.

Furthermore, in Table~\ref{table:data2} we give the host properties of the multi-component radio sources, such as SFR and stellar mass from the counterpart catalogue of \cite{smolcic17b}, and the COSMOS2015 identification number. We also provide classification, AGN or not, based on the spectral energy distribution (SED) fit \citep{delvecchio17,smolcic17b}. 

The SFR and M$_{*}$ used in this analysis were obtained from the study of \cite{delvecchio17}, after fitting the multi-wavelength SEDs of these objects with the SED-fitting code \textsc{magphys} \citep{dacunha08} and the three-component SED-fitting code \textsc{sed3fit} by \cite{berta13}, which accounts for an additional AGN component. In particular they use the plethora of data in the COSMOS2015 catalogue \citep{laigle16}, including Herschel observations to constrain the far-infrared part of the SED for lower redshift galaxies, while for higher redshift they include a large dataset of sub-mm photometry from JCMT/SCUBA-2, LABOCA, Bolocam, JCMT/AzTEC, MAMBO, ALMA and PdBI \citep[see Sec. 2.2 in][ for references and discussion]{delvecchio17}. In addition they make use of the Chandra-COSMOS \citep{elvis09, civano12} and COSMOS-Legacy \citep{civano16} X-ray catalogues.

\subsection{Multi-frequency radio data}
\label{sec:radiodata}

In Table~\ref{table:freqs} we present radio photometry from MHz to GHz frequencies for the multi-component objects at 3 GHz. We have matched the 3-GHz radio positions to the following radio catalogues: 324 MHz from VLA \citep{smolcic14}; 325 MHz from GMRT \citep{tisanic18}; and 1.4 GHz from VLA \citep{schinnerer10}. The matching of the multi-frequency data with the 3-GHz multi-component sources was done by an automatic code which is described in Appendix~\ref{app:matching}. In short, the code takes as prior the positions of the 3-GHz multi-component objects and searches within the area covered by the source, rejecting near-by objects that are not associated with it. The area covered by the source at 3 GHz was estimated by the linear-projected size of the source, which was measured using a semi-automatic method presented in Vardoulaki et al. (in prep.). In some cases more than one blob was matched to a 3-GHz radio source. We thus report in Table ~\ref{table:freqs} the multi-frequency flux densities and whether the source is a multi-component or not at the corresponding radio frequency. In the following paragraphs, we briefly describe the aforementioned radio surveys.

The 324 MHz mosaic \citep{smolcic14} is a shallow survey of the COSMOS field, covering 3.14 deg$^{2}$, with an average $rms$ noise level of 0.5 mJy/beam and an angular resolution of 8".0$\times$6".0. The extracted catalogue includes 182 sources (down to 5.5$\sigma$). Details on the observations can be found in \cite{smolcic14}.

The GMRT radio map at 325 MHz is the result of a single pointing covering the $\sim$2 deg$^{2}$ of the COSMOS field. This was observed for 45 hrs under the project 07SCB01 (P.I.: S. Croft). The 325 MHz catalogue were created from the GMRT maps \citep{tisanic18} using \textsc{blobcat} down to a $5\sigma$ limit, giving 633 radio blobs. At 325 MHz the angular resolution is $\sim$ 10".8$\times$9".5 with a median $rms$ of 87 $\mu$Jy/beam.  

A comparison between 324 and 325 MHz shows a non-negligible difference between the values, which points on the difficulty of measuring homogeneous flux densities for these sources. 
 
 In Table~\ref{table:freqs} we see that the majority of multi-component sources 98\% are single-component sources at 325 MHz, which is a result of the much lower resolution compared to the 3 GHz (10" arcsec).
 
The 2 deg$^{2}$ of the COSMOS field were observed by \cite{schinnerer07} at 1.4 GHz with the VLA, reaching the angular resolution of 1".5$\times$1".4 and a mean $rms$ of 10.5 (15) $\mu$Jy beam in the central 1 (2) deg$^{2}$, yielding $\sim$ 3600 radio sources. From the 67 multi-component radio sources in the 3-GHz VLA-COSMOS map, 28 objects are not classified as such at 1.4 GHz, instead they are single-component radio sources. This number is derived by visual inspecting the 1.4 GHz cutouts. Thus at 3 GHz we find $\sim$ 42\% more multi-component sources than at 1.4 GHz. This is due to the higher resolution of 0.75 arcsec at 3 GHz, compared to 1.5 arcsec at 1.4 GHz. As a result, diffuse radio emission seen at 1.4 GHz might be resolved out, and the observational consequence is that the radio flux distribution splits into several components. In fact, we find that, on average, flux densities at 1.4 GHz are by $\sim$30\% larger than the ones expected on the basis of the 3 GHz flux densities, assuming a simplistic radio spectral index of 0.8. This suggests that at 3 GHz we are missing part of the diffuse emission seen at 1.4 GHz. Furthermore, we find 8 new identifications at the edges of the 3-GHz mosaic (2.6 deg$^{2}$), which were not recovered by the 2 deg$^{2}$ 1.4-GHz observations (10908, 10922, 10924, 10932, 10938, 10941, 10946 \&10950). In Appendix~\ref{sec:resolution} we test how the coarser resolution can change the appearance of the 3-GHz multi-component sources. We note that our classification of multi-component sources differs from the one in \cite{schinnerer07, schinnerer10}, where they deem multi-component objects those that can be fit by multiple Gaussians, thus we refrain from adding their classification in Table~\ref{table:freqs}. Instead we only present the results from matching the 1.4 and 3 GHz data with the matching method described in Appendix~\ref{app:matching}.

%
\begin{sidewaystable*}
\caption{Multi-component Radio Properties}             
\label{table:data}      
\centering          
\begin{tabular}{l l l c c c c c r r c r}   
\hline\hline       
                 
\multicolumn{1}{c}{3-GHz}  &  \multicolumn{1}{c}{} & \multicolumn{1}{c}{R.A.} & 
\multicolumn{1}{c}{Dec.} & \multicolumn{1}{c}{$S_{\rm 3 GHz}$} & \multicolumn{1}{c}{$S_{\rm core~3 GHz}$} & \multicolumn{1}{c}{$S_{\rm VLBA~1.4 GHz}$} &\multicolumn{1}{c}{z} & 1.4-GHz ID& VLBA & radio \\ 
\hline                    
 \multicolumn{1}{c}{ID} & \multicolumn{1}{c}{COSMOSVLA3} & \multicolumn{1}{c}{(h:m:s)}  & \multicolumn{1}{c}{(d:m:s)} & \multicolumn{1}{c}{(mJy)}& \multicolumn{1}{c}{(mJy)} & \multicolumn{1}{c}{(mJy)} & & \multicolumn{1}{c}{COSMOSVLADP} & \multicolumn{1}{c}{ID}& class \\
 \hline                    
 \multicolumn{1}{c}{(1)} & \multicolumn{1}{c}{(2)}  & \multicolumn{1}{c}{(3)} & \multicolumn{1}{c}{(4)}& \multicolumn{1}{c}{(5)}& \multicolumn{1}{c}{(6)} & \multicolumn{1}{c}{(7)}& \multicolumn{1}{c}{(8)}& \multicolumn{1}{c}{(9)}& \multicolumn{1}{c}{(10)} & \multicolumn{1}{c}{(11)}\\
 \hline
10900 & J095908.31+024309.6&09:59:08.319 &+02:43:09.62 &    35.170 &    18.485 &    18.772 &   1.308$^{s}$ &J095908.32+024309.6 &C0686 &AGN-WAT \\
10901 & J095758.04+015825.1& 09:57:58.041 &+01:58:25.18 &    18.160 &     9.555 &    12.099 &   2.239$^{s}$ &J095758.04+015825.2 &C0090 &AGN-SYM   \\
10902 & J095823.31+022628.4 & 09:58:23.310 &+02:26:28.45 &    46.160 &     0.094 &$-$ &   1.168$^{s}$ &J095822.93+022619.8 &$-$ &AGN-SYM \\
10903 & J100208.75+024103.2 & 10:02:08.753 &+02:41:03.29 &     6.950 &     6.480 &     6.096 &   1.213$^{s}$ &J100208.75+024103.3 &C2867 &AGN-CL \\
10904 & J100243.26+015945.0 &10:02:43.266 &+01:59:45.00 &    28.420 &     0.011 &$-$ &   1.206$^{p}$ &J100242.57+015938.7 &$-$ &AGN-SYM \\
10905 & J100229.89+023225.1 & 10:02:29.894 &+02:32:25.15 &     3.170 &     2.086 &     2.016 &   0.432$^{s}$ &J100229.89+023225.2 &C3026 &AGN-SYM\\
10906 & J100212.06+023135.0 &10:02:12.120 &+02:31:35.016 &     8.651 &       $-$ &     0.048 &   0.948$^{p}$ &J100212.06+023134.8 &C2899 &AGN-SYM\\
10907 & J100309.43+022714.1 & 10:03:09.432 &+02:27:14.12 &     2.814 &     2.382 &     0.308 &   1.210$^{p}$ &J100309.43+022714.2 &C3265 &AGN-SYM\\
10908 & J100339.24+015546.6 &10:03:39.240 &+01:55:46.67 &    26.230 &       $-$ &$-$ &   0.921$^{p}$ &$-$ &$-$ &AGN-SYM\\
10909 & J100007.90+024315.3&10:00:07.903 &+02:43:15.34 &     6.828 &     0.212 &     0.191 &   1.438$^{p}$ &J100007.90+024315.4 &C1374 &AGN-XZ\\
& & & & & & & & \&J100008.13+024308.0\\
10910 & J100049.59+014923.7&10:00:49.590 &+01:49:23.71 &     5.867 &     1.306 &     0.835 &   0.530$^{s}$ &J100049.58+014923.7 &C1893 &AGN-WAT\\
10911 & J100114.85+020208.6&10:01:14.858 &+02:02:08.67 &     3.425 &     1.221 &     0.729 &   0.971$^{s}$ &J100114.85+020208.8 &C2214 &AGN-SYM\\
10912 & J095802.10+021540.8& 09:58:02.101 &+02:15:40.87 &     1.220 &     0.954 &     0.923 &   0.943$^{s}$ &J095801.42+021542.3 &C0109 &AGN-CL\\
& & & & & & & & \&J095802.10+021540.9\\
10913 & J100028.28+024103.3& 10:00:28.285 &+02:41:03.37 &    32.090 &     1.057 &     0.833 &   0.349$^{s}$ &J100025.91+024144.0 &C1641 &AGN-WAT\\
& & & & & & & & \&J100028.29+024103.3\\
10914 & J100230.19+020913.2&10:02:30.195 &+02:09:13.25 &     3.327 &     0.611 &     0.431 &   1.437$^{p}$ &J100230.11+020912.4 &C3031 &AGN-XZ\\
10915 & J095959.17+014837.7& 09:59:59.172 &+01:48:37.78 &     3.692 &       $-$ &$-$ &   2.357$^{p}$ &J095959.16+014837.8 &$-$ &AGN-XZ\\
10916 & J100140.12+015129.7& 10:01:40.125 &+01:51:29.76 &     5.438 &     0.050 &$-$ &   0.4594$^{s}$ &J100140.12+015129.9 &$-$ &AGN-SYM\\
& & & & & & & & \&J100140.12+015129.9\\
10917 & J100152.21+024535.3& 10:01:52.216 &+02:45:35.39 &     1.523 &     0.401 &$-$ &   1.446$^{p}$ &J100152.18+024536.0 &$-$ &AGN-BT\\
10918 & J095824.02+024916.1& 09:58:24.021 &+02:49:16.16 &    25.220 &     0.575 &     0.364 &   0.3446$^{s}$ &J095824.02+024916.0 &C0255 &AGN-XZ\\
& & & & & & & & \&J095826.03+024921.9\\
10919 & J100114.13+015444.1& 10:01:14.131 &+01:54:44.17 &     3.157 &     0.262 &     0.157 &   1.483$^{p}$ &J100114.12+015444.3 &C2203 &AGN-XZ\\
10920 & J095839.25+013557.7& 09:58:39.253 &+01:35:57.70 &     1.975 &     0.095 &$-$ &   1.668$^{p}$ &J095839.24+013557.8 &$-$ &AGN-SYM\\
10921 & J095834.09+022703.3& 09:58:34.097 &+02:27:03.31 &     0.981 &     0.458 &     0.047 &   1.318$^{p}$ &J095834.09+022703.4 &C0337 &AGN-SYM\\
10922 & J100343.12+023700.3& 10:03:43.128 &+02:37:00.38 &    24.150 &     0.348 &$-$ &   1.596$^{p}$ &$-$ &$-$ &AGN-SYM\\
10923 & J100303.67+014736.0& 10:03:03.674 &+01:47:36.00 &    13.170 &     0.239 &     0.156 &   1.203$^{p}$ &J100303.66+014736.0 &C3231 &AGN-SYM\\
10924 & J095925.79+030100.5& 09:59:25.797 &+03:01:00.57 &     9.353 &     1.762 &$-$ & $-$ &$-$ &$-$ &AGN-CL\\
10925 & J095741.10+015122.4& 09:57:41.106 &+01:51:22.44 &    18.540 &     0.606 &     0.443 &   0.984$^{p}$ &J095741.10+015122.5 &C0027 &AGN-SYM\\
10926 & J095949.84+015650.3& 09:59:49.848 &+01:56:50.35 &     0.744 &       $-$ &     0.075 &   1.768$^{p}$ &J095949.80+015650.7 &C1152 &AGN-SYM\\
10927 & J100101.98+020511.4& 10:01:01.985 &+02:05:11.46 &     1.055 &     0.013 &$-$ &   0.915$^{p}$ &$-$ &$-$ &AGN-CL\\
10928 & J095822.49+024722.2& 09:58:22.497 &+02:47:22.20 &    11.640 &     0.054 &$-$ &   0.8784$^{s}$ &J095822.30+024721.3 &$-$ &AGN-SYM\\
10929 & J100211.45+015458.0& 10:02:11.451 &+01:54:58.08 &     0.298 &     0.150 &     0.121 &   1.716$^{p}$ &J100211.44+015458.2 &C2896 &AGN-CL\\
& & & & & & & & \&J100211.44+015458.2\\
10930 & J100231.43+015138.1& 10:02:31.435 &+01:51:38.18 &     0.578 &     0.134 &     0.071 &   2.165$^{s}$ &J100231.41+015138.3 &C3043 &AGN-SYM\\
10931 & J095828.64+014407.6& 09:58:28.649 &+01:44:07.69 &     0.867 &     0.147 &$-$ &   0.5947$^{s}$ &J095828.65+014407.7 &$-$ &AGN-WAT\\
               \hline                  
\end{tabular}
\tablefoot{{\bf Column 1 \& 2} is the 3-GHz VLA-COSMOS ID of the multi-component and the IAU name, respectively. {\bf Columns 3 \& 4} give the sexadecimal R.A. and Dec. of the radio position (J2000.0). {\bf Column 5} gives the flux density at 3 GHz in mJy as measured from the map using CASA down to a 2 $\sigma$ level in surface brightness. {\bf Column 6} is the core flux density measured by a gaussian fit on the core position (also down to a 2$\sigma$ level), in mJy. {\bf Column 7} is the matched VLBA flux density at 1.4 GHz which corresponds to the core. {\bf Column 8} gives the redshift from the COSMOS2015 catalogue \citep{laigle16}; the character '$s$' denotes spectroscopic while '$p$' denotes photometric redshift. {\bf Column 9} is the associated VLA source ID at 1.4 GHz from \cite{schinnerer10}, while in {\bf Column 10} is the VLBA ID at 1.4 GHz from \cite{noelia17}. {\bf Column 11} General radio classification: either AGN or SFG based on radio structure at 1.4 and 3 GHz: AGN if it exhibits jets/lobes; AGN-CL for core-lobe and AGN-HT for head-tail AGN (see Sec.~\ref{sec:headtail}); AGN-WAT for wide-angle-tail AGN (see Sec.~\ref{sec:wat}); AGN-XZ for Z-/X-shaped AGN (see Sec.~\ref{sec:restarted}); AGN-BT for bent-tail AGN (see Sec.~\ref{sec:benttail}); AGN-SYM for an AGN with symmetric radio structure (see Sec.~\ref{sec:restAGN}); SFG if radio emission cannot be associated with jets/lobes but rather follows the disk structure of the galaxy; a * denotes that the classification is based on the radio excess (See Sec.~\ref{sec:hosts}).
}
\end{sidewaystable*}
%
%

\addtocounter{table}{-1}

\begin{sidewaystable*}
\caption{Multi-component Radio Properties (continued)}             
\label{table:data}      
\centering          
\begin{tabular}{l l l c c c c r r c r}     
\hline\hline       
                
\multicolumn{1}{c}{3-GHz}  &  \multicolumn{1}{c}{} &  \multicolumn{1}{c}{R.A.} & 
\multicolumn{1}{c}{Dec.} & \multicolumn{1}{c}{$S_{\rm 3 GHz}$} & \multicolumn{1}{c}{$S_{\rm core~3 GHz}$} & \multicolumn{1}{c}{$S_{\rm VLBA~1.4 GHz}$} &\multicolumn{1}{c}{z} &  \multicolumn{1}{c}{1.4-GHz ID}& VLBA & \multicolumn{1}{c}{radio}\\ 
\hline                    
 \multicolumn{1}{c}{ID} & \multicolumn{1}{c}{COSMOSVLA3} & \multicolumn{1}{c}{(h:m:s)}  & \multicolumn{1}{c}{(d:m:s)} & \multicolumn{1}{c}{(mJy)}& \multicolumn{1}{c}{(mJy)} & \multicolumn{1}{c}{(mJy)} & & \multicolumn{1}{c}{COSMOSVLADP} & ID& \multicolumn{1}{c}{class}\\
 \hline                    
 \multicolumn{1}{c}{(1)} & \multicolumn{1}{c}{(2)}  & \multicolumn{1}{c}{(3)} & \multicolumn{1}{c}{(4)}& \multicolumn{1}{c}{(5)}& \multicolumn{1}{c}{(6)} & \multicolumn{1}{c}{(7)}& \multicolumn{1}{c}{(8)}& \multicolumn{1}{c}{(9)}& \multicolumn{1}{c}{(10)}& \multicolumn{1}{c}{(11)} \\
 \hline
 10932 & J095945.81+025924.3& 09:59:45.815 &+02:59:24.35 &     0.816 &     0.446 &$-$ &   1.161$^{p}$ &$-$ &$-$ &AGN-SYM\\
10933 & J100043.18+014607.8& 10:00:43.186 &+01:46:07.87 &    39.300 &     3.349 &     2.525 &   0.346$^{s}$ &J100043.17+014607.9 &C1810 &AGN-XZ\\
10934 & J100047.58+020958.6& 10:00:47.585 &+02:09:58.63 &     0.380 &     0.106 &$-$ &   0.669$^{s}$ &J100047.58+020958.8 &$-$ &AGN-SYM\\
10935 & J095927.25+023729.2& 09:59:27.251 &+02:37:29.28 &     2.382 &     0.105 &$-$ &   0.9544$^{s}$ &J095927.25+023729.2 &$-$ &AGN-XZ\\
10936 & J100028.24+013508.5& 10:00:28.240 &+01:35:08.59 &    10.050 &     0.146 &$-$ &   0.839$^{s}$ &J100028.31+013507.8 &$-$ &AGN-SYM\\
10937 & J095947.83+021023.9& 09:59:47.832 &+02:10:23.98 &     0.230 &       $-$ &$-$ &   1.127$^{p}$ &J095947.83+021024.1 &$-$ &AGN-SYM\\
10938 & J100138.64+030157.7& 10:01:38.640 &+03:01:57.72 &     0.849 &       $-$ &$-$ &   0.913$^{p}$ &$-$ &$-$ &AGN-SYM\\
10939 & J100242.23+013432.1& 10:02:42.232 &+01:34:32.19 &     0.367 &     0.112 &$-$ &   1.519$^{s}$ &J100242.23+013432.3 &$-$ &AGN$^{*}$-CL\\
& & & & & & & & \&J100242.42+013432.9\\
10940 & J095857.36+021315.5 & 09:58:57.360 &+02:13:15.52 &     0.204 &       $-$ &$-$ &   1.024$^{s}$ &J095857.37+021315.2 &$-$ &AGN$^{*}$-CL\\
10941 & J100102.64+012925.9& 10:01:2.6400 &+01:29:25.95 &     0.549 &       $-$ &$-$ &   2.243$^{p}$ &$-$ &$-$ &AGN$^{*}$-CL\\
10942 & J100034.76+014635.7& 10:00:34.760 &+01:46:35.79 &     0.374 &     0.054 &$-$ &   0.7335$^{s}$ &J100034.77+014635.9 &$-$ & SFG$^{*}$\\
10943 & J100104.99+013154.5& 10:01:04.993 &+01:31:54.58 &     0.406 &     0.030 &$-$ &   1.809$^{p}$ &J100105.10+013153.8 &$-$ &AGN-CL\\
& & & & & & & & \&J100105.10+013153.8\\
10944 & J095905.52+023809.9 & 09:59:05.525 &+02:38:09.91 &     0.948 &     0.040 &$-$ &   0.079$^{s}$ &J095905.55+023810.27 &$-$ &SFG\\
10945 & J100124.19+023049.9& 10:01:24.198 &+02:30:49.98 &     0.205 &     0.036 &$-$ &   0.6895$^{s}$ &$-$ &$-$ &AGN-CL\\
10946 & J100144.04+025712.6 & 10:01:44.040 &+02:57:12.67 &     0.116 &       $-$ &$-$ &   0.893$^{p}$ &$-$ &$-$ &SFG\\
10947 & J095918.98+014035.9& 09:59:18.980 &+01:40:35.95 &     0.120 &     0.023 &$-$ &   2.537$^{p}$ &J095919.03+014036.0 &$-$ &AGN-SYM\\
10948 & J100021.78+015959.9& 10:00:21.781 &+01:59:59.97 &     1.943 &     1.100 &     0.552 &   0.219$^{s}$ &J100021.78+020000.2 &C1552 &AGN-SYM\\
10949 & J100124.06+024936.6& 10:01:24.069 &+02:49:36.68 &     2.718 &     1.193 &     0.965 &   0.826$^{s}$ &J100124.09+024936.3 &C2347 &AGN-WAT\\
10950 & J095949.01+025516.3& 09:59:49.015 &+02:55:16.39 &     1.439 &     1.044 &$-$ &   0.1258$^{s}$ &$-$ &$-$ &AGN-WAT\\
10951 & J095917.74+020927.8& 09:59:17.741 &+02:09:27.83 &     0.890 &     0.598 &     0.257 &   1.692$^{p}$ &J095917.70+020923.2 &C0781 &AGN-SYM\\
& & & & & & & & \&J095917.73+020927.9\\
10952 & J100238.68+022152.1& 10:02:38.683 &+02:21:52.19 &     1.687 &     0.548 &     0.205 &   0.827$^{s}$ &J100238.67+022152.1 &C3087 &AGN-WAT\\
10953 & J100018.50+023256.2& 10:00:18.506 &+02:32:56.29 &     1.254 &     0.251 &     0.152 &   0.890$^{s}$ &J100018.50+023256.5 &C1510 &AGN-SYM\\
10954 & J100008.10+024554.5& 10:00:08.106 &+02:45:54.54 &     0.508 &     0.297 &$-$ &   0.029$^{s}$ &J100008.10+024554.5 &$-$ &SFG\\
10955 & J100307.47+023655.8& 10:03:07.477 &+02:36:55.89 &     1.058 &     0.111 &$-$ &   0.370$^{p}$ &J100307.48+023655.9 &$-$ &AGN-HT\\
10956 & J100027.44+022123.2& 10:00:27.440 &+02:21:23.28 &     4.379 &     0.078 &$-$ &   0.2202$^{s}$ &J100027.31+022111.3 &$-$ &AGN-WAT\\
10957 & J100015.55+020731.4& 10:00:15.559 &+02:07:31.40 &     0.514 &     0.069 &$-$ &   0.6612$^{s}$ &J100015.55+020731.5 &$-$ &AGN-HT\\
10958 & J100136.46+022642.0& 10:01:36.464 &+02:26:42.04 &     0.889 &     0.045 &$-$ &   0.123$^{s}$ &J100136.46+022641.8 &$-$ &AGN-XZ\\
10959 & J100245.40+024516.1& 10:02:45.405 &+02:45:16.13 &     8.576 &     0.049 &$-$ &   0.986$^{p}$ &J100245.39+024519.8 &$-$ &AGN-SYM\\
10960 & J100120.64+021816.6& 10:01:20.640 &+02:18:16.60 &     0.188 &       $-$ &$-$ &   0.123$^{s}$ &J100120.60+021817.9 &$-$ &SFG\\
10961 & J095934.59+014924.4& 09:59:34.592 &+01:49:24.47 &     0.107 &     0.001 &$-$ &   0.133$^{s}$ &J095934.57+014923.6 &$-$ &SFG\\
10962 & J102833.60+024248.0& 10:28:33.600 &+02:42:48.07 &    80.250 &       $-$ &$-$ &   0.974$^{p}$ &J100251.11+024248.5 &$-$ &AGN-WAT\\
10963 & J100008.94+024010.9& 10:00:08.941 &+02:40:10.90 &     0.152 &     0.090 &$-$ &   1.599$^{s}$ &J100008.91+024010.3 &$-$ &SFG$^{*}$\\
10964 & J102154.00+013706.8& 10:21:54.000 &+01:37:6.882 &     0.092 &       $-$ &$-$ &   1.592$^{s}$ &J100211.24+013706.8 &$-$ &SFG$^{*}$\\
10965 & J100045.28+013847.4& 10:00:45.286 &+01:38:47.43 &     0.263 &     0.016 &$-$ &   0.2204$^{s}$ &J100045.32+013846.5 &$-$ &SFG\\
10966 & J100106.73+013320.4& 10:01:06.735 &+01:33:20.43 &     0.861 &     0.032 &$-$ &   0.361$^{p}$ &J100106.76+013320.0 &$-$ &AGN-BT\\
   \hline                  
\end{tabular}
\end{sidewaystable*}

\section{A closer look at the multi-component objects}
\label{sec:results}

As explain in Sec.~\ref{sec:sourceid}, we classified objects as AGN if they have clear signs of jets/lobes, and as SFGs if they show other characteristics where the radio emission follows the optical/near-IR morphology. 
The multi-component VLA-COSMOS radio sources at 3 GHz are either AGN (58 out of 67 objects) or star-forming galaxies (9 out of 67). Note that 6 of the 67 objects have uncertain classification on the basis of their radio structure, and they are placed on the AGN or the SFG class using a radio AGN diagnostic from \cite{delvecchio17}; see Sec.~\ref{sec:hosts}. Multi-component AGN exhibit radio lobes and jets, which in most cases extend out to large distances from the host galaxy (e.g. 10900, 10913, 10918 etc). Multi-component SFGs are composed of several radio blobs associated with star-forming regions (10944, 10946, 10954, 10960, 10961, 10965; see Fig.~\ref{fig:sfgs}). Some of the AGN show peculiar radio structure due to strong interaction with the environment or due to internal physical mechanisms (see Sec.~\ref{sec:notes} and Appendix~\ref{app:notes}).

  \begin{figure}[!ht]
   \resizebox{\hsize}{!}{
   \includegraphics[width=0.5cm]{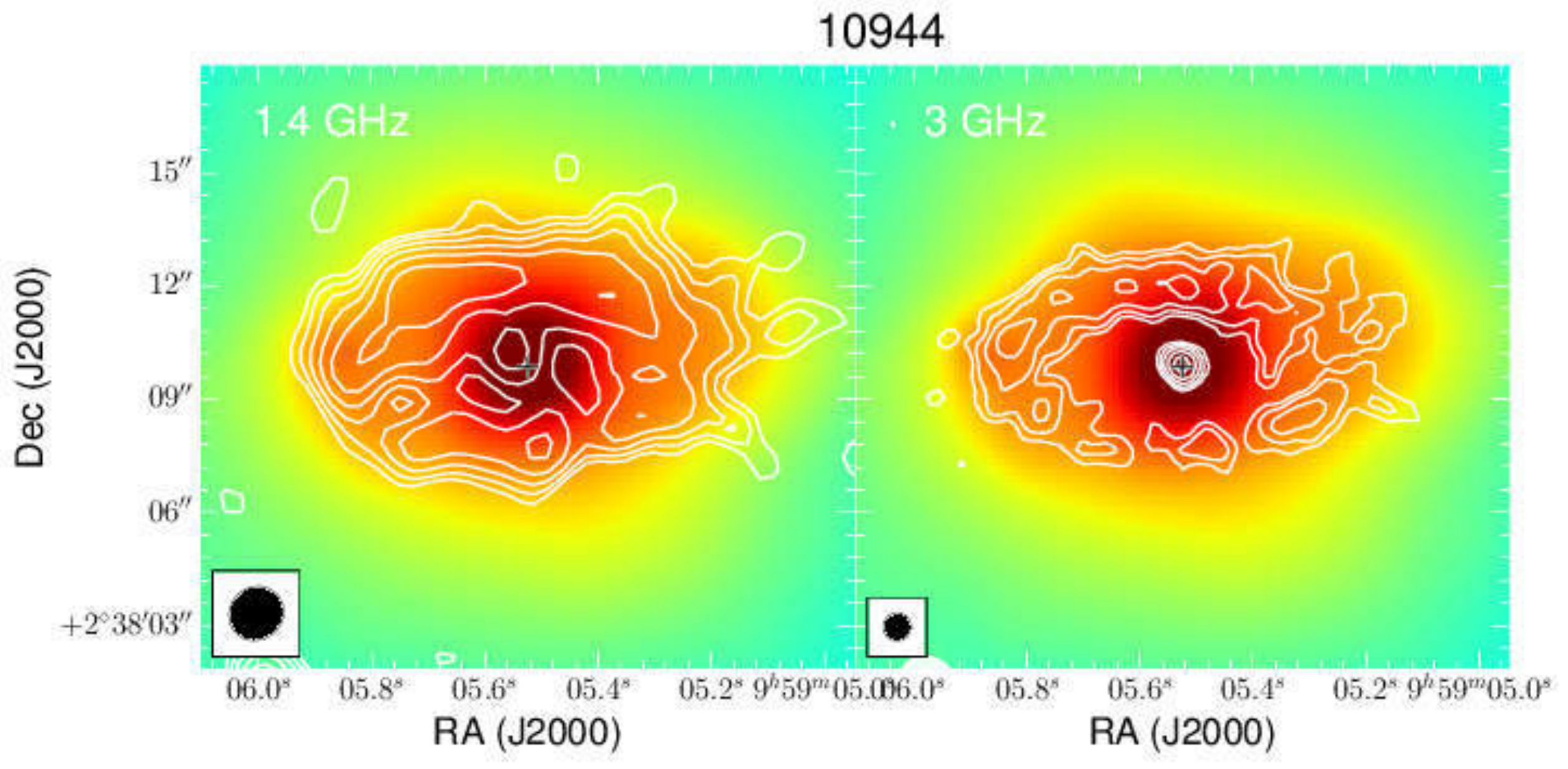}
   }
   \resizebox{\hsize}{!}{
    \includegraphics[width=0.5cm]{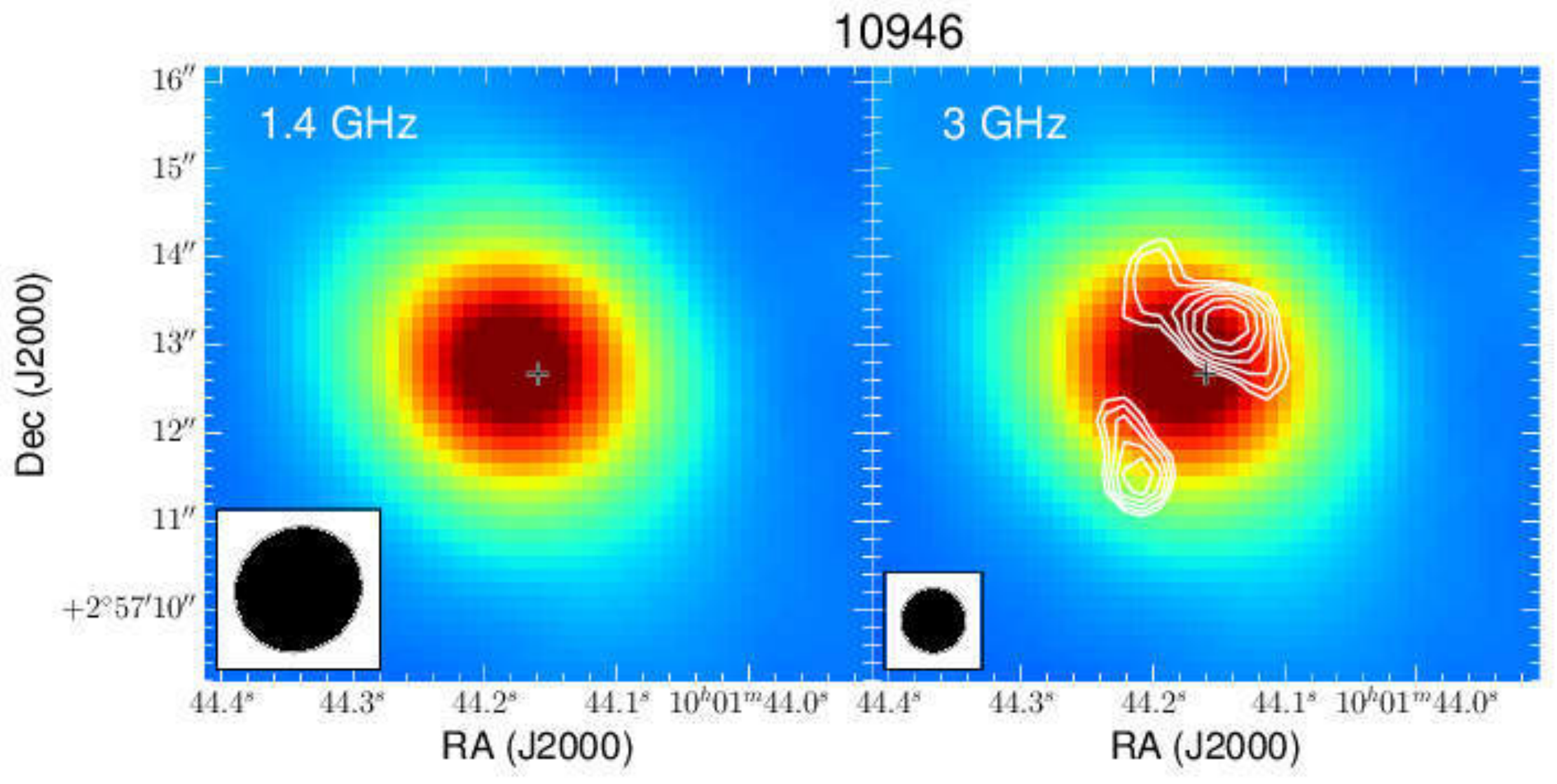}
   }
    \resizebox{\hsize}{!}{
   \includegraphics[width=0.5cm]{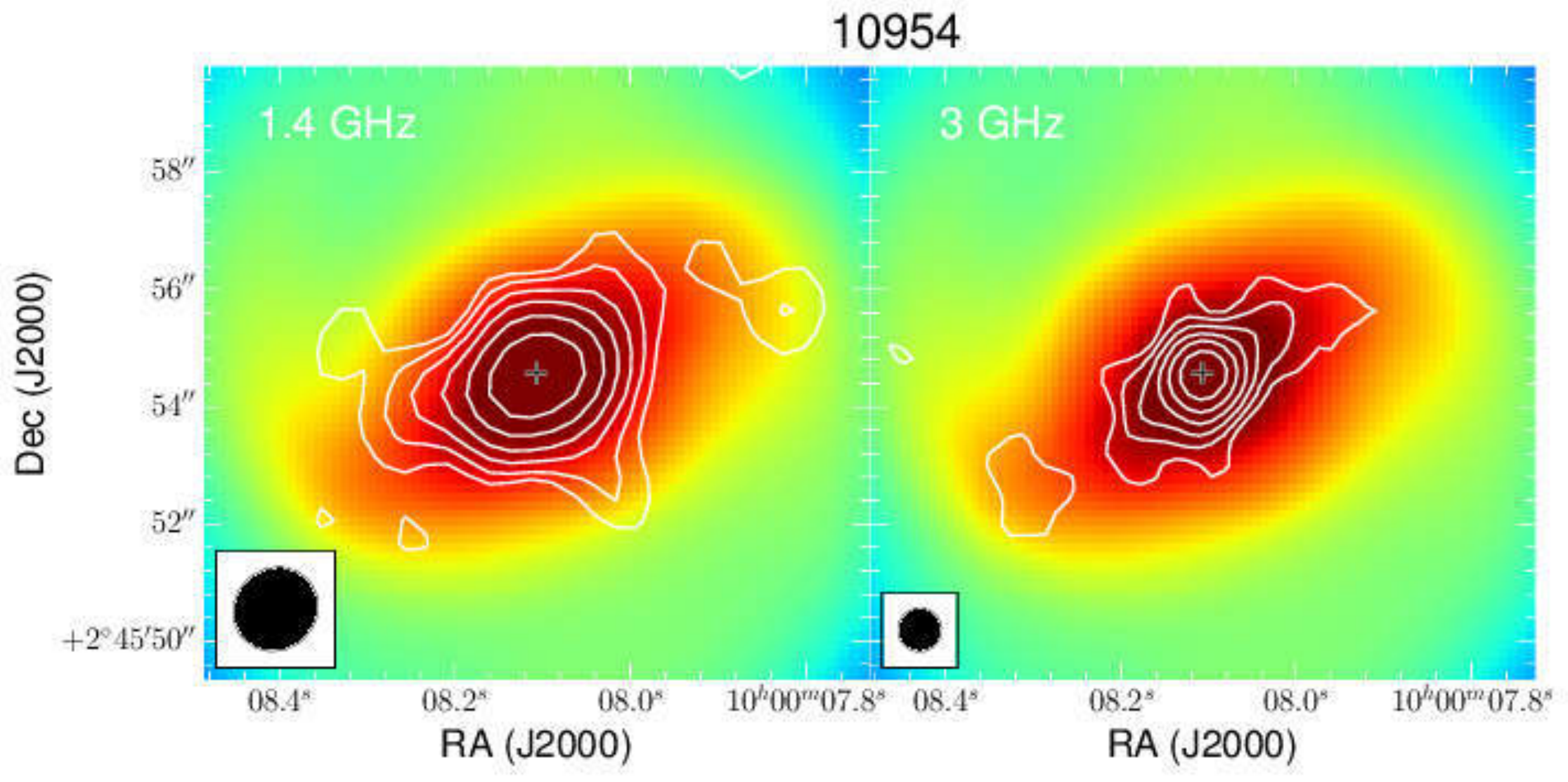}
  }
   \resizebox{\hsize}{!}{
   \includegraphics[width=0.5cm]{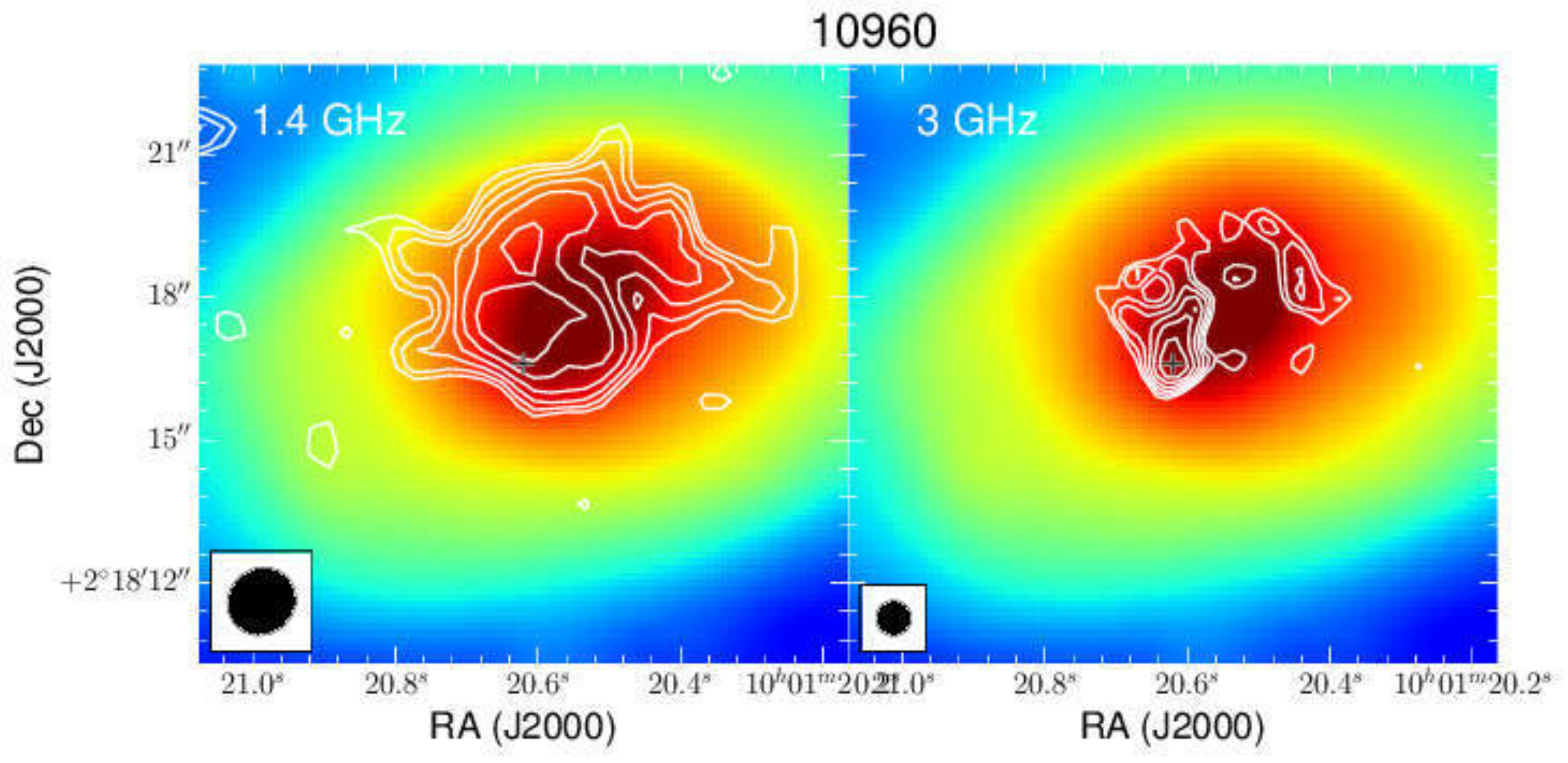}
   }
   \resizebox{\hsize}{!}{
   \includegraphics[width=0.5cm]{JVLA10961-ultrajvlavla.pdf}
  }
   \resizebox{\hsize}{!}{
   \includegraphics[width=0.5cm]{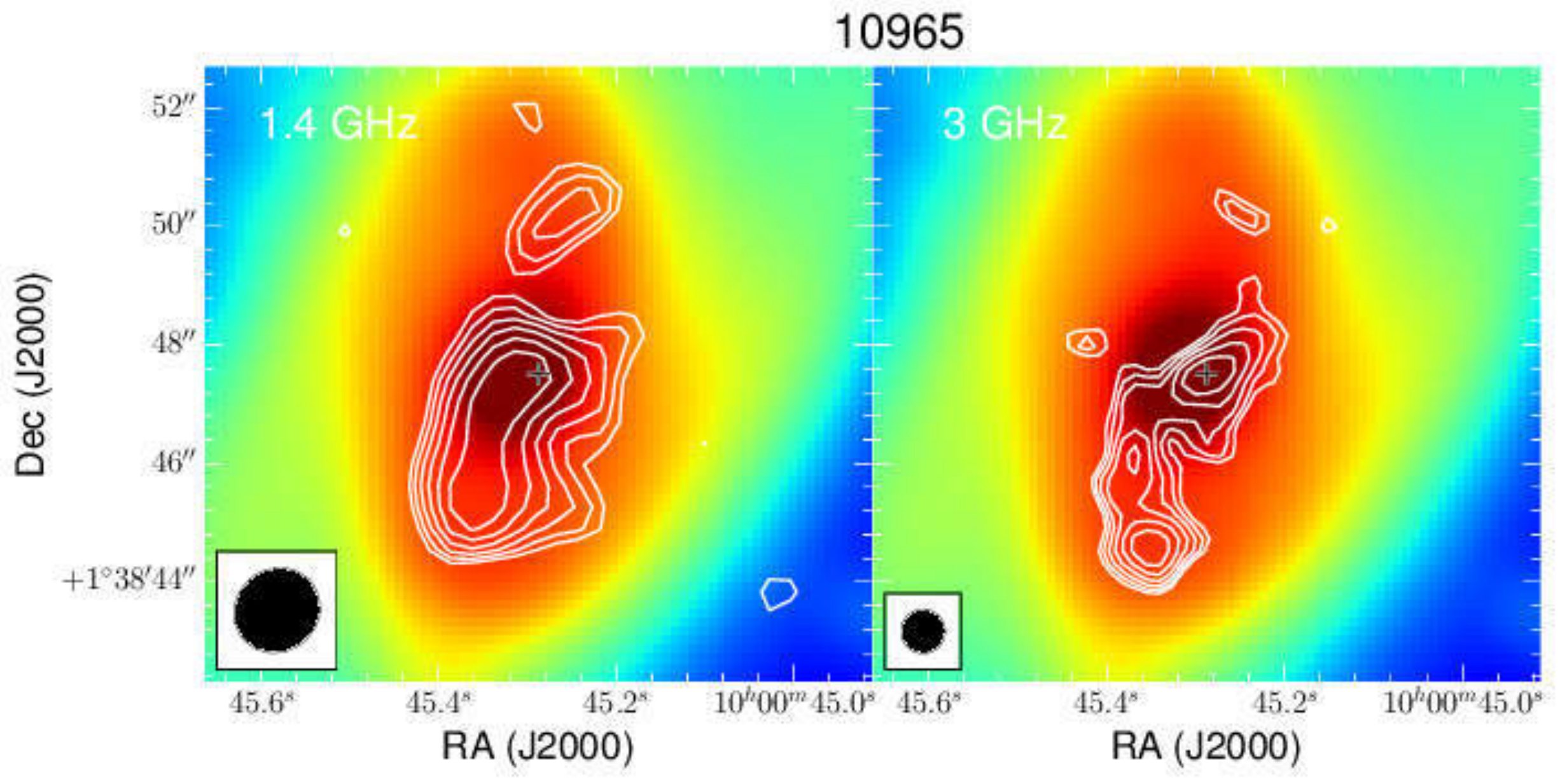}
  }
  \\
      \caption{The 6 star-forming galaxies in our multi-component radio sample at 3 GHz. Images described in Fig.~\ref{fig:radmaps}.
              }
         \label{fig:sfgs}
   \end{figure}

Out the 58 objects with a 1.4-GHz detection, 28 are not multi-component radio sources at 1.4 GHz due to the coarser resolution of 1.5 arcsec (see Fig.~\ref{fig:maps2} in the Appendix). An example is 10966 (see Fig.~\ref{fig:benttail}), where the lobes are separated at 3 GHz but not at 1.4 GHz. Two other examples are 10960 \& 10944 (see Fig.~\ref{fig:sfgs}), which exhibit at least two radio blobs at 3 GHz associated with the host galaxy disk, almost forming a ring around the nucleus; this is not seen at 1.4 GHz where the emission comes from the whole galaxy disk. Additionally, the 3-GHz data reveal in higher detail the sub-structure in the jets of radio AGN (e.g. 10910, 10913, and 10956 in Fig.~\ref{fig:wat}, and 10918 in Fig.~\ref{fig:xshaped}). 

There are objects that do not appear completely separated at 3 GHz (e.g. 10933, 10962; Fig.~\ref{fig:xshaped}~\&~\ref{fig:wat}, respectively) but are included in the multi-component catalogue due to the way \textsc{blobcat} identifies blobs. If the surface brightness falls below 2.5 $\sigma$, then it separates the source into different islands. 10933 and 10962 have small components near-by the main source associated with them, so to include them in the total flux-density measurements we had to re-measure their flux densities. 

In the following sub-sections we present an analysis on the hosts of the multi-component objects in our sample (Sec.~\ref{sec:hosts}), as well as some brief notes on the objects (Sec.~\ref{sec:notes}), with a more detailed description in Appendix~\ref{app:notes}.

   \begin{figure*}[!ht]
    \resizebox{\hsize}{!}
            {\includegraphics{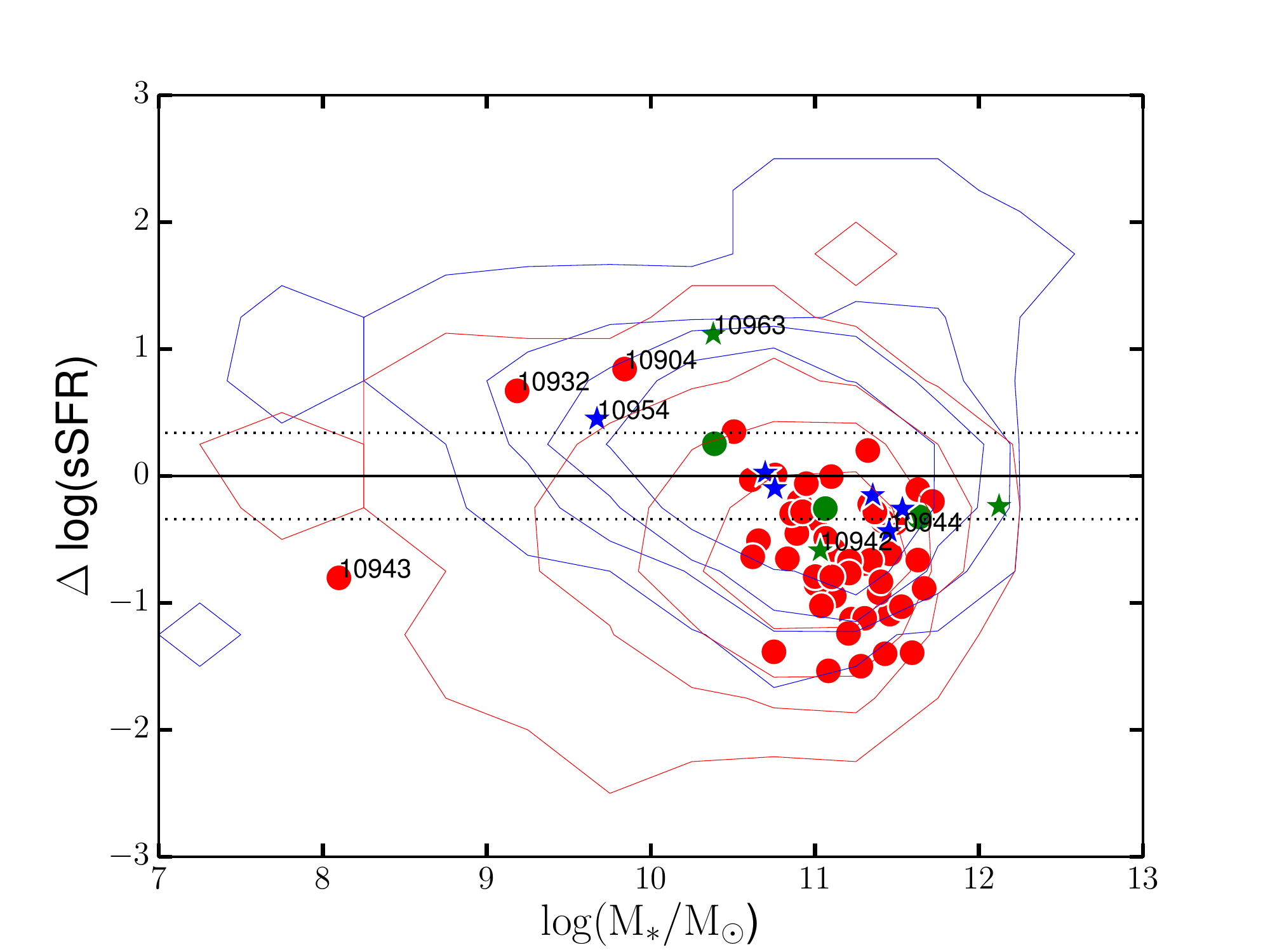}
              \includegraphics{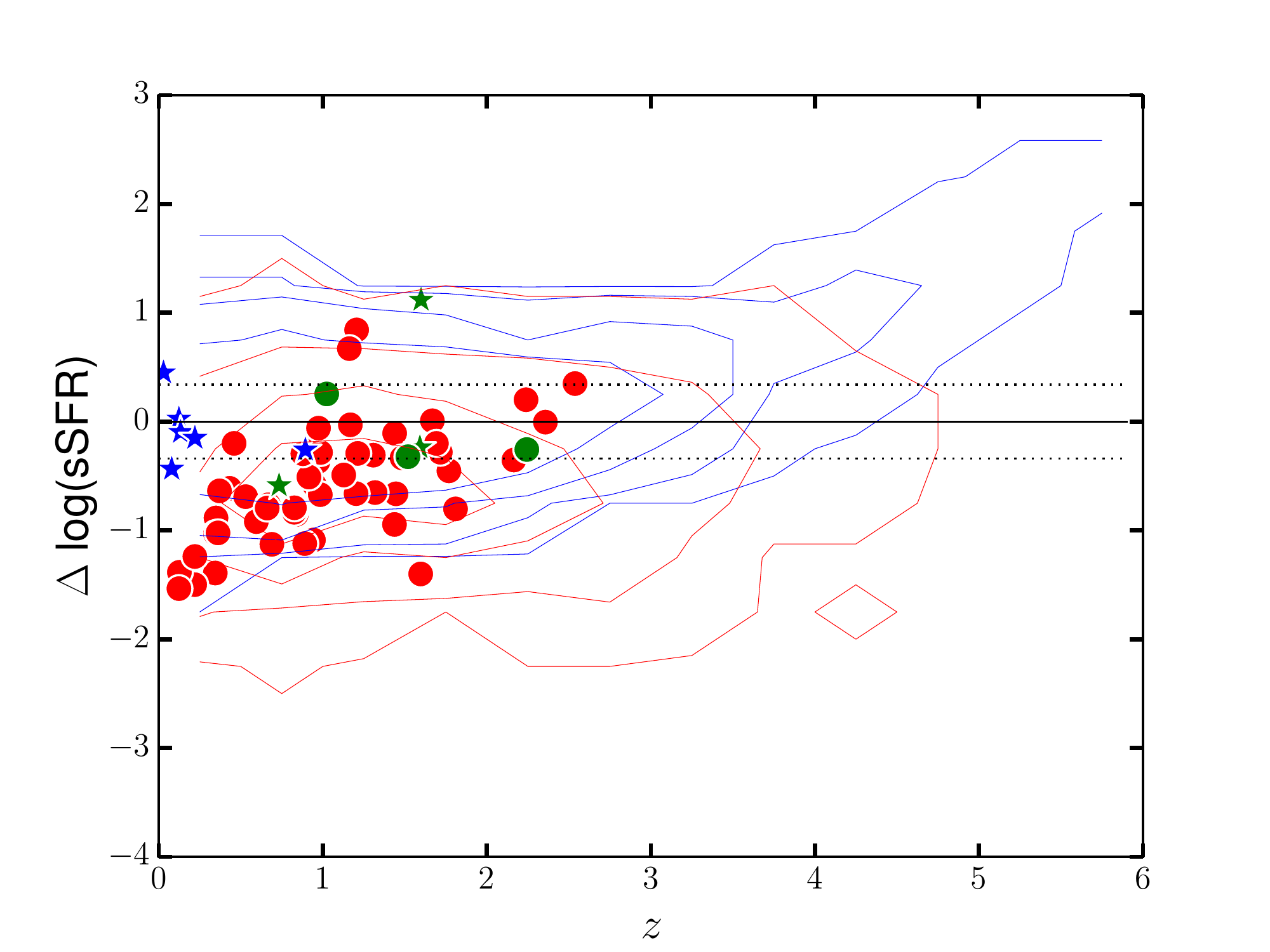}
              
 }

               \caption{$\Delta$sSFR, the logarithmic difference between the specific SFR and the specific SFR of objects lying on the MS and at the same redshift, versus stellar mass M$_{*}$  \citep[see][for details on SFR calculation]{delvecchio17} shown on the $left$, and versus redshift $z$ on the $right$. Symbols: red circles are the multi-component objects classified as AGN at 3-GHz VLA-COSMOS (objects that exhibit jets/lobes), as shown in Table~\ref{table:data}; blue stars multi-component SFGs (see Sec.~\ref{sec:sample} for classification); green circles/stars are multi-component objects with uncertain classification (see Sec.~\ref{sec:hosts}), with circles used for AGN and stars for SFG based on the radio excess flag (see Sec.~\ref{sec:hosts}); the blue contours represent the single-component SFGs \citep[objects that show no radio excess;][]{delvecchio17} and the red contours are for single-component AGN (radio excess objects). The black solid and dashed lines show the main-sequence for star-forming galaxies and its dispersion \citep{whitaker12}.            
   }
              \label{fig:sfr_mstar}%
    \end{figure*}

 \begin{figure}[!ht]
   \resizebox{\hsize}{!}{
   \includegraphics[width=0.5cm]{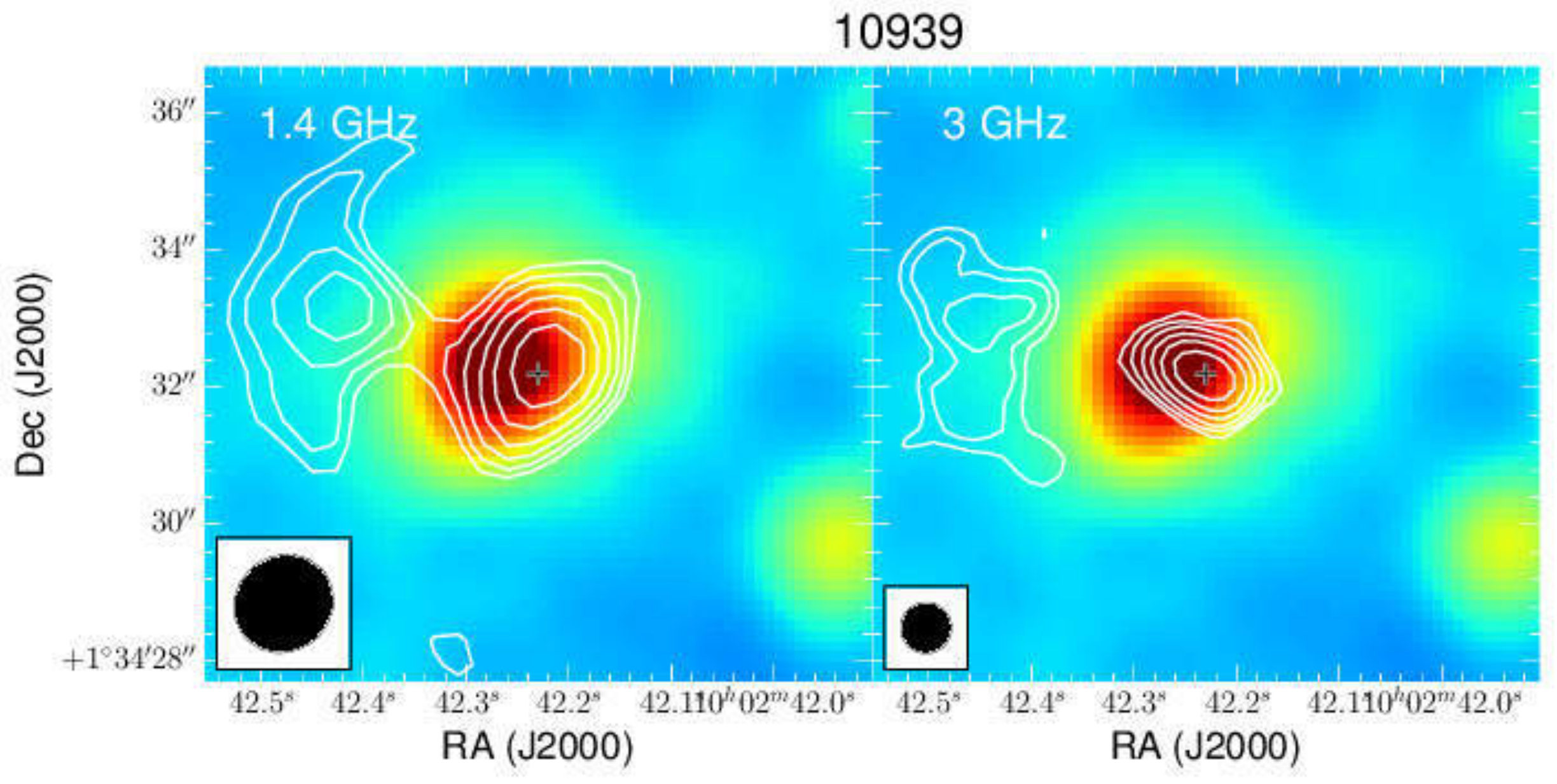}
   }
    \resizebox{\hsize}{!}{
    \includegraphics[width=0.5cm]{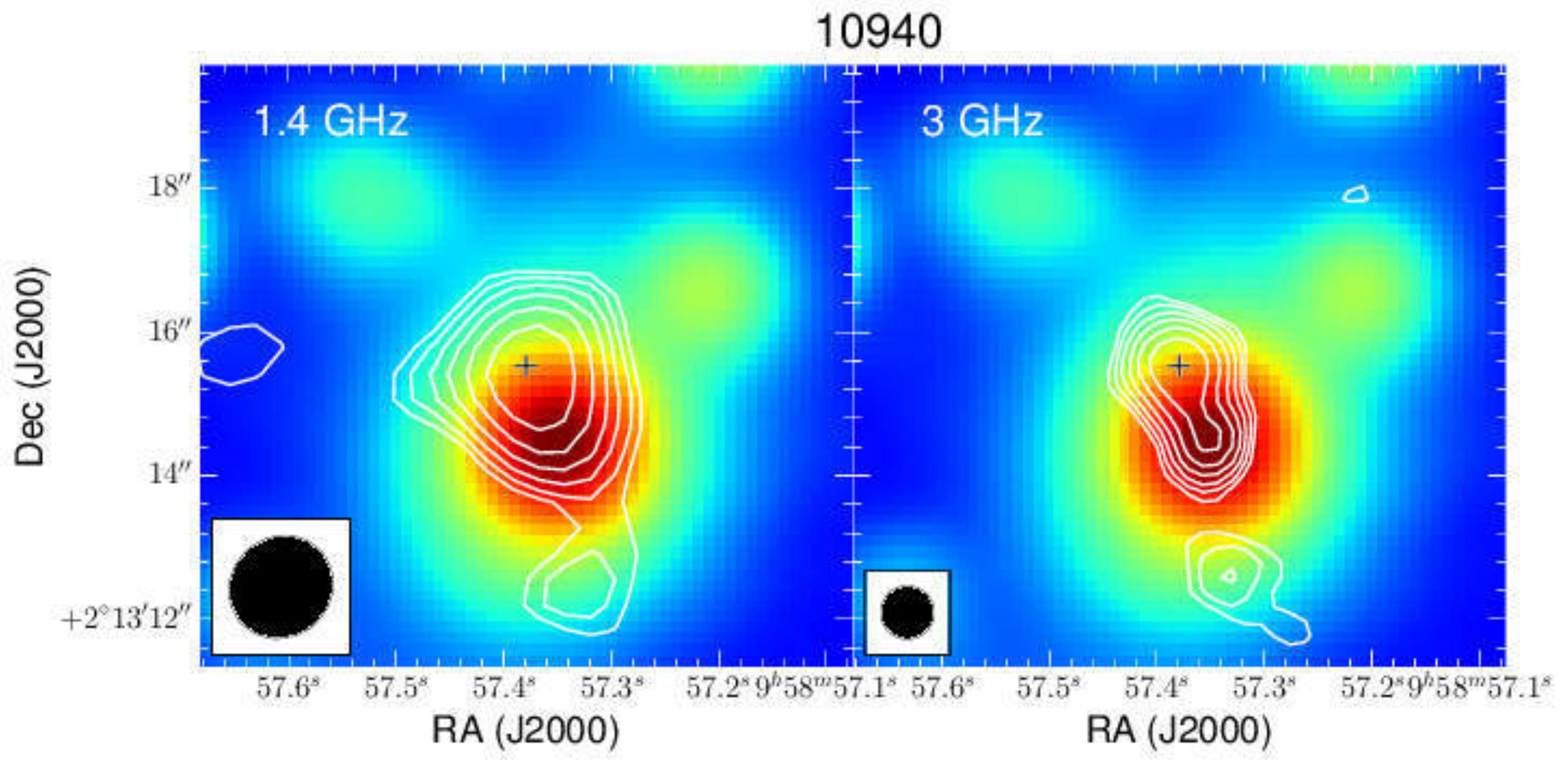}
   }
   
   \resizebox{\hsize}{!}{
   \includegraphics[width=0.5cm]{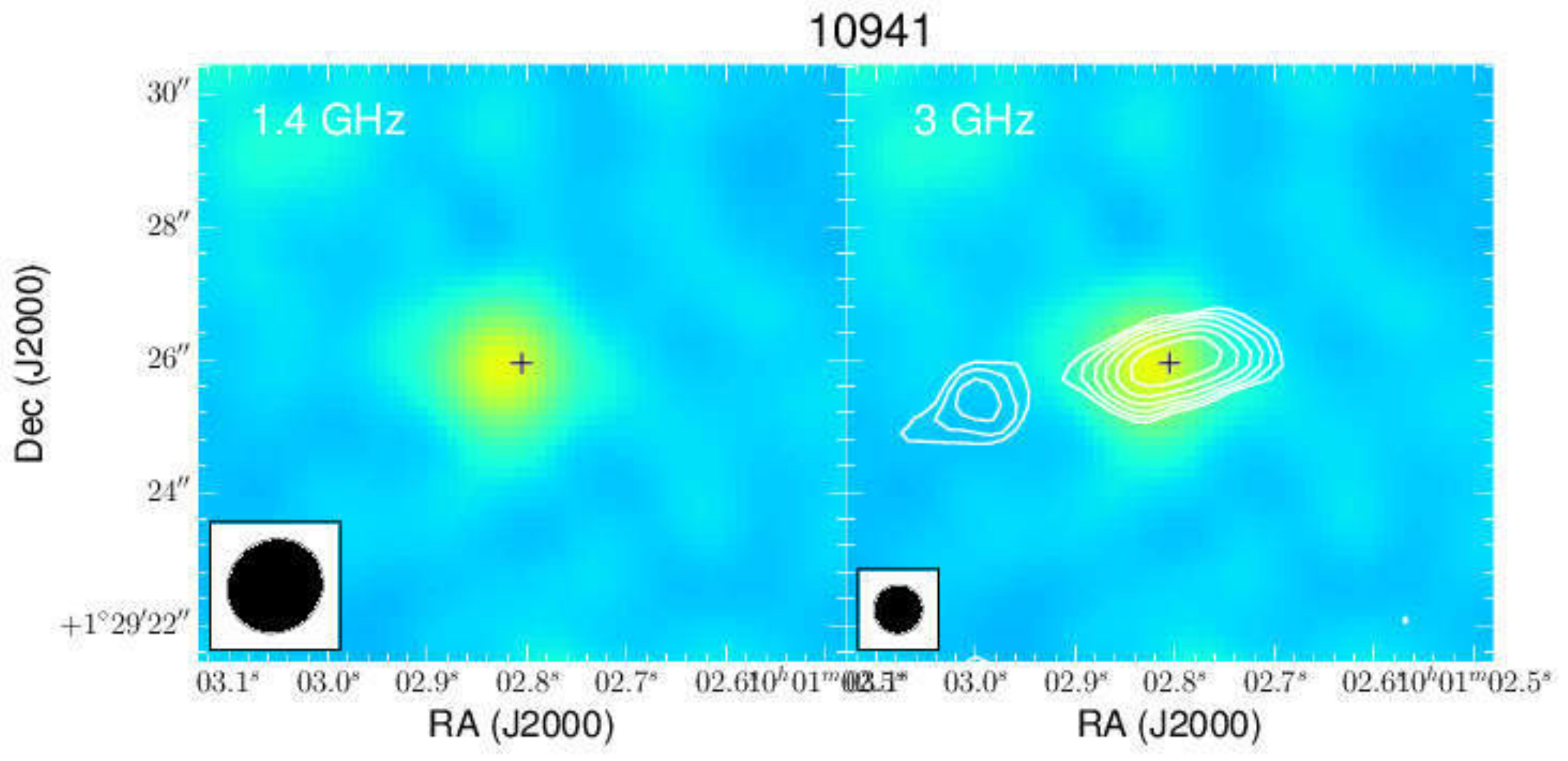}
   }
    \resizebox{\hsize}{!}{
   \includegraphics[width=0.5cm]{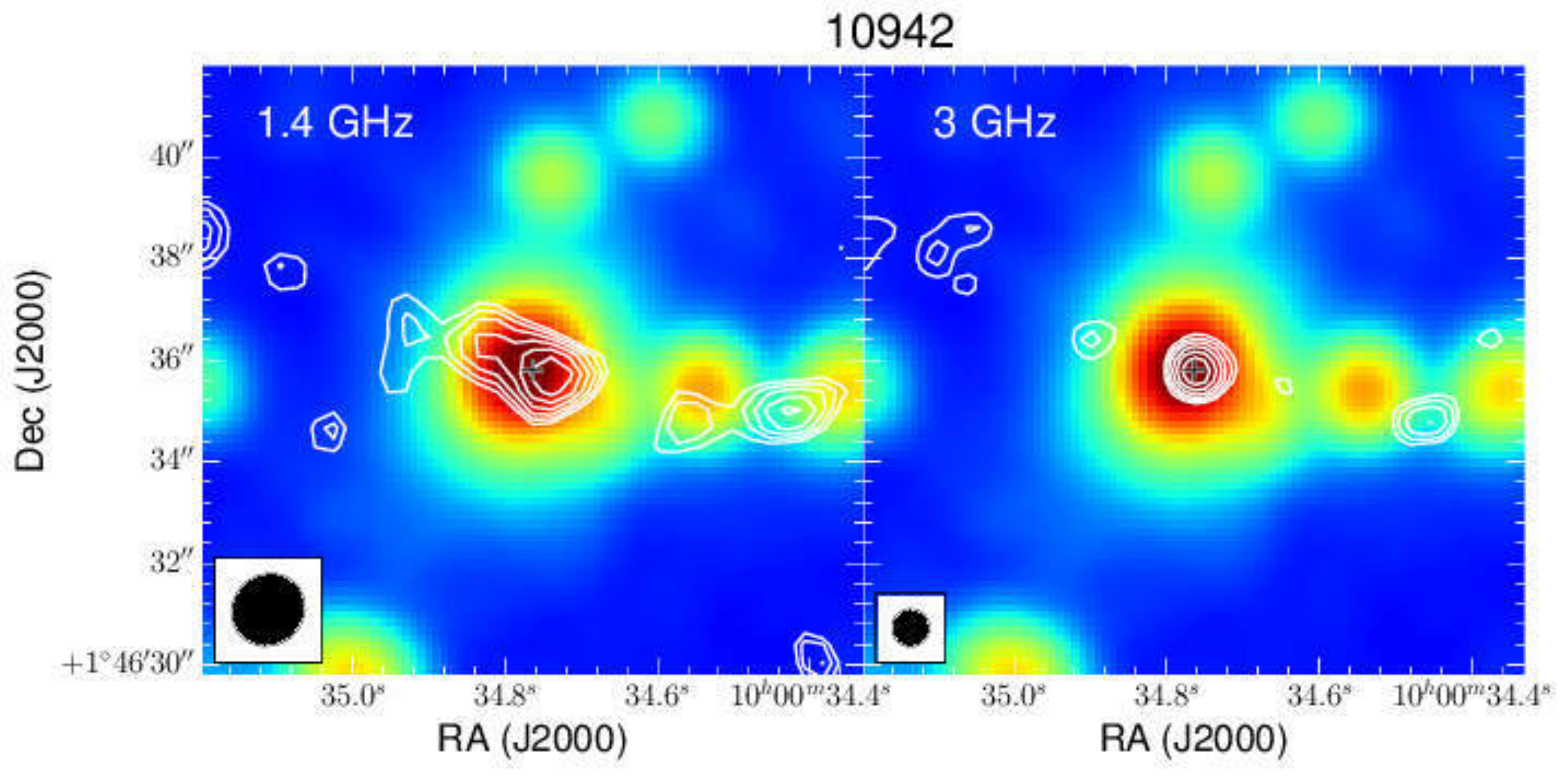}
  }
  
   \resizebox{\hsize}{!}{
   \includegraphics[width=0.5cm]{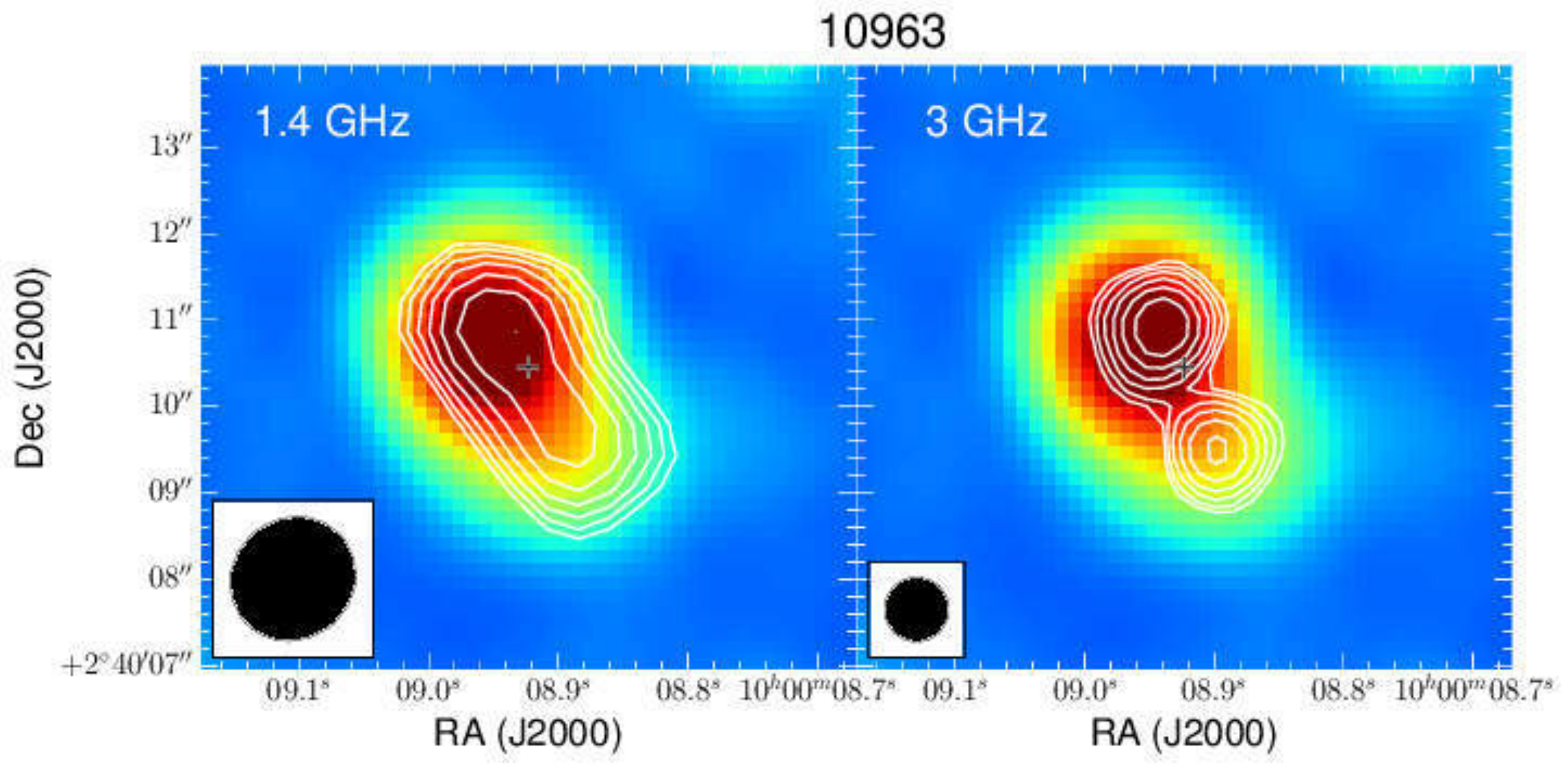}
   }
    \resizebox{\hsize}{!}{
   \includegraphics[width=0.5cm]{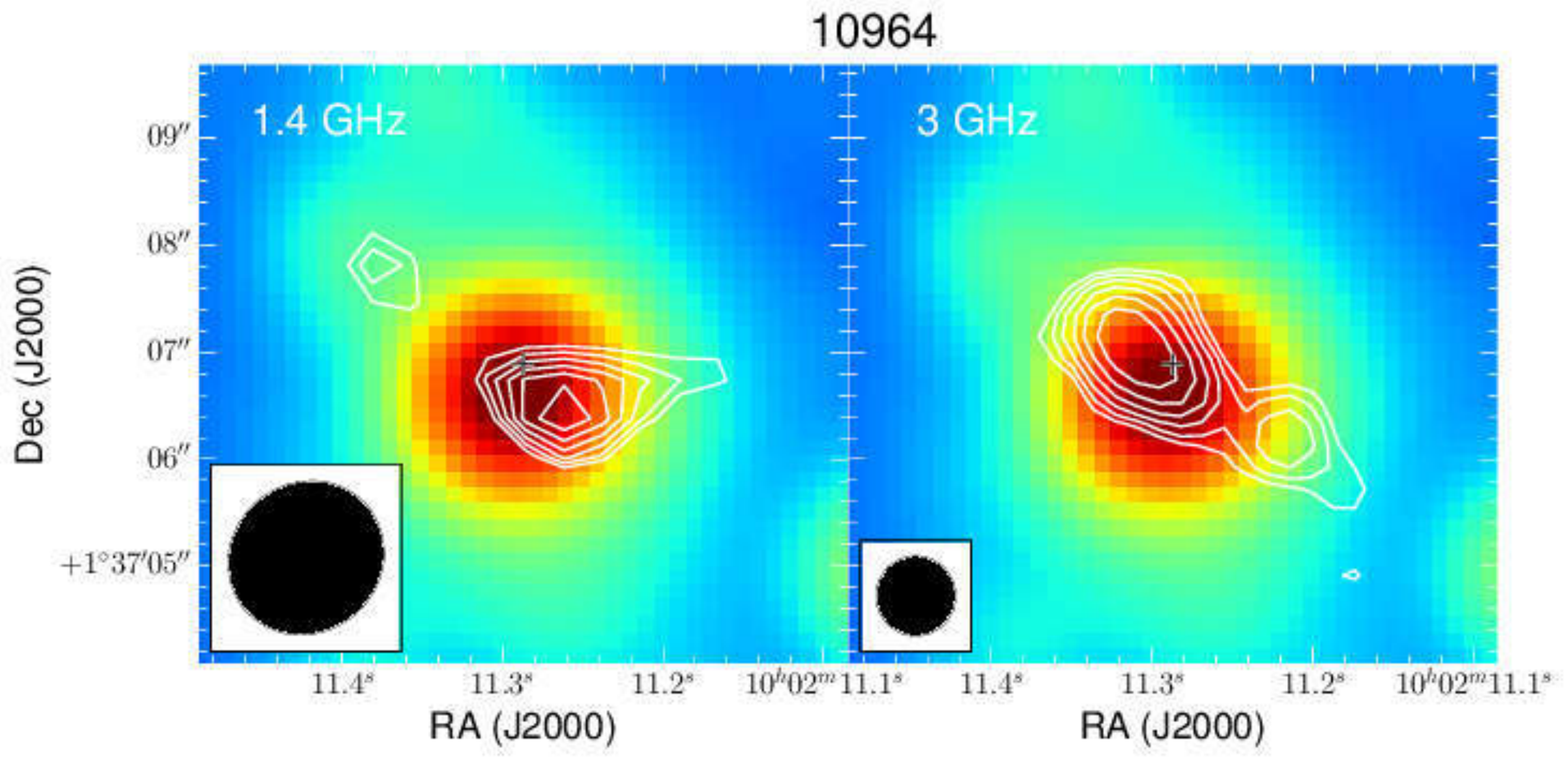}
  }
  \\
      \caption{The 6 multi-component objects at 3 GHz with uncertain classification (see Sec.~\ref{sec:hosts}). Images described in Fig.~\ref{fig:radmaps}. \\ \\
              }
         \label{fig:agnsfg}
   \end{figure}
%

\begin{table}
\caption{Multi-component host properties}             
\label{table:data2}      
\centering          
\begin{tabular}{l r c c l l}     
\hline\hline       
                
\multicolumn{1}{c}{3-GHz}  & SFR$_{\rm IR}$ & $\log_{10}$(M$_{*}$ & SED & \multicolumn{1}{c}{radio}& {\small COSMOS}\\ 
\hline                    
 \multicolumn{1}{c}{ID} & (M$_{\odot}$/yr) & /M$_{\odot}$)&AGN& \multicolumn{1}{c}{class}& \multicolumn{1}{c}{2015} \\
 \hline                    
 \multicolumn{1}{c}{(1)} & \multicolumn{1}{c}{(2)}  & \multicolumn{1}{c}{(3)}& \multicolumn{1}{c}{(4)}& \multicolumn{1}{c}{(5)}& \multicolumn{1}{c}{(6)}\\
 \hline
10900  &   56.29 &   11.47 & T & {\small AGN-WAT} &934339 \\
10901  &  329.01 &   11.32 & F& {\small AGN-SYM} &446143 \\
10902  &   30.51 &   10.61 & T& {\small AGN-SYM} &754369 \\
10903  &   29.25 &   11.01 & T& {\small AGN-CL} &912632 \\
10904  &   91.75 &    9.84 & T & {\small AGN-SYM}&458870 \\
10905  &    6.92 &   11.46 & F& {\small AGN-SYM} &809167 \\
10906  &   11.57 &   11.11 & F & {\small AGN-SYM} &809443 \\
10907  &   16.05 &   11.21 & F & {\small AGN-SYM} &761486 \\
10908  &  $-$ &  $-$ & F& {\small AGN-SYM}& 380833 \\
10909  &  125.21 &   11.63& F & {\small AGN-XZ}& 936454 \\
10910  &    5.59 &   11.30 & F& {\small AGN-WAT}&350495 \\
10911  &   32.35 &   11.49 & T& {\small AGN-SYM}&486067 \\
10912  &    5.63 &   11.46 & F& {\small AGN-CL}&636013 \\
10913  &    4.29 &   11.67 & F& {\small AGN-WAT}&901584 \\
10914  &    9.94 &   11.12 & F& {\small AGN-XZ}&561934 \\
10915  &  174.84 &   11.10 & F& {\small AGN-XZ}&343802 \\
10916  &    8.37 &   10.90 & F& {\small AGN-SYM}&374634 \\
10917  &   24.96 &   11.34 & F& {\small AGN-BT}&960761 \\
10918  &    1.19 &   11.59 & F& {\small AGN-XZ}&996897 \\
10919  &   37.50 &   11.00 & F& {\small AGN-XZ}&407780 \\
10920  &   76.76 &   10.76 & F& {\small AGN-SYM}&210704 \\
10921  &   11.92 &   10.83 & F& {\small AGN-SYM}&759401 \\
10922  &    5.92 &   11.43 & F& {\small AGN-SYM}&1349607 \\
10923  &   26.74 &   11.63 & F& {\small AGN-SYM}&333779 \\
10924  &  $-$ &  $-$ & F& {\small AGN-CL}&351323 \\
10925  &   11.31 &   11.21 & F& {\small AGN-SYM}&372940 \\
10926  &   34.08 &   10.89 & F& {\small AGN-SYM}&429082 \\
10927  &    7.00 &   10.66 & F& {\small AGN-CL}&517689 \\
10928  &   14.10 &   10.86 & F& {\small AGN-SYM}&978441 \\
10929  &   49.87 &   10.92 & F& {\small AGN-CL}&410131 \\
10930  &   96.83 &   11.41 & F& {\small AGN-SYM}&374873 \\
10931  &    4.23 &   11.39 & F& {\small AGN-WAT}&292852 \\
10932  &   24.92 &    9.18 & F& {\small AGN-SYM}&345618 \\
10933  &    2.48 &   11.53 & F& {\small AGN-XZ}&305535 \\
10934  &    5.34 &   11.21 & F& {\small AGN-SYM}&570506 \\
10935  &   35.67 &   11.33 & F& {\small AGN-XZ}&873867 \\
10936  &    4.46 &   11.01 & F& {\small AGN-SYM}&202465 \\
10937  &   17.63 &   11.07 & F& {\small AGN-SYM}&575428 \\
10938  &  $-$ &  $-$ & F& {\small AGN-SYM}&351652 \\
10939  &   83.56 &   11.64 & T & {\small AGN$^{*}$-CL}&195117 \\
10940  &   34.55 &   10.39 & T& {\small AGN$^{*}$-CL}&609017 \\
10941  &   90.12 &   11.06 & F& {\small AGN$^{*}$-CL}&134089 \\
10942  &    7.06 &   11.03 & F& {\small SFG$^{*}$}&323222 \\
10943  &    0.80 &    8.10 & F& {\small AGN-CL}&163557 \\
10944  &    4.89 &   11.45 & F& {\small SFG} &869036 \\
10945  &    2.45 &   11.22 & T & {\small AGN-CL}&801950 \\
10946  &   38.98 &   11.53 & F& {\small SFG}&1124349 \\
10947  &  251.99 &   10.51 & F& {\small AGN-BT}&261526 \\
10948  &    0.44 &   11.28 & F& {\small AGN-SYM}&447542 \\
10949  &    7.82 &   11.40 & F& {\small AGN-WAT}&1003852 \\
10950  &    0.20 &   10.75 & F& {\small AGN-WAT}&1068567 \\
10951  &  140.78 &   11.72 & F& {\small AGN-SYM}&565211 \\
10952  &    4.99 &   11.00 & F& {\small AGN-WAT}&704802 \\

   \hline                  
\end{tabular}
\tablefoot{Host-galaxy properties. {\bf Column 1}: 3-GHz VLA-COSMOS ID; {\bf Columns 2 \& 3}: SFR calculated from the IR SED in M$_{\odot}$/yr \& logarithm of the stellar mass M$_{*}$ in M$_{\odot}$ from the SED fit from \cite{delvecchio17}, respectively;  {\bf Column 4}: AGN based on the SED fit \citep{delvecchio17} or not ('T' for AGN, 'F' no AGN); {\bf Column 5}: Radio class as in Table~\ref{table:data} also presented here as a visual aid to the reader; {\bf Column 6}: COSMOS2015 ID from \cite{laigle16} and counterpart catalogue in \cite{smolcic17b}.}
\end{table}

\addtocounter{table}{-1}

\begin{table}
\caption{Multi-component host properties}             
\label{table:data2}      
\centering          
\begin{tabular}{l r c c l l}     
\hline\hline       
                
\multicolumn{1}{c}{3-GHz}  & SFR$_{\rm IR}$ & $\log_{10}$(M$_{*}$ &SED & \multicolumn{1}{c}{radio}& {\small COSMOS}\\ 
\hline                    
 \multicolumn{1}{c}{ID} & (M$_{\odot}$/yr) & /M$_{\odot}$)&AGN& \multicolumn{1}{c}{class}& \multicolumn{1}{c}{2015} \\
 \hline                    
 \multicolumn{1}{c}{(1)} & \multicolumn{1}{c}{(2)}  & \multicolumn{1}{c}{(3)}& \multicolumn{1}{c}{(4)}& \multicolumn{1}{c}{(5)}& \multicolumn{1}{c}{(6)}\\
 \hline
 10953  &    3.91 &   11.30 & F& {\small AGN-SYM}&826044 \\
10954  &    1.93 &    9.67 & F& {\small SFG}&955856 \\
10955  &    1.65 &   10.62&F & {\small AGN-HT}&869175 \\
10956  &    0.72 &   11.20 & F& {\small AGN-WAT}&689074 \\
10957  &    4.23 &   11.10 & F& {\small AGN-HT}&544105 \\
10958  &    0.24 &   11.08 & F& {\small AGN-XZ}&744655 \\
10959  &   33.94 &   11.37 & F&{\small AGN-SYM} &957772 \\
10960  &    4.75 &   10.70 & F& {\small SFG}&657397 \\
10961  &    4.03 &   10.76 & F& {\small SFG}&342091 \\
10962  &   32.10 &   10.95 & T& {\small AGN-WAT}&931677 \\
10963  &  596.79 &   10.38 & F& {\small SFG$^{*}$}&902320 \\
10964  &  188.52 &   12.12 & T& {\small SFG$^{*}$}&223951 \\
10965  &   11.03 &   11.35 & F& {\small SFG}&234240 \\
10966  &    1.25 &   11.04 & F& {\small AGN-BT}&182559 \\
   \hline                  
\end{tabular}
\tablefoot{ (continued)}
\end{table}

\begin{table}
\caption{Multi-frequency flux densities}             
\label{table:freqs}      
\centering          
\begin{tabular}{l l l l l l   l l l }     
\hline\hline       
                
\multicolumn{1}{c}{3-GHz}  & \multicolumn{4}{c}{$S$  (mJy)} \\
\hline
   ID                                        & 324 & 325 & 1.4 & 3  \\
                                &  \multicolumn{2}{c}{MHz} &  \multicolumn{2}{c}{GHz} \\
 \hline                    
 \multicolumn{1}{c}{(1)} & \multicolumn{1}{c}{(2)} & (3) & (4) & (5)  \\
 \hline
10900 &   155.1 &   131.9 &    60.51&    35.17  \\
10901 &   260.3 &   233.9  &    53.03 &    18.16  \\
10902 &   472.1 &   421.5 &   116.5 &    46.16  \\
10903 &    17.81 &    16.41  &    10.61$^{2m}$ &     6.950  \\
10904 &   205.1 &   182.4  &    64.60$^{2m}$ &    28.42 \\
10905 &    12.40 &    10.43  &     5.673  &     3.170 \\
10906 &    60.33 &    54.33  &    19.21  &     8.651 \\
10907 &    10.90 &     8.862  &     5.271  &     2.814  \\
10908 &   156.3 &$-$ &$-$ &    26.23 \\
10909 &    32.73 &    30.50  &    12.99$^{2m}$ &     6.828  \\
10910 &$-$ &    30.36  &    15.10 &     5.867  \\
10911 &     3.008 &    16.27  &     7.509 &     3.425  \\
10912 &$-$ &     2.777  &     2.200$^{2m}$ &     1.220  \\
10913 &   172.8 &   207.6  &    82.83$^{2m}$ &    32.09  \\
10914 &    21.62 &    19.17 &     7.751 &     3.327  \\
10915 &    46.97 &    42.07  &     9.799 &     3.692  \\
10916 &    32.74 &    34.66  &    12.92$^{2m}$ &     5.438  \\
10917 &     6.394 &     7.087  &     2.658 &     1.523  \\
10918 &   114.0 &   136.4  &    57.26$^{2m}$ &    25.22  \\
10919 &    24.60 &    26.16  &     8.119 &     3.157  \\
10920 &    15.73 &    18.09  &     4.706 &     1.975  \\
10921 &$-$ &     4.486  &     2.422 &     0.981  \\
10922 &   237.4 &$-$  &$-$ &    24.15  \\
10923 &   143.5 &   135.1$^{2m}$  &    37.02 &    13.17  \\
10924 &$-$ &$-$  &$-$ &     9.353  \\
10925 &   146.0 &   144.7  &    47.23 &    18.54  \\
10926 &     9.163 &     9.158  &     2.027  & 0.744\\
10927 &     9.929 &$-$  &     3.428 &     1.055  \\
10928 &    88.64 &    93.43  &    30.12 &    11.64  \\
10929 &$-$ &$-$  &     0.675$^{2m}$ &     0.298  \\
10930 &$-$ &     3.603  &     1.173 &     0.578  \\
10931 &$-$ &     4.458  &     2.275 &     0.867  \\
10932 &$-$ &$-$  &$-$ &     0.816 \\
10933 &   227.4 &   217.2  &    88.17 &    39.30  \\
10934 &$-$ &$-$  &     0.942 &     0.380 \\
10935 &    15.56 &    20.19  &     6.548 &     2.382  \\
10936 &   101.3 &    82.80  &    29.13 &    10.05  \\
10937 &$-$ &     1.122  &     0.432 &     0.230  \\
10938 &$-$ &$-$  &$-$ &     0.849  \\
10939 &$-$ &     3.601  &     0.920$^{2m}$ &     0.367  \\
10940 &$-$ &     0.998  &     0.411 &     0.204  \\
10941 &$-$ &$-$  &$-$ &     0.549  \\
10942 &$-$ &$-$  &     0.107 &     0.374  \\
10943 &$-$ &     4.925  &     0.994$^{2m}$ &     0.406 \\
10944 &$-$ &     6.033  &     2.815 &     0.948  \\
10945 &$-$ &$-$  &$-$ &     0.205  \\
10946 &$-$ &$-$  &$-$ &     0.116  \\
10947 &$-$ &     0.944  &     0.313 &     0.120  \\
10948 &$-$ &     2.982  &     2.625 &     1.943  \\
   \hline                  
\end{tabular}
\tablefoot{Multi-frequency analogues to the 3-GHz multi-component radio sources. Data: 324 MHz from VLA \citep{smolcic14}; 325 MHz from GMRT \citep{tisanic18}; 1.4 GHz from VLA \citep{schinnerer07}; and 3 GHz from VLA \citep{smolcic17a}. The character $m$ denotes the source is multi-component at the corresponding radio frequency, and that the flux densities of the individual components have been added together giving the value presented here; the number in front of it denotes the number of blobs added together. 
}
\end{table}

\addtocounter{table}{-1}

\begin{table}
\caption{Multi-frequency flux densities}             
\label{table:freqs}      
\centering          
\begin{tabular}{l l l l l l l  l  l  }     
\hline\hline       
                
\multicolumn{1}{c}{3-GHz}  & \multicolumn{4}{c}{$S$  (mJy)} \\
\hline
   ID                                        & 324 & 325 & 1.4 & 3 \\
                                &  \multicolumn{2}{c}{MHz} &  \multicolumn{2}{c}{GHz} \\
 \hline                    
 \multicolumn{1}{c}{(1)} & \multicolumn{1}{c}{(2)} & (3) & (4) & (5)  \\
 \hline
 10949 &     7.472 &     6.975  &     5.422 &     2.718  \\
10950 &     3.242 &$-$  &$-$ &     1.439  \\
10951 &$-$ &     2.286  &     1.827$^{2m}$ &     0.890 \\
10952 &$-$ &     8.662  &     4.533 &     1.687  \\
10953 &$-$ &     6.276  &     3.192 &     1.254  \\
10954 &$-$ &     2.306  &     1.387 &     0.508 \\
10955 &$-$ &     4.636  &     6.229 &     1.058  \\
10956 &$-$ &    24.98   &    17.04$^{3m}$ &     4.379  \\
10957 &$-$ &     3.933  &     1.774 &     0.514 \\
10958 &$-$ &     4.354  &     2.305 &     0.889 \\
10959 &   158.3 &   144.1  &    35.14$^{2m}$ &     8.576  \\
10960 &$-$ &$-$  &     0.595 &     0.188  \\
10961 &$-$ &$-$  &     0.334 &     0.107  \\
10962 &   682.6 &   571.9  &   175.5 &    80.25  \\
10963 &$-$ &$-$  &     0.302 &     0.152  \\
10964 &$-$ &$-$  &     0.102 &     0.092  \\
10965 &$-$ &$-$  &     0.526 &     0.263  \\
10966 &$-$ &     6.403  &     2.873 &     0.861  \\
\hline
\end{tabular}
\tablefoot{(continued)}
\end{table}

\subsection{Hosts of 3-GHz multi-components}
\label{sec:hosts}
 
To further investigate the nature of the radio emission from the 67 multi-component objects, we use their SFR and M$_{*}$ quantities derived by \cite{delvecchio17} to explore their offset with respect to the MS of SFGs from \cite{whitaker12}, defined as $\Delta$sSFR-M$_{*}$. In Fig.~\ref{fig:sfr_mstar} we plot the logarithmic difference between the specific star-formation-rate (sSFR) and the specific star-formation-rate of objects on the main-sequence of star-forming galaxies (sSFR$_{\rm MS}$) versus stellar mass M$_{*}$ diagram, and infer trends regarding  the hosts of the multi-component objects. 

As seen on Fig.~\ref{fig:sfr_mstar}, the 6 multi-component SFGs at 3 GHz lie within the 1$\sigma$ dispersion of the MS, while the multi-component AGN have hosts that are spread around the diagram, with some objects lying below the MS, and some within the MS. The AGN hosts are mainly massive galaxies with masses above 10$^{10.5}$ M${_\odot}$, with the exception of 10904, 10932 and 10943. The latter object seems to have been assigned a counterpart in the \cite{smolcic17b} catalogue that is not actually matching to the core of the 3-GHz radio source (see Fig.~\ref{fig:maps2}). Instead it is matching to the western lobe of the source. We will flag this identification as uncertain (see Fig.~\ref{fig:maps2}). Regarding the SFGs (Fig.~\ref{fig:sfgs}), their hosts have masses above 10$^{9.5}$ M$_{*}$. Also, from their radio structure we do not have any indication that their radio emission is powered by an AGN. The multi-wavelength data available for COSMOS support that these are star-forming objects (see also Table~\ref{table:data2}).

Multi-component objects with uncertain radio classification, i.e. objects without clear signs of jets (10939, 10940, 10941, 10942, 10963, 10964; Fig.~\ref{fig:agnsfg}), are mainly located on the MS, with the exception of 10963 which lies in the starburst (SB) region of the $\Delta$sSFR-M$_{*}$ diagram, and of 10942 that lies below the MS. 

Furthermore, we investigate whether the 6 objects with uncertain classification have a VLBA counterpart in COSMOS \citep{noelia17}. A cross-correlation does not give a match, implying that, within the sensitivity of the VLBA observations (10 $\mu$Jy at milli-arcsecond resolution of 16.2$\times$7.3 mas$^{2}$), none of these 6 objects is associated with a bright point-like AGN. Finally, we use the radio excess flag\footnote{This radio excess flag is presented in \cite{delvecchio17}, and can be used to separate AGN from SFGs. Excess radio emission above what is expected from star-formation suggests the object is an AGN.} from \cite{delvecchio17} to further investigate whether these 6 objects with uncertain classification are AGN or SFGs. Based on this criterion, excess radio emission above what is expected from star formation should originate from AGN, thus radio excess can be used to separate AGN from SFGs. Based on that, 10942, 10963 and 10964 are SFGs, while 10939, 10940 and 10941 are AGN as the latter display excess radio emission, and we classify them as such for the rest of our analysis (see also Table~\ref{table:data}). Note that 10942 lies below the MS on the $\Delta$sSFR-M$_{*}$ diagram in the quiescent AGN region, but based on the radio excess parameter and the fit to the infrared SED (see Table~\ref{table:data2}), this is a clear SFG.

The few multi-component AGN in our sample at $z >$ 2 are found within the MS, suggesting their hosts are star-forming galaxies. On the other hand, at $z <$ 2 radio AGN occupy regions just below the scatter in the MS, with some exceptions within the MS and one above it. We also see that at low-z, multi-component AGN lie in their majority in the quiescent region and occupy massive hosts ($>$ 10$^{10.5}$ M$_{*}$).

Compared to the rest of the AGN at 3-GHz VLA-COSMOS shown by the density contours in Fig.~\ref{fig:sfr_mstar}, multi-component AGN lie at lower redshifts (up to $z \sim$ 2.5) but occupy a similar parameter space. On the other hand, we do not find multi-component AGN above $z >$ 2 in the quiescent region. At 3-GHz VLA-COSMOS depth we don't reach the sensitivity to identify diffuse multi-component sources that lie at high redshifts and in the quiescent region, as single-component AGN do. In fact, single-component AGN are located on the MS for SF and the green valley/quiescent region for up to redshifts of 4, as seen in Fig.~\ref{fig:sfr_mstar}. Interestingly, the latter are also found above the MS at very high redshifts.

\subsection{Types of multi-component objects at 3-GHz}
\label{sec:notes}

Here we present the types of multi-component objects we encounter at 3-GHz VLA-COSMOS. These can be separated in two broad classes, AGN or SFGs. A detailed description of the all the multi-component sources presented can be found in Appendix~\ref{app:notes}.

\subsubsection{The AGN multi-component radio sources at 3 GHz}
\label{sec:agn}

Based on our classification, at 3-GHz VLA-COSMOS we find 58 radio AGN. Some display peculiar radio structure that shows interaction with the intergalactic environment. Following visual inspection, we identified different categories for the AGN multi-components in our sample. We note that, as classification depends on the surface brightness and frequency of the survey, as well as resolution, the same objects could be classified differently under different conditions. \\ 

We deem:  \\

\begin{enumerate}
\item Head-tail (HT): One-sided radio AGN that show radio emission in the core and have one-sided jet attached to the core emission (2/58 AGN sources; Sec.~\ref{sec:headtail}).
\item Core-lobe (CL): One-sided radio AGN that show radio emission in the core and have one-sided lobe not attached to the core emission (10/58 AGN sources; Sec.~\ref{sec:headtail}).
\item Wide-angle-tail (WAT): FRI or FRII radio AGN with bent jets/lobes. The bending of the jets/lobes is the result of ram pressure while the source moves through the ICM \citep[e.g.][]{sakelliou00}. These show an inner-jet bent close to the core (9/58 AGN sources; Sec.~\ref{sec:wat}).
\item Z-/X-shaped (XZ): Objects that show signs of Z- or X-shape radio structure. This can be either due to a restarted AGN or axis reorientation (8/58 AGN sources; Sec.~\ref{sec:restarted}).
\item Bent-tail (BT): Objects that show an outer-jet bent which is not a result of movement of the sources through the ICM, but due to the outer part of jet interacting with a denser medium (3/58 AGN sources; Sec.~\ref{sec:benttail}).
\item Symmetric AGN (SYM): Radio AGN in which their jets/lobes form 180 degree angle in respect to each other (26/58 AGN sources; Sec.~\ref{sec:restAGN}).
\end{enumerate}

According to the above classification and the radio structure of the 3-GHz multi-component AGN, we find a large number of peculiar, i.e. disturbed/bent sources (32 out of 58), while 20 of them show intense bending in their radio structure which deviates from a straight line (see Fig.~\ref{fig:maps2}). We discuss this further below in Sec.~\ref{sec:denseenv}. Detailed description of the classification can be found in Appendix~\ref{app:notes}. These sub-classes are also presented in Table~\ref{table:data}, as part of the radio classification.

\subsection{Multi-component SFGs}
\label{sec:sfgs}

Our classification yields 9 SFGs amongst the multi-component objects at 3-GHz VLA-COSMOS. These are presented in Fig.~\ref{fig:sfgs} in Appendix~\ref{app:notes}. The advantage of high-resolution observations is that we are able to resolve out star-forming regions within the galaxy disk (e.g. 10951 \& 10965, see Fig.~\ref{fig:sfgs} in Appendix~\ref{app:notes}). The most striking object in this sub-sample is 10944. This is the only object in COSMOS that shows a well resolved radio ring and a compact core radio emission. Nevertheless, it is not associated with an AGN, based on multi-wavelength data. We discuss this in detail in Sec.~\ref{sec:ring} in Appendix~\ref{app:notes}.

Furthermore, 10965 is quite an interesting object when it comes to disentangling radio AGN from SFGs. Classifying this object solely by looking at its radio map can be misleading. Depending on one's point of view, this can look either as an AGN where the jets are bent due to interaction with the environment, or as a disk of a star-forming galaxy.


\section{Discussion}
\label{sec:discuss}

In this section we discuss the effects of high resolution (0.75 arcsec) and sensitivity ($rms$ = 2.3 $\mu$Jy/beam) of the VLA-COSMOS Large Project at 3 GHz in the context of multi-component sources.

\subsection{Hosts of multi-component sources at 3-GHz VLA-COSMOS}
\label{sec:disc_hosts}

In Sec.~\ref{sec:hosts} we demonstrated that multi-component objects in our sample occupy a similar parameter space as single-component 3-GHz sources in the $\Delta$sSFR vs M$_{*}$ diagram (Fig.~\ref{fig:sfr_mstar}). They also have similar sSFRs to single-component AGN and SFGs, but lie at lower redshifts. A small fraction of the multi-component objects are SFGs ($\sim$ 13\%) and are found within the MS. The rest are AGN ($\sim$ 87\%) and are preferentially located below the main sequence.  

AGN are believed to be responsible for quenching in situ star-formation in star-forming galaxies lying in the MS through feedback \citep[e.g.][]{hopkins08, combes15}, due to their jet emission (radio-mode feedback) or in the form of winds \citep[quasar-mode feedback;][]{combes13, combes14}. Hydrodynamical simulations show that AGN feedback transforms spiral galaxies to ellipticals \citep{dubois13}. So objects are expected to transition from the MS to the red-and-dead AGN region of the SFR-M$_{*}$ plane passing through the green valley \citep[e.g.][]{leslie16}. This scenario is supported by the high fraction of AGN within the green valley in studies such as 3D-HST/CANDELS \citep[0.5 $< z <$ 2.5, ][]{gu18}. Nevertheless, studies in the COSMOS field from X-ray detected AGN with far-IR emission \citep{mahoro17, lanzuisi17}, which lie in the green valley, show that AGN feedback plays a role in star-formation, but not necessarily by quenching star formation. \cite{silk14} discuss how AGN outflows can enhance star formation, through compression of dense clouds within the ISM. In our sample, all multi-component AGN display clear signs of jets or lobes by definition. The ones located in the green valley could suggest that AGN radio-mode feedback is in place, but it is not necessarily negative.

At very low-$z$ ($\sim$ 0.5), in particular, we have multi-component AGN in the quiescent region in their majority all lying in massive hosts ($>$ 10$^{10.5}$ M$_{*}$). Although this picture agrees with the scenario of passive evolution of galaxies, read-and-dead galaxies are an indirect evidence for AGN feedback \citep[see][]{fabian12}.

AGN feedback is not necessarily the only mechanism responsible for quenching star-formation. For example, \cite{schawinski14} propose a scenario for late-type spiral galaxies where once the cosmological inflow of gas has stopped, the galaxy leaves the MS while still converting gas and dust to stars, but at slower rates, until it becomes a red-and-dead late-type spiral, several Gyr later. We suspect that something similar is happening to the multi-component SFGs in our sample, 10942 and 10944, which lie below the MS. These objects are presented in detail in Appendix~\ref{app:notes}. 

In conclusion we see that multi-component AGN at 3-GHz in COSMOS occupy in their majority the green valley and quiescent region of Fig.~\ref{fig:sfr_mstar}, which suggests they can be responsible for star-formation quenching. Also, the multi-component AGN seem to be representative of the general AGN population. Regarding the multi-component SFGs, they mainly occupy the MS for SF.

\begin{table*}
\caption{Comparison of 3-GHz VLA-COSMOS and ATLAS multi-components per flux-density bin}             
\label{table:binflux}      
\centering          
\begin{tabular}{l l l l l l l l l l l  l l l}     
\hline\hline       
Sample    & area  & $rms$&{\small resolution} & $N_{\rm m}/N_{\rm tot}$   &   \multicolumn{9}{c}{$N$ in $S_{1.4 \rm GHz}$ bins}   \\
\hline
& (deg$^{2})$& ($\mu$Jy/& (arcsec) &  &\multicolumn{1}{c}{{\small 0.01-0.1}} & \multicolumn{1}{c}{{\small 0.1-0.6}}  & \multicolumn{1}{c}{{\small 0.6 -1}} & \multicolumn{1}{c}{{\small 1-6}} & \multicolumn{1}{c}{{\small 6-10}} & \multicolumn{1}{c}{{\small 10-60}} &  \multicolumn{1}{c}{{\small 60-100}} & \multicolumn{1}{c}{{\small 100-160}} & \multicolumn{1}{c}{{\small $>$ 160}}  \\
& & beam) &  &   & \multicolumn{9}{c}{ (mJy) }     \\
 \hline 
{\small 3-GHz}   		& 2.6 & 2.3 & 0.75& {\small 67/10,830}& 0 & 11 & 6 & 25 & 4 & 17 & 3 & 1 & 0    \\
{\small VLA-COSMOS} &  &  & & &  & &     &   &   &  &       &  \\
{\small ATLAS-CDFS}       & 3.7& 30& 10&  {\small 41/726}  &0 & 0 & 1 & 13 & 7 & 14 & 3 & 1 &  2      \\
 \hline

\hline
\end{tabular}
\tablefoot{Comparison of 3-GHz VLA-COSMOS multicomponent sources to the multi-component sources in the ATLAS 3.7 deg$^{2}$ Chandra Deep Field-South (CDFS) field \citep{norris06}. ATLAS multi-components are identified from the flag CID in Table 6 of \cite{norris06}. 3 GHz flux densities for COSMOS multi-component sources have been converted to 1.4 GHz fluxes using a typical steep spectrum index $\alpha$ = 0.8. N$_{tot}$ is the total number of radio sources in the survey and N$_{m}$ the number of multi-component sources. The rest of the columns show the number of multi-component sources per flux-density bin, from 10 $\mu$Jy to 2000 mJy. 
}
\end{table*}

\subsection{Investigating the high number of bent/disturbed sources at 3-GHz VLA-COSMOS}
\label{sec:denseenv}

We have reported a large number (32 out of 58) of peculiar sources associated with AGN within the multi-component sample. These either show interaction with their IGM (e.g. 10950) or with the ICM (e.g. 10913, 10956). In Appendix~\ref{app:notes} we describe the shapes of these objects in more detail. The important question that emerges is why do we have so many bent sources within 2.6 deg$^{2}$ of COSMOS, down to very small linear sizes.

In order to understand the reason behind this, we match the 3-GHz multi-component AGN sample (58 objects) to the X-ray group-sample (247 groups at 0.08 $\leq z <$ 1.53) of \cite{gozaliasl18} from Chandra/XMM-Newton data in COSMOS \citep[see also][]{george11}. By X-ray groups we refer to a set of galaxies with a common dark matter halo \citep{george11}. The X-ray groups have halo masses ($M_{200}$) of the order of 10$^{12-14}$ M$_{\odot}$ \citep[see][]{gozaliasl18}. We use a search radius defined by the virial radius ($R_{200}$) of each group and the corresponding redshift of the source (see Table~\ref{table:data}), with $\Delta z = \pm 0.007\times(1+z_{\rm xgroup})$ to match the photometric redshift accuracy \citep{laigle16}.

We find 12 out of 58 ($\sim$ 21\%) multi-component AGN (10902, 10910, 10912, 10913, 10918, 10933, 10948, 10950, 10953, 10956, 10958, 10966) within an X-ray group in COSMOS. From these objects, 3 are symmetric AGN (10902, 10948, 10953). We note that 10956 is member of 3 X-ray groups, which is already known from the study of \cite{smolcic07}, who report that we could be seeing the formation of a large cluster which is being assembled by several smaller groups. Thus, 9 out of 58 ($\sim$ 16\%) multi-component AGN that lie within X-ray groups in COSMOS exhibit peculiarities in their radio structure. 

The rest of the AGN multi-component objects (46 out of 58; $\sim$ 79\%) are not within the \cite{gozaliasl18} X-ray groups in COSMOS, thus we assume they lie in the field, or they are at redshifts not probed by the current X-ray data. Still, they might lie in mass halos below $< 1.5(1+z)\times 10^{13}$ M$_{\odot}$ not probed by our current X-ray data \citep[see Fig. 4 of][]{gozaliasl18}. 

From the AGN outside X-ray groups, 23 out of 46 ($\sim$50\%) show peculiarities in their radio structure. These belong to one of the sub-classes described in Section~\ref{sec:agn}  (excluding symmetric AGN). We note that from the 23, 4 lie at redshifts above 1.53 (10915, 10929, 10941, 10943) and are not in the current X-ray group catalogue, and for one (10924) we have no redshift measurement. This leaves us with 18 multi-component AGN with peculiarities\footnote{In detail, there are 5 CL (10903, 10927, 10939, 10940, 10945), 2 HT (10955, 10957), 5 WAT (10900, 10931, 10949, 10952, 10962), 4 Z-/X-shaped (10909, 10914, 109019, 10935) and 2 BT (10917, 10947).} that do not lie within the X-ray groups in COSMOS of \cite{gozaliasl18}. 

As we discuss in Sec.~\ref{sec:restarted}, the Z/X shape can be a result of several scenarios, which involve galaxy merging and restarting of AGN activity \citep[e.g.][]{{Gopal-Krishna12}}. WATs are the result of ram pressure on the source moving through the ICM \citep[e.g.][]{sakelliou00}, as described in Sec.~\ref{sec:wat} (e.g. 10956). Also, WATs are known to trace clusters of galaxies \citep[e.g.][]{smolcic07}. Thus these objects can be used to identify groups of galaxies with smaller halo masses than probed by the current X-ray observations in COSMOS. Core-lobe sources that don't lie within the known X-ray groups in COSMOS might be old AGN that started to fade away, or they lie within smaller groups not identified at X-rays (e.g. 10903, 10924). The rest of the objects outside X-ray groups are the one-sided head-tail sources and the bent-tail sources (10917 (BT), 10947 (HT), 10955 (HT)). We  believe the reason for the interruption of the jet path in these 3 bent sources  is a dense immediate environment outside the galaxy, the circum-galactic medium. 

We note that within the X-ray groups of \cite{gozaliasl18} lies the multi-component SFG of our sample, 10965.

To summarise, from the 32 AGN with peculiarities and bents in their radio structure we find 10 inside X-ray groups and 18 outside of groups, while 4 are above the redshift range of the X-ray group catalogue, and 1 has no redshift. The fraction of AGN disturbed objects outside X-ray groups to the total number of AGN within groups (18/12) is significantly higher than the fraction of disturbed AGN within groups to the total AGN within groups (9/12). 
The large number of bent sources outside groups suggests two things: 
\begin{enumerate}
\item We are probing density environments that cannot be probed by the current X-ray observations in COSMOS, i.e. less dense groups with halo masses below $< 1.5(1+z)\times 10^{13}$ M$_{\odot}$. These objects can be used to identify small halo mass groups not currently identified by X-ray observations. An interesting question that arises is how the IGM affects the radio structure of AGN.
\item The COSMOS field is filled with a large number of bent radio sources associated with AGN. The latter can either be small or large in their size, or have diffuse radio structure. We have described a plethora of peculiarities in this group of objects. This is a result of the 0.75 arcsec resolution, which allows to dive in the substructure of these sources and reveal objects that could not have been identified in surveys with poorer resolution. This is evident in Table~\ref{table:binflux} where we compare the ATLAS survey in the Chandra Deep Field-South (CDFS) field \citep{norris06}, which has an angular resolution of 10 arcsec and contains 41 multi-component sources. At 3-GHz we have a larger number of multi-component sources than in ATLAS CDFS, and we identify these as deep as 100 $\mu$Jy, in contrast to ALTAS CDFS.

\end{enumerate}

\subsection{Multi-component AGN and SFG disentangling, and future surveys}
\label{sec:disentangling}

Deep radio surveys such as the 3-GHz VLA-COSMOS, despite their advantage in detecting of the order of ten thousand objects in a very small patch of sky (2.6 deg$^{2}$), impose challenges when it comes to disentangling AGN and SFG populations of radio sources, which can have implications on how one can automatically identify sources in a radio survey. Here we briefly discuss the difficulties faced by current automatic algorithms in properly identifying such sources and disentangling the plethora of radio structures, given the upcoming radio surveys such as SKA and precursors. 

As we have demonstrated with the example of 10965 (see Fig.~\ref{fig:maps2}), classifying AGN and SFGs solely by their radio images can lead to mis-classifications. 10965 resembles a bent-tail radio AGN at 3 GHz, when in fact what we observe is emission from the face-on disk of the SFG. An AGN diagnostic that is purely based on the radio structure can be misleading, especially for surveys that are dominated by SFGs at faint flux densities, as the 3-GHz VLA-COSMOS \citep[see Fig. 13 of][]{smolcic17b}. This is the reason we use the UltraVISTA $\chi^{2}$ map in combination to the radio maps to match the blobs in Sec.~\ref{sec:sourceid}. 

High angular resolution (milli-arcsec) radio observations such as very long baseline interferometry (VLBI) are being used often in order to disentangle AGN and SF. This is a good diagnostic, but it also has its limitations. Although a source needs to have brightness temperature $\sim$ 10$^{6}$ K to be observed with VLBI, there are cases of SF objects reaching those high values \citep{kaviraj15, noelia17}. Indeed, many studies support that not even VLBI observations can disentangle AGN and SF in some cases, for example at higher redshifts \citep{noelia17}, or in extreme objects such as Arp 220 \citep{lonsdale06}, and not every VLBI core can be associated to an AGN \citep{kaviraj15}.

The difficulty in disentangling AGN and SFGs has implications regarding automatic classification methods. As we have already discussed in Sec.~\ref{sec:sourceid}, the current codes available for automatic identification of radio blobs or islands in radio mosaics do not perform a matching of the blobs, which is what is needed to identify multi-component objects. We performed a visual inspection to identify the objects that are composed of two or more blobs, combined them into a single parent source, and then update the catalogue. Visually inspecting 10,899 blobs is inefficient, subjective and time consuming. The method we describe in Sec.~\ref{sec:sourceid} is efficient and relatively fast, but has disadvantages when it comes to disentangling AGN from SFGs: if the objective is to select AGN, faint or diffuse sources with S/N and extent below the cut will not be included in the selection.
 
As \cite{hopkins15} stress, there is no automated source finder for large surveys that can produce 100\% completeness below the 10$\sigma$ detection level; there is always a compromise when it comes to reliability and completeness of the catalogue produced by the available codes, which is the case with automatic algorithms. Attempting to fit extended sources with an automatic algorithm is a rather complicated task. If the algorithm uses gaussian fits to find the extent of the source (e.g. \textsc{pyBDSF}), this can be proven difficult for large or diffuse sources. Thus, available codes provide much better results on compact sources after exclusion of extended/diffuse objects. Development of source finders is necessary for upcoming radio surveys, i.e. algorithms that can handle both compact and extended sources, perform de-blending, blob association, eliminate spurious sources and artefacts, and can be reliable when it comes to completeness and detection levels. Current precursors for the Square Kilometre Array \citep[SKA;][]{norris13, prandoni15} are investing on the development of automatic techniques and testing their effectiveness on large data-sets (e.g. the Canadian Initiative for Radio Astronomy Data Analysis - CIRADA\footnote{CIRADA is dedicated to the analysis of radio data from SKA and precursors. http://www.dunlap.utoronto.ca/instrumentation/cirada-canadian-initiative-for-radio-astronomy-data-analysis/.}).

For an all-sky survey like the Very Large Array Sky Survey (VLASS\footnote{https://science.nrao.edu/science/surveys/vlass}; Lacy et al. in prep.) for example,  or the Evolutionary Map of the Universe survey \citep[EMU\footnote{http://www.atnf.csiro.au/people/Ray.Norris/emu/index.html. EMU is being observed by the Australian Square Kilometre Array Pathfinder (ASKAP) telescope.};][]{norris11}, which will be able to observe millions of radio sources down to 69 and 10 $\mu$Jy/beam at angular resolution of 2.5 and 10 arcsec, respectively, the challenge of matching radio blobs into their parent source will be a major issue. The expected number of sources for VLASS is 5$\times$10$^{6}$ within 33,885 deg$^{2}$, and around 7$\times$10$^{7}$ sources for EMU over the $\sim$75\% of the sky. Given that a survey like the 3-GHz VLA-COSMOS revealed $\sim$ 11,000 sources in 2.6 deg$^{2}$ out of which 67 are multi-component, all-sky surveys should expect to find hundreds of thousands of multi-component sources. 

Assuming VLASS would reach a resolution similar to 3-GHz VLA-COSMOS, the expected number of multi-component sources of VLASS would be $\sim$ 10$^{5}$, given the depth and area coverage. But for a factor of 3 worse resolution, the expected number of multi-component sources is of the order of 10$^{4}$, making it not practical to visually inspect such a large number of sources. For EMU this number is smaller, of the order of 10$^{3}$, due to the much poorer resolution compared to 3-GHz VLA-COSMOS. 

We believe that the COSMOS field is an excellent laboratory for testing new tools due to the plethora of multi-wavelength observations that can be used to cross-check whether an automatic algorithm performs well, i.e. provides enough information for a good identification of the source. Such codes can take into account the physical mechanisms responsible for the emission in a multi-frequency and multi-resolution parameter space, and not only rely on the appearance of the source at a specific frequency. The next step would be for developed algorithm to be used for future radio surveys, for which we do not have all the multi-wavelength information available. Along these lines, there are already efforts for development of machine learning algorithms for classification by using self-organising maps (SOM) in LOFAR\footnote{http://www.lofar.org} for example \citep[e.g.][and Michiel Brentjens priv. comm.]{shimwell17}, as implemented in the PINK software by \cite{polsterer15}. But there are also methods that take into account the physical properties of objects and convolutional networks, as for example in Galaxy Zoo (Dennis Turp \& Kevin Schawinski priv. comm.) and the EMU collaboration (priv. comm).

\section{Conclusions}
\label{sec:conc}

In this data paper we present the multi-component radio sources of the VLA-COSMOS Large Project at 3 GHz \citep{smolcic17a}, i.e. the radio sources composed of two or several radio blobs. These were identified by selecting the brightest and most extended objects in the catalogue, and were verified by visual inspection using a multi-wavelength approach. 

The 3-GHz VLA-COSMOS survey demonstrates the importance of high resolution and sensitivity in identifying the substructure of radio sources within a field. Our study shows that we are probing populations that occupy lower density environments than what is probed by X-ray studies in COSMOS.

The multi-component objects can be either associated to AGN or to star-forming galaxies. Our results are summarised as follows:

\begin{enumerate}
\item We find 28 new multi-component sources that were not identified as such at 1.4 GHz by \cite{schinnerer07}, due to the higher resolution of the 3-GHz observations. 
\item We identify 58 AGN and 9 star-forming-galaxies, based on the 1.4- \& 3-GHz radio data and the UltraVISTA map for COSMOS.
\item From the 58 multi-component AGN we find 2 head-tail, 10 core-lobe, 9 WAT, 8 double Z-/X-shaped, 3 bent-tail, and 26 symmetric AGN.
\item Due to the high resolution we are able to resolve the sub-structure of SFGs to smaller star-forming regions.
\item At the high-mass end ($\rm > 10^{10.5} M_\odot$), AGN lie in their majority below the MS for SF, i.e. in the green valley. 
\item We find a large number (32/58) of disturbed/bent radio AGN, 18 of which do not lie within X-ray groups in COSMOS \citep[0.08$\leq z <$ 1.53][]{gozaliasl18}.
\item The use of small, diffuse, bent radio AGN within COSMOS can pinpoint to the location of small groups within COSMOS with halo masses  $< 1.5(1+z)\times 10^{13}$ M$_{\odot}$, not yet identified by COSMOS X-ray studies.
\item Disentangling AGN and SFGs using solely radio observations can be misleading, especially at the depths reached by 3-GHz VLA-COSMOS.
\end{enumerate}

We believe that future radio surveys will benefit from the development of automatic algorithms that not only perform identification of radio blobs, but also match them in a single radio identification when these belong to the same parent source. These should include multi-wavelength information and use of AGN diagnostics to disentangle AGN and SF, as they are both present at the low end of the radio luminosity function. COSMOS, with the plethora of auxiliary multi-wavelength data available, can be a laboratory for the development and testing of machine learning codes, which can be used for future radio surveys (SKA and precursors) in order to identify radio sources and eventually classify them based on type, without the need for extensive prior information.

\begin{acknowledgements}
      We would like to thank the anonymous referee for suggestions that significantly improved the manuscript. We would also like to thank Mattia Vaccari for very useful comments. EV and SL acknowledge funding from the DFG grant BE 1837/13-1. Support for BM was provided by the DFG priority program 1573 "The physics of the interstellar medium". EV, EFJA, AK, BM and FB acknowledge support of the Collaborative Research Center 956, subproject A1 and C4, funded by the Deutsche Forschungsgemeinschaft (DFG). KPM is a Jansky Fellow (NRAO, Caltech). VS, MN, KT and JD acknowledge support from the European Union's Seventh Frame-work program under grant agreement 337595 (ERC Starting Grant, `CoSMass').
\end{acknowledgements}

\bibliographystyle{aa} 

\begin{thebibliography}{}
  
  	\bibitem[Antonucci(1993)]{antonucci93} Antonucci, R., 1993, ARA\&A, 31, 473
	
	\bibitem[Becker et al.(1994)]{becker94} Becker, R. H., White, R. L., \& Helfand, D. J. 1994, ASPC, 61, 165
	
	\bibitem[Berta et al.(2013)]{berta13} Berta, S., Lutz, D., Santini, P., et al. 2013, A\&A, 551, A100
	
	\bibitem[Best et al.(2014)]{best14} Best, P. N., Ker, L. M., Simpson, C., Rigby, E. E., Sabater, J., 2014, MNRAS, 445, 955
  
  	\bibitem[Burns(1990)]{burns90} Burns, J. O., 1990, AJ, 99, 14
	
	\bibitem[Cielo et al.(2017)]{cielo17} Cielo, S., Antonuccio-Delogu, V., Silk, J., Romeo, A. D., 2017, MNRAS, 467, 4526
	
	\bibitem[Civano et al.(2012)]{civano12} Civano, F., Elvis, M., Brusa, M., et al. 2012, ApJS, 201, 30
	
	\bibitem[Civano et al.(2016)]{civano16} Civano, F., Marchesi, S., Comastri, A., et al. 2016, ApJ, 819, 62
	
	\bibitem[Chabrier(2003)]{chabrier03} Chabrier, G. 2003, ApJ, 586, L133
  	
	\bibitem[Combes(2015)]{combes15} Combes, F., 2015, IAUS, 309, 182
	
	\bibitem[Combes et al.(2014)]{combes14} Combes, F.; Garc\'{i}a-Burillo, S.; Casasola, V., et al. 2014, A\&A, 564, 1
	
	\bibitem[Combes et al.(2013)]{combes13} Combes, F., Garc\'{i}a-Burillo, S., Casasola, V., et al. 2013, A\&A, 558, A124
	
	\bibitem[Condon et al.(1998)]{condon98} Condon, J. J., Cotton, W. D., Greisen, E. W., Yin, Q. F., Perley, R. A., Taylor, G. B., \& Broderick, J. J. 1998, AJ, 115, 1693
	
	\bibitem[Condon(1992)]{condon92} Condon, J. J., 1992, ARA\&A, 30, 575
		
	\bibitem[da Cunha et al.(2008)]{dacunha08} da Cunha, E., Charlot, S., \& Elbaz, D. 2008, MNRAS, 388, 1595
	
	\bibitem[De Young(2010)]{deyoung10} De Young, D. S., 2010, ApJ, 710, 743
		
  	\bibitem[Delvecchio et al.(2017)]{delvecchio17} Delvecchio, I., Smol\v{c}ic, V., Zamorani, G., et al., 2017, A\&A, A3
	
	\bibitem[Dubois et al.(2013)]{dubois13} Dubois, Y., Gavazzi, R., Peirani, S., Silk, J., 2013, MNRAS, 433, 3297
	
	\bibitem[Elvis et al.(2009)]{elvis09} Elvis, M., Civano, F., Vignali, C., et al. 2009, ApJS, 184, 158
	
	\bibitem[Fabian(2012)]{fabian12} Fabian, A. C., 2012, ARA\&A, 50, 455
		
	\bibitem[Fan et al.(2015)]{fan15} Fan, Dongwei, Budavari, Tamas, Norris, Ray P., Hopkins, Andrew M., 2015, MNRAS, 451, 1299
	
  	\bibitem[Fanaroff \& Riley(1974)]{fr74} Fanaroff, B. L., Riley, J. M., 1974, MNRAS, 167, 31
		
	\bibitem[George et al.(2011)]{george11} George, M. R., Leauthaud, A., Bundy, K., et al. 2011, ApJ, 742, 125
	
	\bibitem[Gopal-Krishna et al.(2012)]{Gopal-Krishna12} Gopal-Krishna, Biermann, P. L., Gergely, L\'{a}szl\'{o} \'{A}., Wiita, P. J., 2012, RAA, 12, 127
	
	\bibitem[Gopal-Krishna et al.(2003)]{Gopal-Krishna03} Gopal-Krishna, Biermann, P. L., Wiita, P. J., 2003, ApJ, 594L, 103
	
	\bibitem[Gozaliasl et al.(2018)]{gozaliasl18} Gozaliasl, G., Finoguenov, A., Tanaka, M., et al. 2018, MNRAS, in press
	
	\bibitem[Gu et al.(2018)]{gu18} Gu, Y., Fang, G., Yuan, Q., Cai, Z., Wang, T., 2018, arXiv:1802.00550
	
	\bibitem[Herrera Ruiz et al.(2017)]{noelia17}Herrera Ruiz, N., Middelberg, E., Deller, A., et al. 2017, A\&A, 607, 132
	
	\bibitem[Hales et al.(2012)]{hales12} Hales, C. A., Murphy, T., Curran, J. R., et al. 2012, MNRAS, 425, 979
	
	
	\bibitem[Hopkins et al.(2015)]{hopkins15} Hopkins, A. M.; Whiting, M. T.; Seymour, N.; Chow, K. E.; Norris, R. P.; Bonavera, L.; Breton, R.; Carbone, D.; Ferrari, C.; Franzen, T. M. O.; et al., 2015, PASA, 32, 37
	
	\bibitem[Hopkins et al.(2008)]{hopkins08} Hopkins, P. F., Hernquist, L., Cox, T. J., et al., 2008, ApJS, 175, 356
	
	
	\bibitem[Ishibashi et al.(2014)]{ishibashi14} Ishibashi, W.; Auger, M. W.; Zhang, D.; Fabian, A. C., 2014, MNRAS, 443, 1339
	
	
	\bibitem[Kaviraj et al.(2015)]{kaviraj15} Kaviraj, S., Shabala, S. S., Deller, A. T., Middelberg, E., 2015, MNRAS, 452, 774
	
	
	\bibitem[Koekemoer et al.(2007)]{koekemoer07} Koekemoer et al. 2007, ApJS 172, 196
	
	
	\bibitem[Laigle et al.(2016)]{laigle16} Laigle, C., McCracken, H. J., Ilbert, O., et al. 2016, ApJS, 224, 24
	
	\bibitem[Lanzuisi et al.(2017)]{lanzuisi17} Lanzuisi, G., Delvecchio, I., Berta, S., Brusa, M., Comastri, A., Gilli, R., Gruppioni, C., Marchesi, S., Perna, M., Pozzi, F., et al., 2017, A\&A, 602A, 123
	
	\bibitem[Leahy \& Williams(1984)]{leahy84} Leahy, J. P., Williams, A. G., 1984, MNRAS, 210, 929
	
	\bibitem[Leslie et al.(2016)]{leslie16} Leslie, S. K., Kewley, L. J., Sanders, D. B., Lee, N., 2016, MNRAS , 455L, 82
	
	
	\bibitem[Lonsdale et al.(2006)]{lonsdale06} Lonsdale C. J., Diamond P. J., Thrall H., Smith H. E., Lonsdale C. J., 2006, ApJ, 647, 185
	
	\bibitem[Mahoro et al.(2017)]{mahoro17} Mahoro, A., Povi\'{c}, M., Nkundabakura, P., 2017, MNRAS, 471, 3226
	
	\bibitem[Marchesi et al.(2016)]{marchesi16} Marchesi, S., Lanzuisi, G., Civano, F., et al., 2016, ApJ, 830, 100
	
	
	\bibitem[Mao et al.(2010)]{mao10} Mao, M. Y., Sharp, R., Saikia, D. J., Norris, R. P., Johnston-Hollitt, M., Middelberg, E., Lovell, J. E. J., 2010, MNRAS, 406, 2578
	
	\bibitem[Mao et al.(2009)]{mao09} Mao, M. Y., Johnston-Hollitt, M., Stevens, J. B., Wotherspoon, S. J., 2009, MNRAS, 392, 1070
	
	
	\bibitem[McMullin et al.(2007)]{McMullin07} McMullin, J. P., Waters, B., Schiebel, D., Young, W., \& Golap, K. 2007, Astronomical Data Analysis Software and Systems XVI (ASP Conf. Ser. 376), ed. R. A. Shaw, F. Hill, \& D. J. Bell (San Francisco, CA: ASP), 127
	
	
	\bibitem[Miley et al.(1972)]{miley72} Miley, G. K., Perola, G. C., van der Kruit, P. C., van der Laan, H., 1972, Nature, 237, 269
	
	\bibitem[Mohan \& Rafferty(2015)]{mohan15} Mohan \& Rafferty, 2015, ASCL, 1502.007
	
	\bibitem[Moissev et al.(2011)]{moiseev11} Moiseev, A. V., Smirnova, K. I., Smirnova, A. A., Reshetnikov, V. P., 2011, MNRAS, 418, 244
	
	\bibitem[Murphy(2009)]{murphy09} Murphy, E. J., 2009, ApJ, 706, 482
	
	
	\bibitem[Netzer(2015)]{netzer15} Netzer, H., 2015, ARA\&A, 53, 365
	
	\bibitem[Norris et al.(2013)]{norris13} Norris, R. P., Afonso, J., Bacon, D., et al. 2013, PASA, 30, 20
	
	\bibitem[Norris et al.(2011)]{norris11} Norris, Ray P., Hopkins, A. M., Afonso, J., Brown, S., Condon, J. J., Dunne, L., Feain, I., Hollow, R., Jarvis, M., Johnston-Hollitt, M., et al. 2011, PASA, 28, 215
	
	\bibitem[Norris et al.(2006)]{norris06} Norris, Ray P., Afonso, Jos\'{e}, Appleton, Phil N., Boyle, Brian J., Ciliegi, Paolo, Croom, Scott M., Huynh, Minh T., Jackson, Carole A., Koekemoer, Anton M., Lonsdale, Carol J., et al., 2006, AJ, 132, 2409
	
	\bibitem[Oklop\v{c}i\'{c} et al.(2010)]{oklopcic10} Oklop\v{c}i\'{c}, A., Smol\v{c}i\'{c}, V., Giodini, S., et al., 2010, ApJ, 713, 484
		
	\bibitem[Owen \& Rudnick(1976)]{owen76} Owen, F. N., Rudnick, L., 1976, ApJ, 205L, 1
	
	\bibitem[Owen \& Laing(1989)]{owen89} Owen, Frazer N., Laing, Robert A., 1989, MNRAS, 238, 357
	
	\bibitem[Padovani(2016)]{padovani16} Padovani, P., 2016, A\&ARv, 24, 13
	
	\bibitem[Prandoni \& Seymour(2015)]{prandoni15} Prandoni, I., \& Seymour, N. 2015, in Proc. Advancing Astrophysics with the
Square Kilometre Array (AASKA14), 67
	
	\bibitem[Polsterer et al.(2015)]{polsterer15} Polsterer, K. L., Gieseke, F., Igel, C., 2015, ASPC, 495, 81
	
	\bibitem[Prestage \& Peacock(1988)]{prestage88} Prestage, Richard M., Peacock, John A, 1988, MNRAS, 230, 131
	
	
	\bibitem[Rodriguez et al.(2006)]{rodriguez06} Rodriguez, C., Taylor, G. B., Zavala, R. T., Peck, A. B., Pollack, L. K., Romani, R. W., 2006, ApJ, 646, 49
	
	\bibitem[Rudnick \& Owen(1976)]{rudnick76} Rudnick, L., Owen, F. N., 1976, ApJ, 203, 107
	
	
	\bibitem[Saikia \& Jamrozy(2009)]{saikia09} Saikia, D. J., Jamrozy, M., 2009, BASI, 37, 63
	
	\bibitem[Sakelliou \& Merrifield(2000)]{sakelliou00} Sakelliou, Irini, Merrifield, Michael R., 2000, MNRAS, 311, 649

	\bibitem[Sanders et al.(2007)]{sanders07} Sanders, D. B., Salvato, M., Aussel, H., et al. 2007, ApJS, 172, 86
	
	\bibitem[Schinnerer et al.(2007)]{schinnerer07} Schinnerer, E.; Smolcic, V.; Carilli, C. L.; Bondi, M.; Ciliegi, P.; Jahnke, K.; Scoville, N. Z.; Aussel, H.; Bertoldi, F.; Blain, A. W.; et al., 2007, ApJS, 172, 46
	
	\bibitem[Schinnerer et al.(2010)]{schinnerer10} Schinnerer, E., Sargent, M. T., Bondi, M., et al. 2010, ApJS, 188, 38
	
	\bibitem[Schoenmakers et al.(2000)]{schoenmakers00} Schoenmakers, Arno P., de Bruyn, A. G., R\"{o}ttgering, H. J. A., van der Laan, H., Kaiser, C. R., 2000, MNRAS, 315, 371
	
	\bibitem[Seymour et al.(2009)]{seymour09} Seymour, N., Huynh, M., Dwelly, T., Symeonidis, M., Hopkins, A., McHardy, I. M., Page, M. J., Rieke, G., MNRAS, 398, 1573
	
	\bibitem[Silk et al.(2014)]{silk14} 2014, Proceedings of the International School of Physics 'Enrico Fermi' Course 186 'New Horizons for Observational Cosmology' Vol. 186, p. 137-187, arXiv:1312.0107
	
	\bibitem[Simpson(2017)]{simpson17} Simpson, C., 2017, RSOS, 470522

	\bibitem[Shimwell et al.(2017)]{shimwell17} Shimwell, T. W.; R\"{o}ttgering, H. J. A.; Best, P. N.; Williams, W. L.; Dijkema, T. J.; de Gasperin, F.; Hardcastle, M. J.; Heald, G. H.; Hoang, D. N.; Horneffer, A.; Intema, H.; et al. 2017, A\&A, 598A, 104


	\bibitem[Smol\v{c}i\'{c} et al.(2007)]{smolcic07} Smol\v{c}i\'{c}, Schinnerer, E., Finoguenov, A., et al. 2007, ApJSS, 172, 313
	
	\bibitem[Smol\v{c}i\'{c} et al.(2014)]{smolcic14} Smol\v{c}i\'{c}, V., Ciliegi, P., Jeli\'{c}, V., et al., 2014, MNRAS, 443, 2590
	
	\bibitem[Smol\v{c}i\'{c} et al.(2017a)]{smolcic17a} Smol\v{c}i\'{c}, V., Novak, M., Bondi, M., et al. 2017a, A\&A, 602A, 1
	
	\bibitem[Smol\v{c}i\'{c} et al.(2017b)]{smolcic17b} Smol\v{c}i\'{c}, V., Delvecchio, I., Zamorani, G., et al. 2017b, A\&A, 602A, 2
	
	\bibitem[Smol\v{c}i\'{c} et al.(2017c)]{smolcic17c} Smol\v{c}i\'{c}, V., Novak, M.; Delvecchio, I.; Ceraj, L.; Bondi, M.; Delhaize, J.; Marchesi, S.; Murphy, E.; Schinnerer, E.; Vardoulaki, E.; Zamorani, G., et al. 2017c, A\&A, 602A, 6
		
	\bibitem[Schawinski et al.(2014)]{schawinski14} Schawinski, K., Urry, C. M., Simmons, B. D., et al. 2014, MNRAS, 440, 889
	
	\bibitem[Symeonidis et al.(2013)]{symeonidis13} Symeonidis, M.; Kartaltepe, J.; Salvato, M.; Bongiorno, A.; Brusa, M.; Page, M. J.; Ilbert, O.; Sanders, D.; van der Wel, A., 2013, MNRAS, 433, 1015
	
	\bibitem[Taniguchi et al.(2015)]{taniguchi15} Taniguchi, Y., Kajisawa, M., Kobayashi, M. A. R., et al. 2015, PASJ, 67, 104
	
	\bibitem[Taniguchi et al.(2007)]{taniguchi07} Taniguchi, Y., Scoville, N., Murayama, T., et al. 2007, ApJS, 172, 9
	
	\bibitem[Tisani\'{c} et al.(2018)]{tisanic18} K. Tisani\'{c}, V. Smol\v{c}i\'{c}, J. Delhaize, M. Novak, H. Intema, I. Delvecchio, E. Schinnerer, G. Zamorani, M. Bondi, and E. Vardoulaki, 2018, A\&A, in press, arXiv:1812.03392
		
	\bibitem[Williams \& R\"{o}ttgering(2015)]{williams15} Williams, W. L., R\"{o}ttgering, H. J. A., 2015, MNRAS, 450. 1538
	
	\bibitem[Willott et al.(1999)]{willott99} Willott, C. J.. Rawlings, S., Blundell, K. M., Lacy, M., MNRAS, 309, 1017
    	
	\bibitem[Whitaker et al.(2012)]{whitaker12} Whitaker, K. E., van Dokkum, P. G., Brammer, G., Franx, M., 2012, ApJ, 754L, 29
	
	\bibitem[Zhao et al.(1989)]{zhao89} Zhao, Jun-Hui, Burns, Jack O., Owen, Frazer N., 1989, AJ, 98, 64
      
      

    
   \end{thebibliography}

\appendix

\section{Testing the effect of coarser resolution to the 3-GHz sources}
\label{sec:resolution}

To test how the coarser resolution can change the appearance of the 3-GHz multi-component sources, we degraded the resolution of the 3-GHz mosaic to match that at 1.4 GHz, and re-examined the objects. We find that $\sim$55\% (37 out of 67) of the 3-GHz multi-component objects become single-component in the coarser resolution map, i.e. the radio emission does not break between the components and the sources appear to be composed of a single radio blob in the 1.5 arcsec resolution 3-GHz mosaic\footnote{Multi-component objects at 3 GHz that become single-component sources after convolving the mosaic to 1.5 arcsec resolution: 10906, 10907, 10908, 10909, 10915, 10917, 10922,  10924, 10928, 10930, 10932, 10934, 10935, 10937, 10938, 10939, 10940, 10941, 10943, 10944, 10945, 10946, 10947, 10948, 10949, 10950, 10952, 10953, 10954, 10955, 10957, 10960, 10961, 10962, 10965, 10966}. Images are shown in Fig.~\ref{fig:maps2} in the Appendix, next to the actual 3-GHz resolution images for comparison.

Finally, in order to check whether we fully recover the flux from the source we cross-match the 3-GHz radio positions to the NVSS \citep[angular resolution 45 arcsec][]{condon98} and FIRST \citep[angular resolution 5 arcsec][]{becker94} radio-source surveys. We find 29 matches with NVSS within a 15-arcsec search-radius and 20 in FIRST. As a search radius for FIRST we use 20 arcsec, except for the following sources, due to their larger extent: 10901 (40"), 10910 (40"), 10913 (90"), 10925 (35"), 10928 (30"), 10936 (40"). The resulting flux densities are given in Table~\ref{table:nvss}. We find that NVSS and FIRST flux densities are on average $\sim$ 29\% and 43\%, respectively, smaller than the 1.4-GHz COSMOS flux densities. To compare the FIRST and 3-GHz flux densities we use a typical steep radio spectral index of 0.8. We find that FIRST fluxes are on average $\sim$ 19\% larger than fluxes at 3 GHz. We do not compare the NVSS to 3-GHz VLA-COSMOS flux densities due to the large difference in angular resolution. These differences can be a consequence of a simplistic assumption of the spectral index but also an effect of different resolution. We also note that, in cases, FIRST flux densities are much lower than the NVSS, or the 1.4 and 3 GHz. For example in 10935, only the north lobe is identified by FIRST, while the south lobe, which is diffuse, is not observed. Also, in 10910 the north lobe is not identified by FIRST resulting in the low flux density value in Table~\ref{table:nvss}. Note that sources without 1.4-GHz flux entry in Table~\ref{table:nvss} lie outside the coverage of the survey.

\section{Matching 3-GHz multi-component radio sources with their multi-frequency analogues}
\label{app:matching}

In this section we describe how we matched the multi-frequency radio catalogues for COSMOS to the 3-GHz multi-component sources presented in this paper. We use the data for COSMOS available from 4 catalogues, 324 MHz \citep{smolcic14}, 325 MHz \citep{tisanic18} and 1.4 GHz \citep{schinnerer07, schinnerer10}. In order to match the 3-GHz multi-component COSMOS objects we had to take into account: i) the different angular resolution of each map the data were extracted from, and 2) near-by sources that could blend with the parent source in cases of poorer resolution. This method was developed to be able to rely solely on catalogue information, rather than carrying out a full visual inspection of the radio maps. Due to the high resolution of the 3-GHz data of 0".75 we are able to distinguish the substructure of these multi-component sources and disentangle them from near-by objects. But, in the case of the 325 MHz map, with angular resolution of 10 arcsec, we need to be careful when performing the matching not to include blended sources or mis-matches. For that reason the following automatic technique was applied. 

The code is written in IDL and takes as priors the radio positions of the 3-GHz multi-component sources as reported in Table~\ref{table:data}. The purpose is to search within the area occupied by the multi-component source. For this we need the size of the source and the area it occupies. The linear projected sizes of these objects were measured by a semi-automatic machine learning technique that i) identifies extended blobs by using the skeleton of the source, ii) joins blobs that belong to the same parent source via a clustering algorithm, and iii) measures the linear projected size of the source in arcsec. The method is described in Vardoulaki et al. (in prep.; credit E. Jimenez-Andrade). 

Now we can search within the area the source occupies. The actual search radius was set to size/1.5 to account for asymmetries in the lobe-core distance of some objects. But just using a search radius does not remove near-by objects not associated with the parent source. For that reason a mask was applied to the 3-GHz stamps, and then the algorithm searches within the radius given by the linear size of the source. 

The 3-GHz stamps were convolved to the coarser resolution of 10 arcsec\footnote{There is no need to regrid as we are interested in creating a mask and using it in combination to the size of the source.}, and all pixels below 3$\sigma$ of the local $rms$ were flagged. This mask was then used to approve/reject sources within the search radius. So the algorithm gives a match when it satisfies two conditions: 1) radio position within the mask, and 2) radio position within the search radius (size/1.5). 

\begin{figure}[!ht]
   \resizebox{\hsize}{!}{
   \includegraphics[trim={3cm 8cm 1.5cm 10cm},clip]{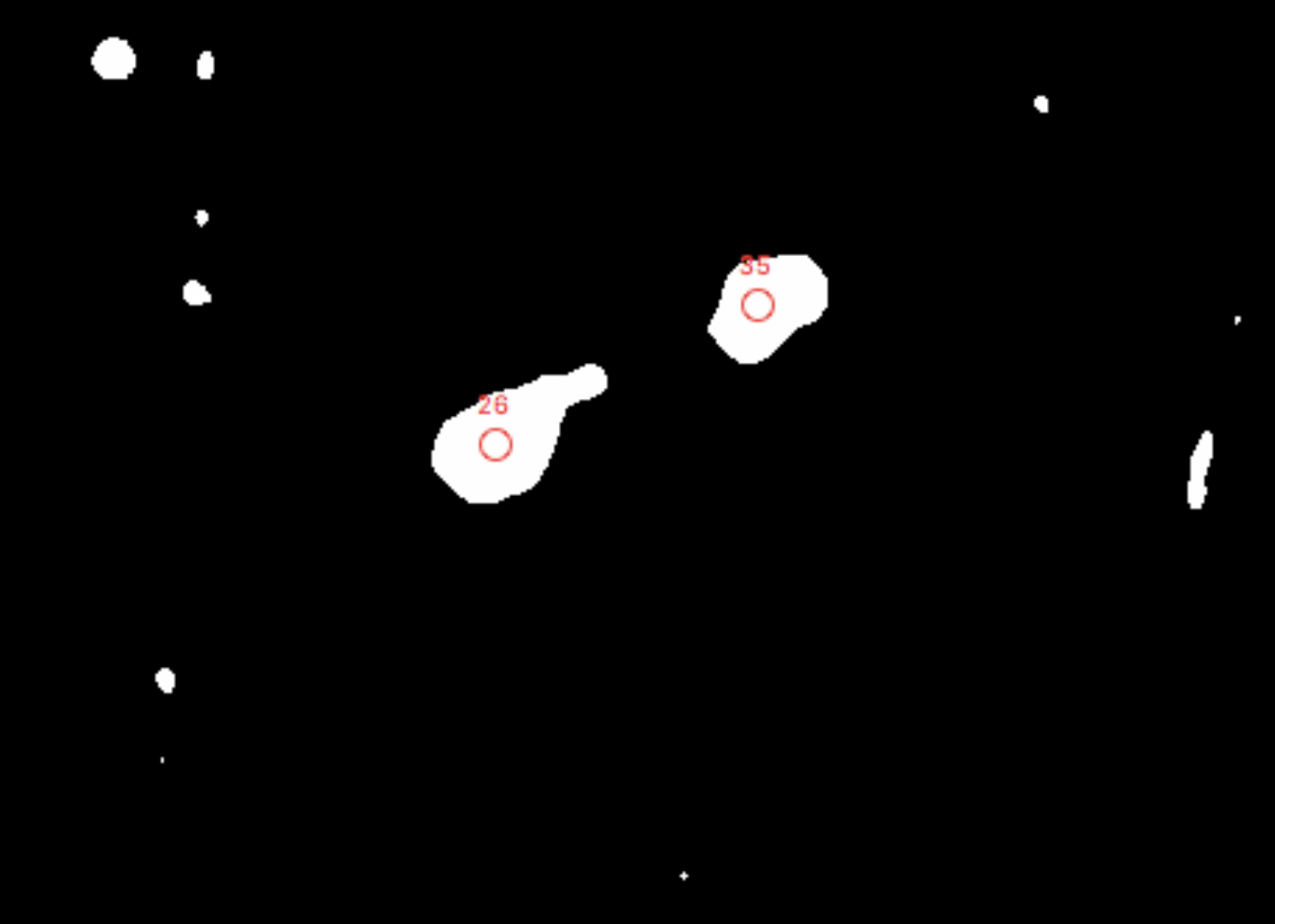}
   \\
 }

      \caption{Example of 3-GHz stamp convolved to 10" with 325-MHz radio positions within the stamp. The mask is shown in black \& white, where flux-densities above 3$\sigma$ at 3 GHz are shown as white. Red circles mark the 325-MHz blob radio positions. Only circles inside the white area occupied by the 3-GHz source were matched to the 3-GHz multi-component source. 
              }
         \label{fig:10923}
   \end{figure}

In some cases at 325 MHz, the 3-GHz multi-component sources were matched with more than one 325-MHz blob. These 325 MHz multi-component sources were joined automatically by the code and their flux densities were added together. An example is shown in Fig.~\ref{fig:10923}. The results of the matching are shown in Table~\ref{table:freqs}. Multi-component sources that were joined together are marked with $m$.

\begin{table}
\caption{FIRST \& NVSS matches to 1.4- \& 3-GHz VLA-COSMOS}             
\label{table:nvss}      
\centering          
\begin{tabular}{l l l l l l l  l l l }     
\hline\hline       
                
\multicolumn{1}{c}{3-GHz}  & \multicolumn{4}{c}{$S$  (mJy)} \\
\hline
   ID                                       & 1.4-GHz & 1.4-GHz  & 1.4-GHz  & 3-GHz  \\
                                &   NVSS & FIRST & COSMOS & COSMOS\\
 \hline                    
 \multicolumn{1}{c}{(1)} & \multicolumn{1}{c}{(2)} & \multicolumn{1}{c}{(3)} & \multicolumn{1}{c}{(4)} & \multicolumn{1}{c}{(5)}\\
 \hline
10900 &  59.4 &  55.91$^{3m}$ &  60.51 &  35.17 \\
10901 &    52.2& 43.91$^{3m}$ &  53.03 & 18.16  \\
10902 &   $-$ & $-$ &  116.5  & 46.16 \\
10903 &    11.6 & 10.01&  10.61 &   6.950 \\
10904 &    58.7 & 53.52$^{2m}$ &     64.60 &    28.42  \\
10905 &  8.2  &2.66 &    5.673 &     3.170 \\
10906 &  17.4  & 16.28$^{2m}$&    19.21 &     8.651  \\
10907 &    5.2 & 4.43&     5.271 &     2.814  \\
10908 &   50.3 & 49.97$^{2m}$  &$-$ &    26.23  \\
10909 &  12.0 &  9.98$^{2m}$&   12.99 &     6.828  \\
10910 & 12.4 &  5.48$^{2m}$&    15.10 &     5.867  \\
10911 &  6.2 &  4.77$^{2m}$&    7.509 &     3.425  \\
10912 &$-$ & $-$ &     2.200 &     1.220  \\
10913 & 73.1$^{2m}$ & 52.51$^{3m}$ &    82.83 &    32.09  \\
10914 &   6.3 & 6.40$^{2m}$ &  7.751 &     3.327  \\
10915 &  8.4 & 7.69&    9.799 &     3.692  \\
10916 &  11.1 &  7.84 &    12.92 &     5.438  \\
10917 &   2.6 & 2.25&      2.658 &     1.523  \\
10918 & 50.9 & 34.43$^{2m}$ &     57.26 &    25.22  \\
10919 &  6.2 &  4.99$^{2m}$&      8.119 &     3.157  \\
10920 &   4.9 &  2.55&    4.706 &     1.975  \\

   \hline                  
\end{tabular}
\tablefoot{Matches with FIRST \citep{becker94} and NVSS \citep{condon98} radio surveys within a 20" and 15" arcsec radius, respectively. Note that for the following sources we used a larger radius due to their entent: 10901 (40"), 10910 (40"), 10913 (90"), 10925 (35"), 10928 (30"), 10936 (40"). Also for 10913 we used a 90" search radius in NVSS to recover the lobes. The surveys' archives (FIRST: http://sundog.stsci.edu/cgi-bin/searchfirst \& NVSS: http://www.cv.nrao.edu/nvss/NVSSlist.shtml) were utilised for the search. The character $m$ denotes multi-component source, and that the fluxes of the components have been added together, while a number if front of it indicates the number of blobs matched.
}
\end{table}

\addtocounter{table}{-1}

\begin{table}
\caption{FIRST \& NVSS matches to 1.4- \& 3-GHz VLA-COSMOS}             
\label{table:nvss}      
\centering          
\begin{tabular}{l l l l l l l  l l l  }     
\hline\hline       
                
\multicolumn{1}{c}{3-GHz}  & \multicolumn{4}{c}{$S$  (mJy)} \\
\hline
    ID                                       & 1.4-GHz & 1.4-GHz  & 1.4-GHz  & 3-GHz  \\
                                &   NVSS & FIRST & COSMOS & COSMOS\\
 \hline                    
 \multicolumn{1}{c}{(1)} & \multicolumn{1}{c}{(2)} & (3) & (4)  & (5) \\
 \hline
 10921 &$-$ &  $-$ &     2.422 &     0.981  \\
10922 & 58.0 & 52.04$^{2m}$ &  $-$ &    24.15 \\
10923 &  $-$ & $-$ &    37.02 &    13.17 \\
10924 & 27.7 & 16.49$^{2m}$& $-$ &     9.353  \\
10925 &  43.2 & 31.17$^{3m}$ &  47.23 &    18.54 \\
 10926 &   $-$ & 1.77 &     2.027 &     0.744 \\
10927 &    $-$ & $-$ &     3.428 &     1.055  \\
10928 &  22.7 & 17.36$^{2m}$ &    30.12 &    11.64  \\
10929 &  $-$ & $-$&     0.675 &     0.298 \\
10930 &   $-$ & $-$&      1.173 &     0.578  \\
10931 &  $-$ & $-$&     2.275 &     0.867  \\
10932 &  $-$ & $-$ & $-$ &     0.816  \\
10933 &  82.3 &  77.24$^{2m}$ &    88.17 &    39.30  \\
10934 & $-$ & $-$ &     0.942 &     0.380  \\
10935 &   5.9 &  1.94&    6.548 &     2.382  \\
10936 &   26.6 & 14.65$^{2m}$ &    29.13 &    10.05  \\
10937 & $-$ & $-$  &     0.432 &     0.230 \\
10938 & $-$ & $-$ & $-$ &     0.849  \\
10939 & $-$ & $-$ &    0.920 &     0.367 \\
10940 & $-$ & $-$ &      0.411 &     0.204  \\
10941 &$-$ & $-$ & $-$ &     0.549  \\
10942 &$-$ & $-$ &   0.107 &     0.374  \\
10943 &$-$ & $-$  &   0.994 &     0.406  \\
10944 &$-$ & $-$  &     2.815 &     0.948  \\
10945 &$-$ &$-$ & $-$ &     0.205  \\
10946 &$-$ &$-$ & $-$ &     0.116  \\
10947 &$-$ &  $-$ &     0.313 &     0.120  \\
10948 &$-$ & $-$ &    2.625 &     1.943  \\
10949 &   3.9 & 3.39 &    5.422 &     2.718  \\
10950 &  $-$ & 1.87 &    $-$ &     1.439  \\
10951 & 4.2 & $-$&  1.827 &     0.890 \\
10952 & 4.0 & $-$&  4.533 &     1.687  \\
10953 & $-$ & $-$&    3.192 &     1.254  \\
10954 &$-$ &  $-$ &  1.387 &     0.508  \\
10955 &$-$ & $-$ &    6.229 &     1.058  \\
10956 &$-$ & $-$  &  17.04 &     4.379  \\
10957 &$-$ & $-$ &      1.774 &     0.514  \\
10958 &$-$ & $-$  &  2.305 &     0.889 \\
10959 &  28.2 & 20.00$^{2m}$ &  35.14&     8.576  \\
10960 &$-$ & $-$&      0.595 &     0.188 \\
10961 &$-$ & $-$&       0.334 &     0.107  \\
10962 &  $-$ &$-$ & 175.5 &    80.25  \\
10963 & $-$ & $-$&    0.302 &     0.152 \\
10964 &$-$ & $-$& 0.102 &     0.092  \\
10965 &$-$ & $-$&       0.526 &     0.263  \\
10966 &$-$ & 1.57  &    2.873 &     0.861  \\
\hline
\end{tabular}
\tablefoot{(continued)}
\end{table}

\newpage

\section{Notes on the objects}
\label{app:notes}

Here we present a detailed description of the multi-component sources at 3-GHz VLA-COSMOS. These are separated in two major classes, AGN or SFGs. We further separate AGN in classes based on their radio structure.

\subsection{ The AGN multi-component radio sources at 3 GHz}
\label{sec:notesagn}

Based on our classification, at 3-GHz VLA-COSMOS we find 58 multi-component radio AGN. Some display peculiar radio structure that shows interaction with their  environment. Following visual inspection, we further identified different categories for the multi-component sources: head-tail (2) and core-lobe (10) sources (Sec.~\ref{sec:headtail}); wide-angle-tail sources (9 sources; Sec.~\ref{sec:wat}); Z-/X-shaped radio galaxies (8 sources; Sec.~\ref{sec:restarted}); bent-tail (3 sources; Sec.~\ref{sec:benttail}), and 26 symmetric AGN, where the lobes form an angle of approximately 180 degrees in respect to each other (Sec.~\ref{sec:restAGN}). These sub-classes are also presented in Table~\ref{table:data}, as part of the radio classification.

  \begin{figure*}[!ht]
 \resizebox{\hsize}{!}{
   \includegraphics[width=0.5cm]{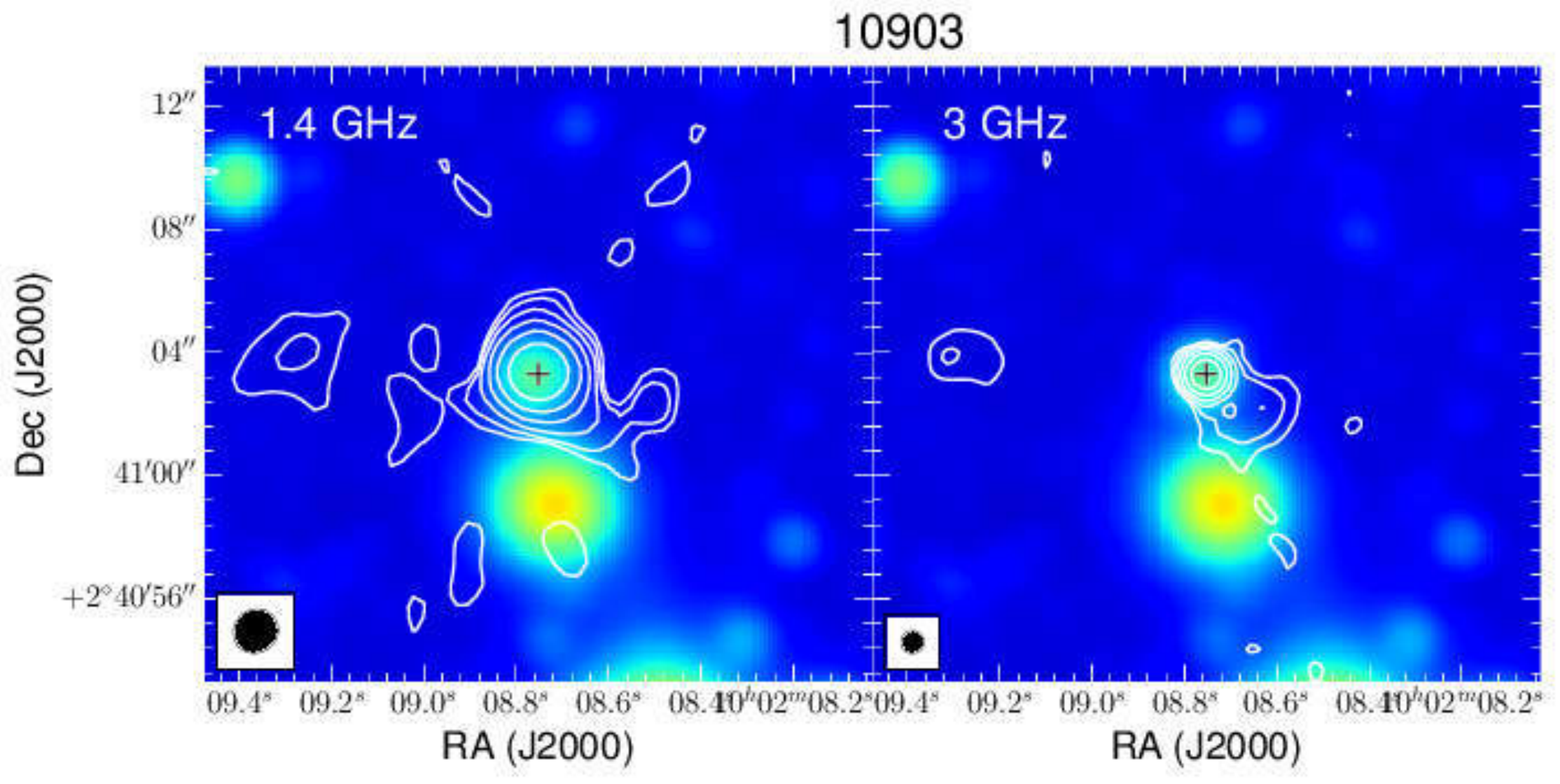}
    \includegraphics[width=0.5cm]{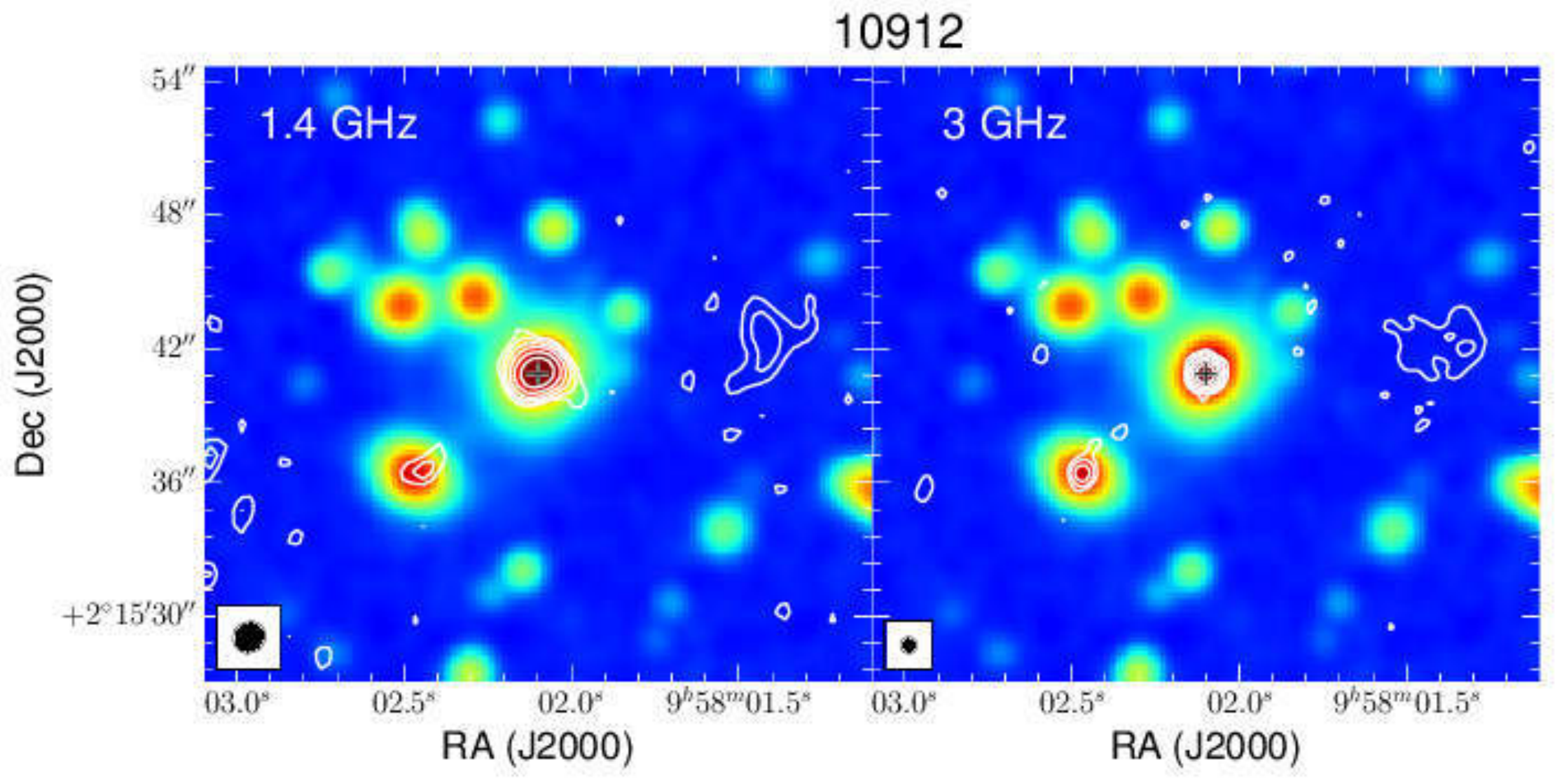}
   }
    \resizebox{\hsize}{!}{
   \includegraphics[width=0.5cm]{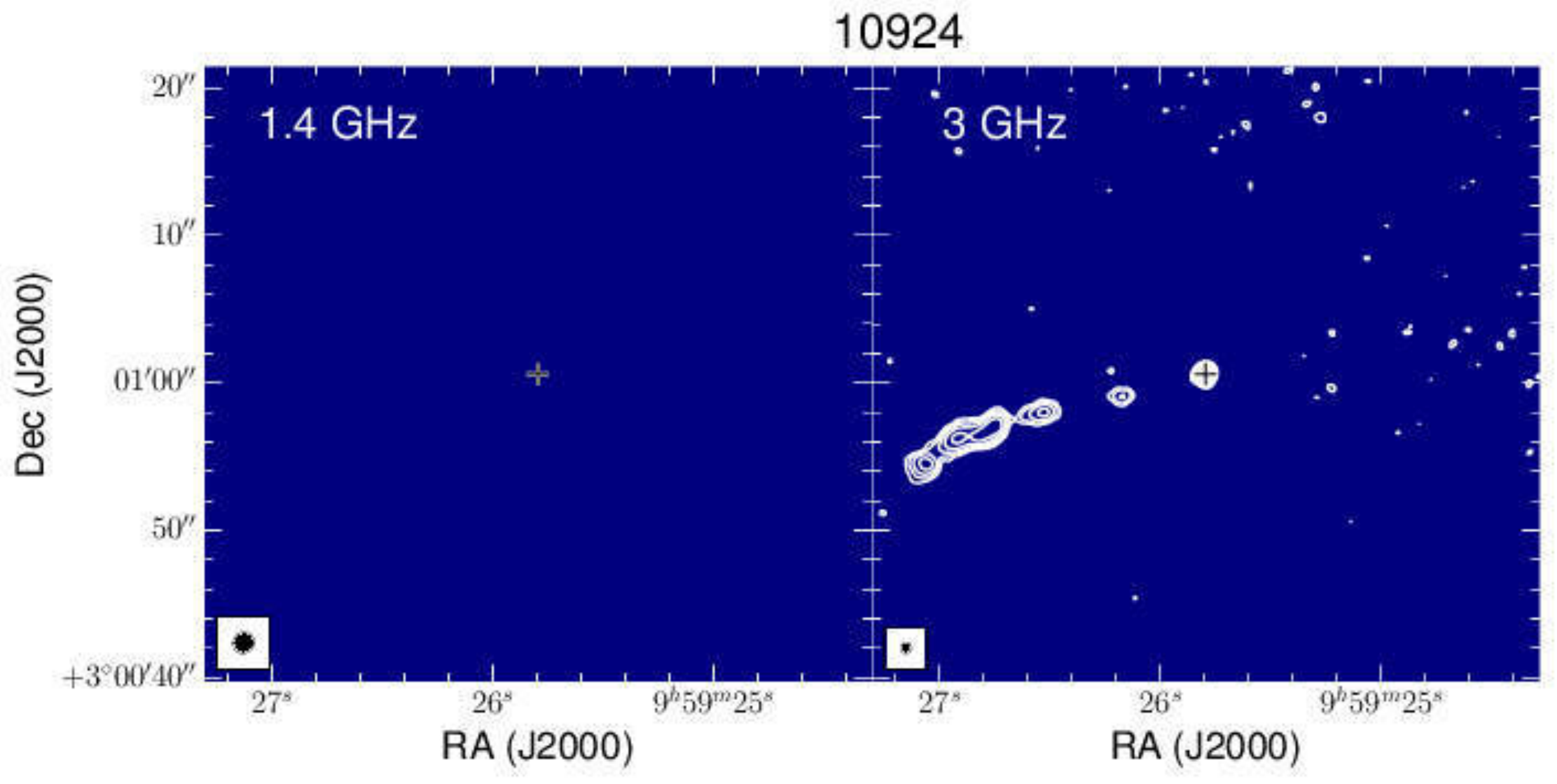}
    \includegraphics[width=0.5cm]{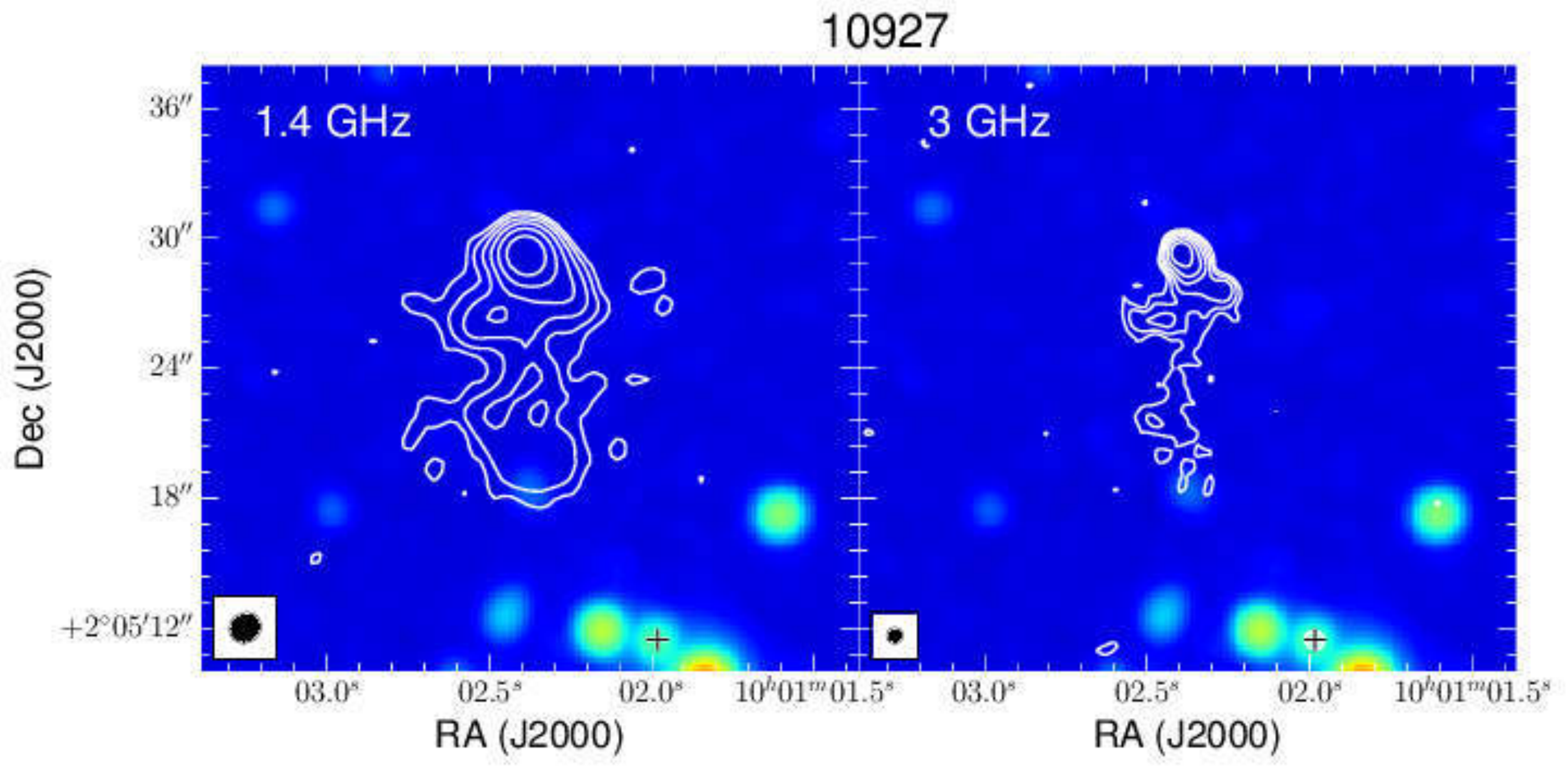}
   }
   \resizebox{\hsize}{!}{
   \includegraphics[width=0.5cm]{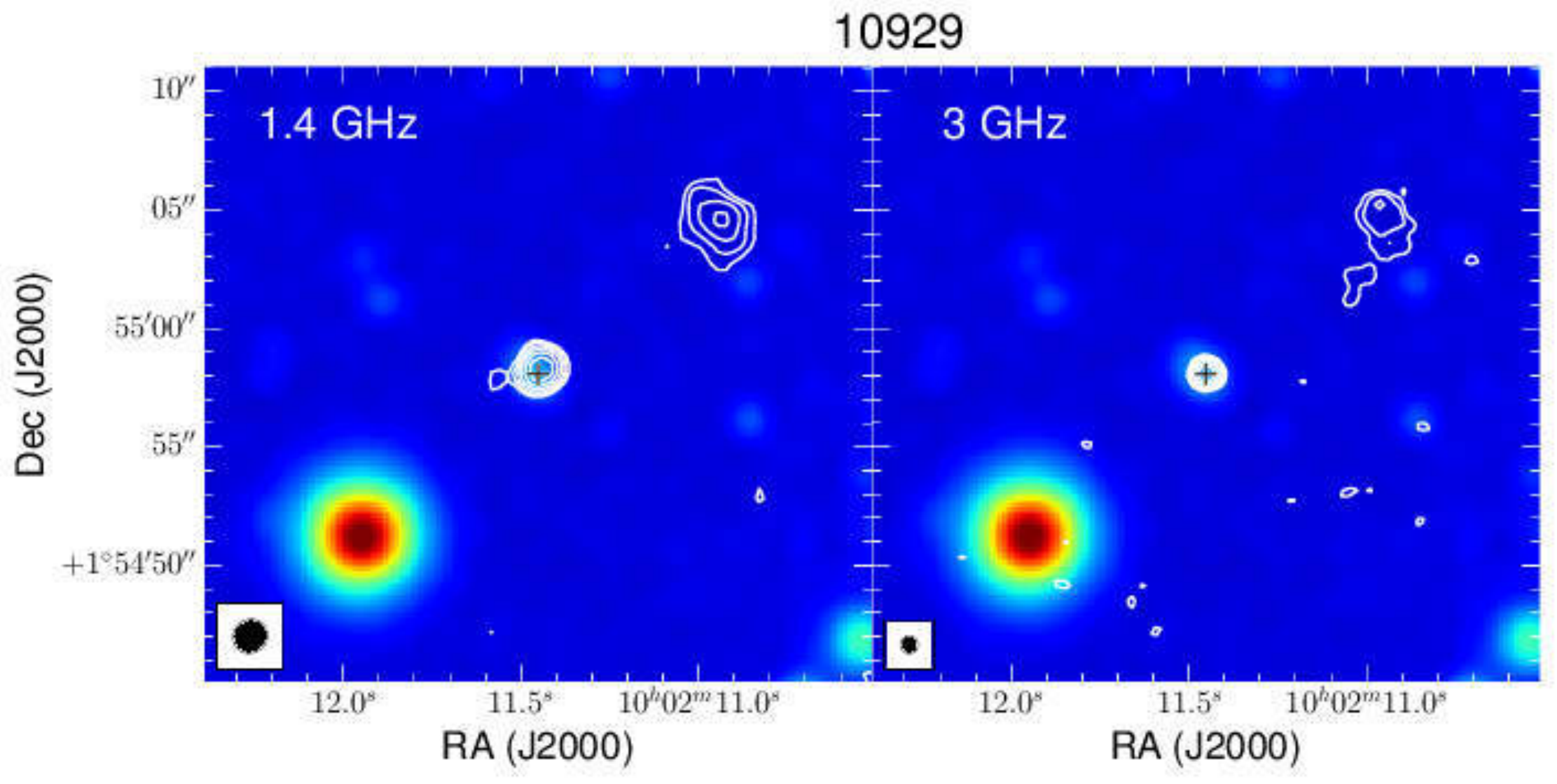}
    \includegraphics[width=0.5cm]{JVLA10939-ultrajvlavla.pdf}
  }
     \resizebox{\hsize}{!}{
 \includegraphics[width=0.5cm]{JVLA10940-ultrajvlavla.pdf}
  \includegraphics[width=0.5cm]{JVLA10941-ultrajvlavla.pdf}
}
   \resizebox{\hsize}{!}{
 \includegraphics[width=0.5cm]{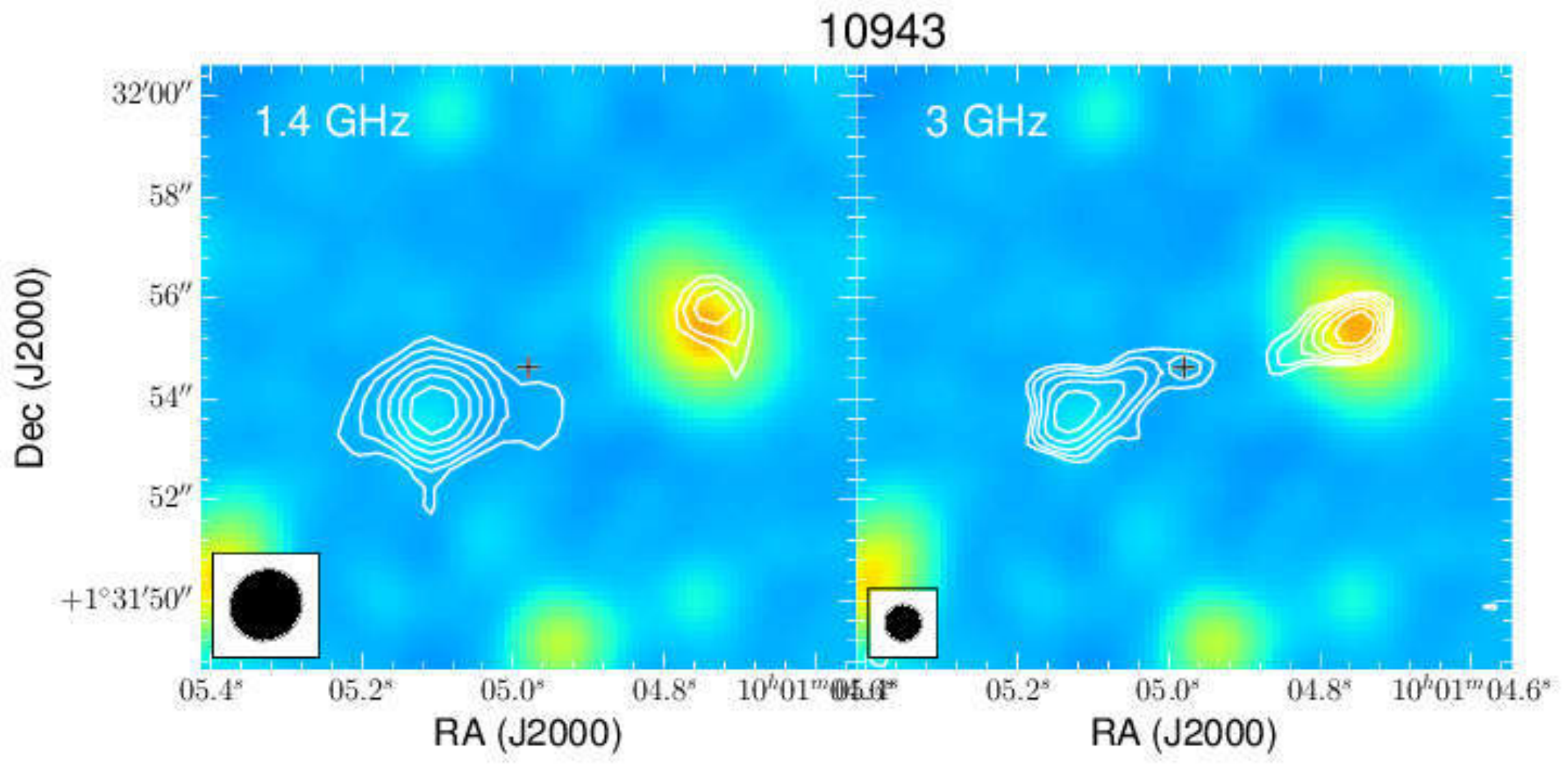}
  \includegraphics[width=0.5cm]{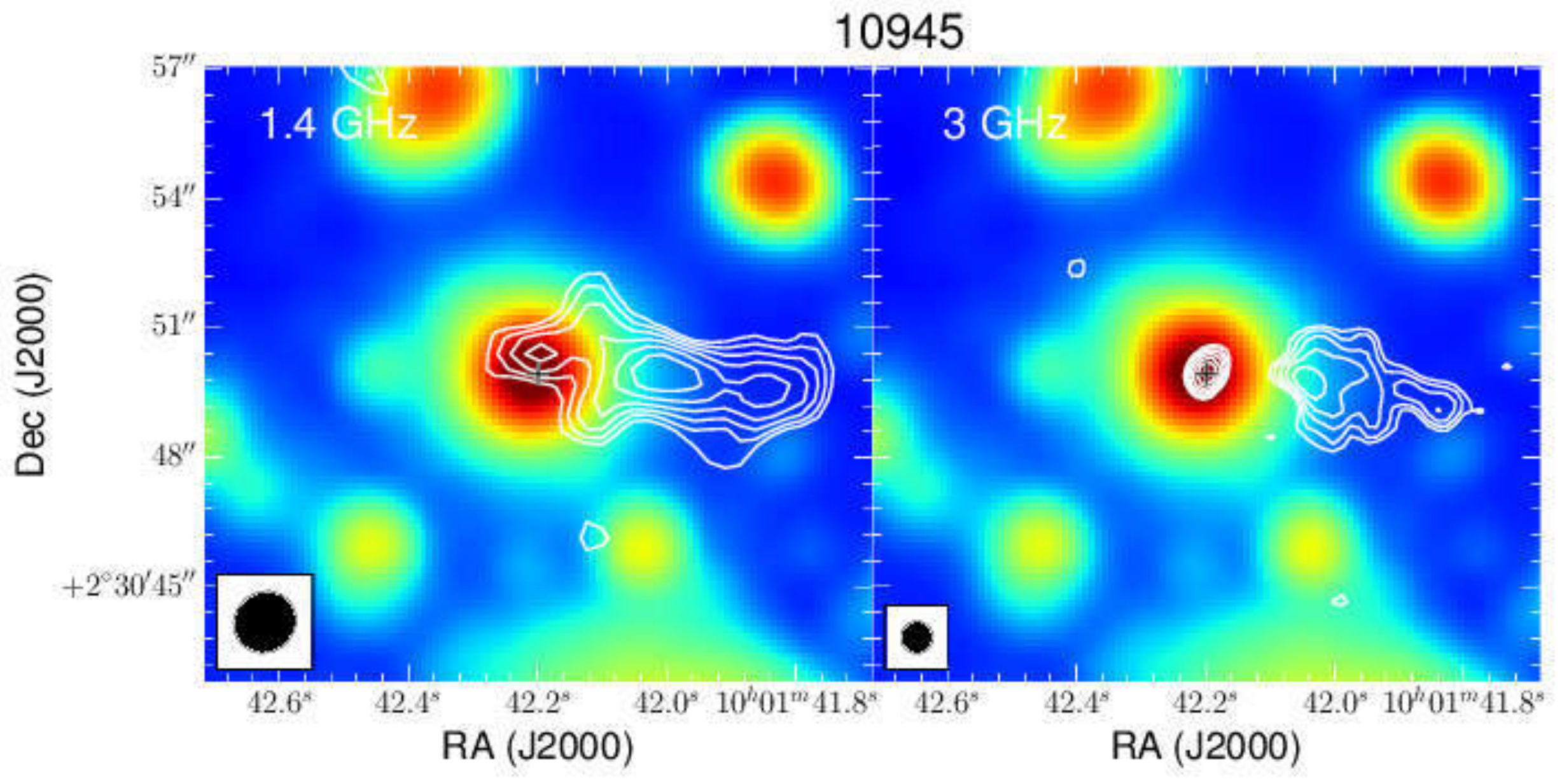}
}
   \resizebox{\hsize}{!}{
    \includegraphics[width=0.5cm]{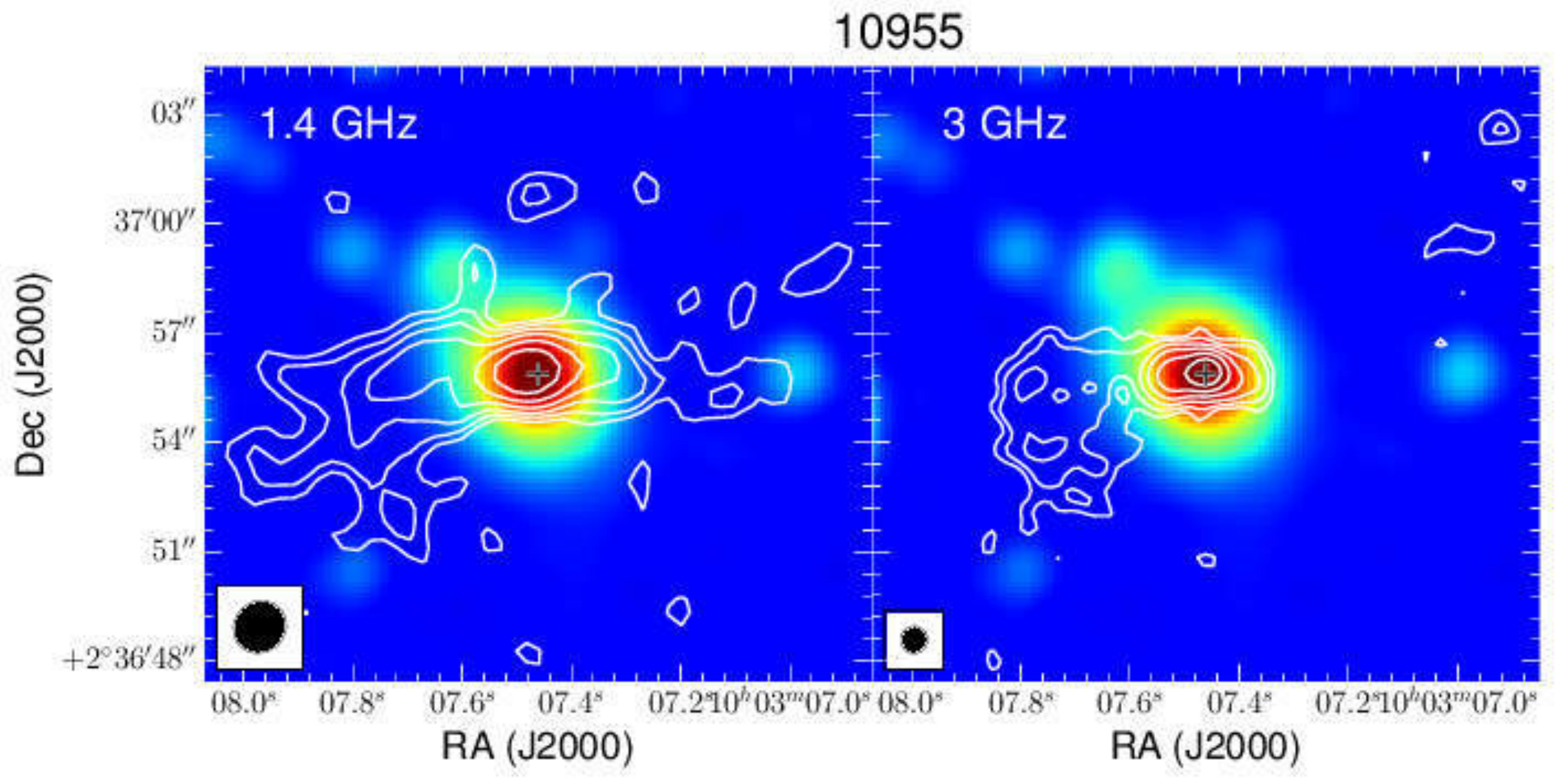}
 \includegraphics[width=0.5cm]{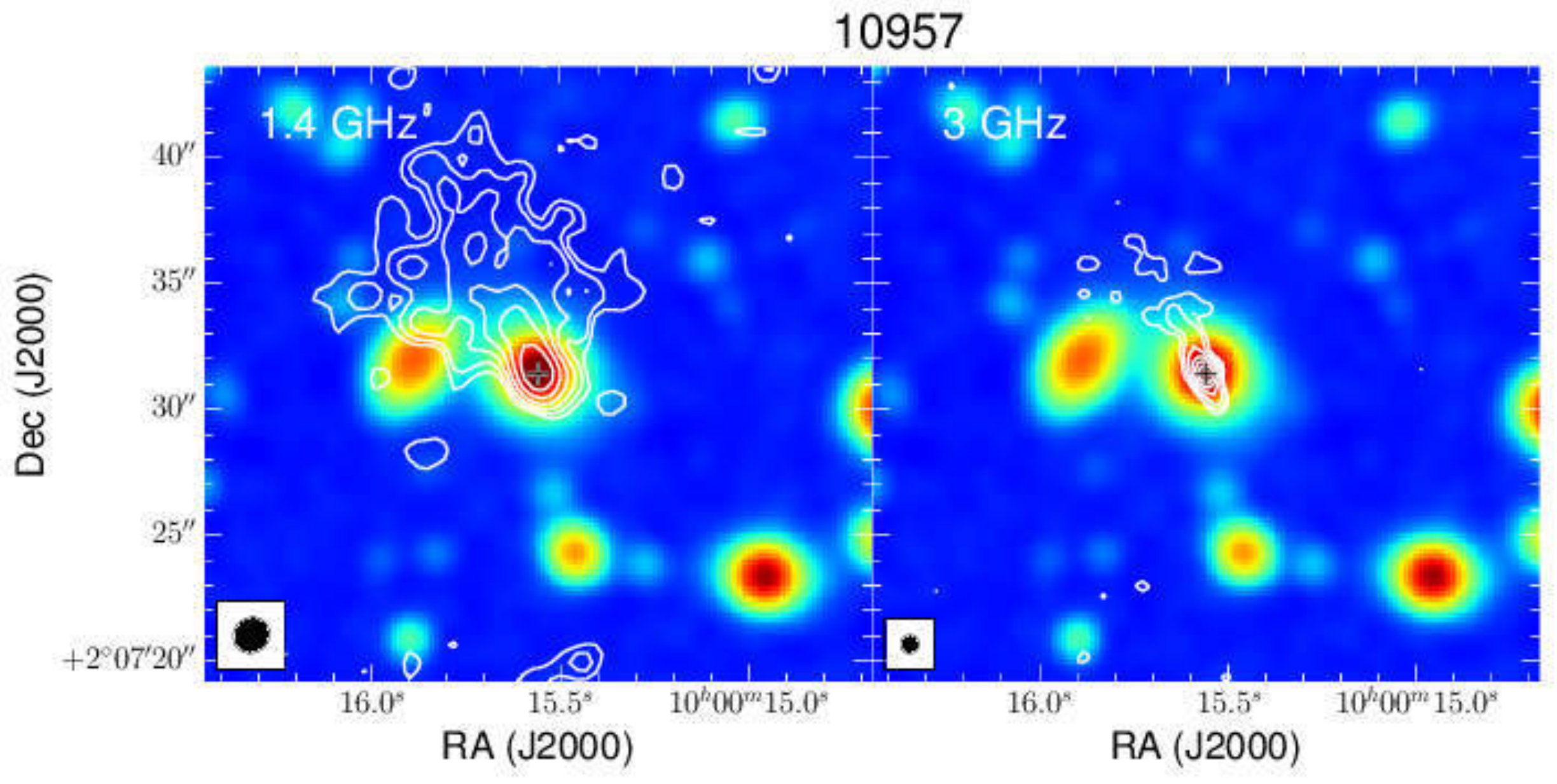}
 }
      \caption{The 10 core-lobe (10903, 10912, 10924, 10927, 10929, 10939, 10940, 10941, 10943, 10945) \& 2 head-tail (10955, 10957) multi-component 3-GHz radio sources in COSMOS. Images described in Fig.~\ref{fig:maps2}.
              }
         \label{fig:headtail}
   \end{figure*}

\subsubsection{ Head-tail and core-lobe multi-component radio sources at 3 GHz}
\label{sec:headtail}

This sub-class includes multi-component AGN radio sources which display one-sided emission. Either core emission and a single lobe (10903, 10912, 10924, 10927, 10929, 10939, 10940, 10941, 10943, 10945), or a one-sided radio jet called head-tail (10955, 10957). The core-lobe and head-tail objects in our sample are presented in Fig.~\ref{fig:headtail}. 

In more detail, 10903 shows a lobe at the east and core emission that slightly extends to the south-west. 

10902 displays core emission which is unresolved at 3 GHz and a lobe at the west of the source. 

10924 lies in a rather noisy region close to the edge of the 3-GHz map. It shows core emission and a jet to the east. We classify it as a possible core-lobe source. 

10927 displays a core and a single lobe, with the host identified to be on the south of the lobe $\sim$ 7 arcsec way (see also counterpart catalogue, \cite{smolcic17b}). We note that this object could instead be a narrow-angle-tail radio source \citep{miley72}, i.e. a WAT seen edge-on, and lies within an X-ray group in COSMOS from the catalogue of \cite{gozaliasl18}. We classify it as a possible a core-lobe source. 

10929 is also core-lobe source. 

10939 was initially in the uncertain class, but since it displays radio excess (see Sec.~\ref{sec:hosts}) we classify it as AGN; it also has a core-single lobe radio structure. 

10940 was classified as AGN based on its radio excess. It shows extended core emission, probably a blend of core and jet emission, and a lobe to the south. 

10941 was also classified as AGN based on its radio excess and is similar to 10940, with the lobe emission to the east. 10943 also shows slightly extended core emission and a radio lobe to the east. 

10945 shows a core-lobe radio structure at 3 GHz, while at 1.4 GHz is a head-tail. The enhanced resolution and sensitivity of the 3 GHz have resulted in this object being a multi-component at 3 GHz and classified as a core-lobe. 

10955 is one-sided at 3 GHz (head-tail), showing bending towards the edge of the lobe/jet-like radio structure. 

10957 is a diffuse head-tail radio source.

\subsubsection{Wide-angle-tail sources at 3 GHz}
\label{sec:wat}

Here we present the multi-component wide-angle-tail radio sources at 3-GHz VLA-COSMOS. In our sample, the most striking case of a WAT is 10956, known with the name CWAT-01 \citep{smolcic07}. A huge ($\sim$ 200kpc long), bent, C-shaped radio structure, that has been previously studied by \cite{smolcic07} in great detail. They perform a Voronoi tesselation analysis and find that  10956 lies in a supercluster, where smaller clusters are actively forming a larger one with a total mass 20\% of the Coma cluster. Other WAT multi-component radio sources at 3-GHz VLA-COSMOS are: 10900, 10910, 10913, 10931, 10949 (?), 10950, 10952 and 10962. 

10949 has a very small bent and is rather diffuse, so we caution the reader about the WAT. 

10962 is not a typical WAT. Although it is symmetric and the jets and lobes form an angle in respect to each other, the lobes seem to go backwards. This could probably be a projection effect, although we cannot rule out the case where the jet encounters a denser medium causing it to flow backwards towards the point of ejection \citep[e.g.][]{cielo17}. 

10900 has a diffuse south lobe. 

10910 shows a bent jet on the south, leading to a diffuse lobe. This source was studied a few years ago by \cite{oklopcic10}, who refer to it as CWAT-02. 

In 10913 the jets seem dragged in the ICM forming an angle in respect to the original jet-emission path. The lobes themselves seem bent and slightly diffuse, suggesting the radio source is fading. 10931 is a very diffuse WAT. 

10950 is a small size WAT. 10952 has a diffuse north lobe.

This C-shaped radio structure in WATs is caused by strong interaction of the jets of the source with the environment \citep[e.g.][]{owen76, owen89, zhao89, burns90, smolcic07, mao10}. Ram pressure distorts the radio structure and forces the jets to bend backwards as the source is moving through the ICM. For that reason, they are often used as probes or tracers of clusters of galaxies \citep[e.g.][]{prestage88, owen89, burns90, smolcic07}. These types of objects show a bent on the inner part of the jets (inner-jet bent-radio sources), in contrast to objects showing bents on the part of the jet further away from the core (outer-get-bent radio sources). The bent in the jet in the latter case is due to the jet interacting with a denser environment and not due to ram pressure from the movement of the source in its ICM. An example is object 10966 (not a WAT), where the outer parts of the jets are bent to almost 90 degrees in respect to the jet ejection direction. This is actually classified as BT (see Sec.~\ref{sec:benttail}).

WAT radio sources are often found inside clusters of galaxies and can be the result of the galaxy's movement inside a hot and dense intracluster medium \citep{rudnick76, mao09}. The dense medium is what causes jets of the source to bend backwards due to ram pressure \citep{mao09}. This mechanism explains the bent shapes of WAT sources within our sample.

   \begin{figure*}[!ht]
   \resizebox{\hsize}{!}{
   \includegraphics[width=0.5cm]{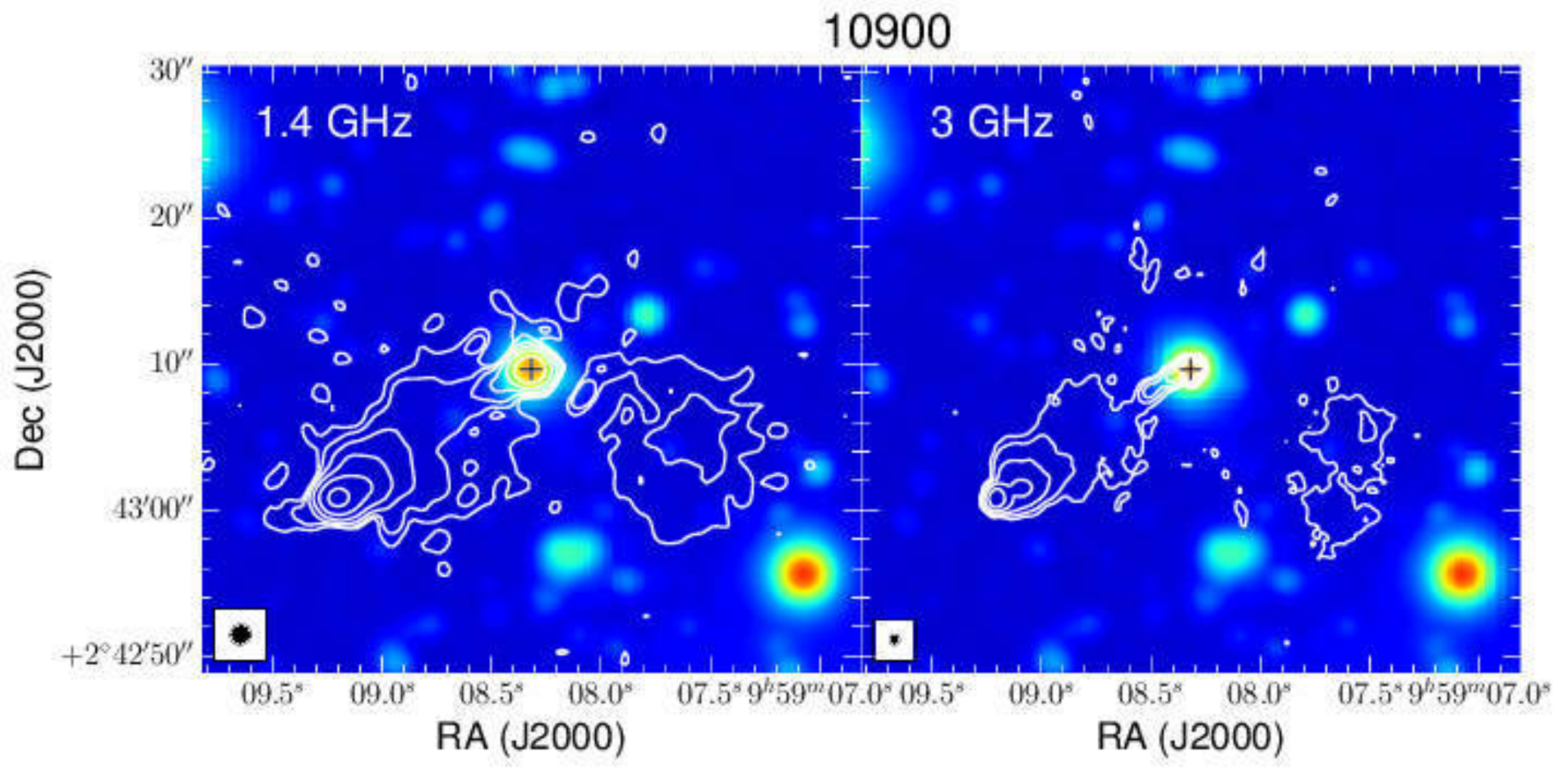}
    \includegraphics[width=0.5cm]{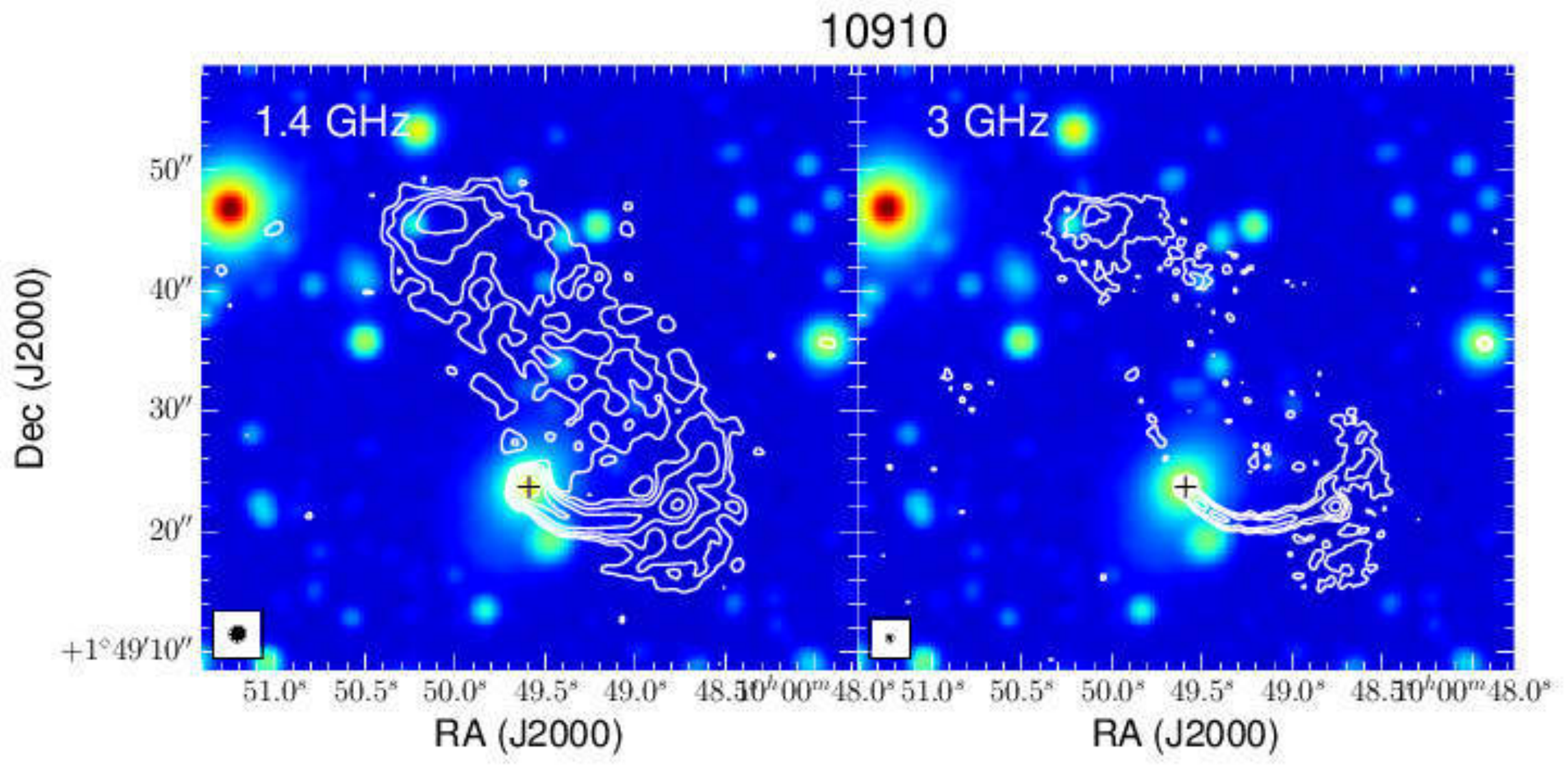}
   }
    \resizebox{\hsize}{!}{
     \includegraphics[width=0.5cm]{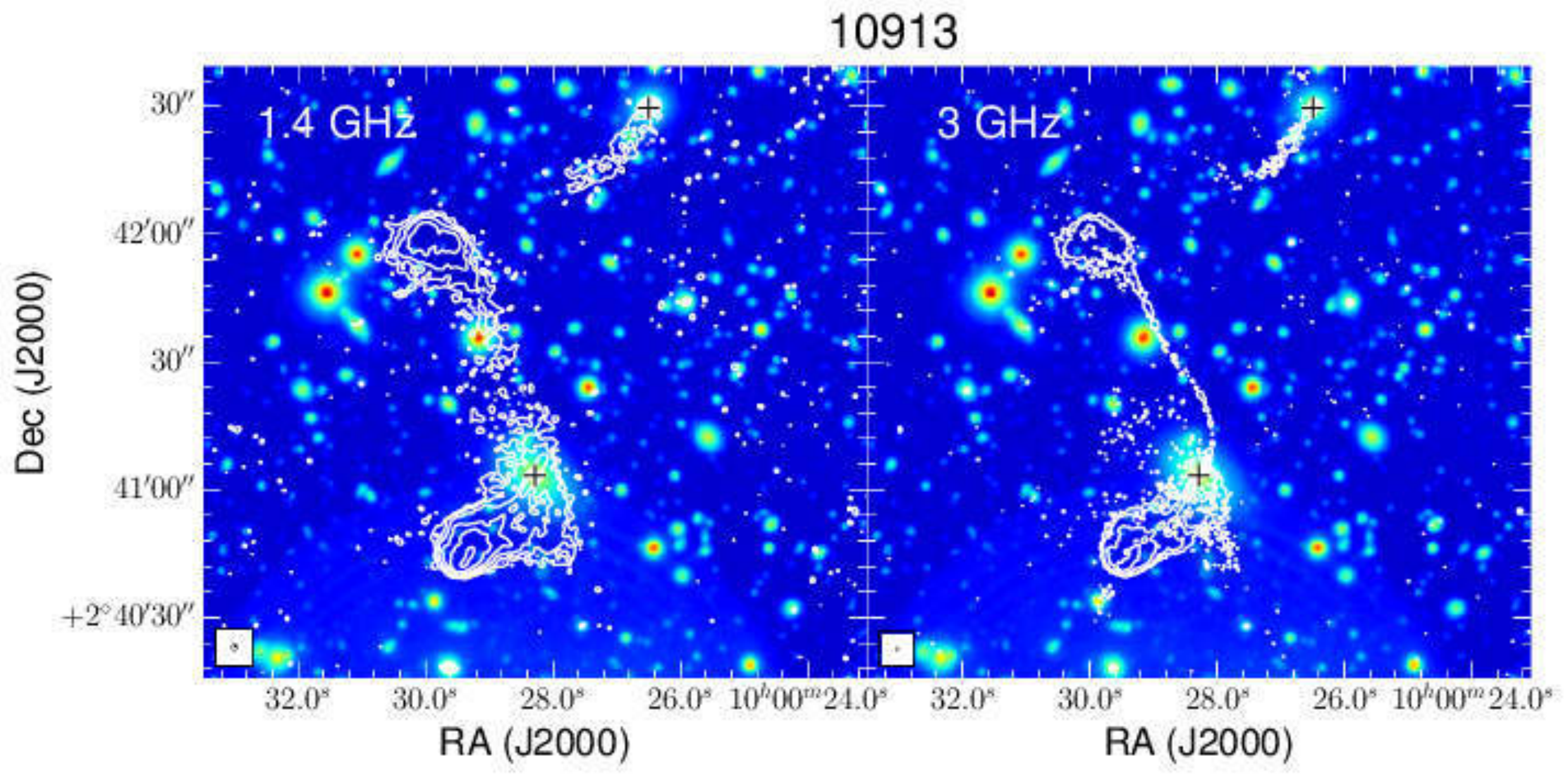}
   \includegraphics[width=0.5cm]{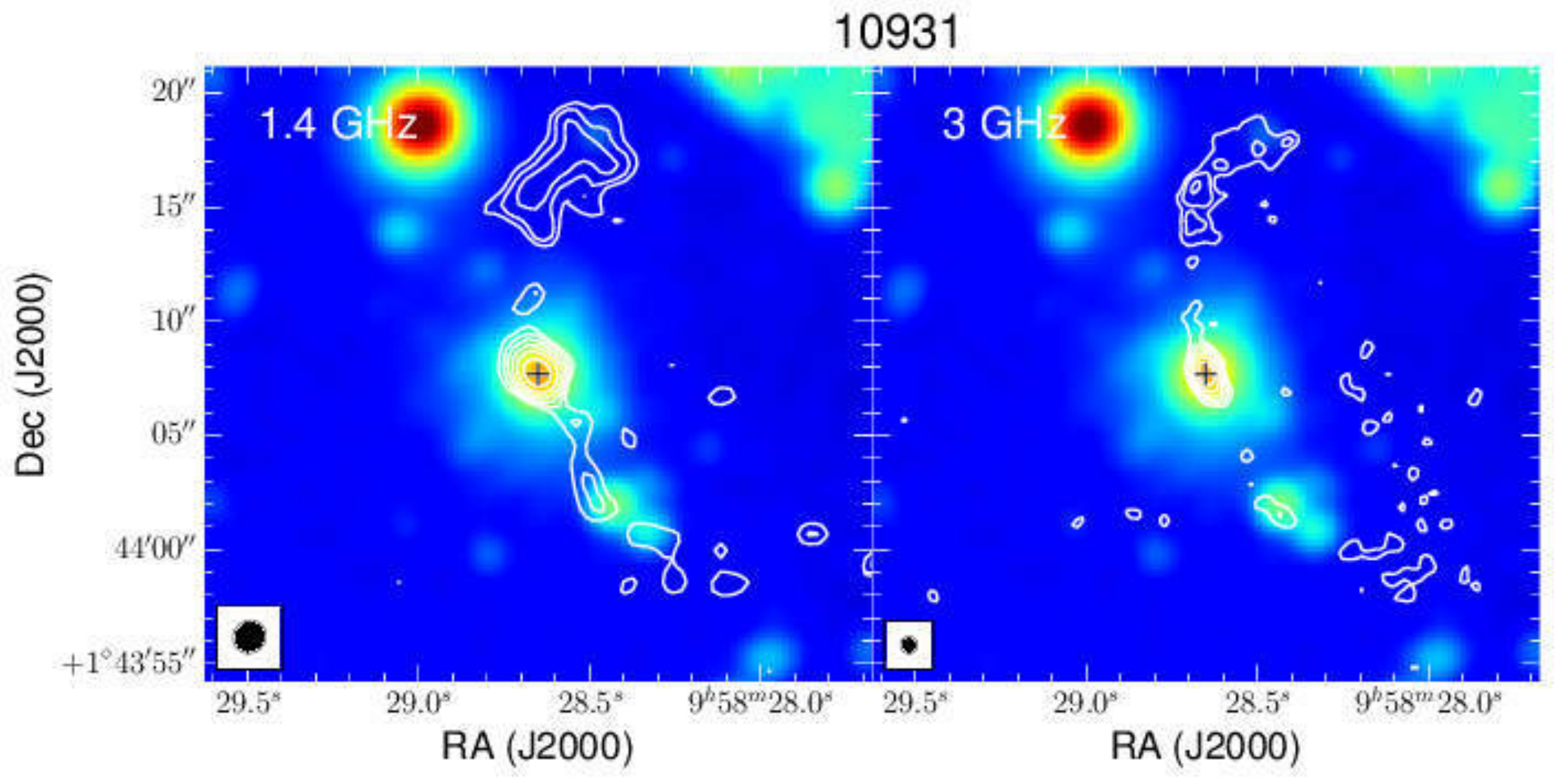}
   }
   
   \resizebox{\hsize}{!}{
    \includegraphics[width=0.5cm]{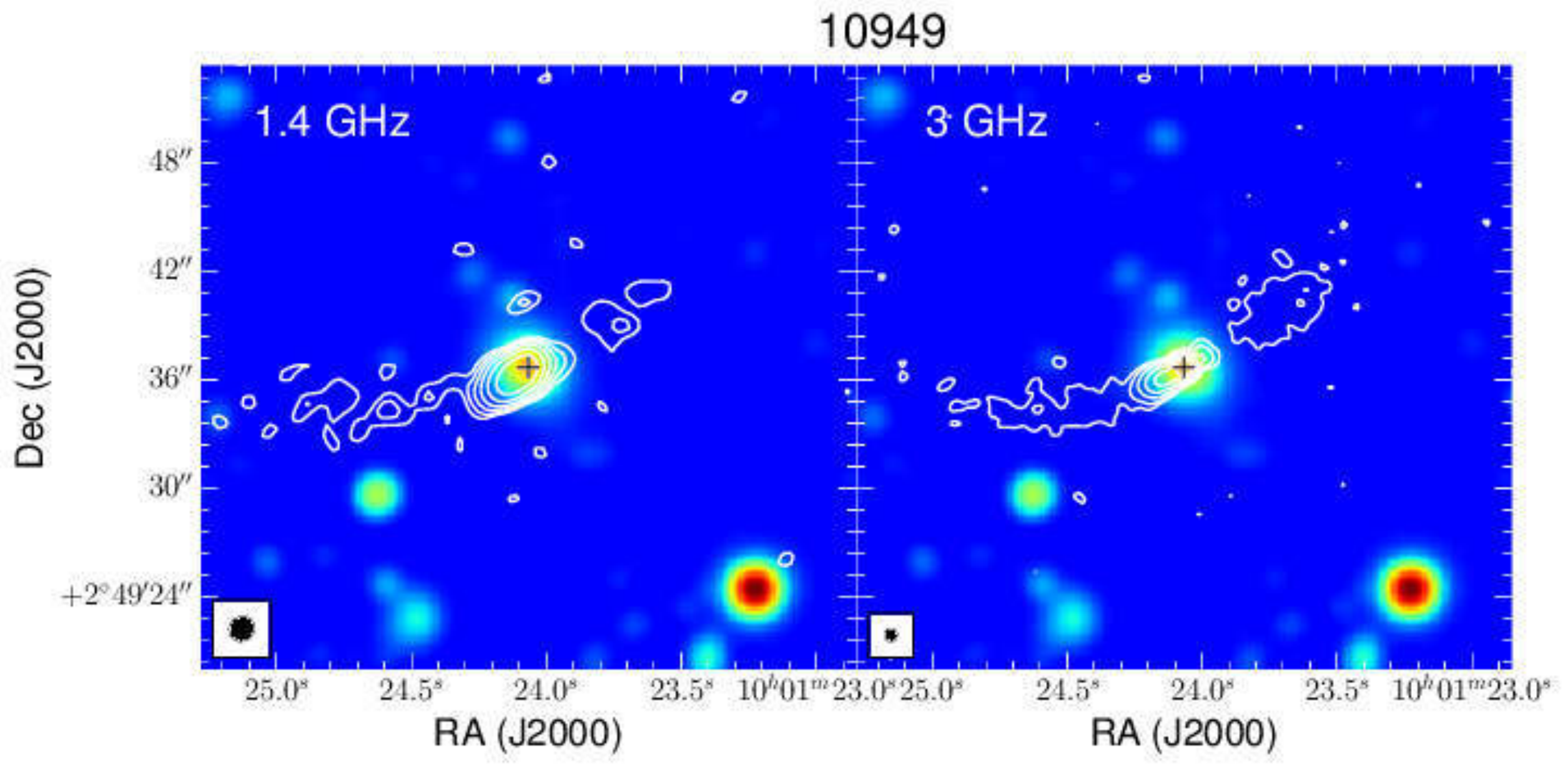}
   \includegraphics[width=0.5cm]{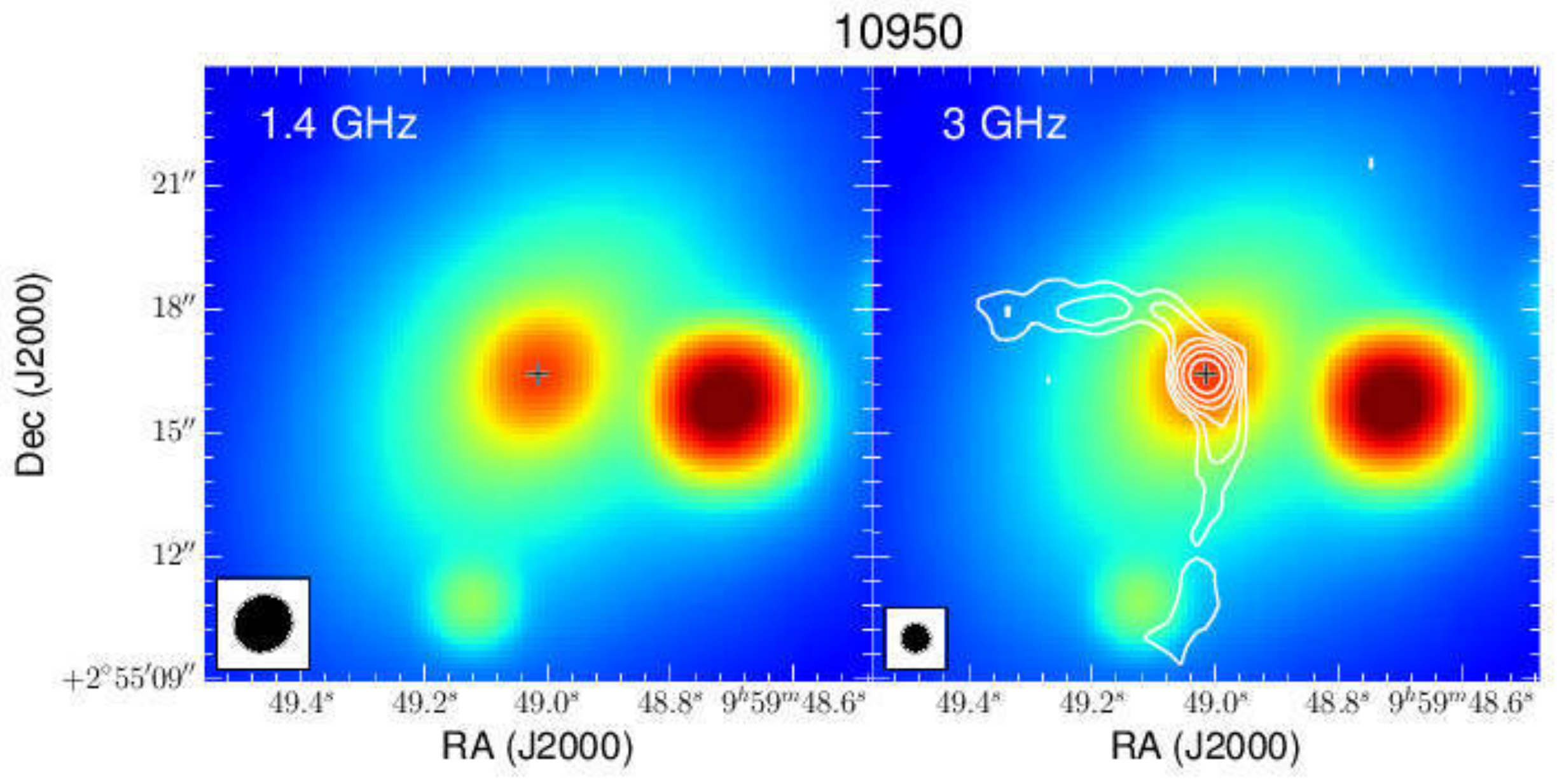}
   }
    \resizebox{\hsize}{!}{
  \includegraphics[width=0.5cm]{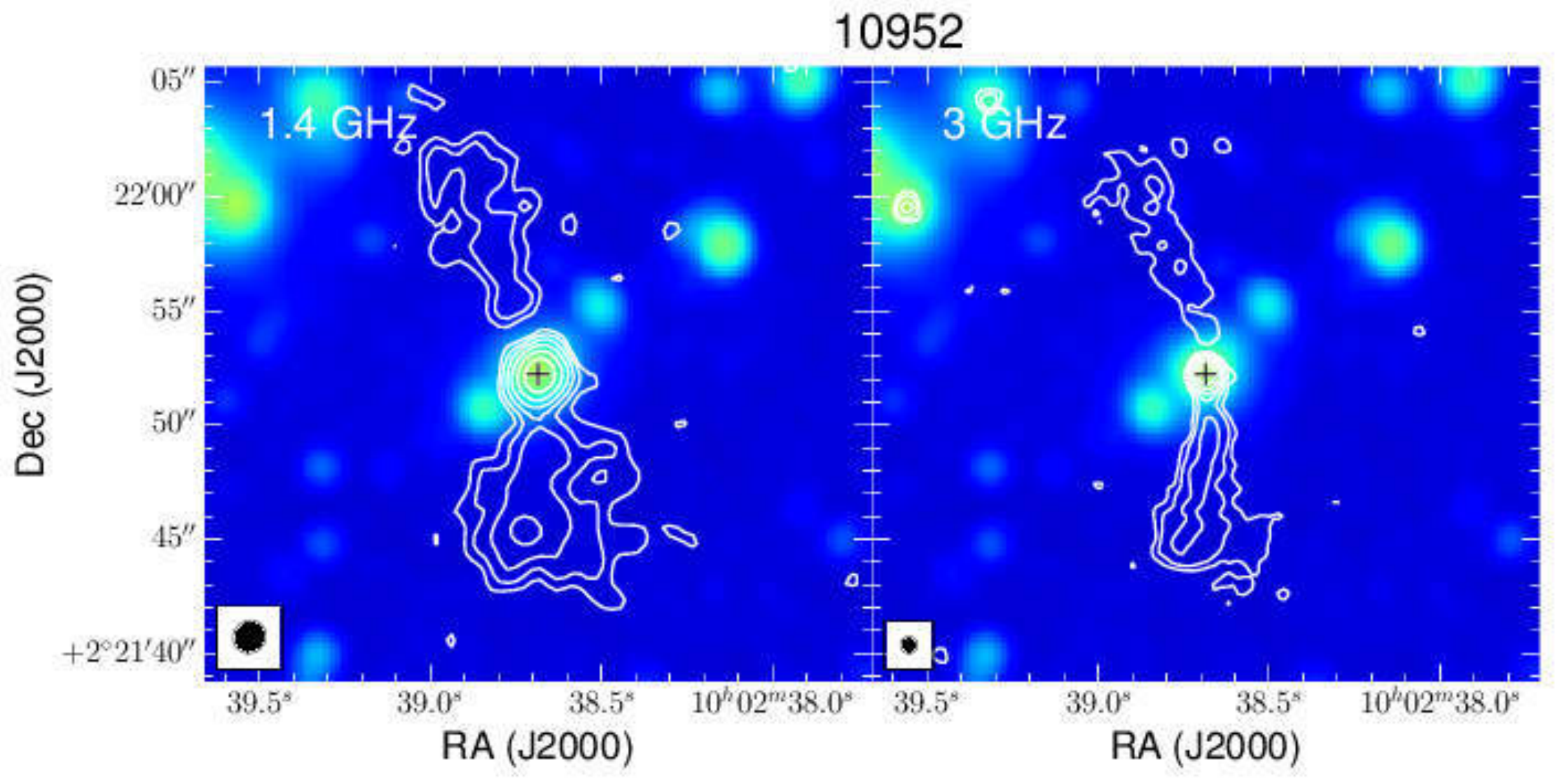}
  \includegraphics[width=0.5cm]{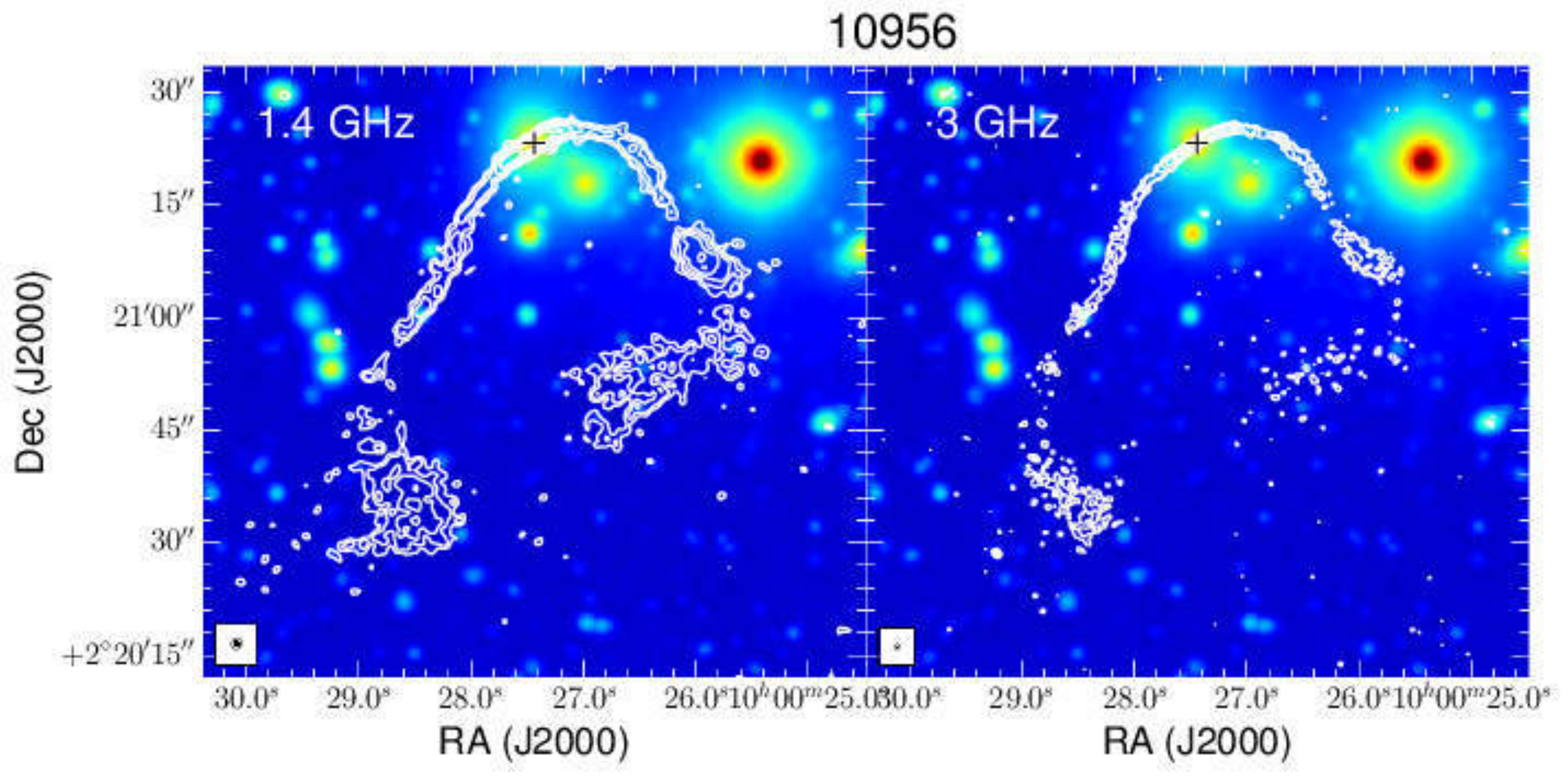}
   }
    \resizebox{\hsize}{!}{
   \includegraphics[width=0.08cm]{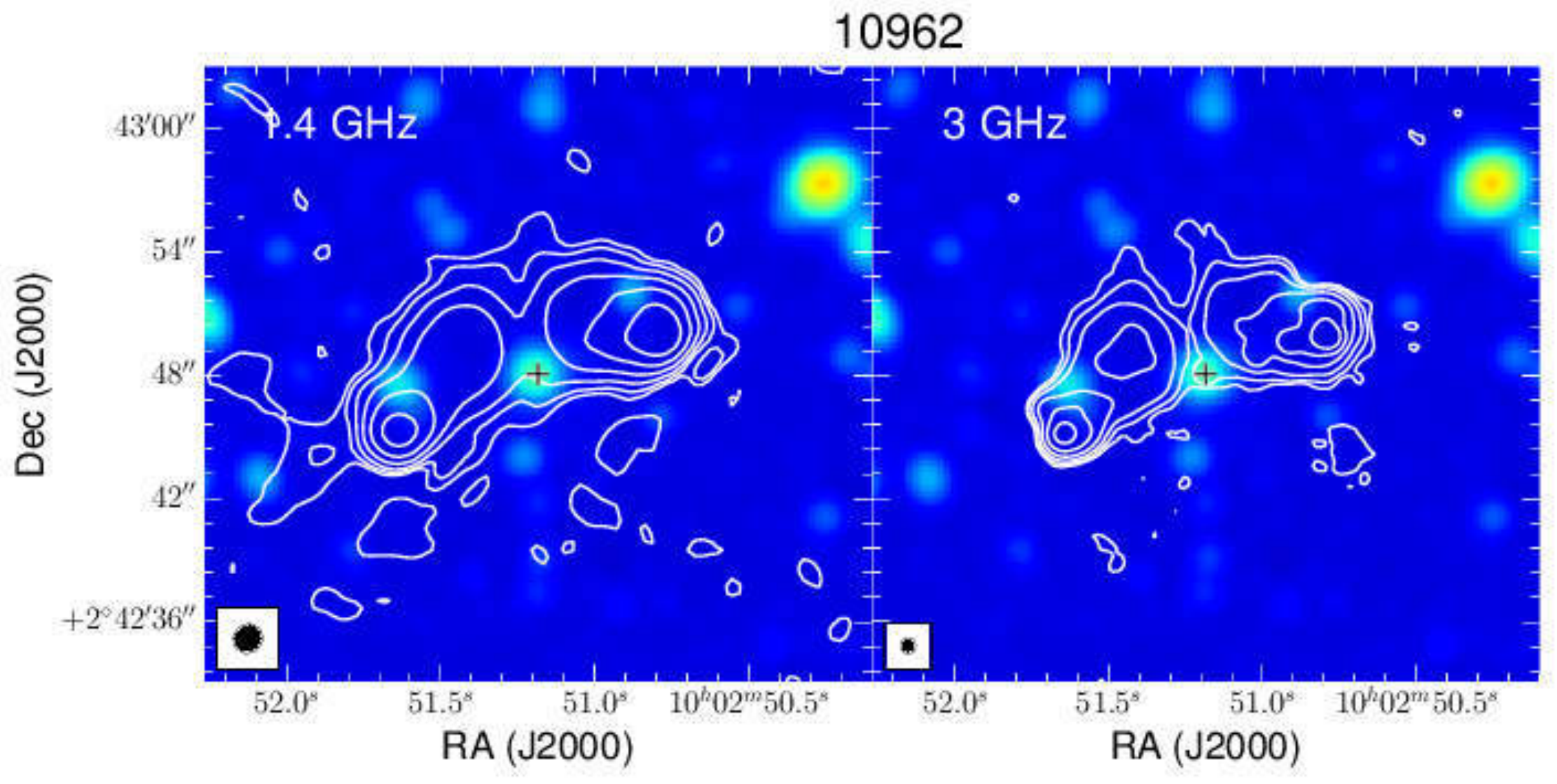}
 
  }
      \caption{The 9 wide-angle-tail multi-component objects at 3 GHz. Images described in Fig.~\ref{fig:maps2}.
              }
         \label{fig:wat}
   \end{figure*}

\subsubsection{Double-double, Z-/X-shaped, and restarted AGN at 3 GHz}
\label{sec:restarted}

\begin{figure*}[!ht]
   \resizebox{\hsize}{!}{
   \includegraphics[trim={0 8.05cm 0 7.1cm},clip,width=0.95cm]{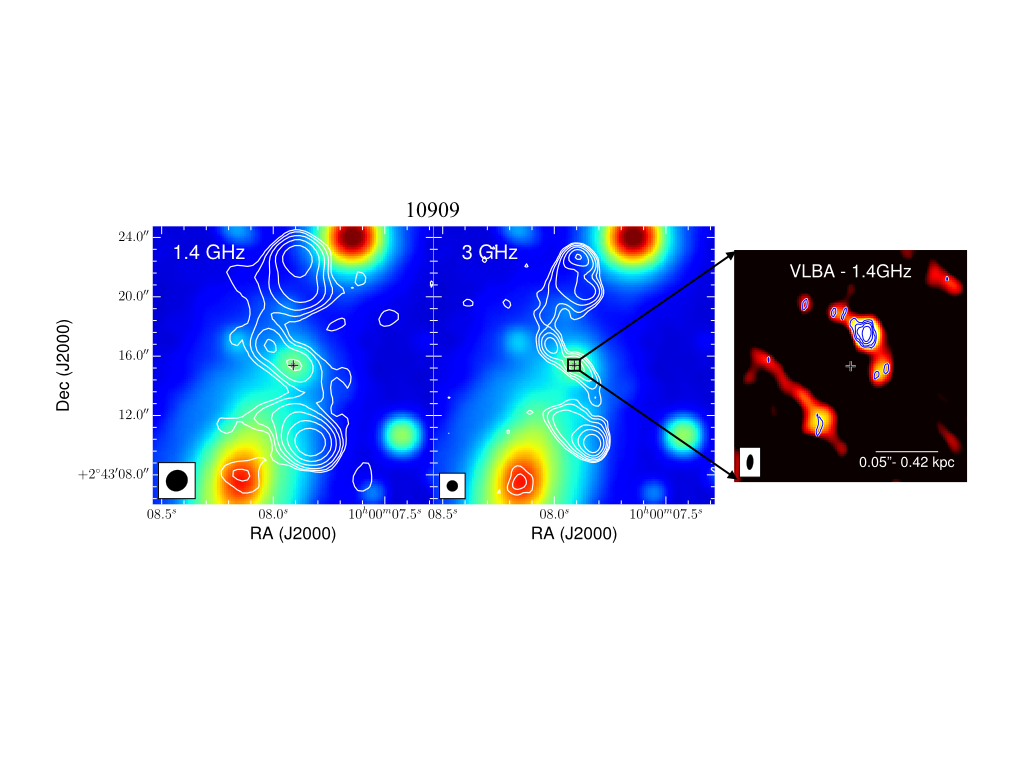}
     \includegraphics[trim={0 8.05cm 0 7.1cm},clip,width=0.95cm]{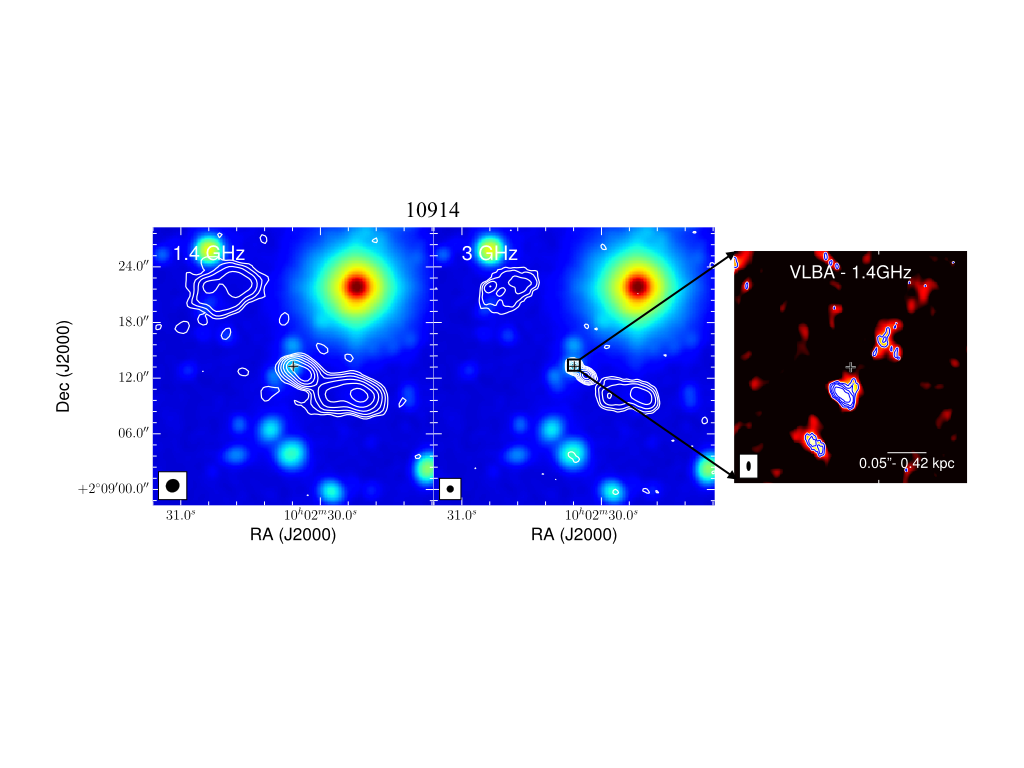}
   }

   \resizebox{\hsize}{!}{
   \includegraphics[clip,width=0.95cm]{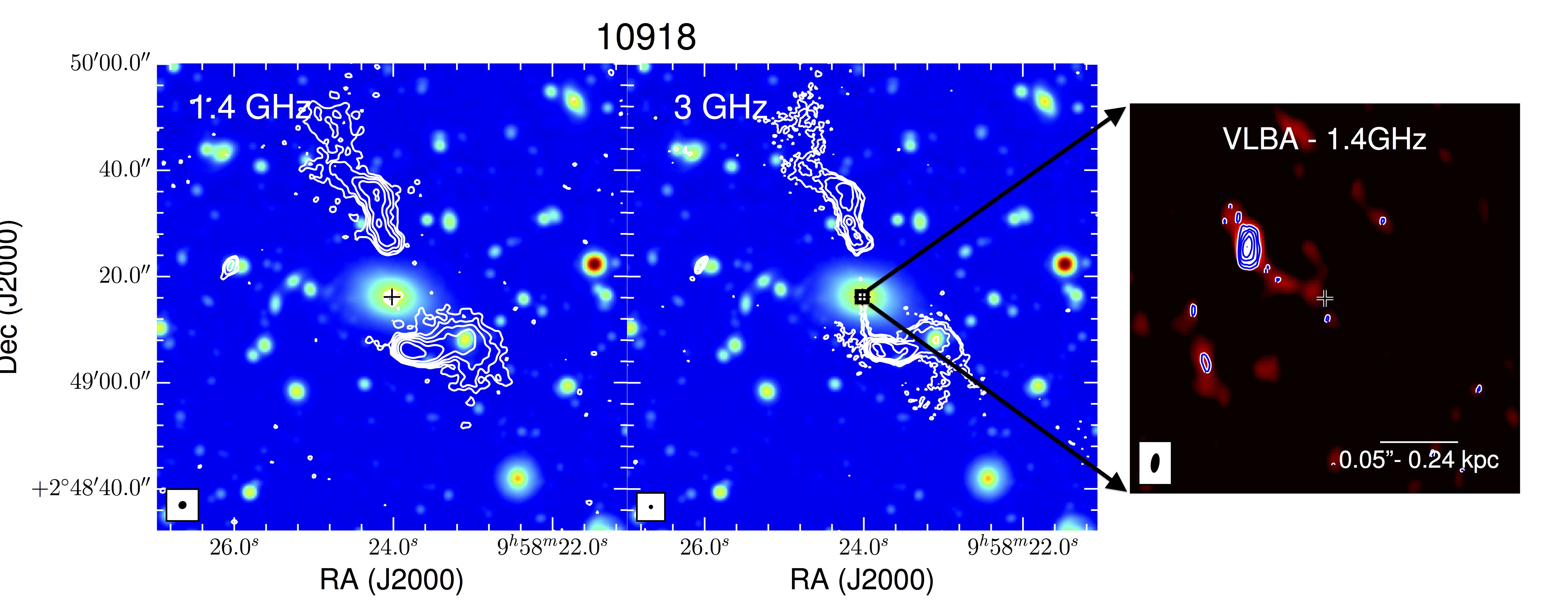}
    \includegraphics[trim={0 8.05cm 0 7.1cm},clip,width=0.95cm]{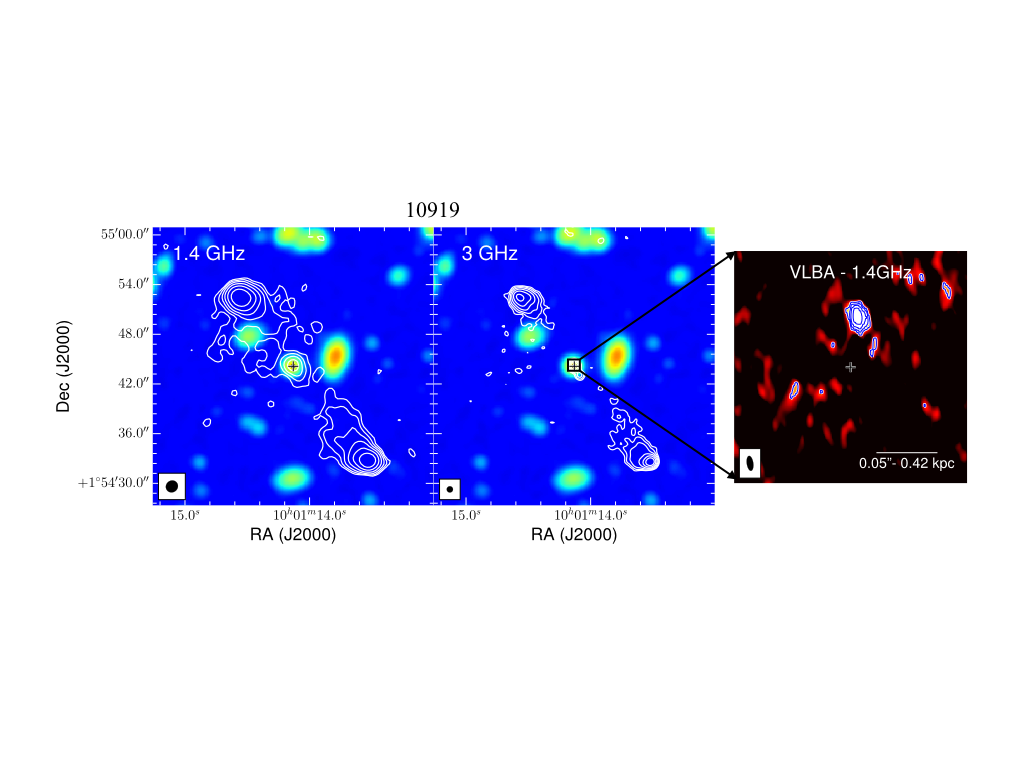}
   }

   \resizebox{\hsize}{!}{
   \includegraphics[trim={0 8.05cm 0 7.1cm},clip,width=0.95cm]{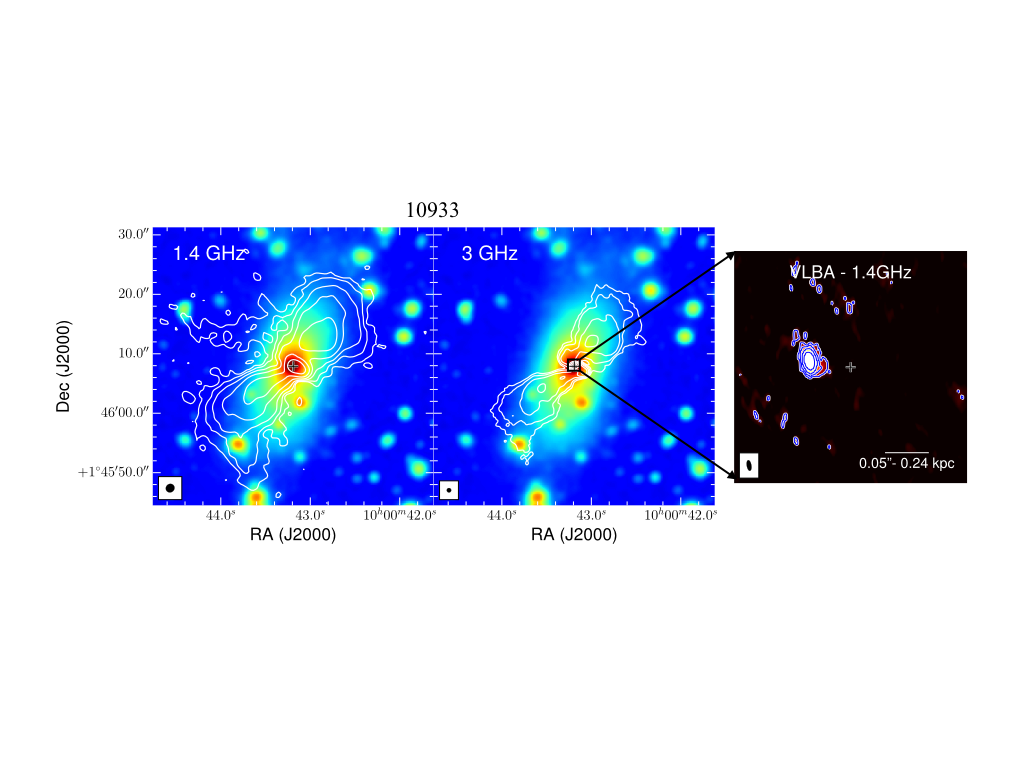}
    \includegraphics[width=0.8cm]{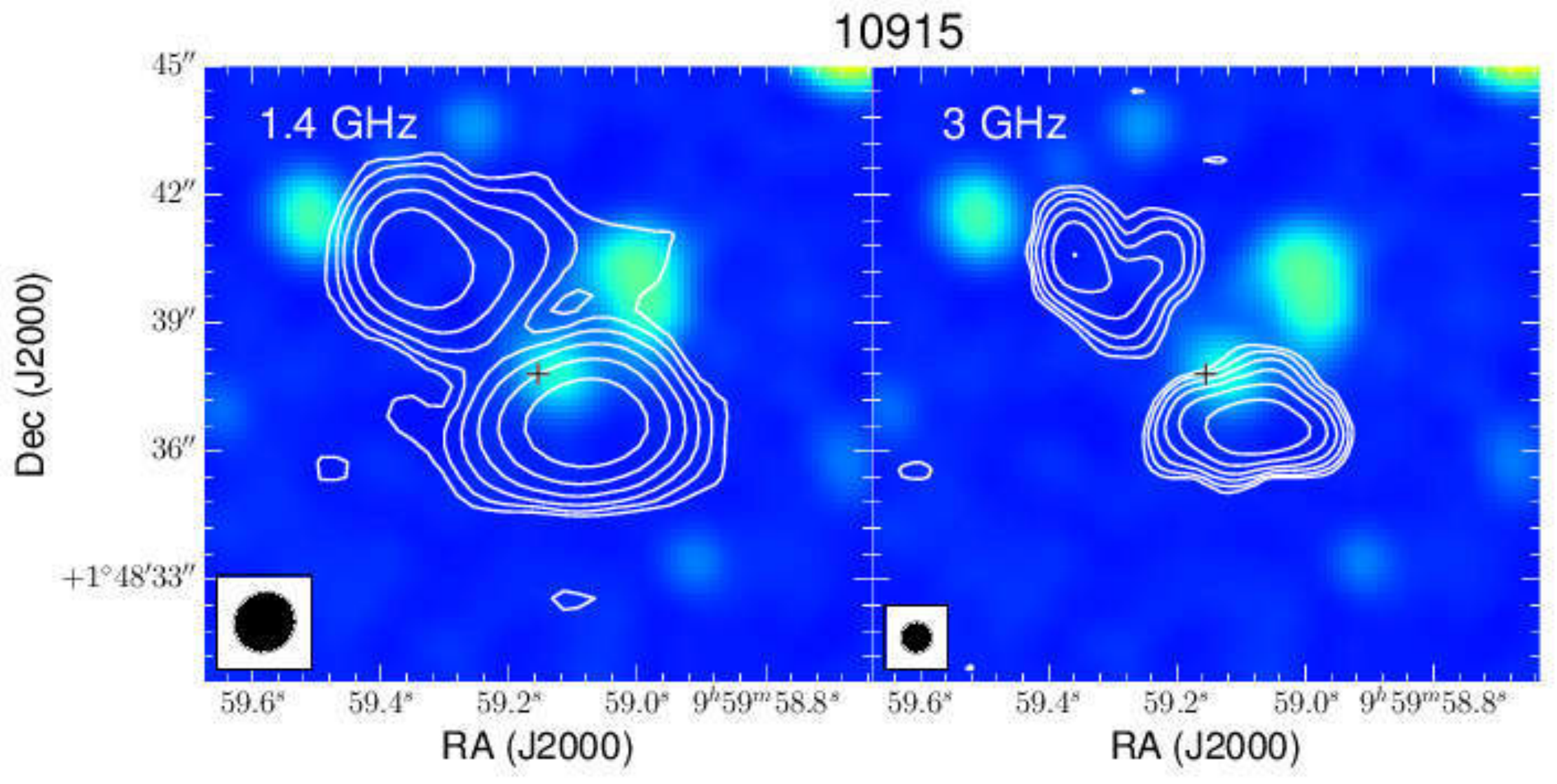}
   }
     \resizebox{\hsize}{!}{
   \includegraphics[width=0.95cm]{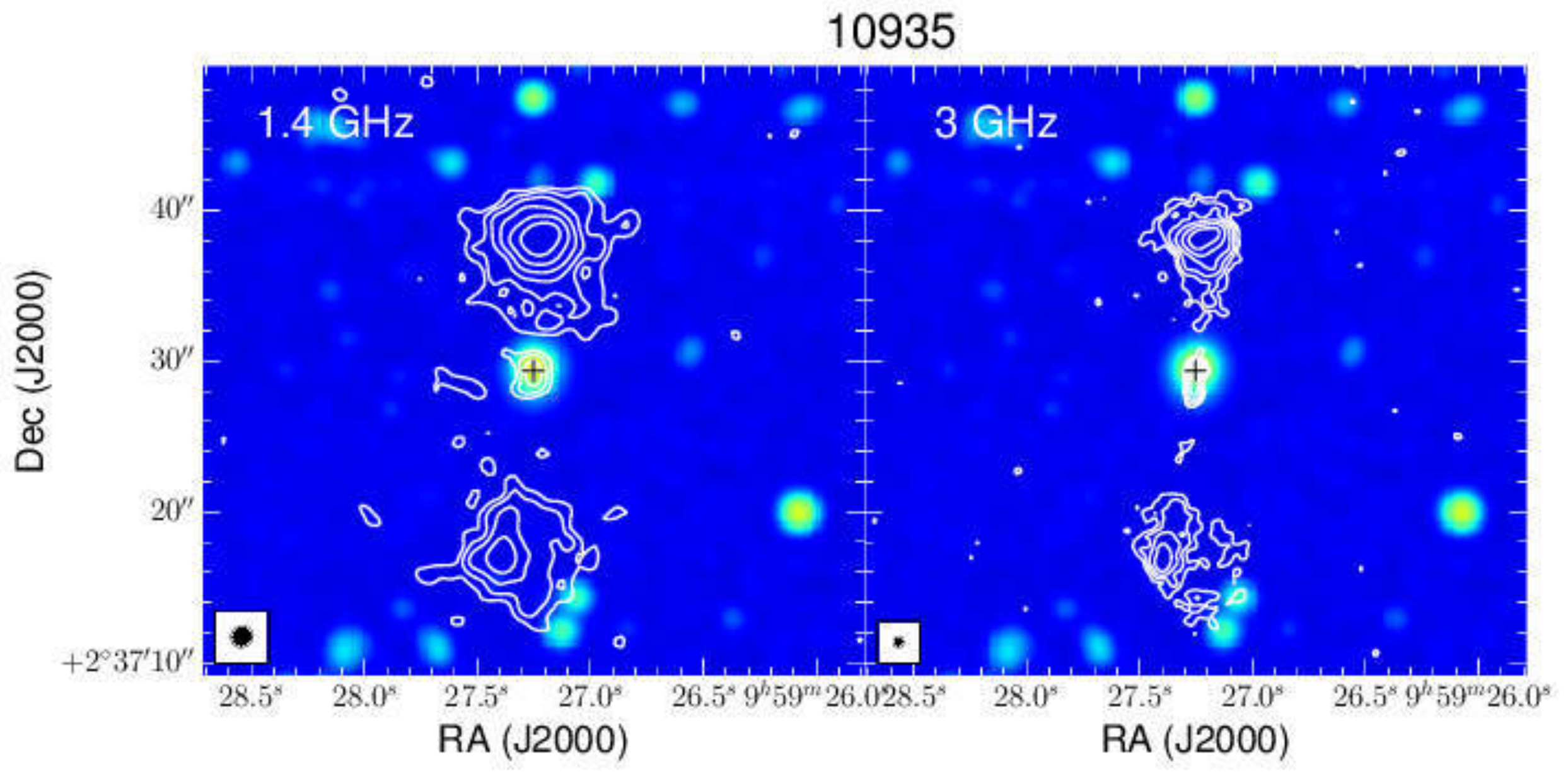}
    \includegraphics[width=0.95cm]{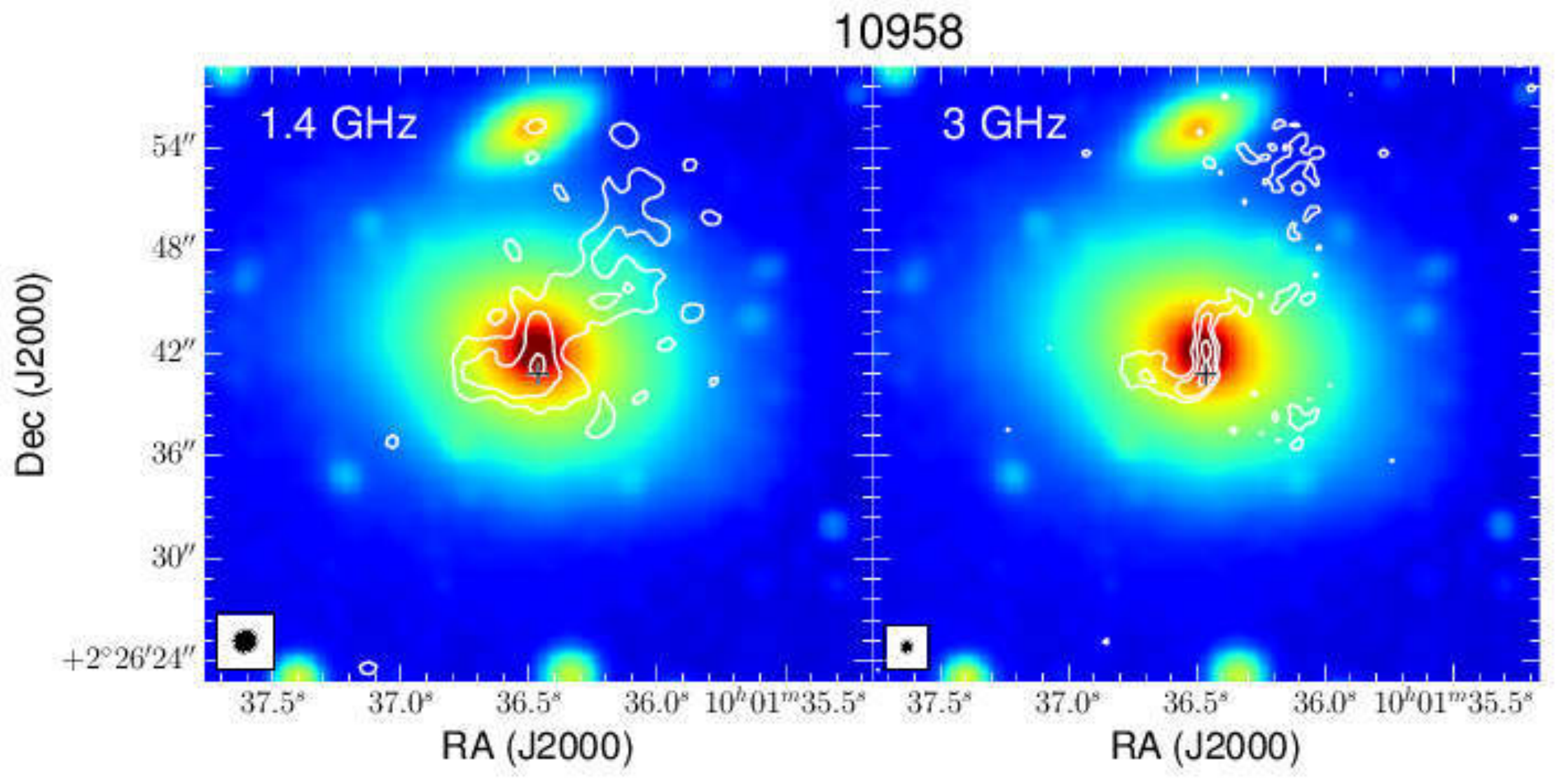}
   }
    
      \caption{ The 8 double-double, Z-/X-shaped multi-component objects at 3 GHz. The top panels show the 5 sources described in Sec.~\ref{sec:restarted} with double-double, Z-/X-shaped radio structure and VLBA observations. Each set of images gives the 1.4-GHz VLA ($left$), the 3-GHz VLA ($middle$) and the 1.4-GHz VLBA ($right$). The $left$ and $middle$ panels are similar to the ones in Fig.~\ref{fig:radmaps}, while the $right$ panels show the VLBA stamp corresponding to the nucleus; the corresponding radio contours start at 3 $\sigma$ increasing logarithmically, and the beam is 16.2$\times$7.3 mas$^{2}$. The VLBA stamps of these objects do not show signs for double core system that could verify the existence of a binary SMBH in the centre of these objects, and which would give rise to a double-double radio structure. 
      The 3 sources at the bottom lack VLBA identification. The radio/infrared overlays for all objects are described in Fig.~\ref{fig:maps2}.
          }
              
         \label{fig:xshaped}
   \end{figure*}

In the multi-component sample we identify 7 double-double or Z-/X-shaped AGN. These display double-double radio structure: 10909, 10914, 10915 (?), 10918, 10919, 10933, 10935 and 10958, as shown in Fig.~\ref{fig:xshaped}. Within the sources in our multi-component sample the following objects fall in the restarted AGN category due to their radio structure at 3 GHz: 10909, 10915, 10933, 10935. 

10909 is a double lobed radio AGN with the furthest away lobes being the oldest, while closer to the core it exhibits another double-lobed structure. This twisted radio structure is most probably not the result of movement of the source in the ICM as in that case we would see a WAT or head-tail source, depending on orientation. This strong resemblance to an X-shaped source might be due to a restarted AGN phase due to jet interruption or axis re-orientation, rather than a projection effect. 

In 10915 we see only two unevenly shaped lobes and no core emission at 3 GHz. 

10918 displays a bend in the south lobe probably caused by interaction with a denser medium. The 3-GHz map shows in great detail the south jet leading to the bent lobe structure, which is not seen at 1.4 GHz. The bent in the north lobe is towards the opposite direction, thus it is placed in this Z-shaped class. 

10933 displays a rotation of the jet/lobe structure close to the centre of emission (marked by the black cross in Fig.~\ref{fig:xshaped}). This does not look like a projection effect as the two jet/lobe structures look alike and symmetric. Thus the radio structure is most probably a result of axis reorientation. 

10935 is a symmetric source that displays a double-double radio structure, which is based on the observation that at both 1.4 and 3 GHz the lobes are very diffuse, but there is a clear jet at 3 GHz, which could indicate restarted activity. 10958 is a rather peculiar object, which is S-shaped. The jets bent to opposite directions and seem to do that within the galaxy itself. 

Z-/X-shaped radio sources, i.e. radio sources that have a double-double, Z- or X-like radio structure, have been reported in past radio studies and are likely linked to recurrent AGN activity \citep[see][for a review]{Gopal-Krishna12}. Episodic mass accretion onto a supermassive black hole results in two pairs \citep[e.g.][]{schoenmakers00} or even three pairs of radio lobes \cite[e.g.][]{saikia09}. These can be well aligned \cite[e.g.][]{schoenmakers00}, or can be X-shaped \cite[e.g.][]{leahy84} or Z-shaped \citep[e.g.][]{Gopal-Krishna03}. The mechanisms behind this recurrent AGN activity can be either axis reorientation, also known as 'spin-flip' \citep[e.g.][and references therein]{Gopal-Krishna12}, backflow, or jet activity interruptions \citep[e.g.][]{schoenmakers00}. In some cases the jet-shell interaction model, which does not require a spin-flip or axis reorientation, has been proposed \citep{Gopal-Krishna12}.

Z- or X-shaped radio sources can be probes of binary black-hole systems in the galaxy, as a product of a galaxy merger. \cite{Gopal-Krishna03} suggest that Z-shaped radio sources are at early stages of interaction with a dense environment, that was the result of the concentration of gas at a distance of $\sim$ 10 kpc from the nucleus due to the galaxy merger. This interaction with the gas causes the jets to bend symmetrically when expelled from the source. The only observationally confirmed example of a binary black-hole galaxy system is the study of \cite{rodriguez06} in the radio galaxy 0402+379, with a projected separation between the black holes of 7.3 pc. X-shaped sources are assumed to be later stages of that interaction \citep{Gopal-Krishna03}. A striking example in the multi-component sources is 10933, where we see this Z-shaped symmetric radio structure close to the nucleus (see Figs.~\ref{fig:xshaped}~\&~\ref{fig:maps2}).

To identify possible signatures of binary black-hole systems that give rise to this double-double radio structure, cross-matched these sources with the 1.4-GHz VLBA data available for the COSMOS field \citep{noelia17}.  

We find 5 matches between X-shaped 3-GHz objects and VLBA data: 10909, 10914, 10918, 10919, 10933 and 10958. The observations can be seen in Fig.~\ref{fig:xshaped}. We also report the core flux densities from VLBA in Table~\ref{table:data}. In all 5 sources the 1.4-GHz VLBA core is offset from the 3-GHz VLA core, by about 0.04 arcsec, which is not significant; it is within the uncertainty of the astrometry in the 3-GHz mosaic. 

As seen in Fig.~\ref{fig:xshaped}, the radio cores are resolved and show some slight extended features (e.g. 10933). 10909 presents to strong emission peaks at VLBA separated by 0.05 arcsec. These lie on top of the fringes in the image, thus we refrain from making a strong statement about a double core. 10914, at $z_{\rm phot}$ = 1.437, displays a feature on the south-west of the core, but this could be due to band-width smearing, and the orientation of the jet is perpendicular to the direction of the 3-GHz emission.

\subsubsection{Bent-tail multi-component AGN}
\label{sec:benttail}

\begin{figure}[!ht]
    
     \resizebox{\hsize}{!}{
   \includegraphics[width=0.55cm]{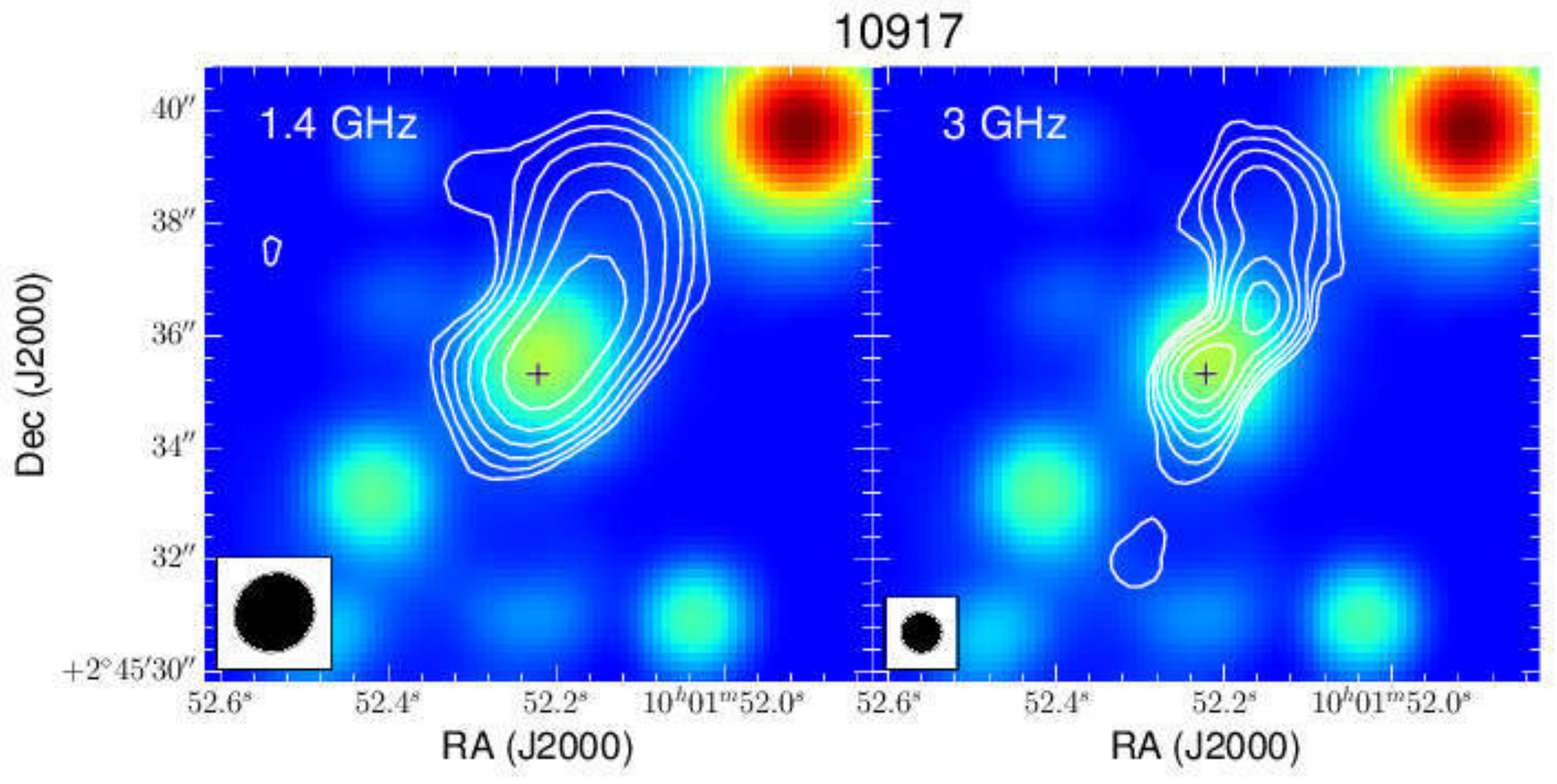}
   }
      \resizebox{\hsize}{!}{
 \includegraphics[width=0.55cm]{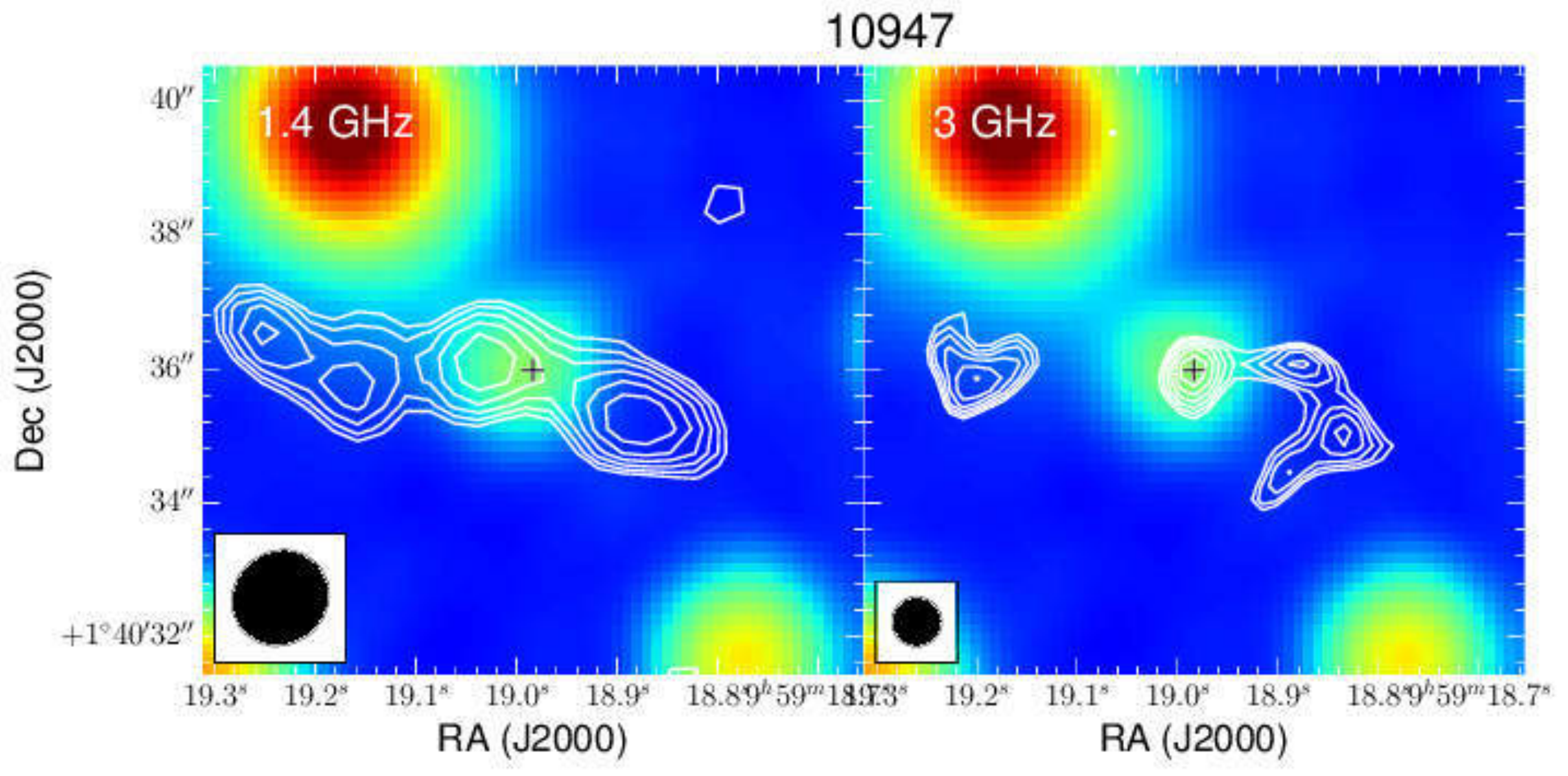}
 }

   \resizebox{\hsize}{!}{
 \includegraphics[width=0.55cm]{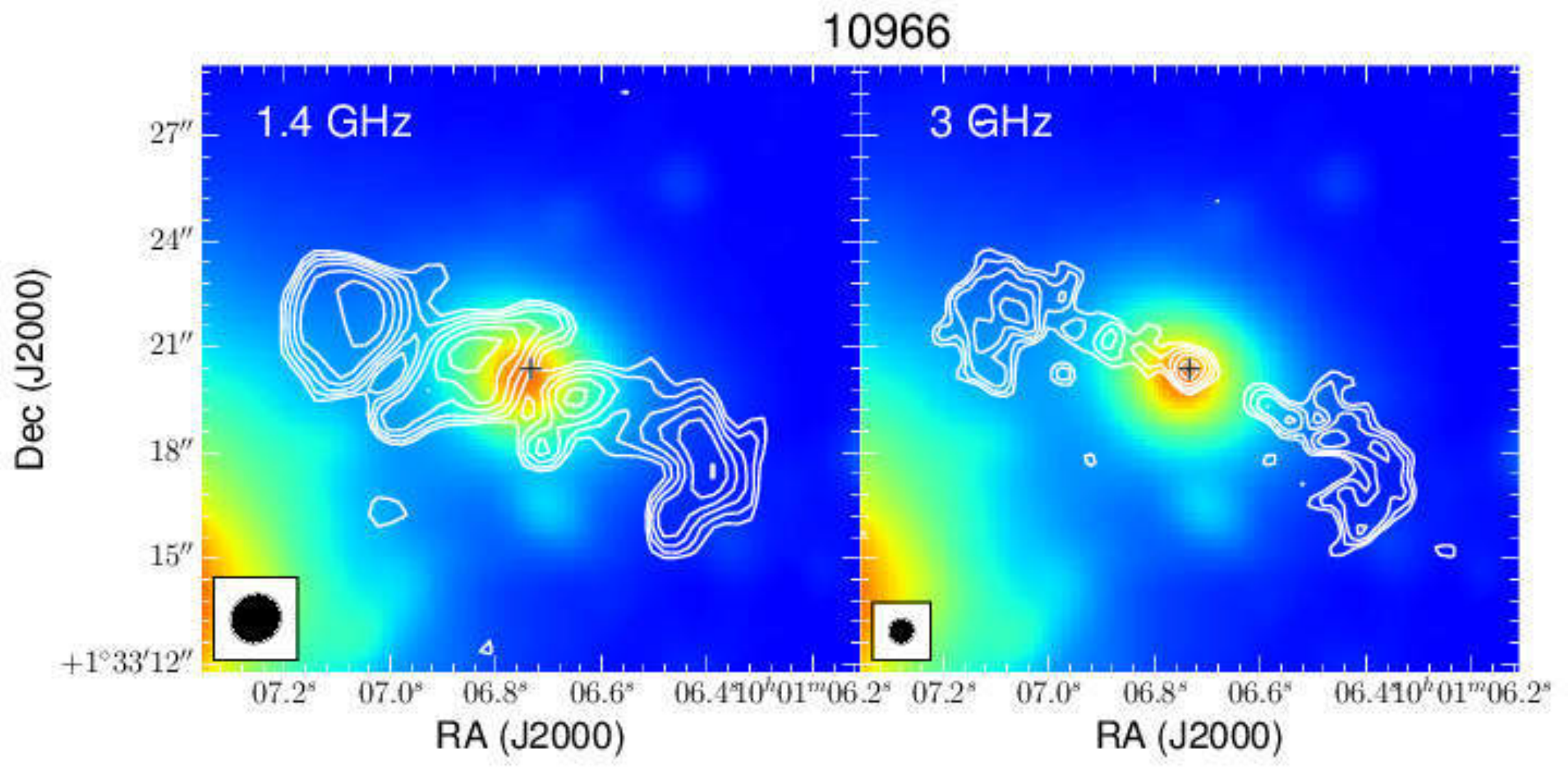}
 }
      \caption{The 3 bent-tail 3-GHz radio sources in COSMOS, show interaction with a dense environment. Images described in Fig.~\ref{fig:maps2}.
              }
         \label{fig:benttail}
   \end{figure}

In this Section we present objects that show bending in their radio structure and in particular in the outer jets, in contrast to WAT sources that show bents on both jets close to the central galaxy. We identify 3 such objects (10917, 10947 and 10966) in our sample of multi-component radio sources (see Fig.~\ref{fig:benttail}). We note that these are not all the bent radio sources at 3-GHz VLA-COSMOS, as there are also single-component bent sources not included in the multi-component sample.

10917 displays a core and single lobe to the north and a faint blob to the south, which was included in the estimate of the flux of this object. Its lobe-like structure to the north, is bent probably due to interaction with the surrounding environment. 

10947 displays a core and single lobe to the west and a faint blob to the east, which was included in the flux of this object. Its structure looks quite different though from the 1.4 GHz analogue, where the bend on the west jet is not so pronounced. 

10966 is an example of outer-jet bent. It exhibits bending towards the edge of both its jets, suggesting interaction with the ICM at distances far from the core of the radio source.

\subsubsection{Symmetric multi-component AGN}
\label{sec:restAGN}

This section lists the multi-component AGN that their lobes/jets form a 180 degree angle to each other. We classify these sources as symmetric multi-component AGN. We identify 26 such objects within the multi-component sample: 10901, 10902, 10904, 10905, 10906, 10907, 10908, 10911, 10916, 10920, 10921, 10922, 10923, 10925, 10926, 10928, 10930, 10932, 10934, 10936, 10937, 10938, 10948, 10951, 10953 and 10959. These can be found in Fig.~\ref{fig:maps2}. All of these display 2 lobes/jets. The exception is 10905, where the east lobe is very diffuse. Similarly, 10911 has a well defined north lobe but the south lobe is very diffuse at 3 GHz. 10937 is the smallest double source in our sample, with largest angular size of $\sim$ 3 arcsec. Finally, in 4 sources both lobes are diffuse (10907, 10921, 10948 and 10951).

\subsection{Multi-component SFGs}
\label{sec:sfgs}

Our classification yields 9 SFGs amongst the multi-component objects at 3-GHz VLA-COSMOS. These are presented in Fig.~\ref{fig:sfgs}, and the most striking case which shows a radio ring, 10944, is discussed in Sec.~\ref{sec:ring}. The rest of the objects show star-forming regions associated with the galaxy disk (shown in the infrared in the stacked Ultra-VISTA map). Below we give a brief description of the radio structure of the SFGs in our sample.

10942 is classified as SFG based on the absence of radio-excess (see Sec.~\ref{sec:hosts}). Multi-wavelength AGN diagnostics also suggest this is a SFG (from the publicly available counterpart catalogue of \cite{smolcic17b}; see also Table~\ref{table:data2}). The map shows several radio blobs at 3 GHz, the strongest of which is coinciding spatially with the galaxy in the infrared.

10946 is classified as SFG due to its radio structure at 3 GHz. There are no indications of a radio jet. The source is composed by 2 radio blobs at 3 GHz which lie on the galaxy disk. It lies outside the 1.4-GHz coverage of the COSMOS field.

10954 is classified as star-forming galaxy due to the radio structure at 3 GHz which has 2 radio blobs (one larger and one smaller) associated with the disk of the galaxy seen in the Ultra-VISTA map. 

10960 displays several blobs of different size and has a ring-like radio structure at 3 GHz. This is why it is classified as SFG. At 1.4 GHz it is a single-component complex source.

10961 is composed of several small approximately beam-sized blobs at 3 GHz, some larger some smaller, that coincide spatially with the disk of the infrared galaxy seen in the Ultra-VISTA map. This is why it is classified as SFG. The 1.4-GHz map shows an S-shaped source and several smaller ones.

10963 is a double source associated with the disk of the infrared galaxy shown in Fig.~\ref{fig:agnsfg}. This initially was in the uncertain class, but we classify it as SFG due to the lack of radio excess (see Sec.~\ref{sec:hosts}).

10964 is a double source at 3 GHz, with radio structure that is different from the one at 1.4 GHz. We classify this as SFG due to the lack of radio excess (see Sec.~\ref{sec:hosts}). It is also SFG based on the SED fit (see Table~\ref{table:data2}).

10965 displays a large twisted blob at 3 GHz, which follows the spiral arm of the galaxy shown in infrared in Fig.~\ref{fig:sfgs}, and also has several smaller blobs associated with the galaxy disk. This is why we classify it as SFG.

  \begin{figure*}[!ht]
    \resizebox{\hsize}{!}{
   \includegraphics[width=0.5cm]{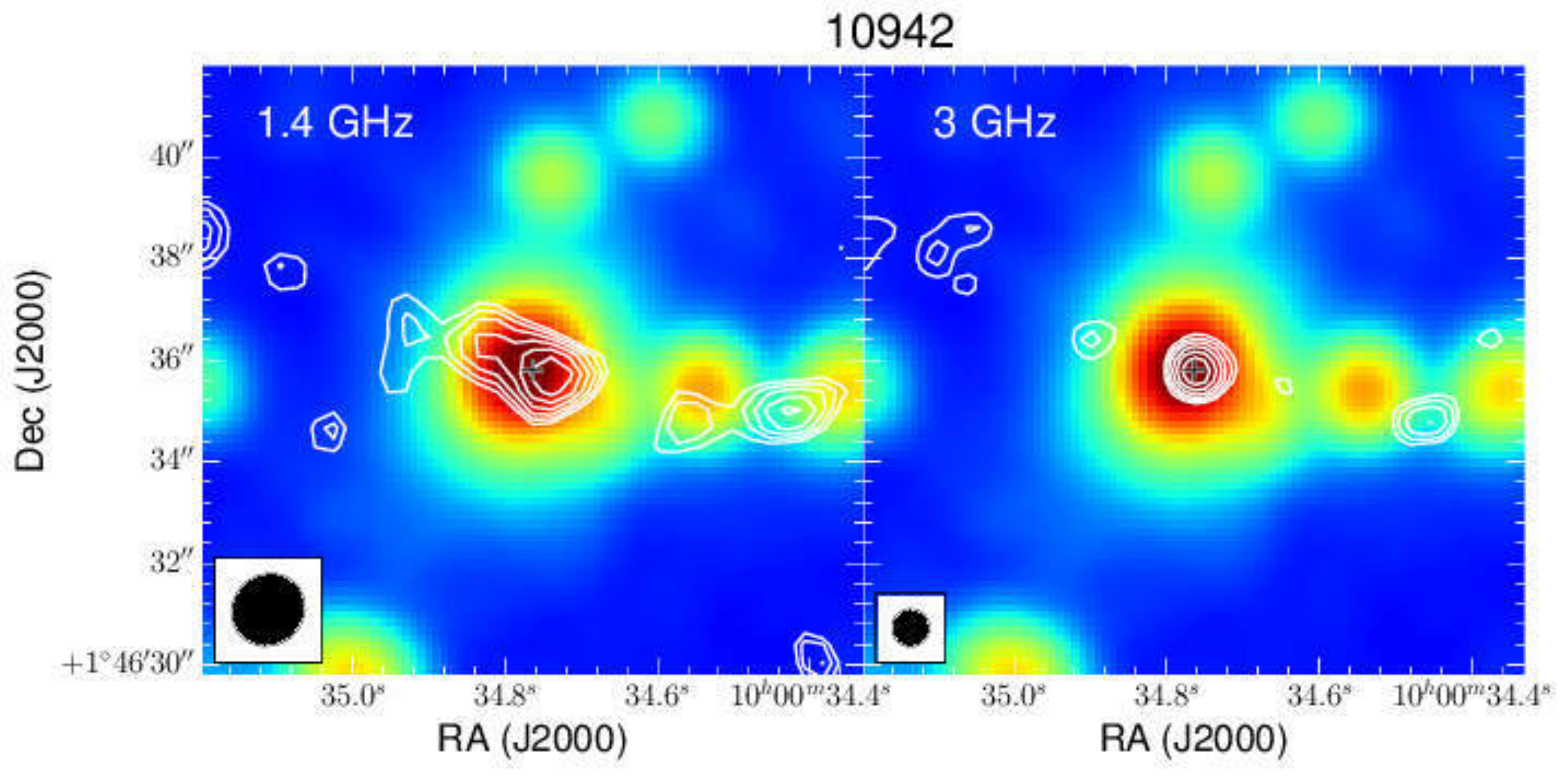}
    \includegraphics[width=0.5cm]{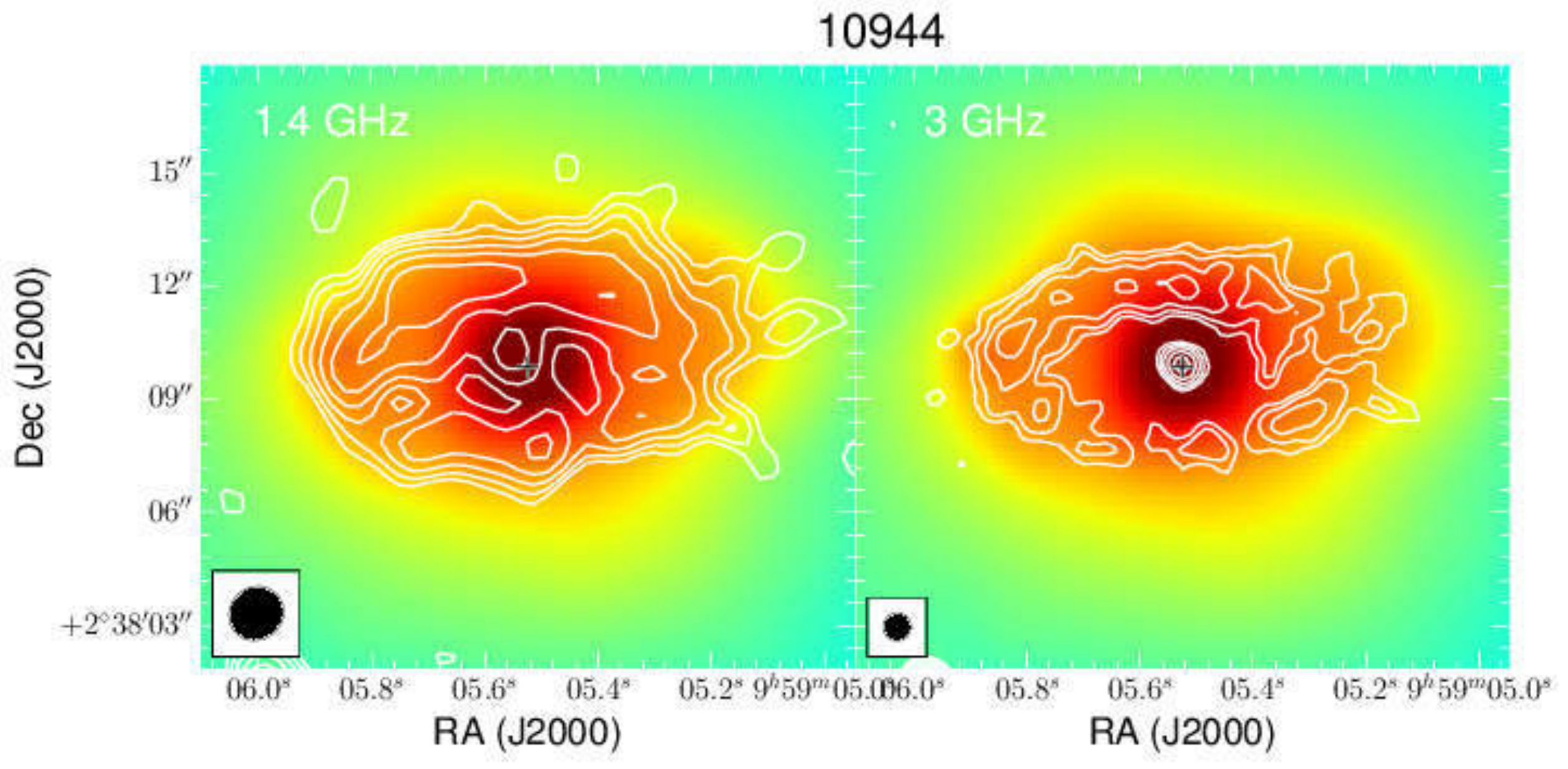}
   }
   \resizebox{\hsize}{!}{
   \includegraphics[width=0.5cm]{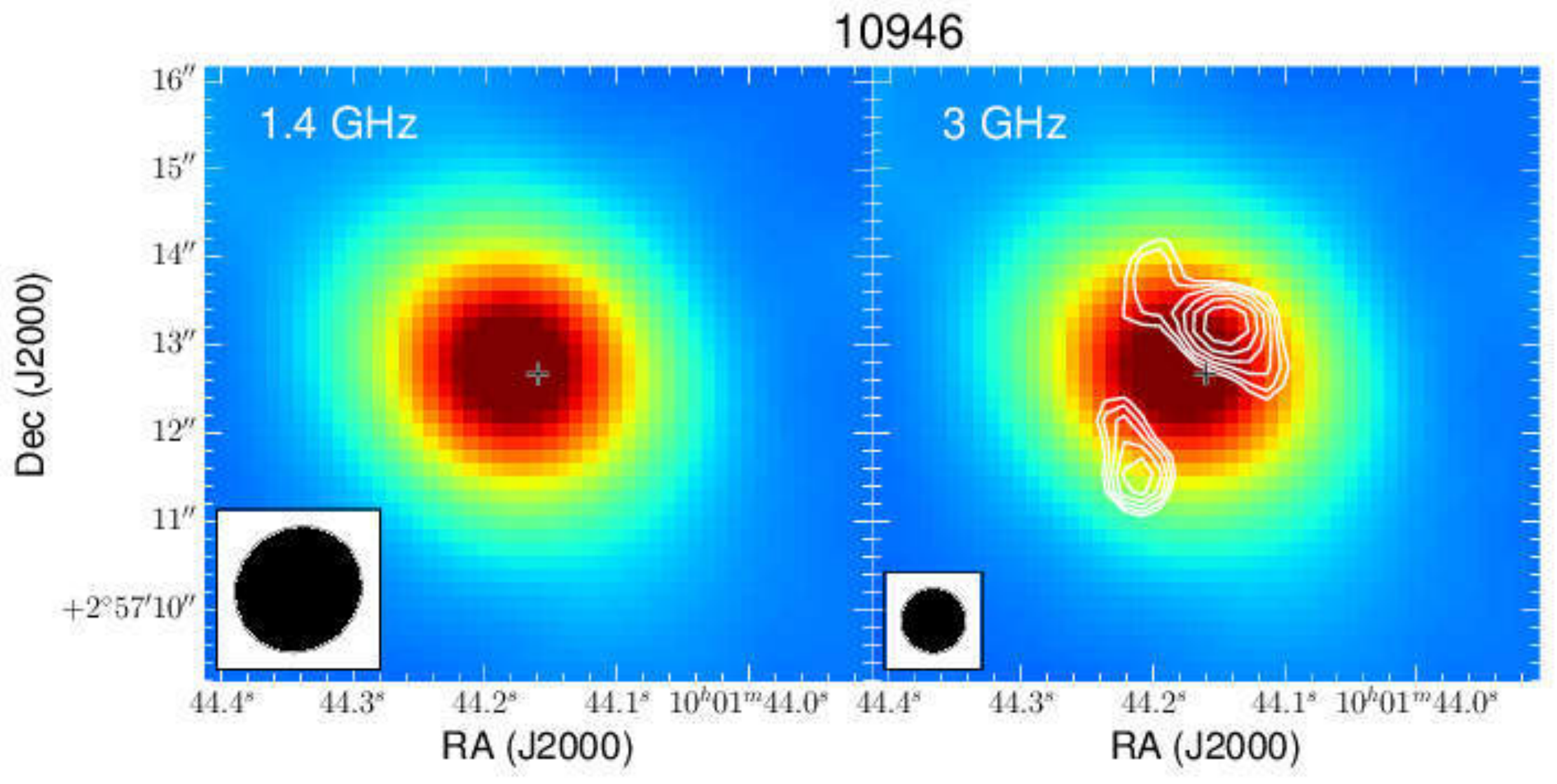}
    \includegraphics[width=0.5cm]{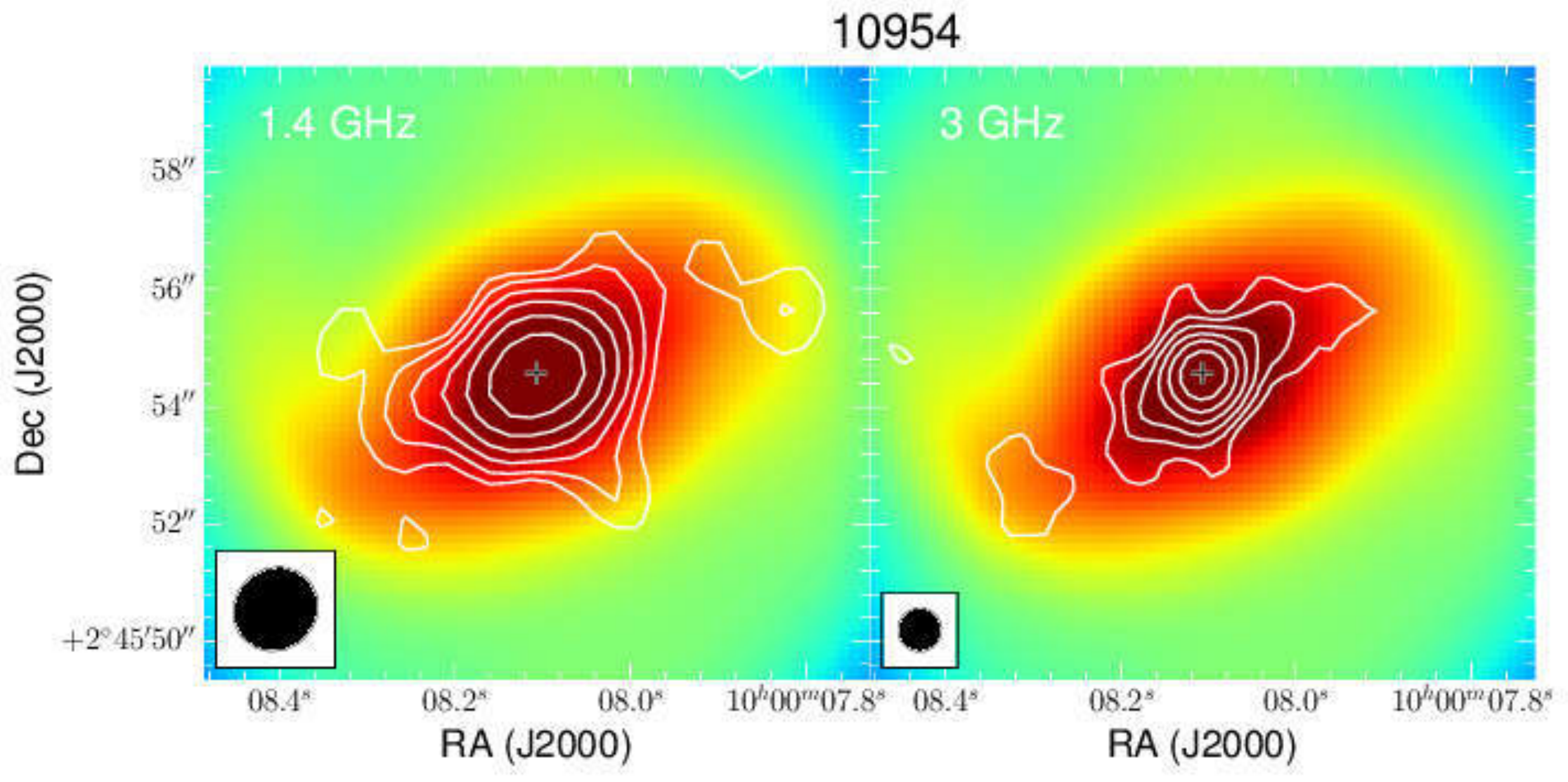}
   }
   \resizebox{\hsize}{!}{
   \includegraphics[width=0.5cm]{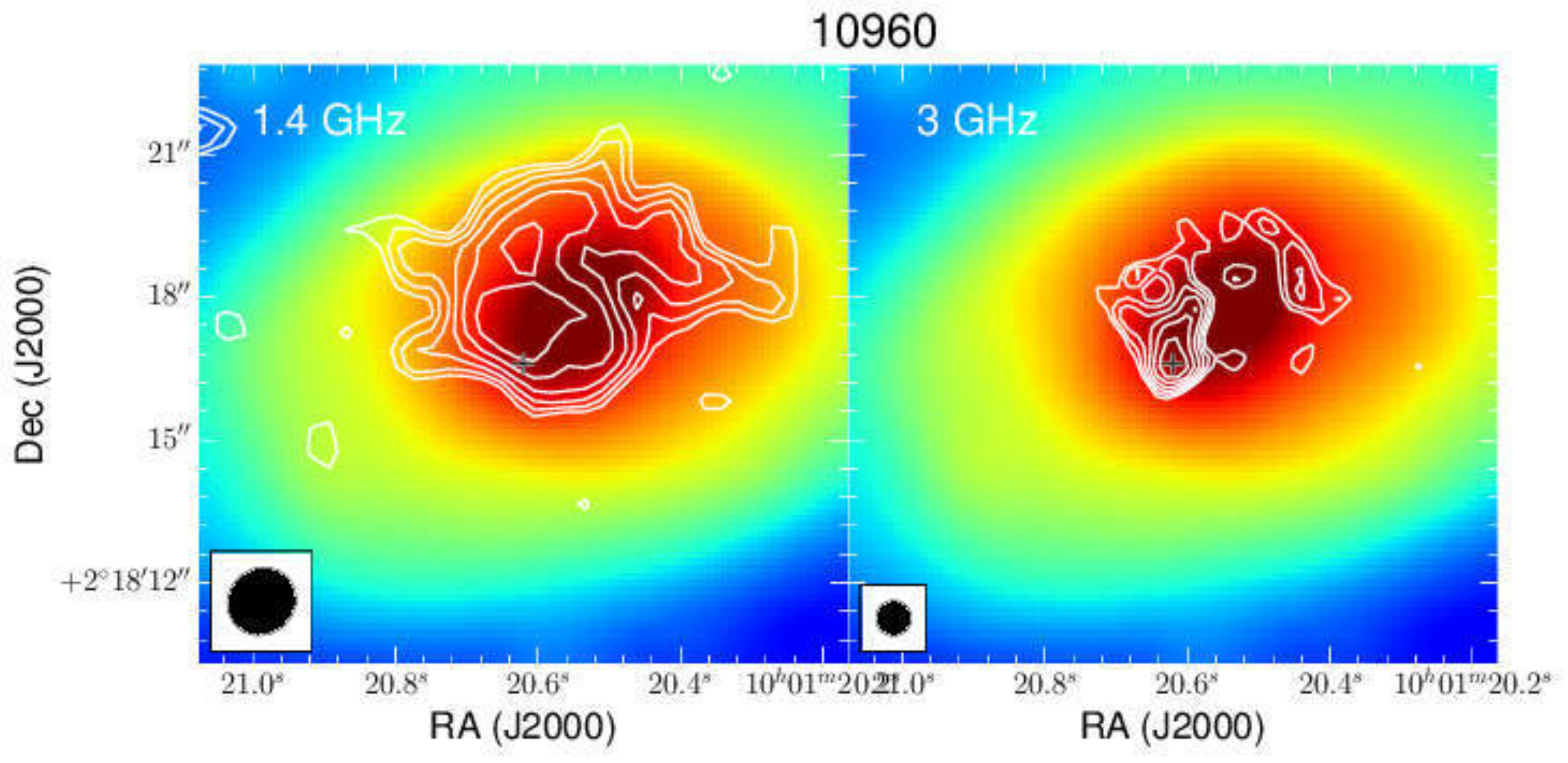}
    \includegraphics[width=0.5cm]{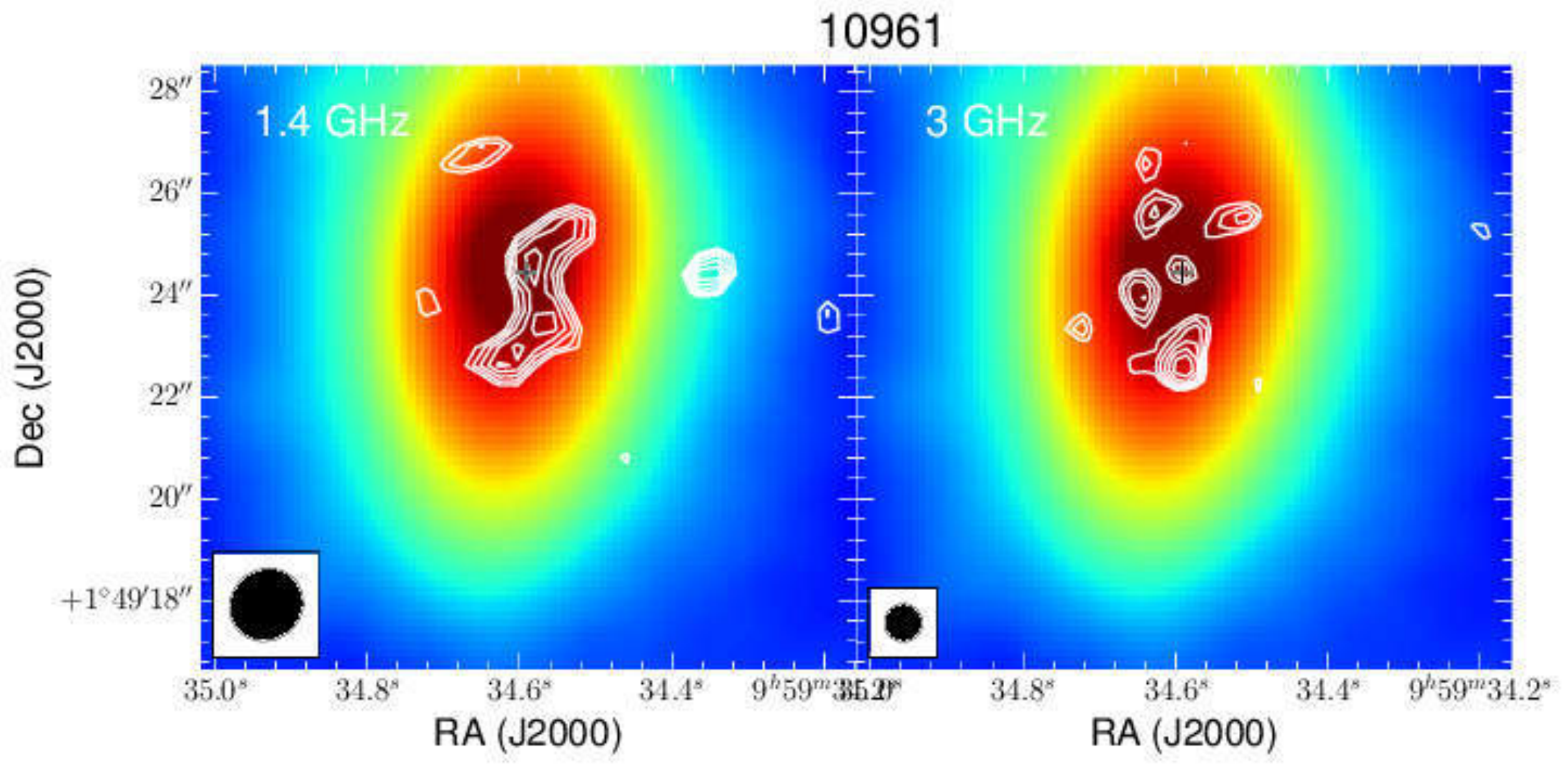}
    }
    \resizebox{\hsize}{!}{
   \includegraphics[width=0.5cm]{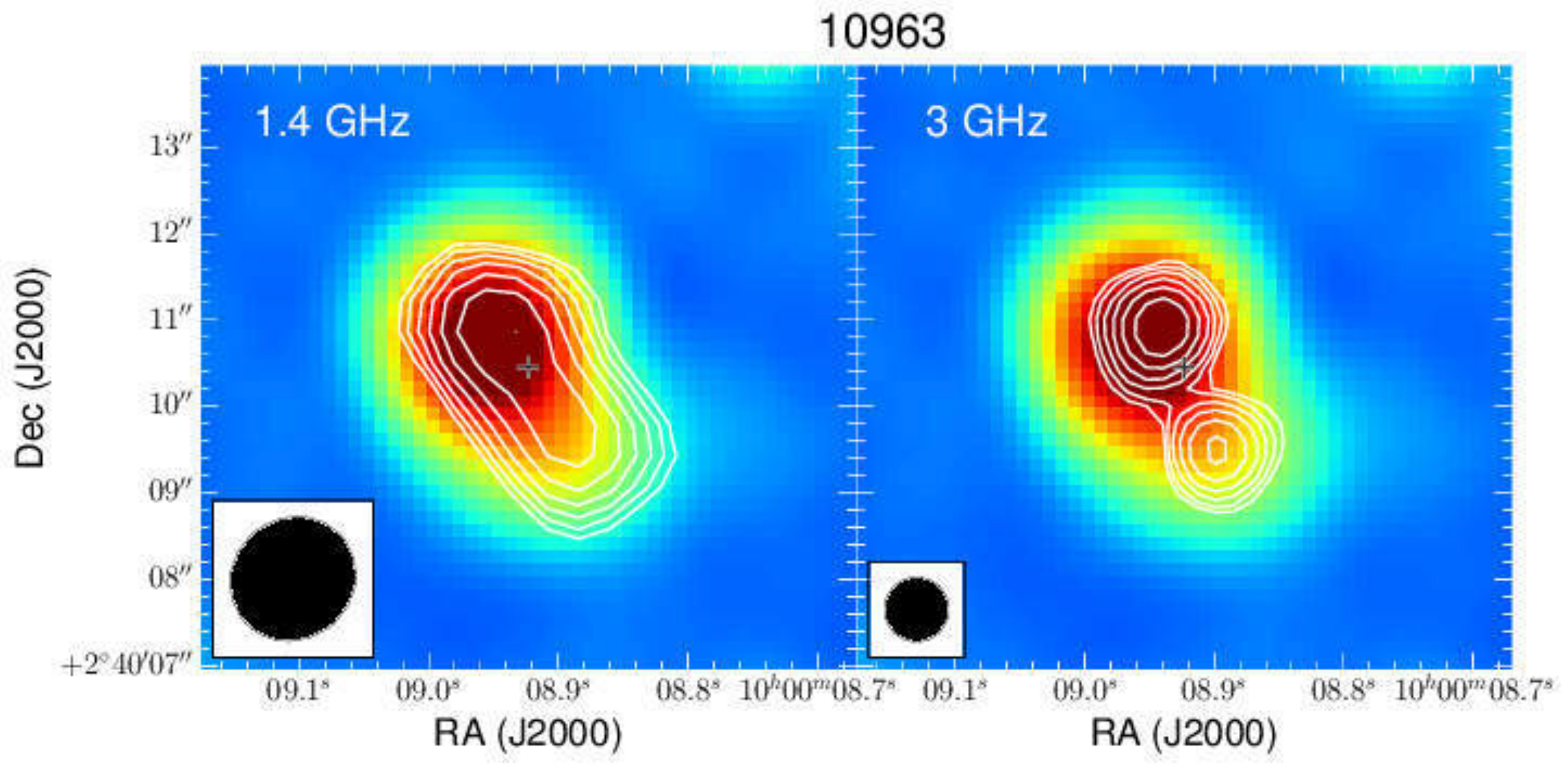}
    \includegraphics[width=0.5cm]{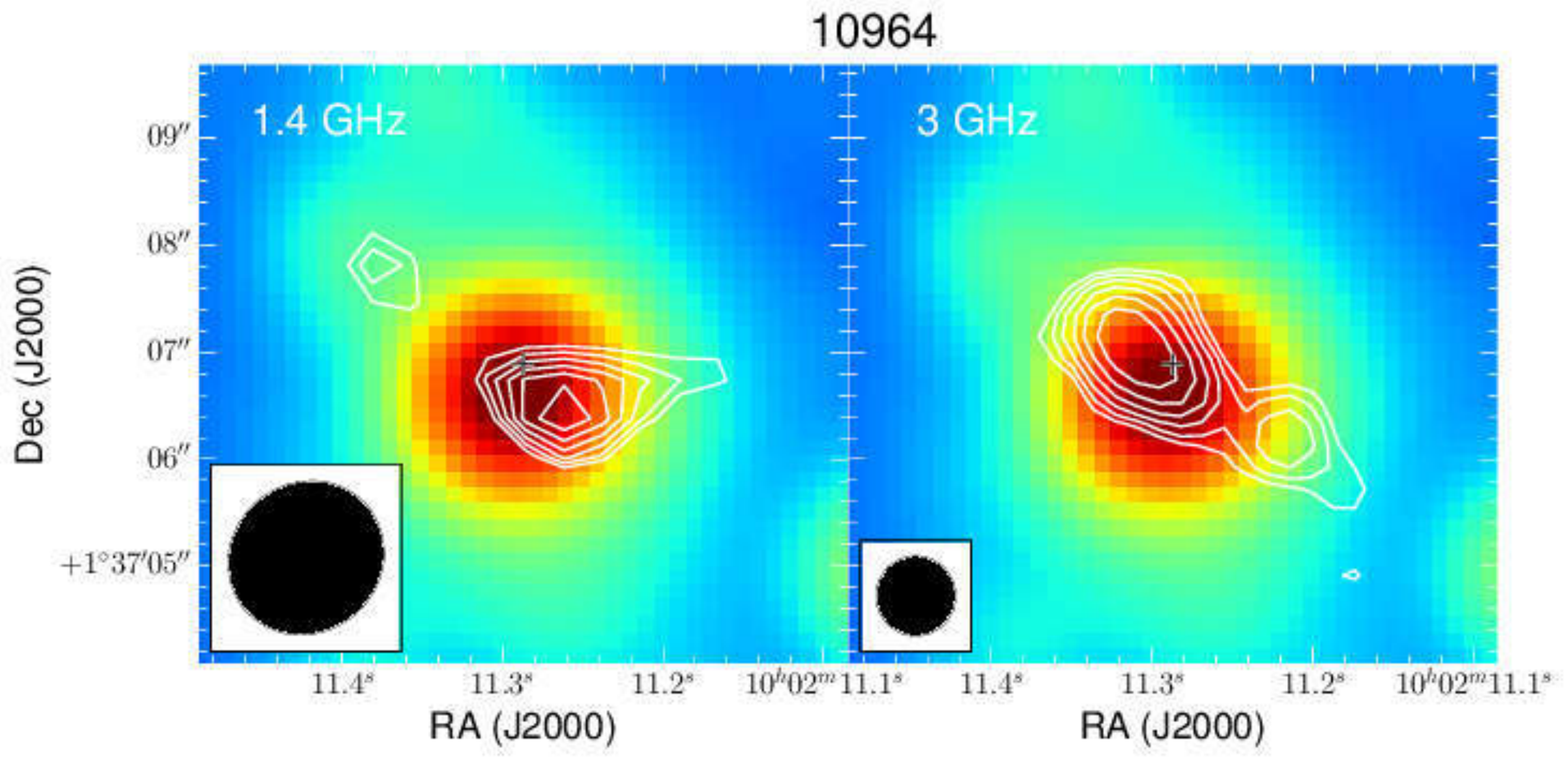}
    }
   \resizebox{\hsize}{!}{
   \includegraphics[width=0.088cm]{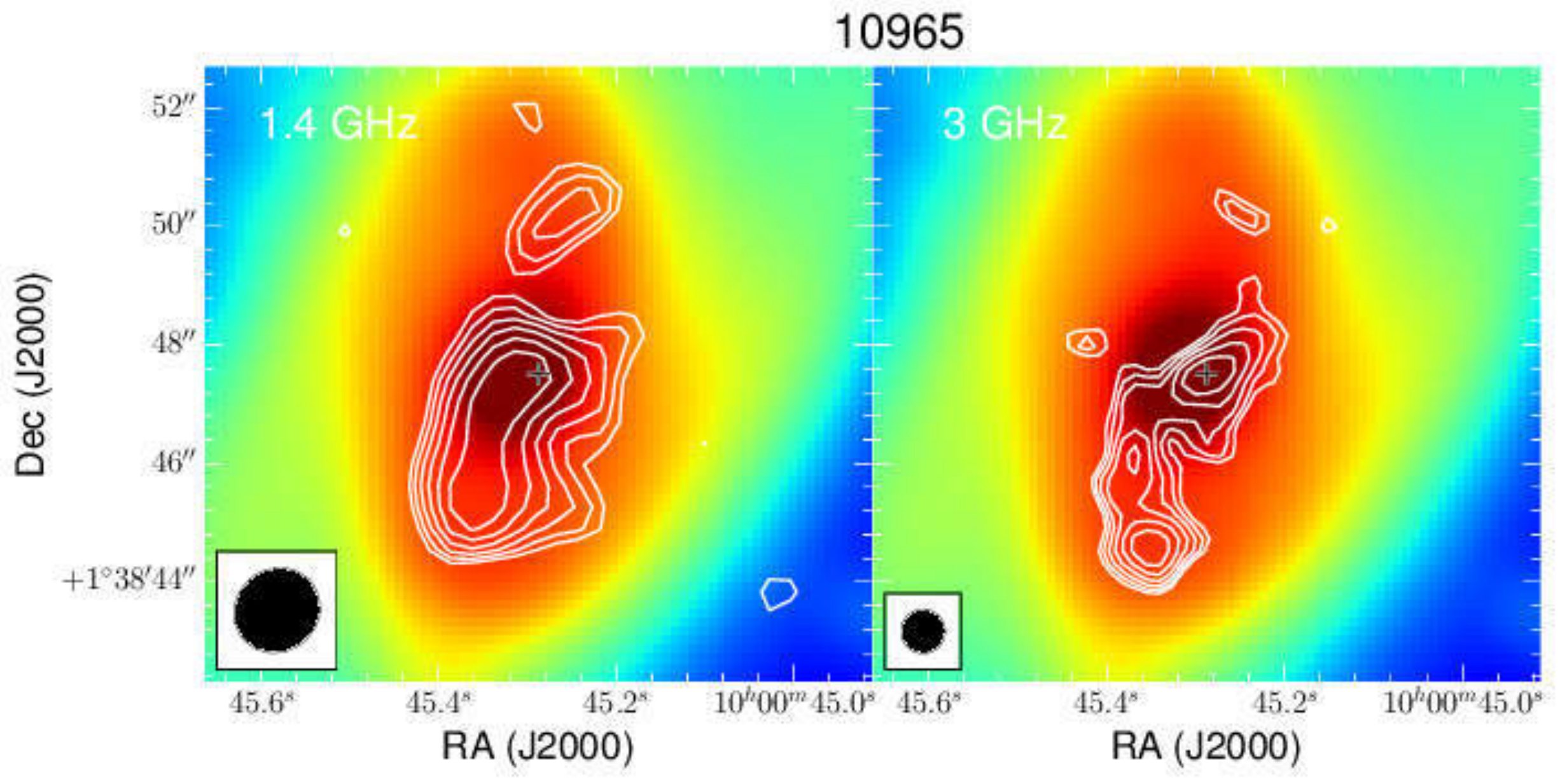}
  }
 
      \caption{The 9 Star-forming-galaxies in our multi-component radio sample at 3 GHz, including uncertain object (marked as SFG* in Table~\ref{table:data}). The latter have been placed in this class as they are lacking radio excess (see Sec.~\ref{sec:hosts}). Images described in Fig.~\ref{fig:maps2}.
              }
         \label{fig:sfgs}
   \end{figure*}

\subsubsection{The "eye" of COSMOS: a star-forming radio galaxy-ring}
\label{sec:ring}

\begin{figure*}[!ht]
   \resizebox{\hsize}{!}{
   \includegraphics[trim={2cm 0cm 2cm 0cm},clip,angle=90]{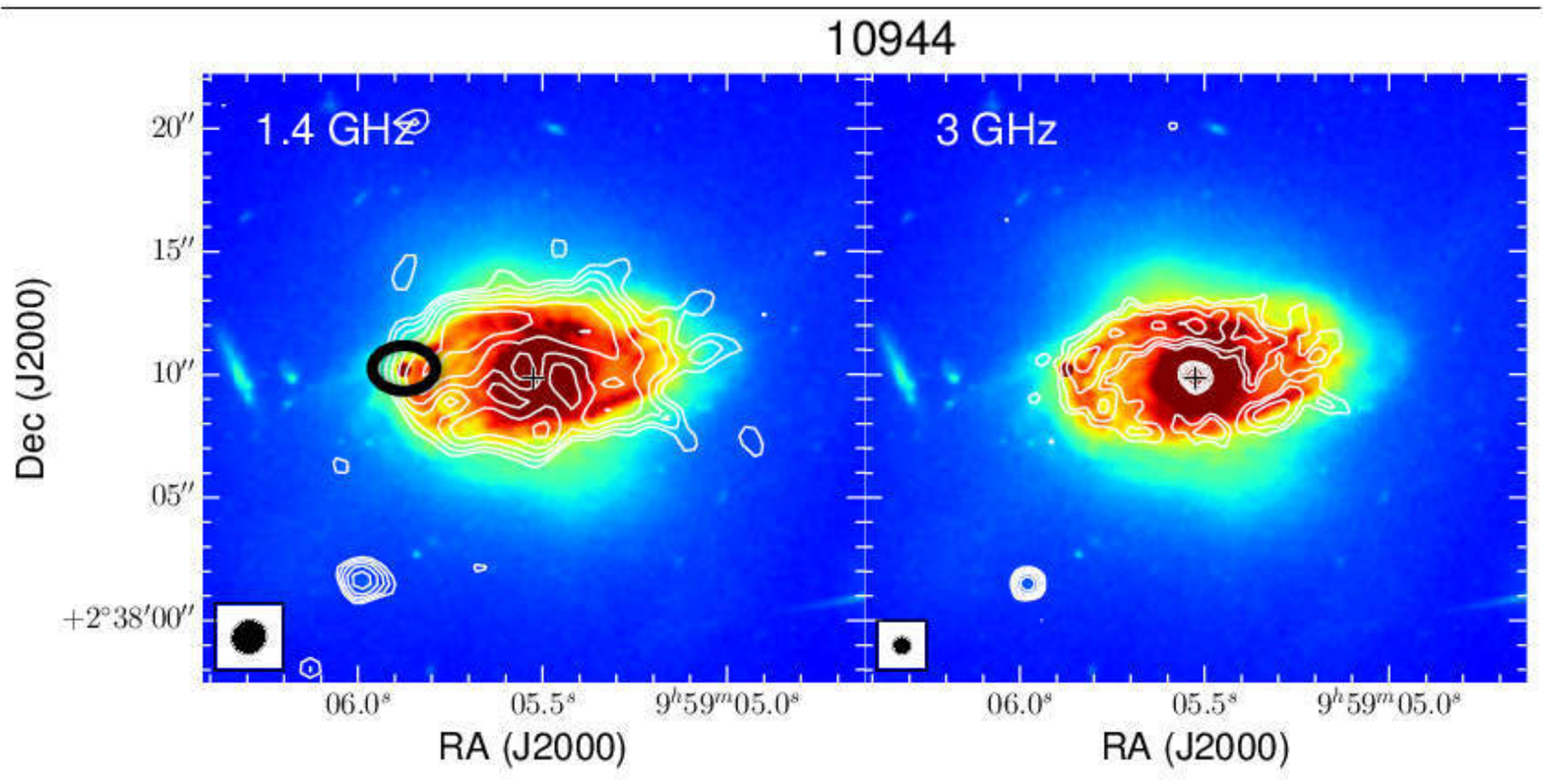}
 }
     \\ 
      \caption{10944 radio contours at 1.4 GHz ($left$) and 3 GHz ($right$) overlaid on the HST map (colourscale). The contours start at the 3$\sigma$ level and rise logarithmically. The HST map is from \cite{koekemoer07}. The black circle marks the knot discussed in the text.
              }
         \label{fig:10944}
   \end{figure*}

Based on the 3-GHz radio map (Fig.~\ref{fig:10944}), 10944 is a star-forming radio ring with a compact radio core in the centre, associated with the nucleus of the galaxy. This ring-like radio synchrotron emission follows well the spiral arms of this face-on galaxy, as seen in the overlay image of radio and HST i-band image \citep{koekemoer07}. From the HST image we classify this particular galaxy as a polar-ring galaxy \citep[e.g.][]{moiseev11}, because it exhibits an elongation along in the NW-SE direction, almost perpendicular to the disk. Polar-ring galaxies are the result of galaxy mergers, and this would indicate that 10944 is the result of a galaxy merger. This the only object in COSMOS that shows a radio ring and a compact radio core.

The unprecedented high resolution of 0.75 arcsec of the 3-GHz VLA-COSMOS data has revealed the substructure in the ring which is not seen at 1.4 GHz. Besides the ring, we also see a small unresolved core associated with the centre of the galaxy. To investigate whether this is due to star-formation or an AGN, we cross-match it to the VLBA data available for the COSMOS field from \cite{noelia17}, but we find no match: this source has been observed but not detected by VLBA (Herrera Ruiz priv. comm.). The host galaxy is a SF galaxy at low redshift, $z_{\rm spec}$ = 0.07905, with no signs of AGN at any observed wavelengths, and it does not display any radio excess over the star-formation-rate.  Our multi-wavelength AGN diagnostics do not show any evidence for the existence of an AGN. It is not a mid-IR AGN nor an X-ray AGN and is classified as a SFG \citep{smolcic17b}. 

Only 4\% of the flux is located in the "core", while the rest is in the ring structure. It has a SFR$_{\rm IR}$ = 4.9 $\rm M_{\odot} / yr$ and a stellar mass of 2.8$\times$10$^{11}$ M$_{\odot}$ and falls just below the 1$\sigma$ dispersion of the main-sequence for star-forming galaxies \citep[e.g.][]{whitaker12} at the redshift of the source. It is a sub-luminous infrared galaxy (sub-LIRG) with total infrared luminosity $L_{\rm IR}$ = 4.9$\times10^{10}$ L$_{\odot}$.

The radio "core" emission at 3 GHz is barely resolved at 3 GHz, giving an upper limit for the size of the radio emitting area at 1.12 kpc (FWHM). We also note that the peak "core" emission is slightly offset at 1.4 GHz from that at 3 GHz, by 0.57 arcsec. This could be attributed to differences in the beam and pixel size between the two maps, or could simply be an effect of noise.

Given that there is no indication for an AGN from the multi-wavelength AGN diagnostics available for COSMOS, we conclude that this is a circum-nuclear star-forming region close to the nucleus of this galaxy and the radio emission is not a result of an AGN. If indeed there were a low-luminosity AGN, this would be Compton-thick and heavily obscured by dust surrounding the nuclear region. 

The bright knot at the eastern part of the ring (marked with a black circle) can be seen in the 1.4 GHz, 3 GHz and HST maps in Fig.~\ref{fig:10944} but it does not spatially coincide, and it lies in a distance of $\sim$7 kpc from the 3-GHz "core". The distance between the HST and 3-GHz local peak flux density is 0.9 arcsec and between the HST and 1.4-GHz local flux density 3 arcsec. Probably this is a star-forming region associated with the spiral arm of the galaxy. Finally, the radio ring has a radius of $\sim$6.5 kpc (major axis) and $\sim$2.8 kpc (minor axis). 

\clearpage
\newpage

\begin{figure*}[!ht]
  \resizebox{\hsize}{!}
 {\includegraphics{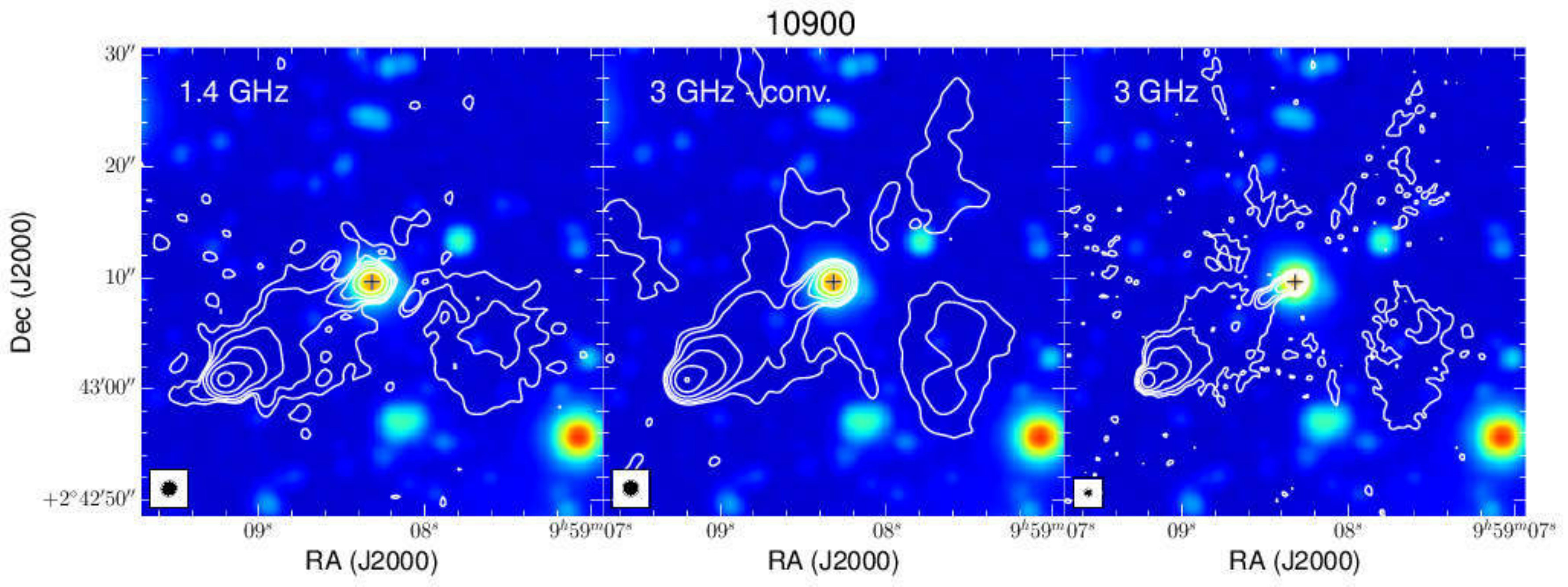}
 \includegraphics{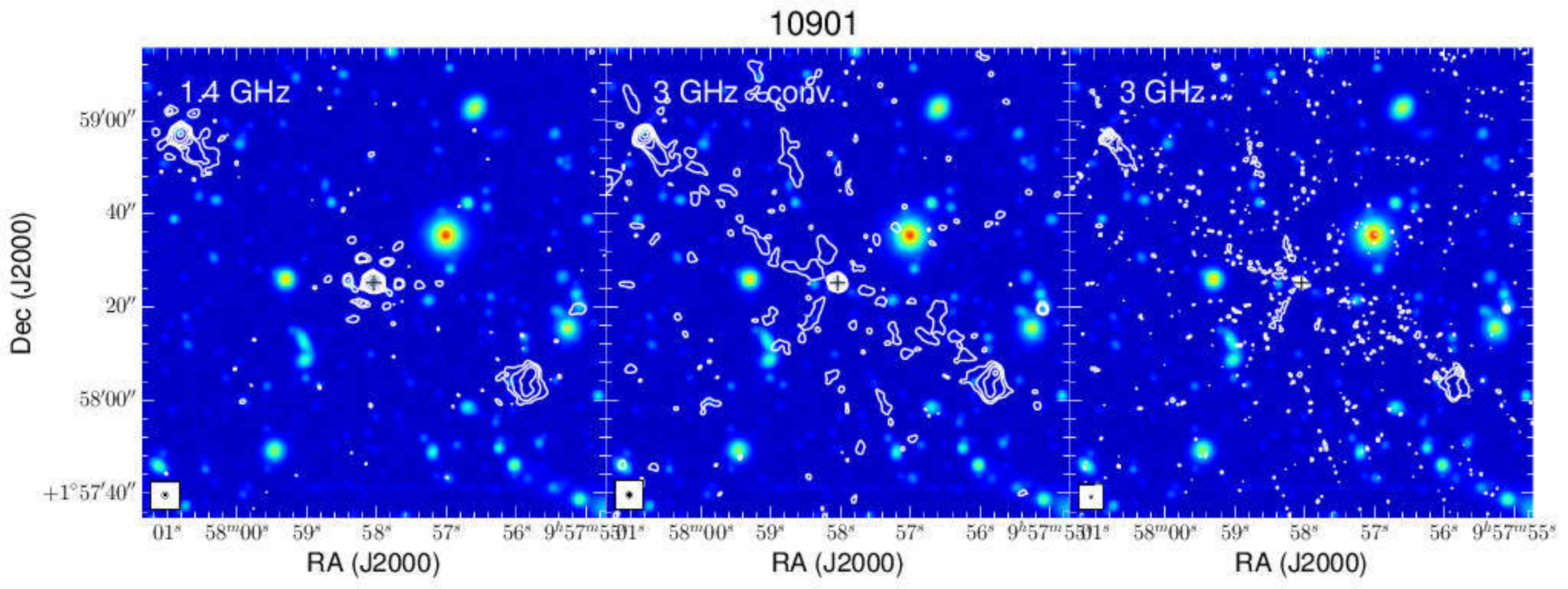}
            }
             \\ \\
      \resizebox{\hsize}{!}
       {\includegraphics{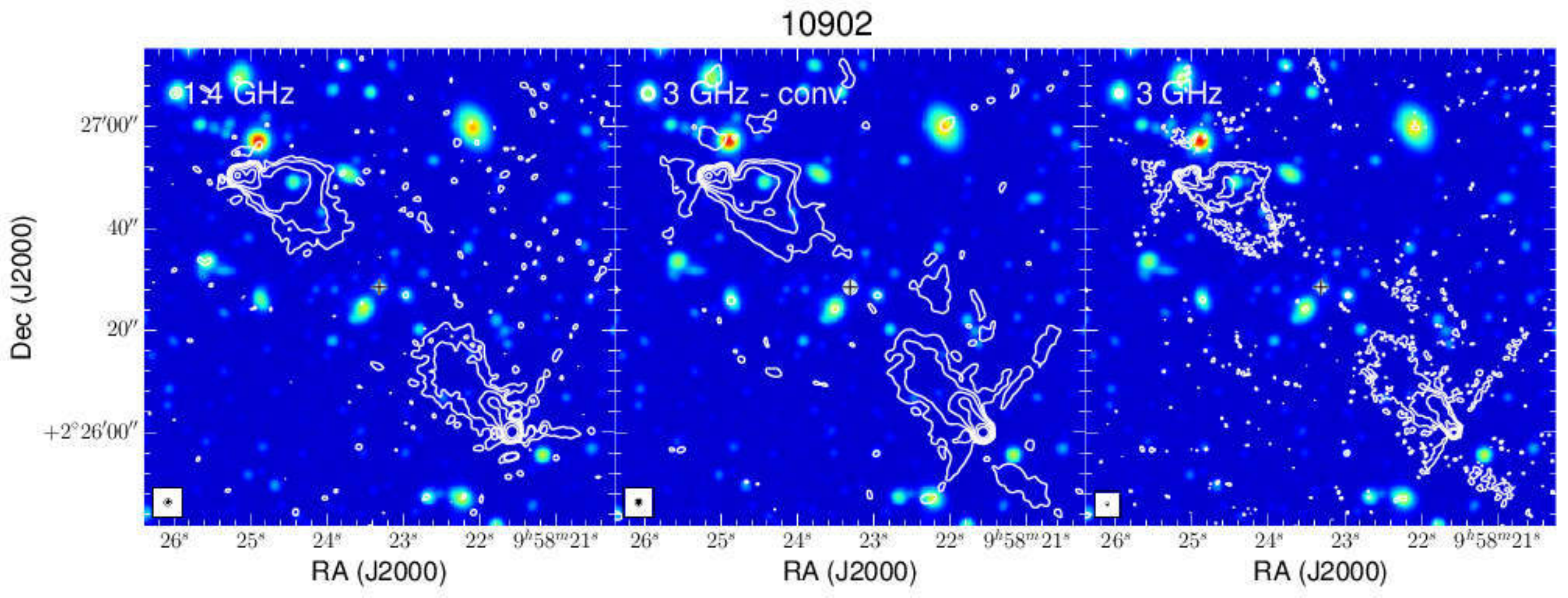}
        \includegraphics{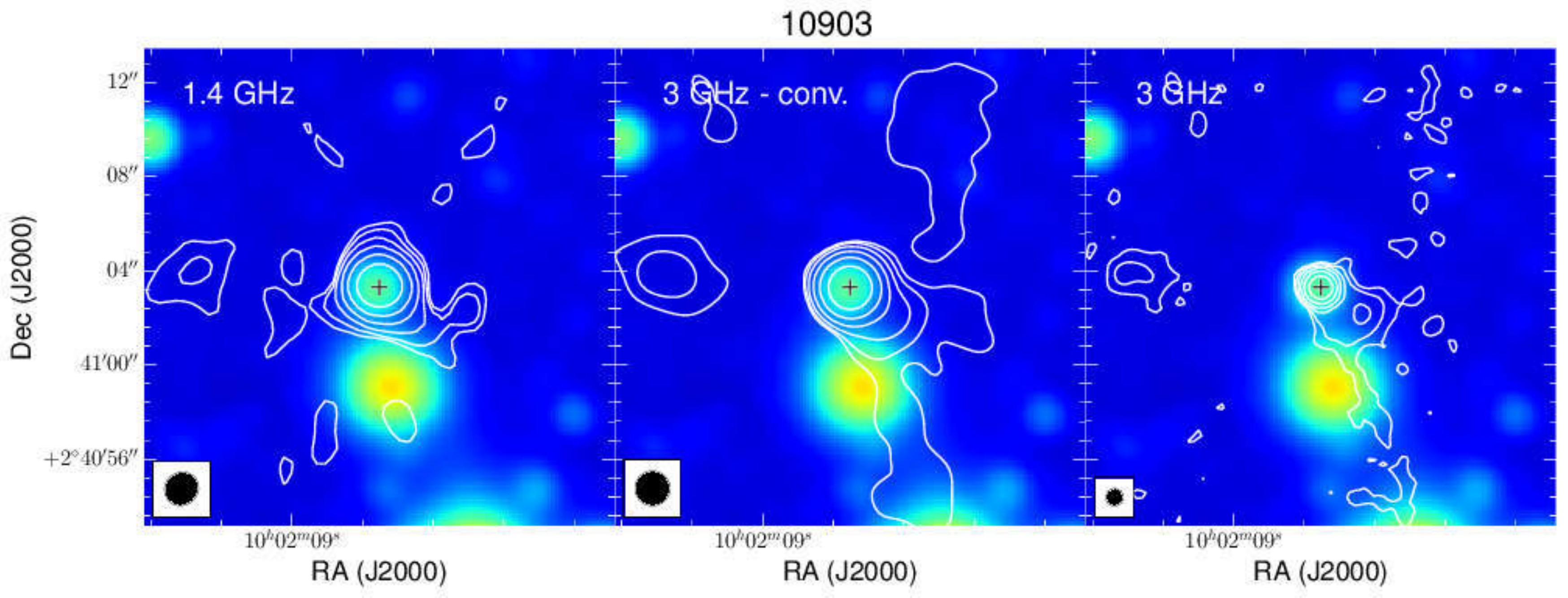}
            }
 \\ \\
 \resizebox{\hsize}{!}
{\includegraphics{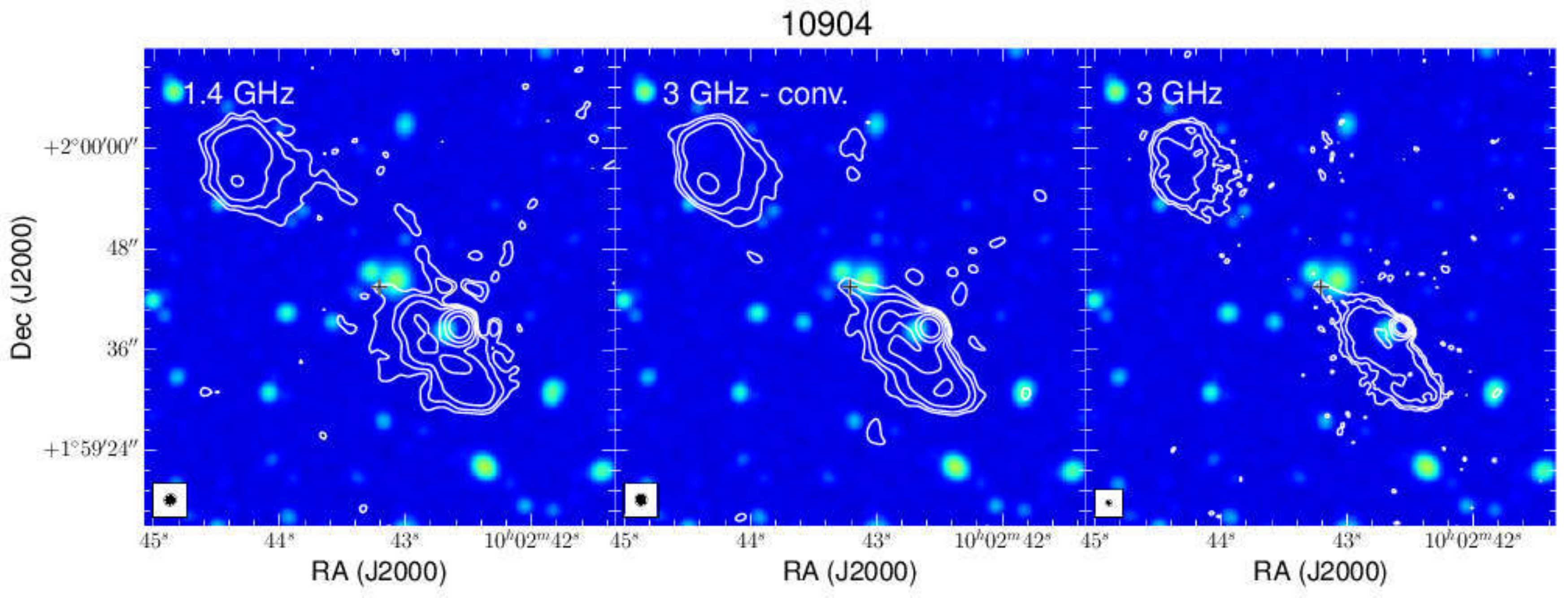}
 \includegraphics{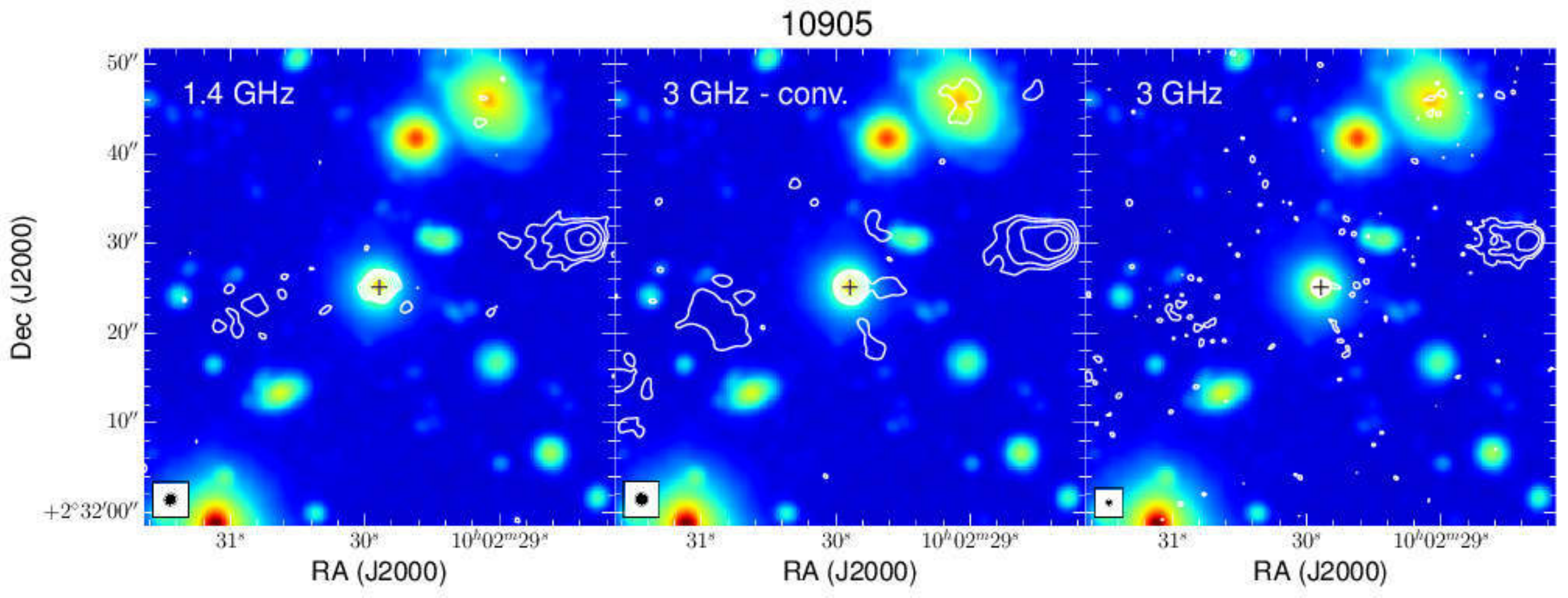}
            }
            \\ \\
  \resizebox{\hsize}{!}
 {\includegraphics{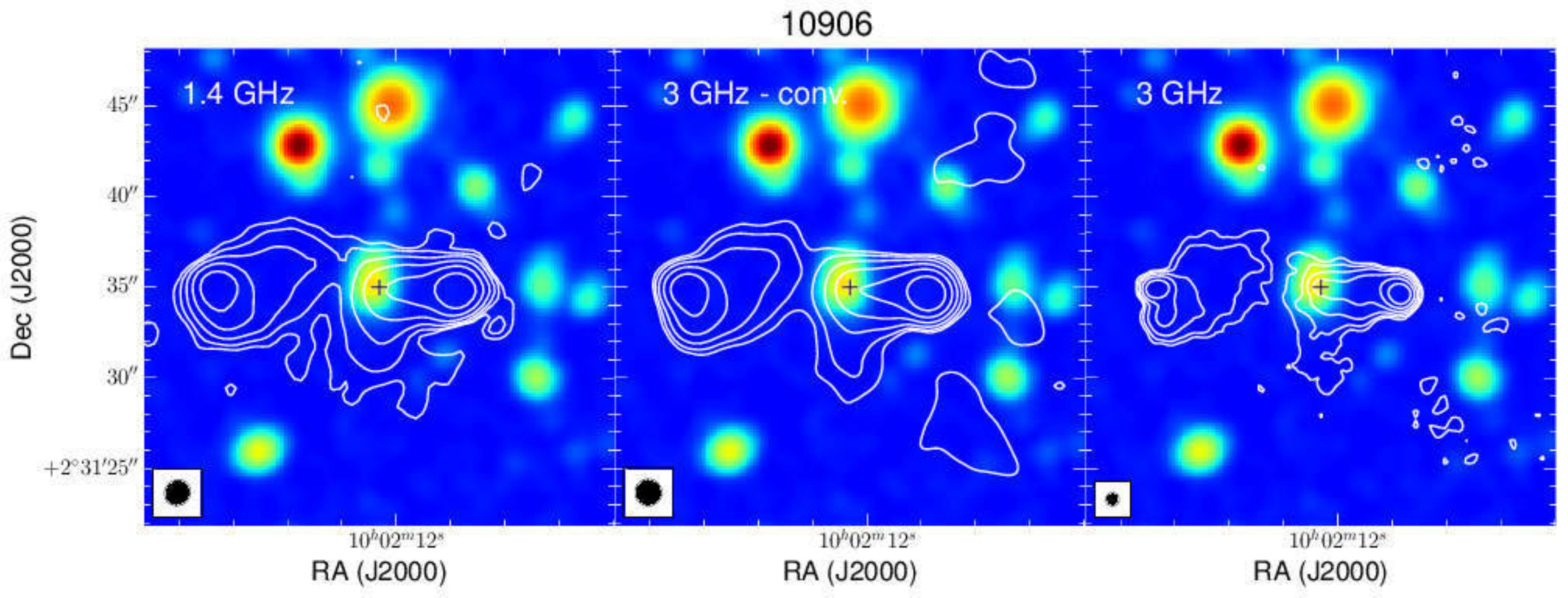}
    \includegraphics{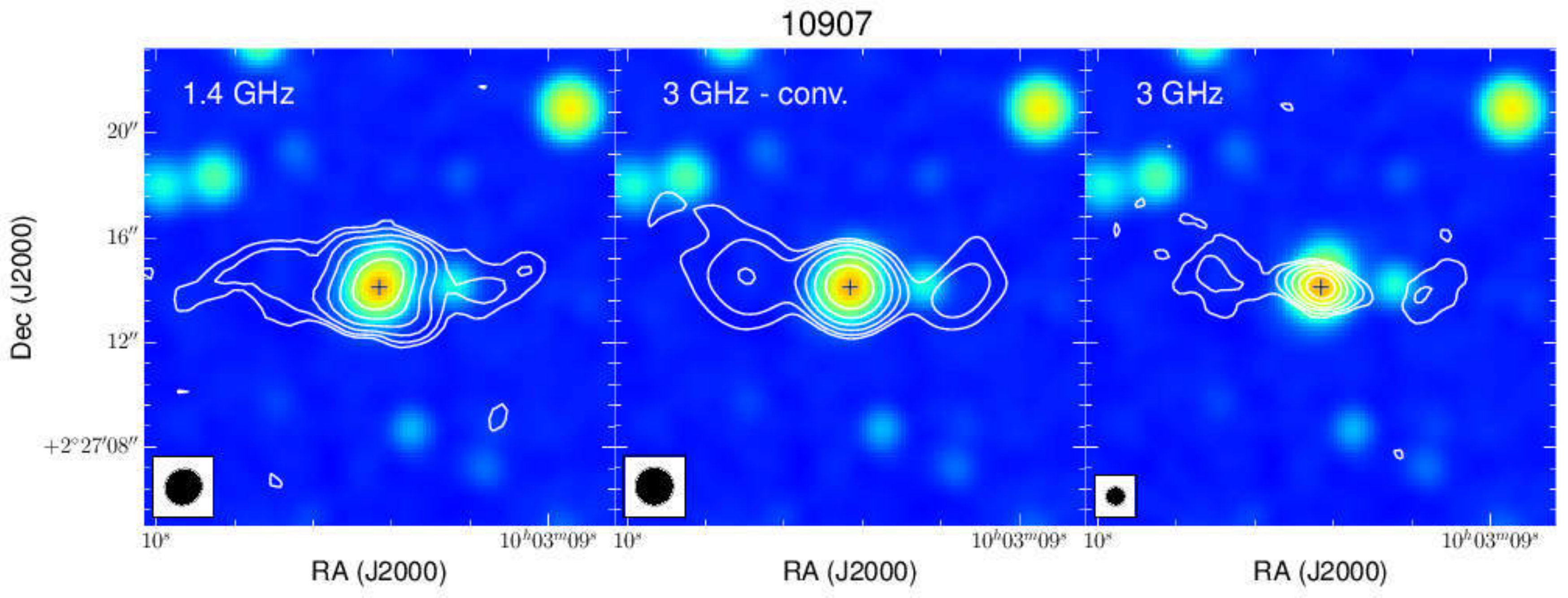}
            }
             \\ \\
      \resizebox{\hsize}{!}
       {\includegraphics{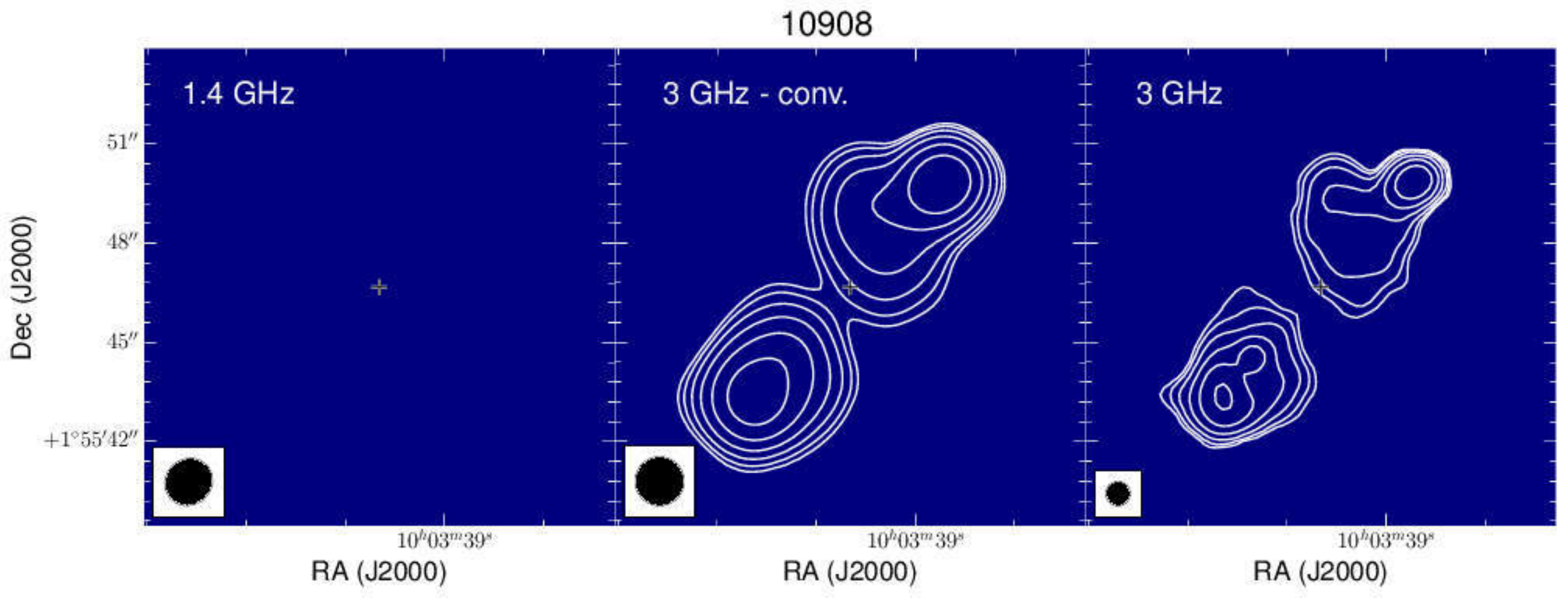}
        \includegraphics{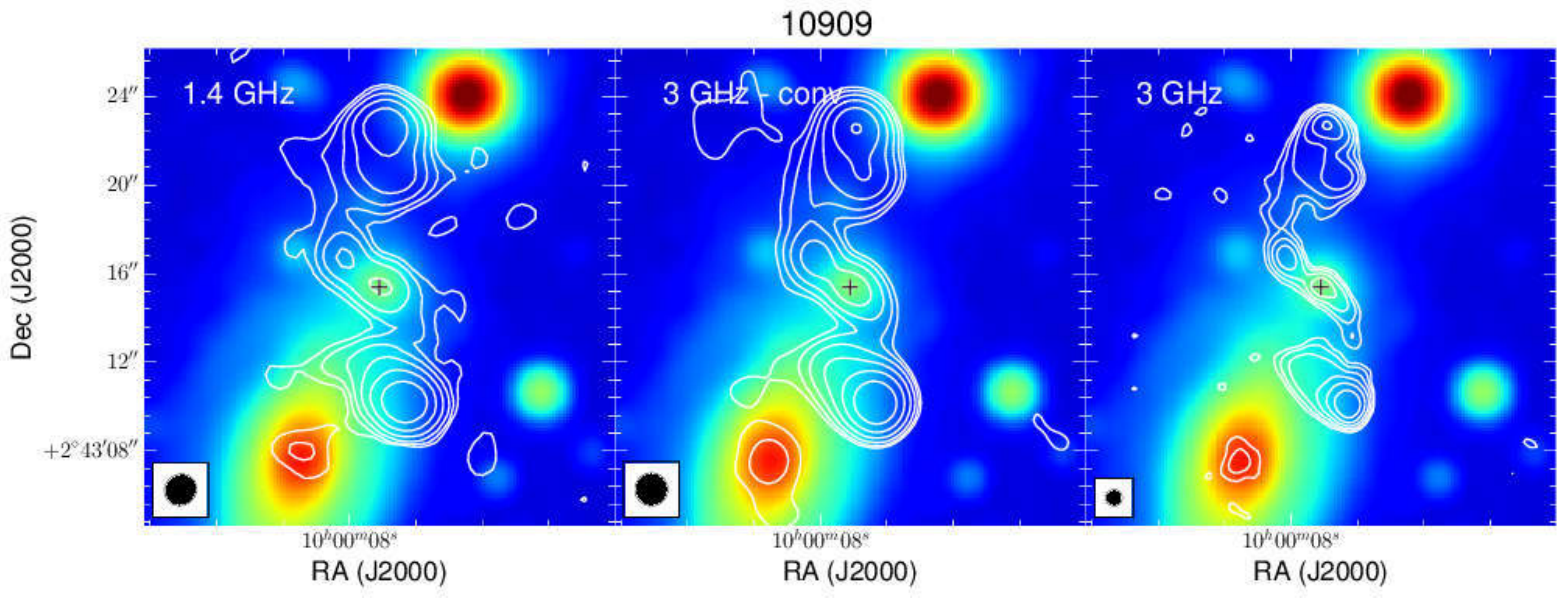}
            }
   \caption{Set of VLA 1.4-GHz and 3-GHz VLA-COSMOS stamps for the multi-component objects. For each object we give the 1.4-GHz map ($left$), the 3-GHz map at the original resolution of 0.75 arcsec ($right$), and the convolved 3-GHz map at 1.5 arcsec resolution ($middle$), to match the one at 1.4 GHz. The contour levels are equally spaced on a log-scale, where the lowest is set at 3 $\sigma$ and the highest at the maximum peak flux-density of the radio structure, and are overlaid on the UltraVISTA stacked image (see Sec.~\ref{sec:sample}). The 1.4- and 3-GHz beams are shown on the bottom left. Objects without a colour background are located in masked areas in the UltraVISTA coverage. Some objects also lack 1.4-GHz contours, as they lie outside the 1.4-GHz coverage.
   }
              \label{fig:maps2}%
    \end{figure*}
\addtocounter{figure}{-1}
\begin{figure*}[!ht]
    \resizebox{\hsize}{!}
       {\includegraphics{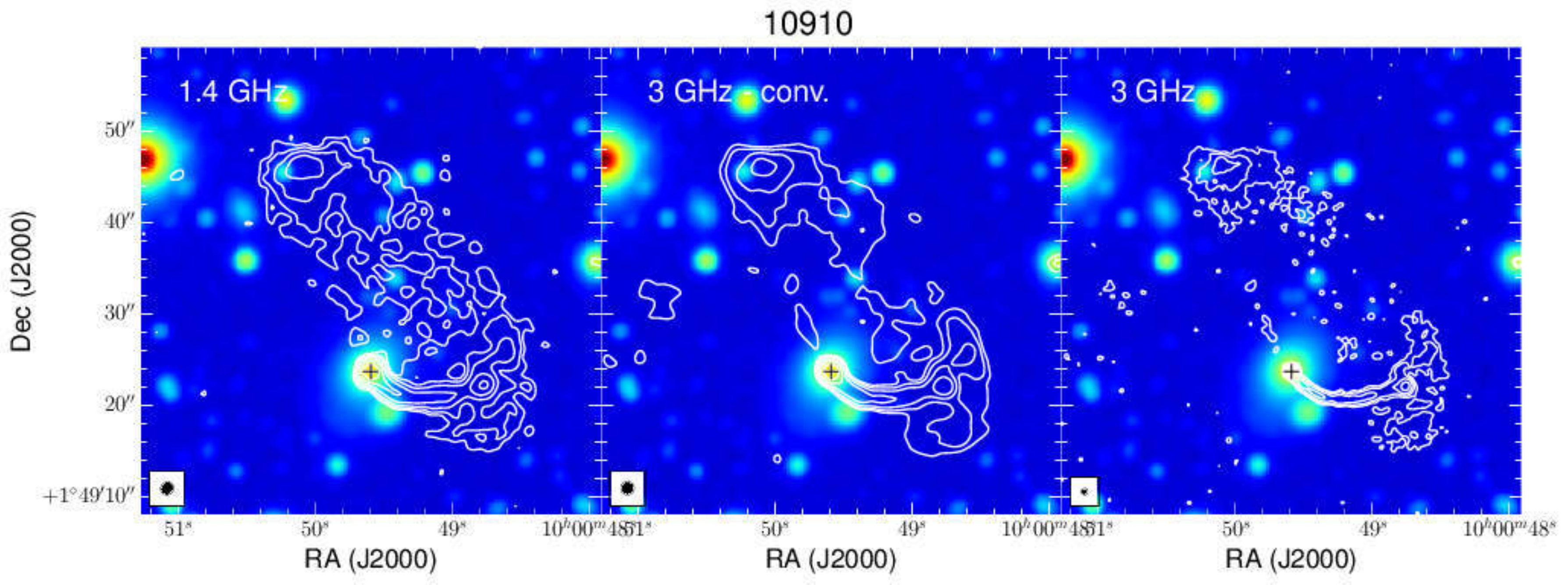}
        \includegraphics{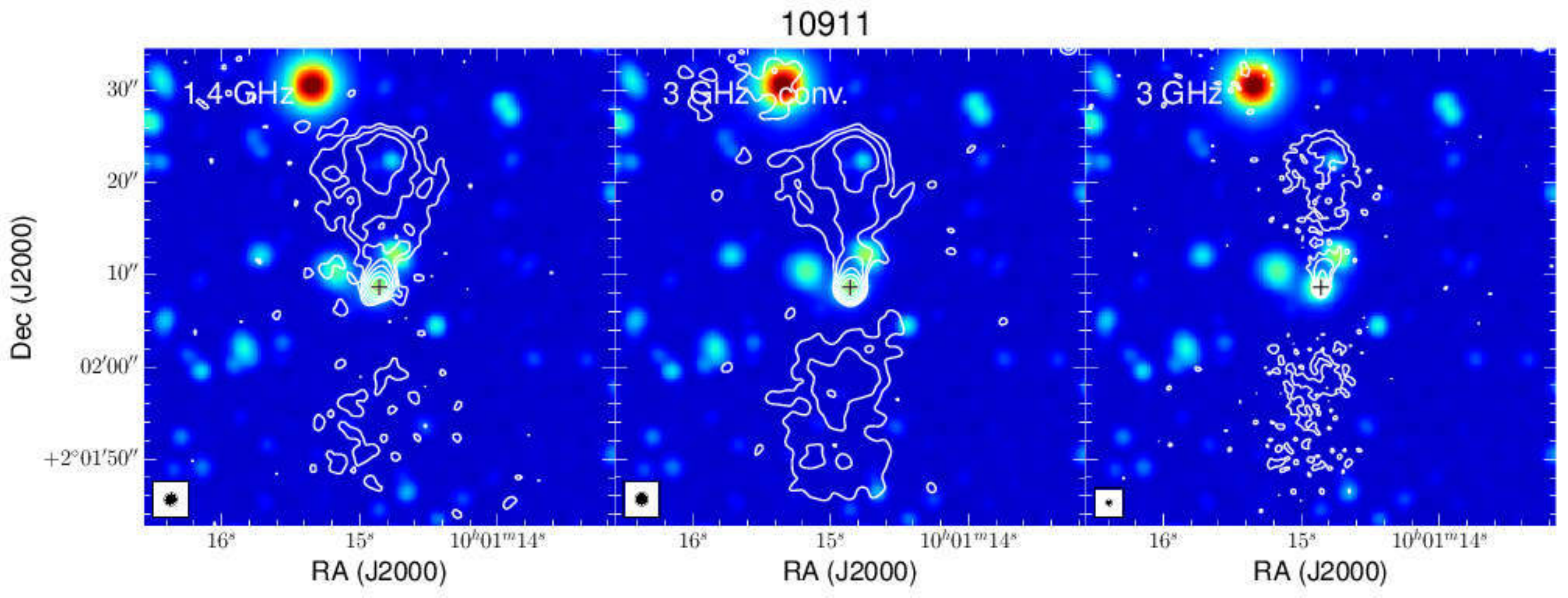}
            }
            \\ \\
 \resizebox{\hsize}{!}
{\includegraphics{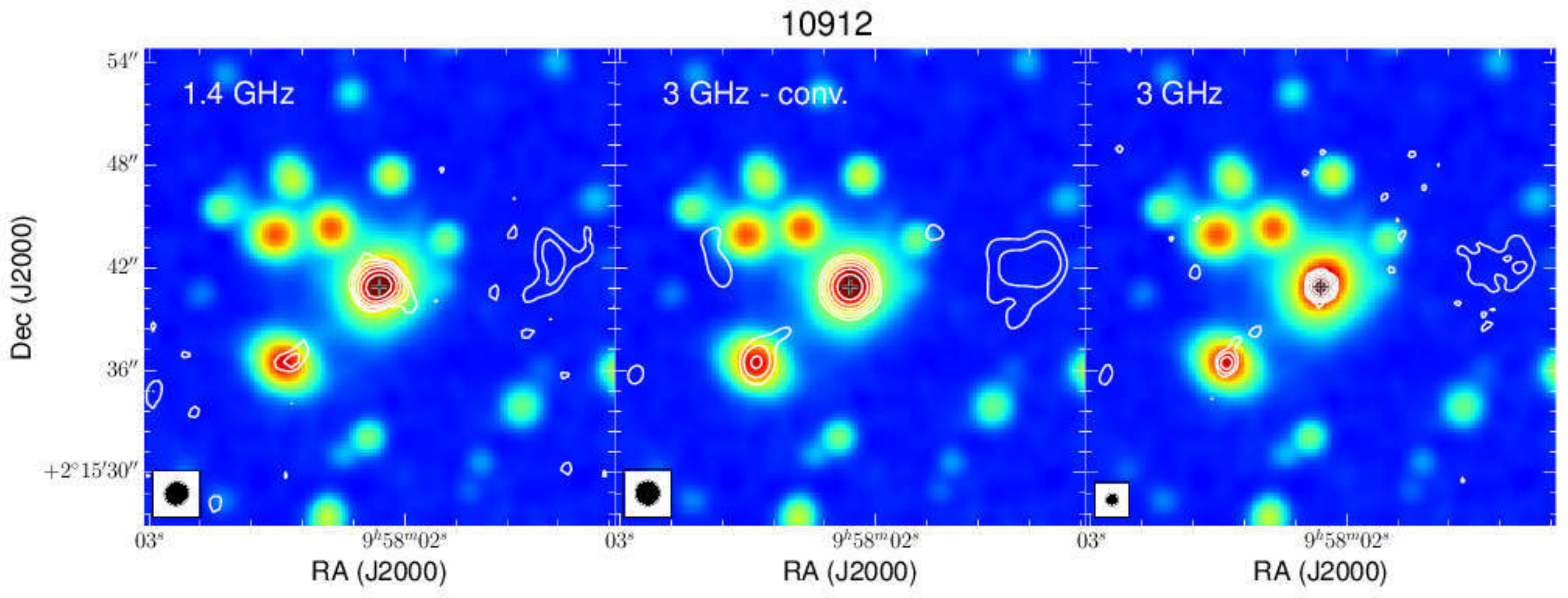}
 \includegraphics{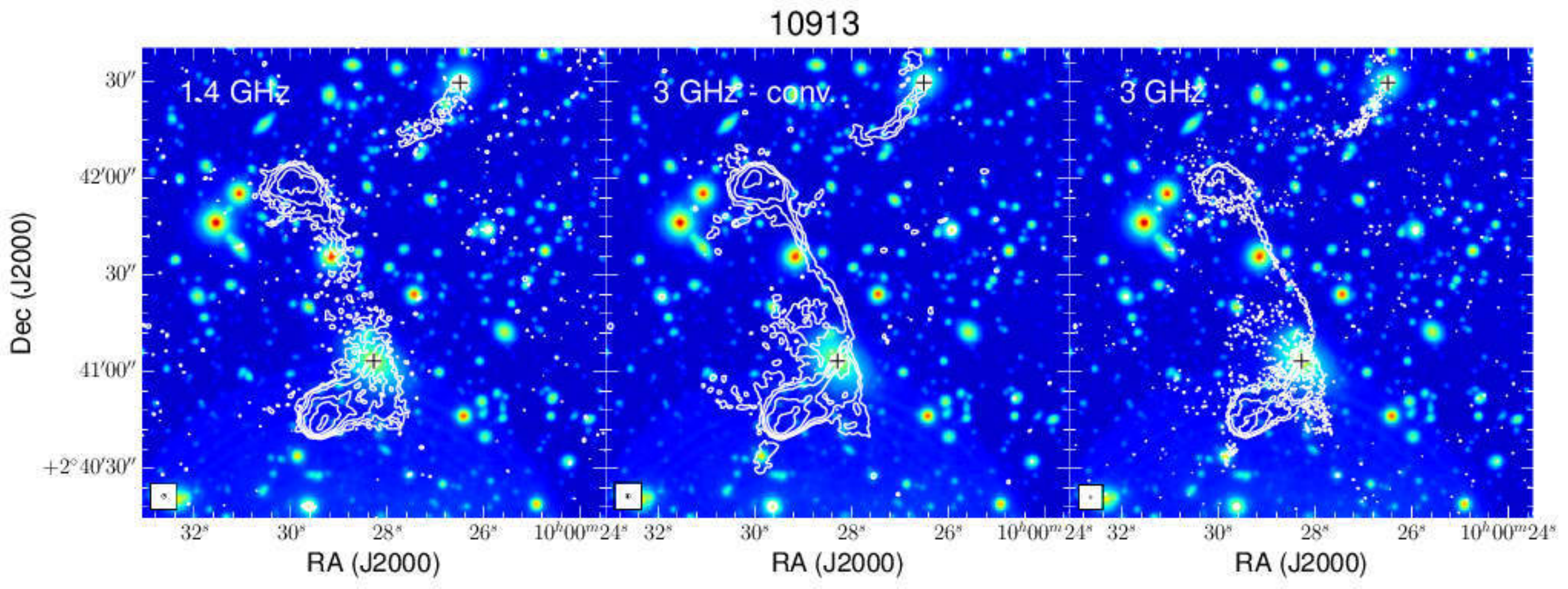}
            }
            \\ \\
  \resizebox{\hsize}{!}
 {\includegraphics{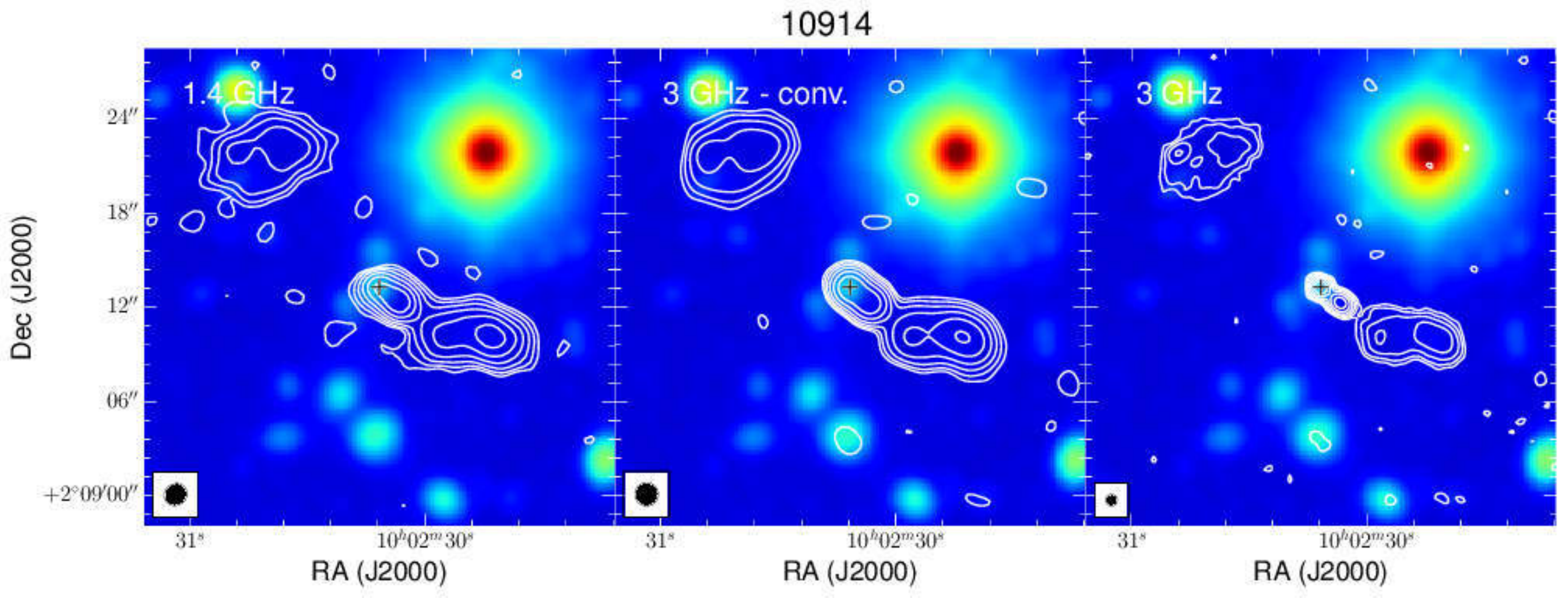}
    \includegraphics{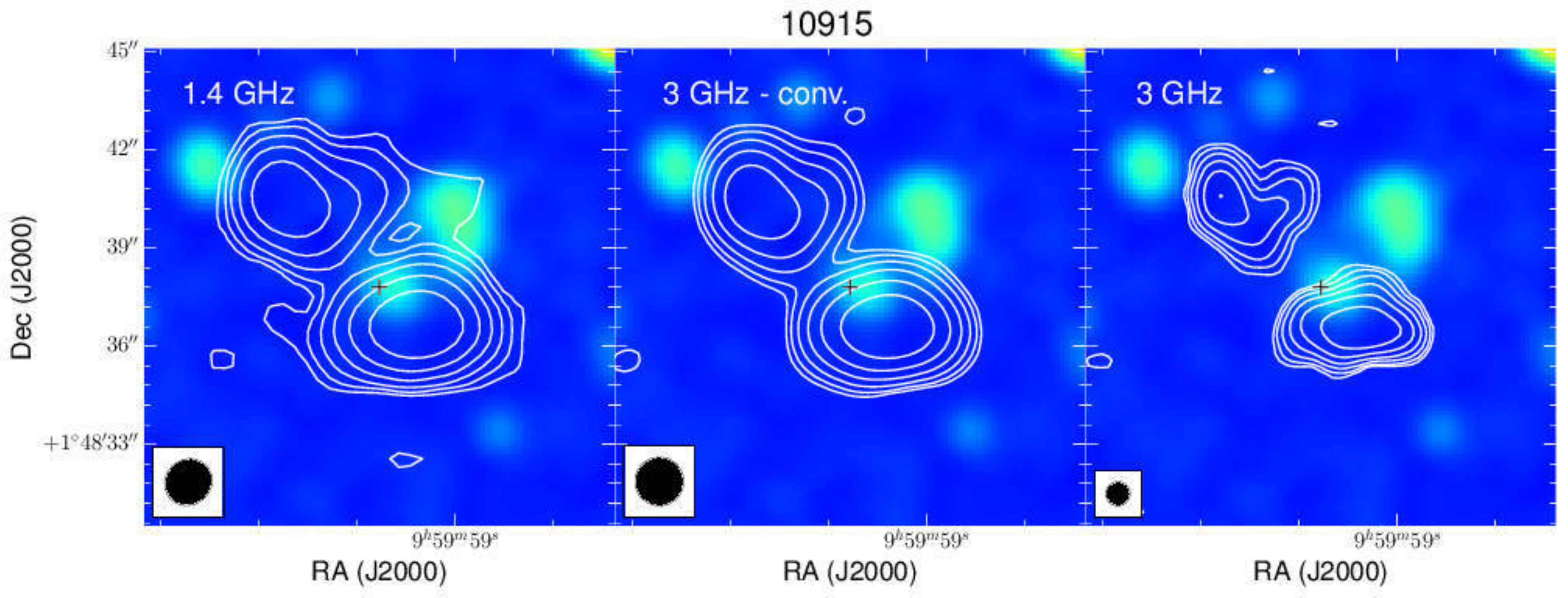}
            }
             \\ \\ 
      \resizebox{\hsize}{!}
       {\includegraphics{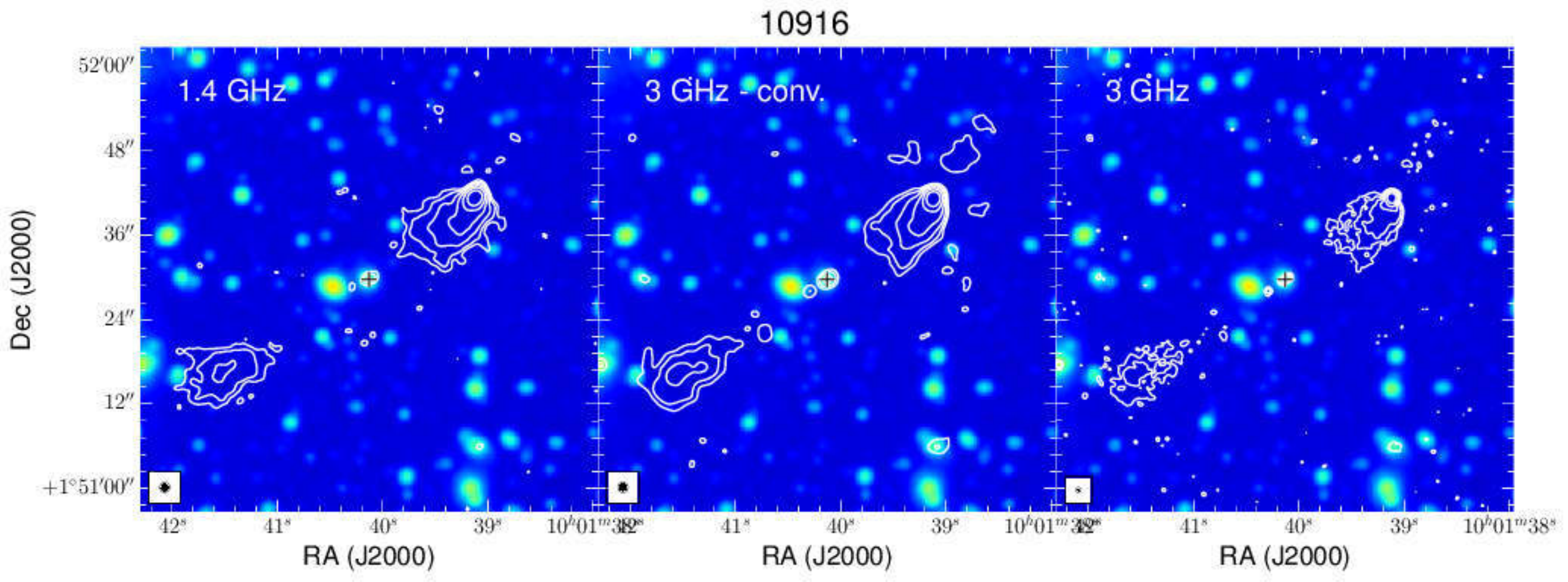}
        \includegraphics{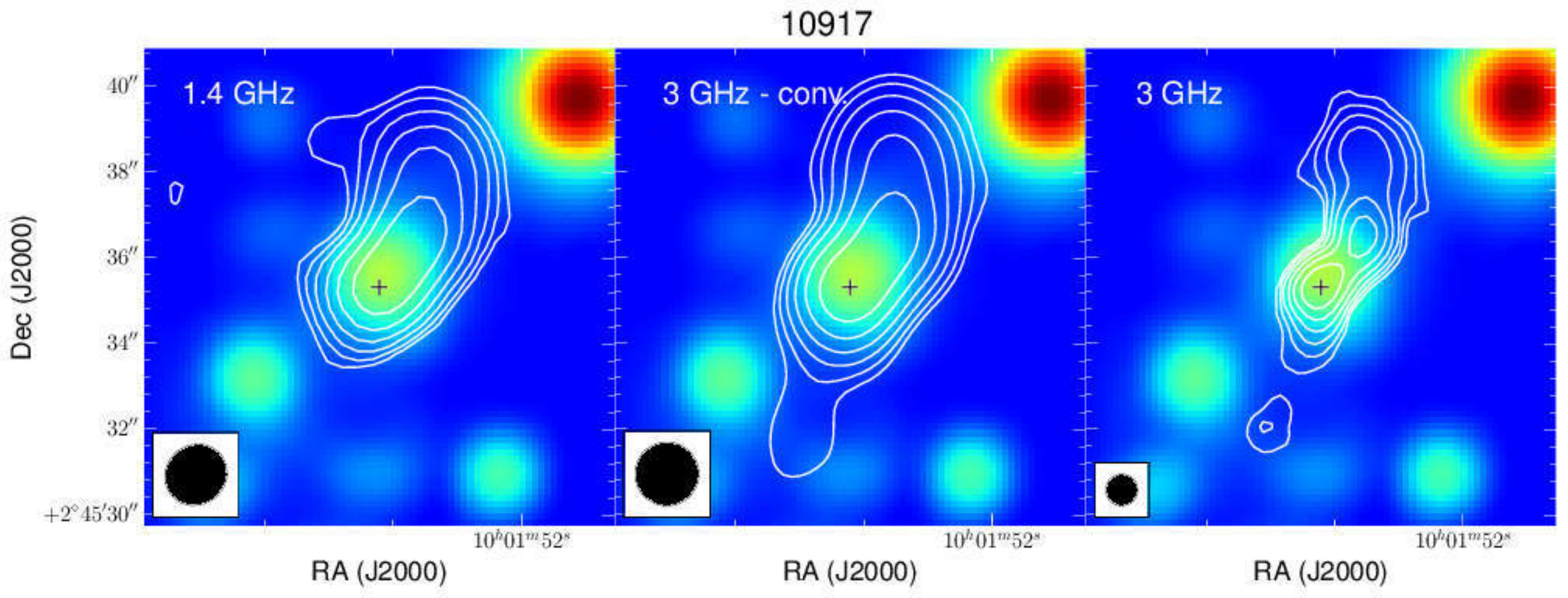}
            }
\\ \\
 \resizebox{\hsize}{!}
{\includegraphics{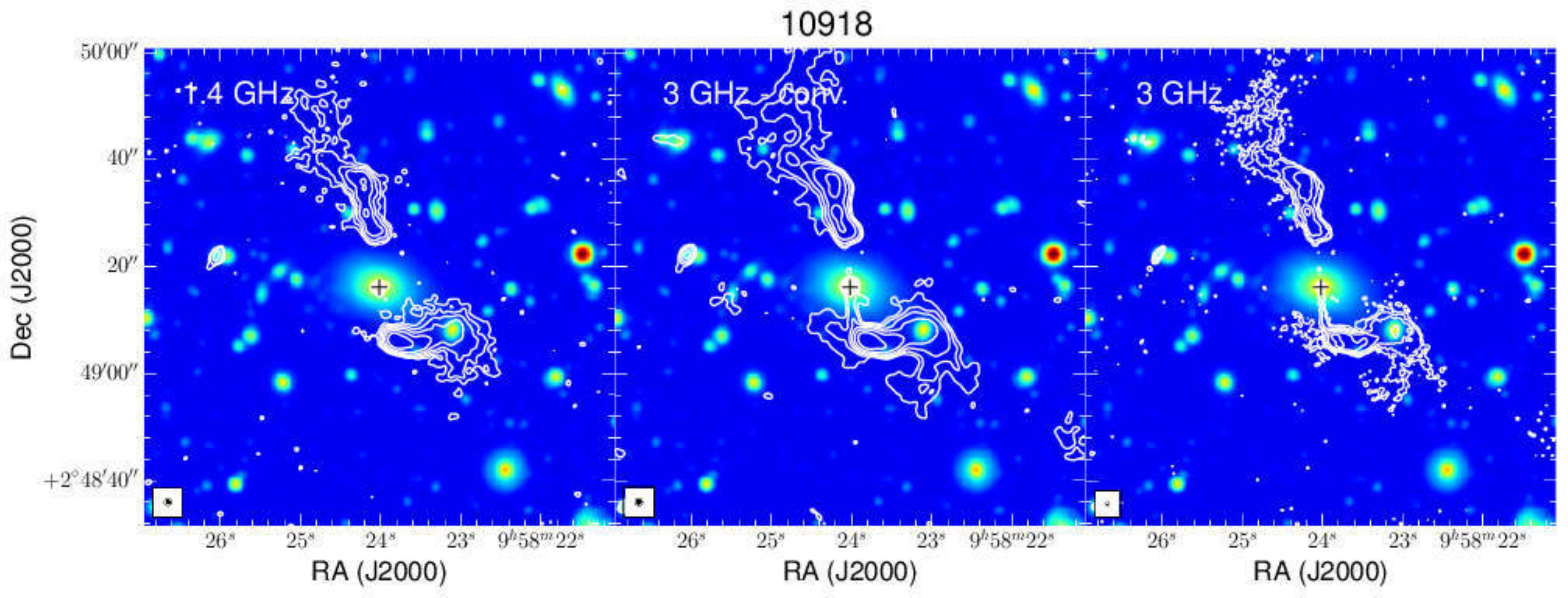}
 \includegraphics{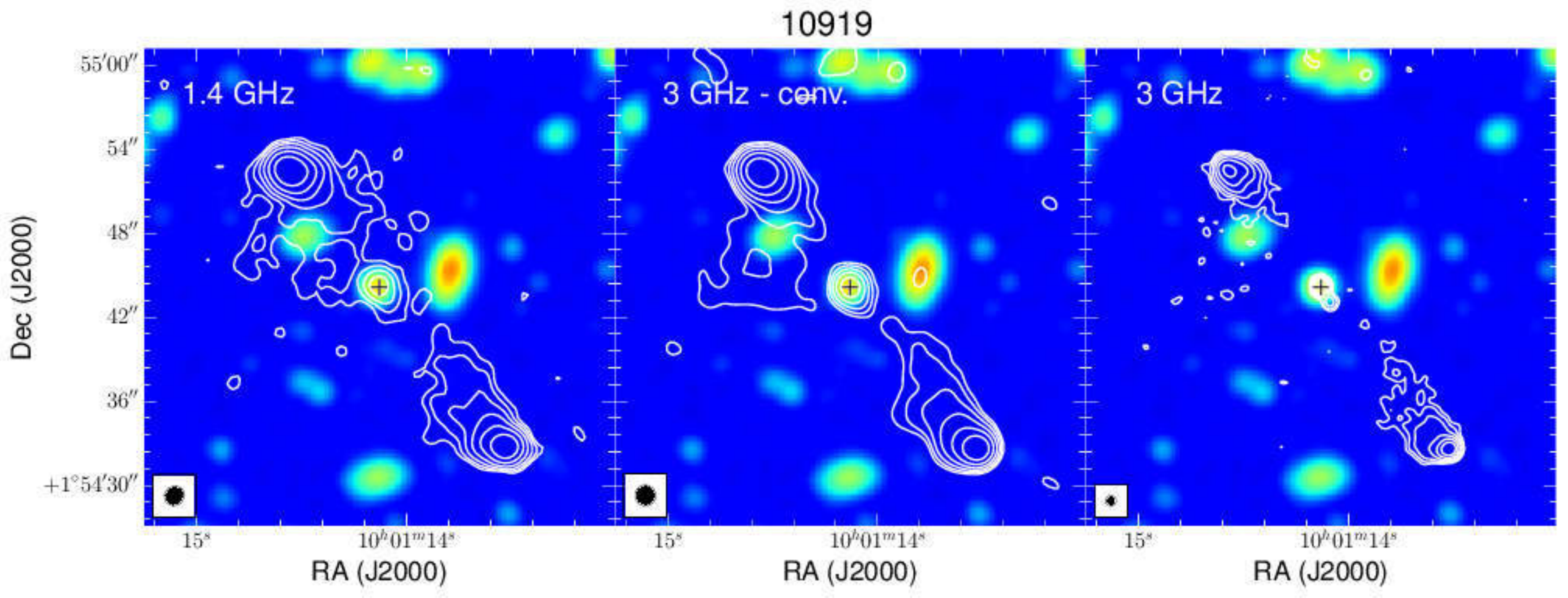}
            }
   \caption{(continued)
   }
              \label{fig:maps2}%
    \end{figure*}
\addtocounter{figure}{-1}
\begin{figure*}[!ht]
    \resizebox{\hsize}{!}
       {\includegraphics{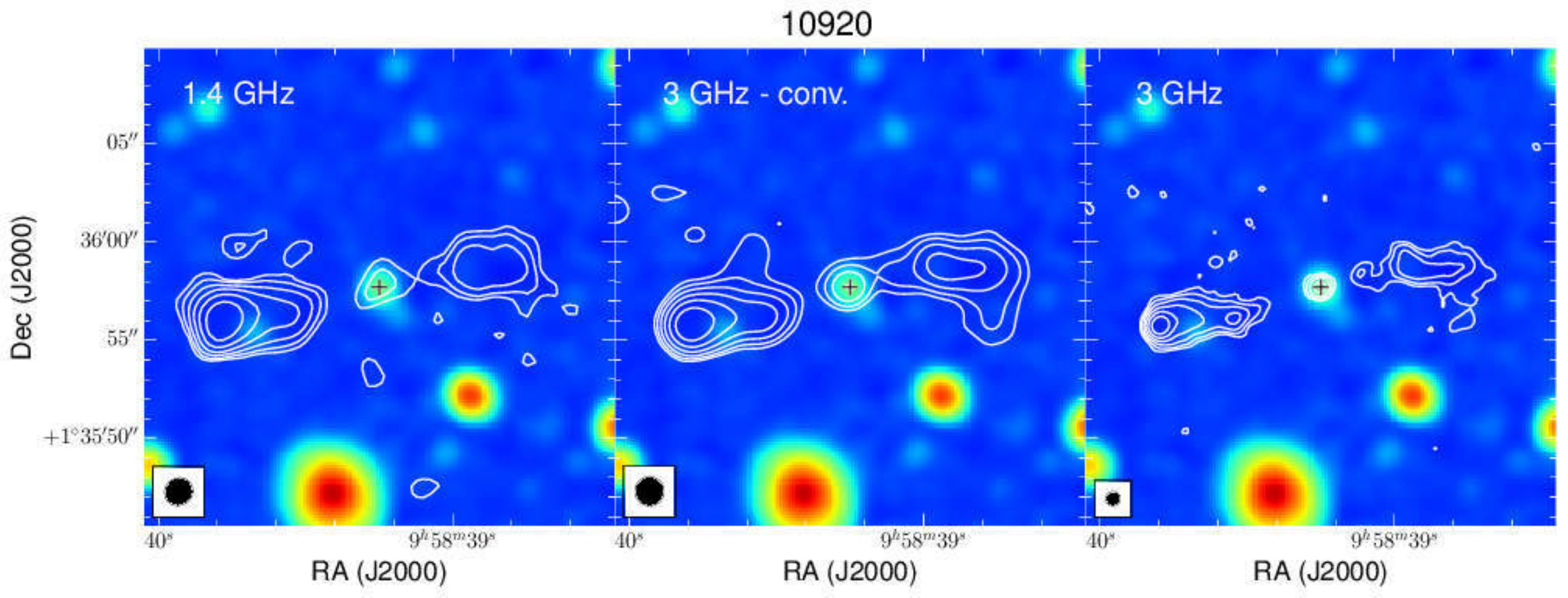}
        \includegraphics{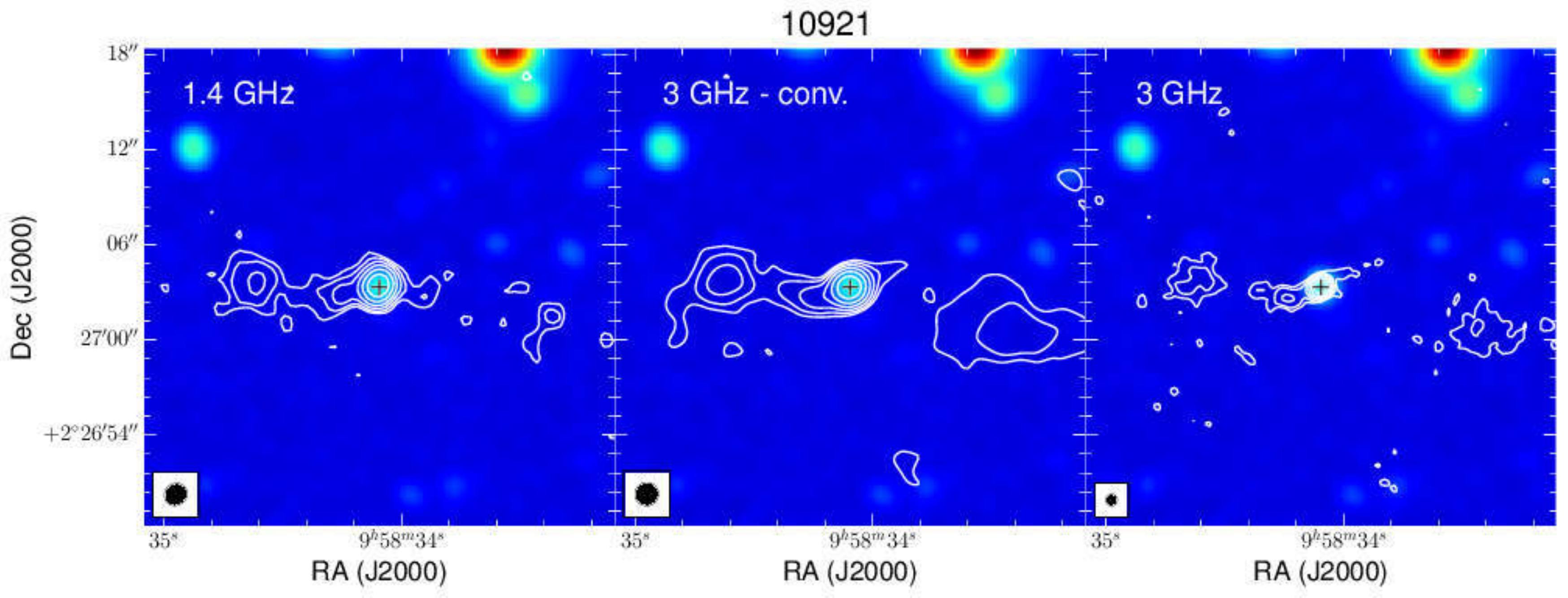}
            }
            \\ \\
 \resizebox{\hsize}{!}
{\includegraphics{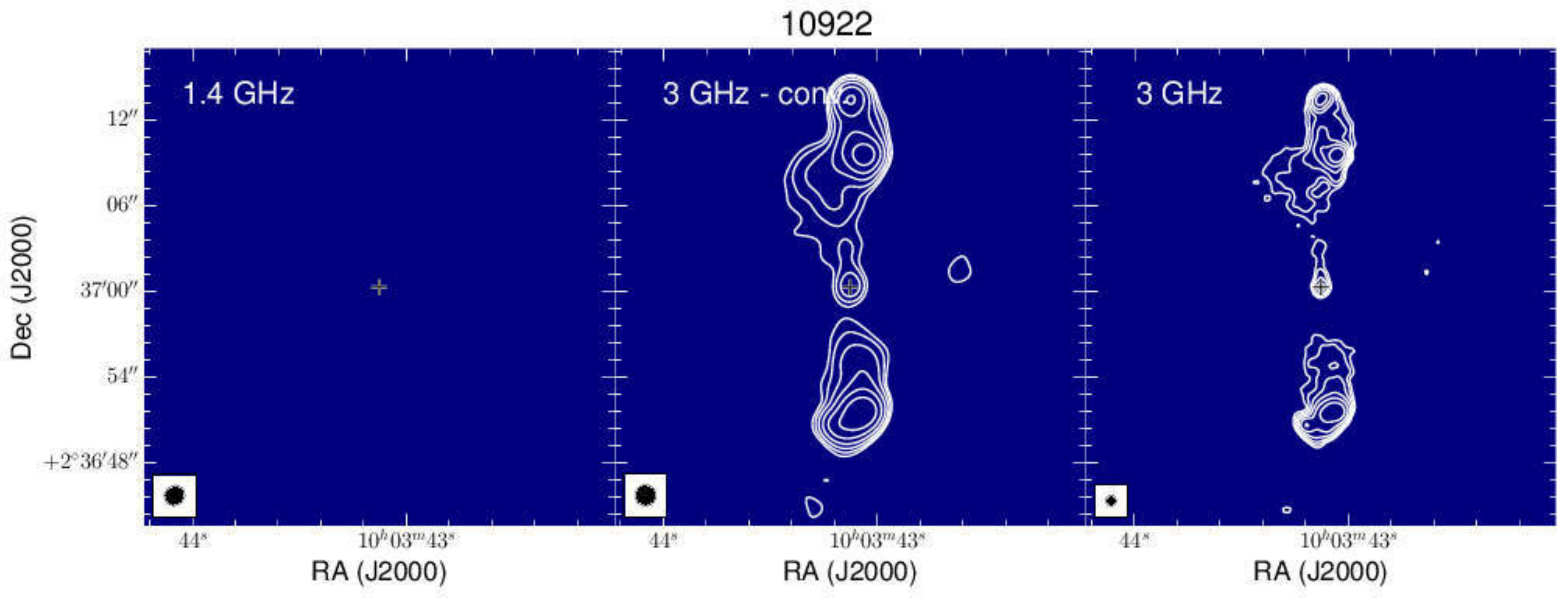}
 \includegraphics{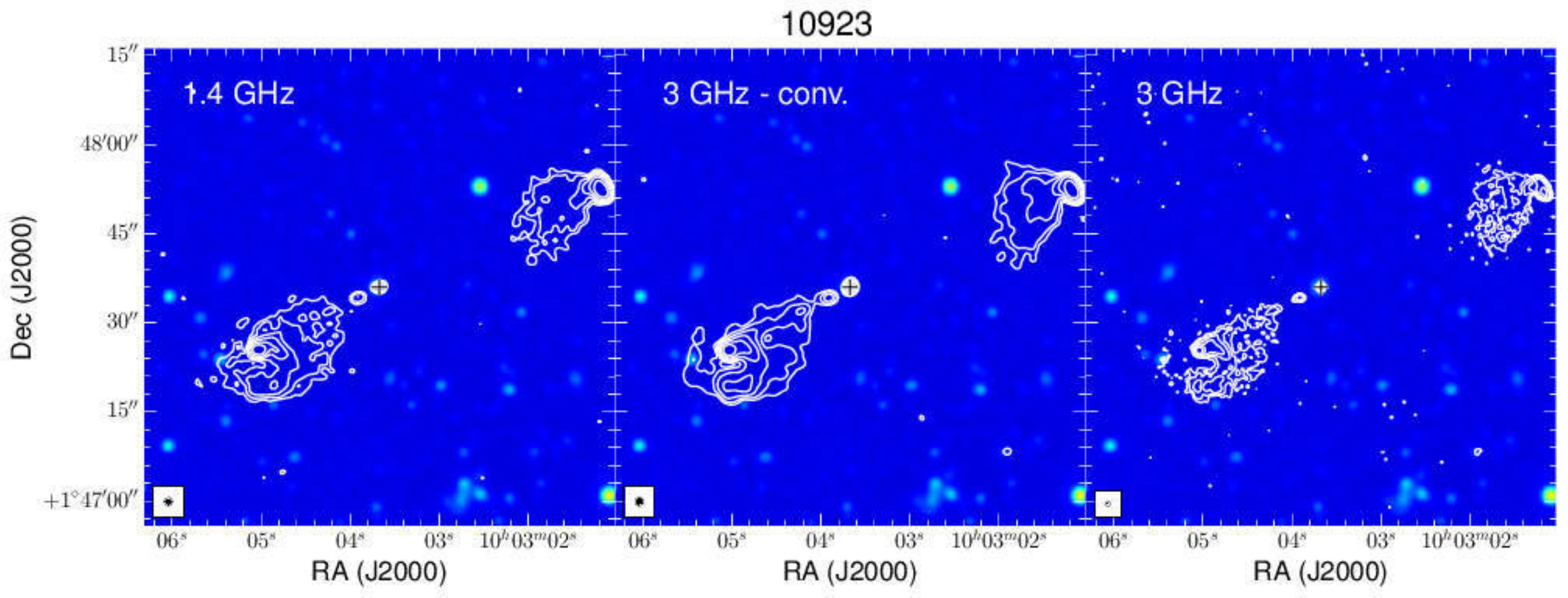}
            }
            \\ \\
  \resizebox{\hsize}{!}
 {\includegraphics{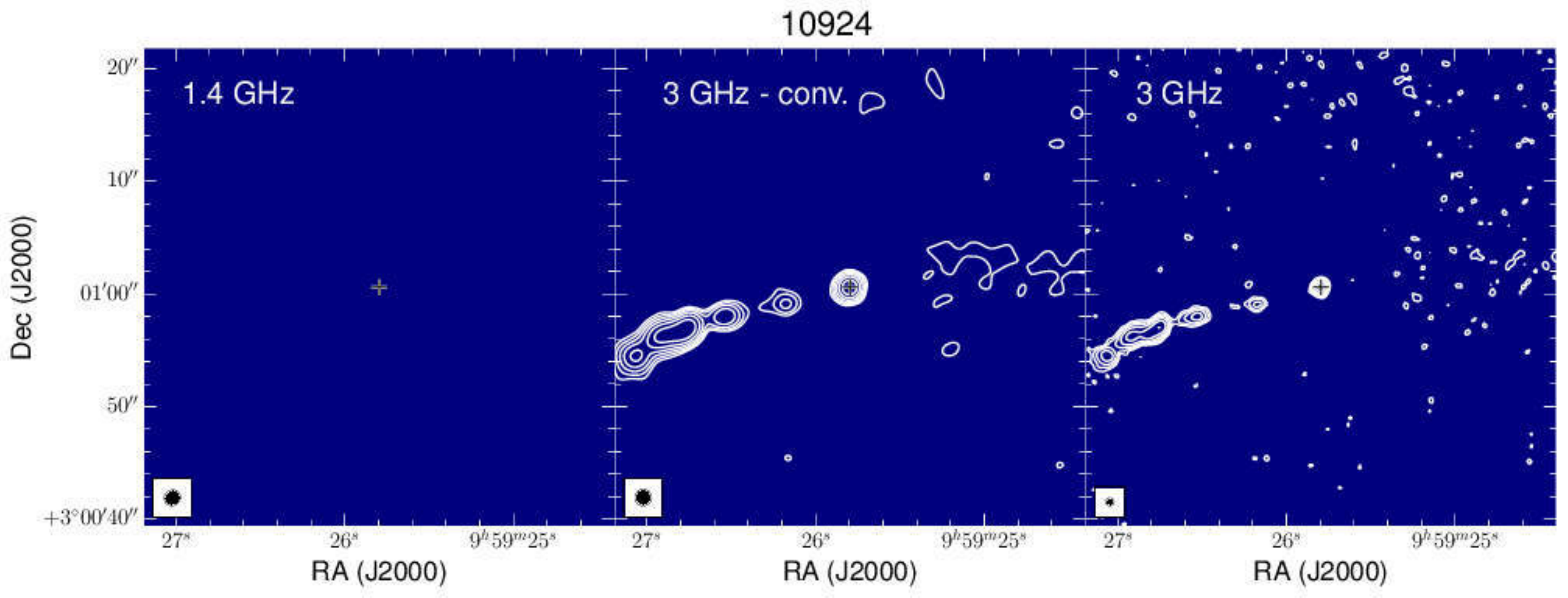}
    \includegraphics{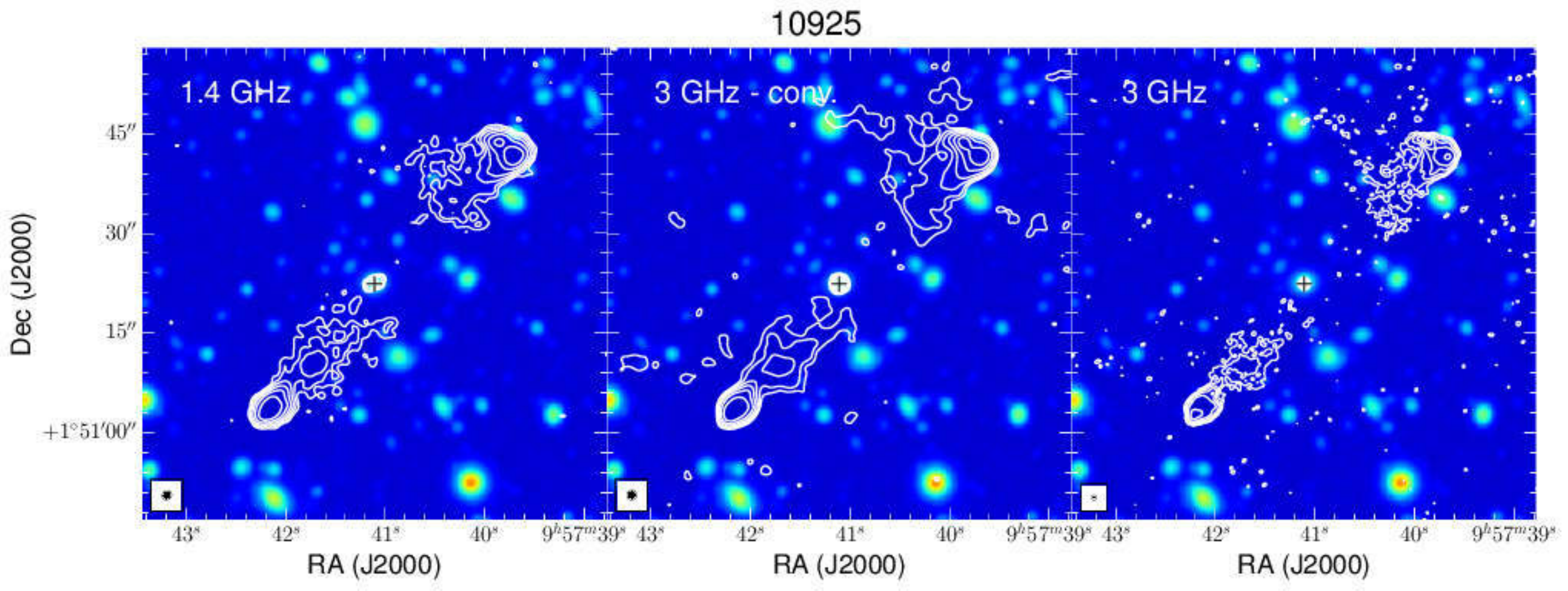}
            }
             \\ \\ 
      \resizebox{\hsize}{!}
       {\includegraphics{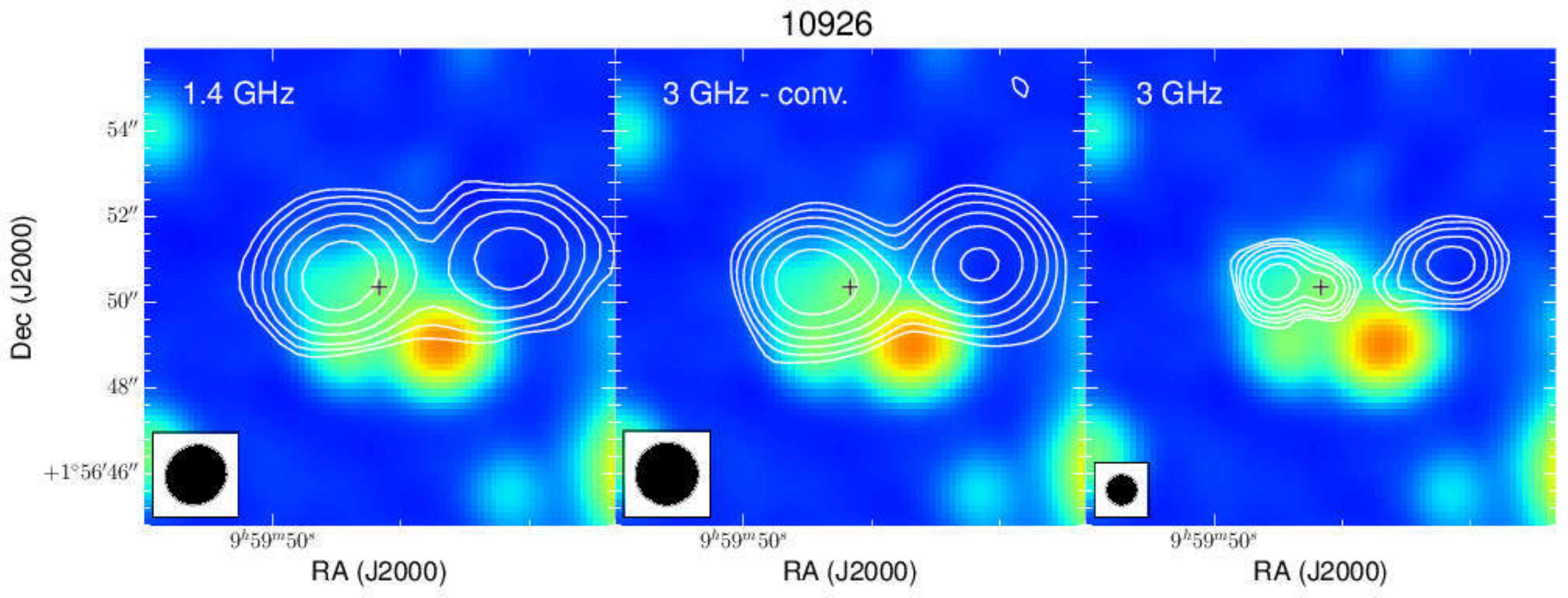}
        \includegraphics{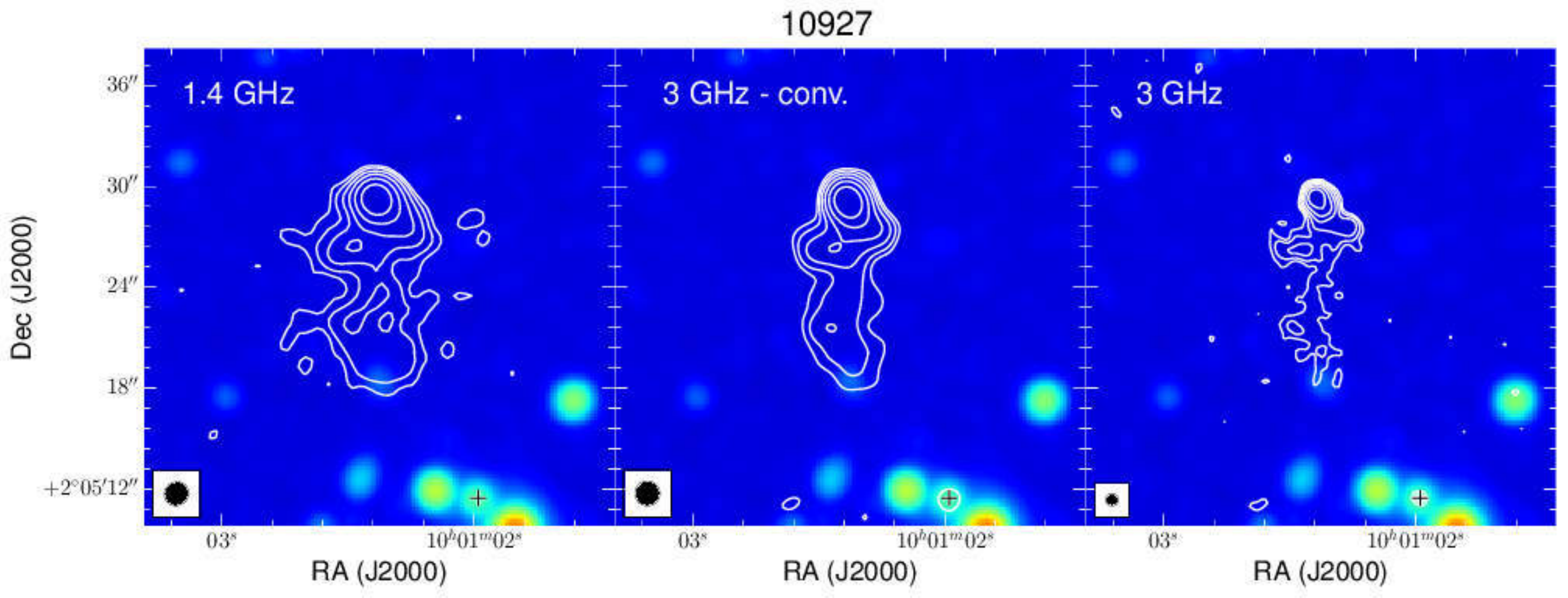}
            }
\\ \\
 \resizebox{\hsize}{!}
{\includegraphics{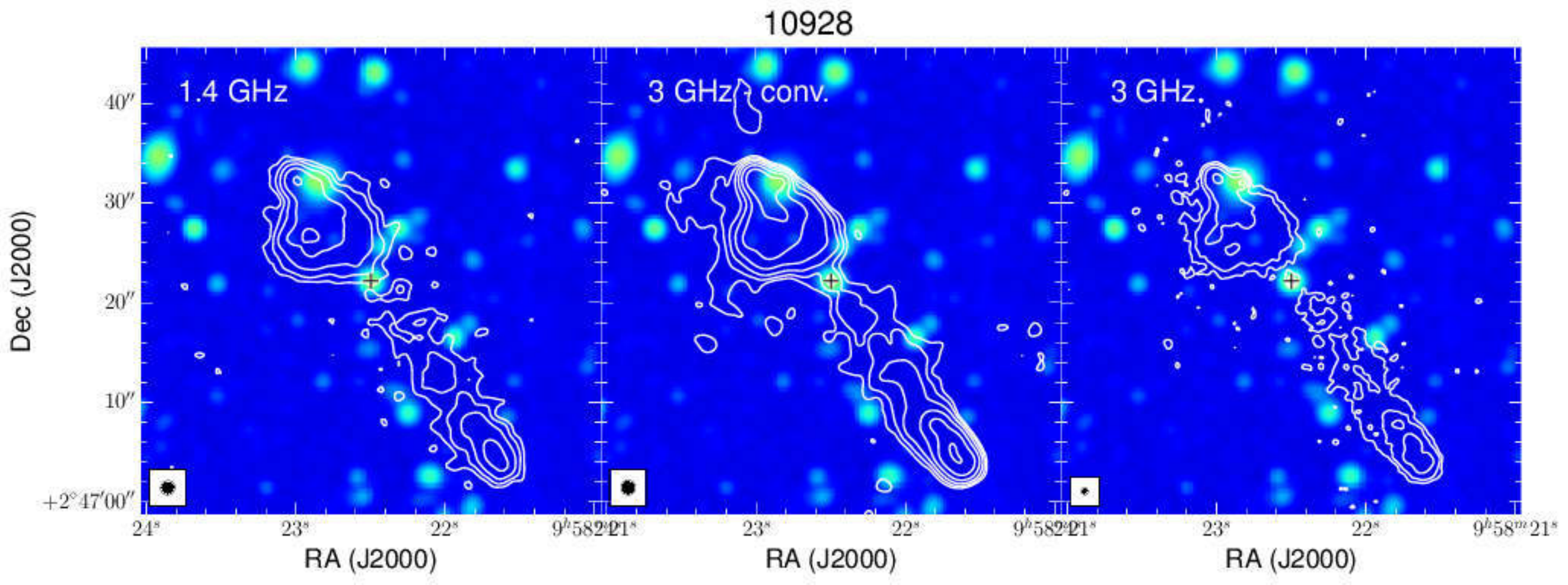}
 \includegraphics{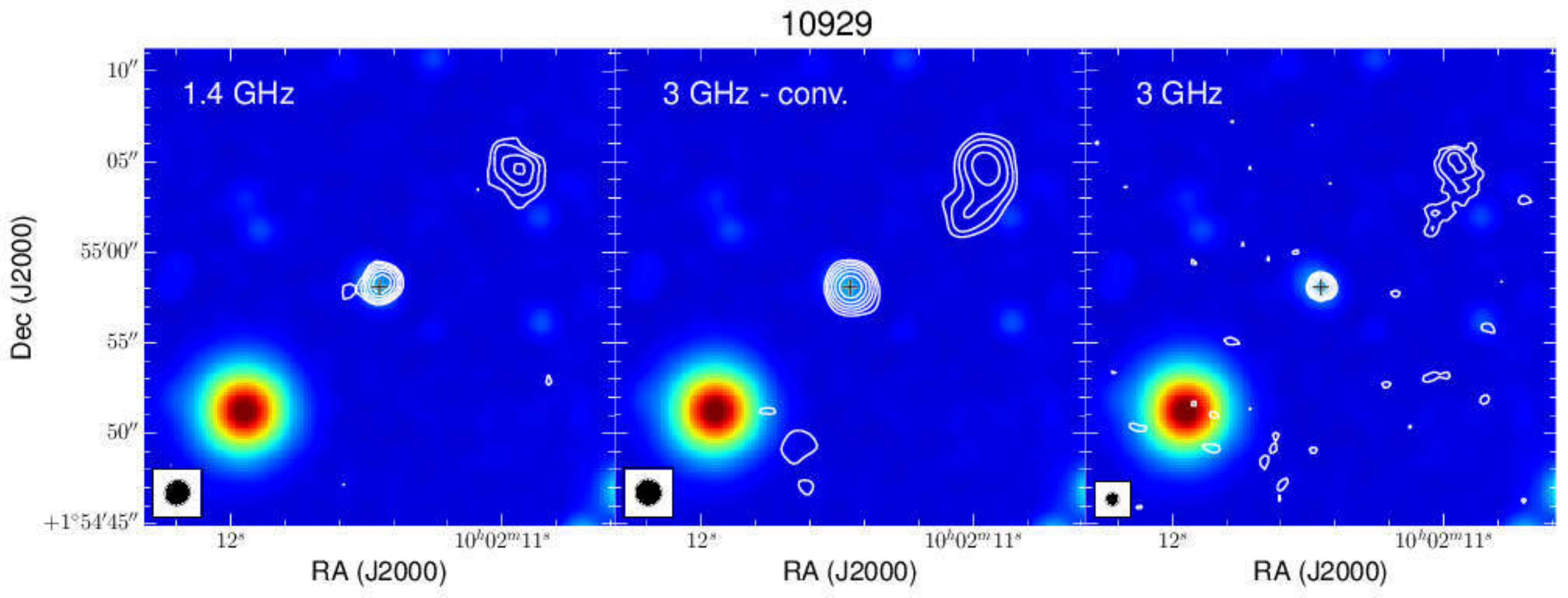}
            }
   \caption{(continued)
   }
              \label{fig:maps2}%
    \end{figure*}
\addtocounter{figure}{-1}
\begin{figure*}[!ht]
    \resizebox{\hsize}{!}
       {\includegraphics{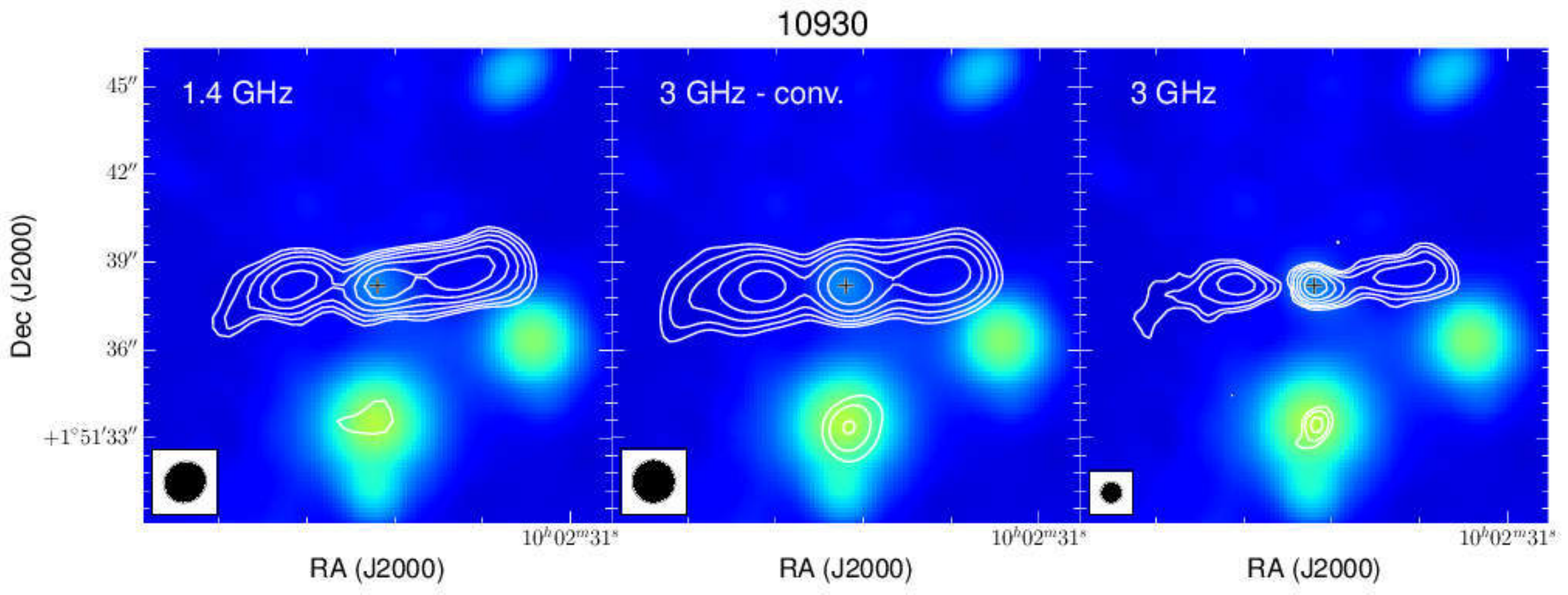}
        \includegraphics{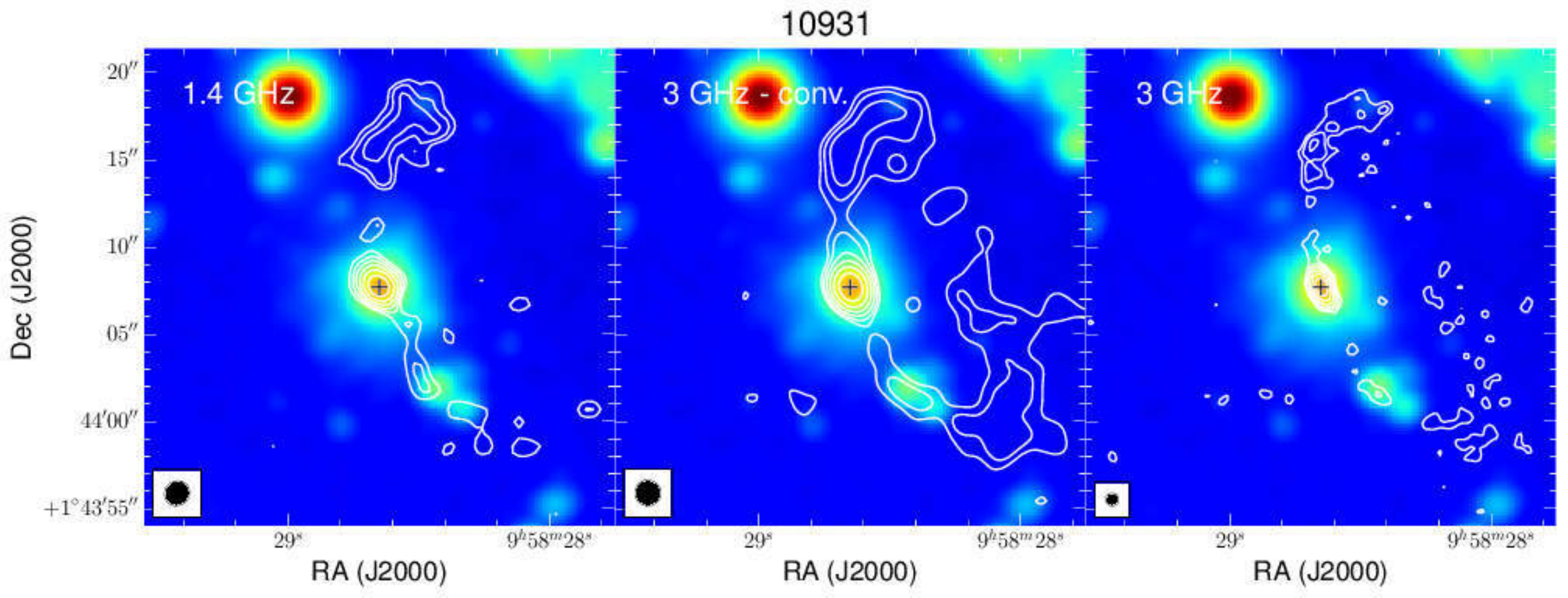}
            }
            \\ \\
 \resizebox{\hsize}{!}
{\includegraphics{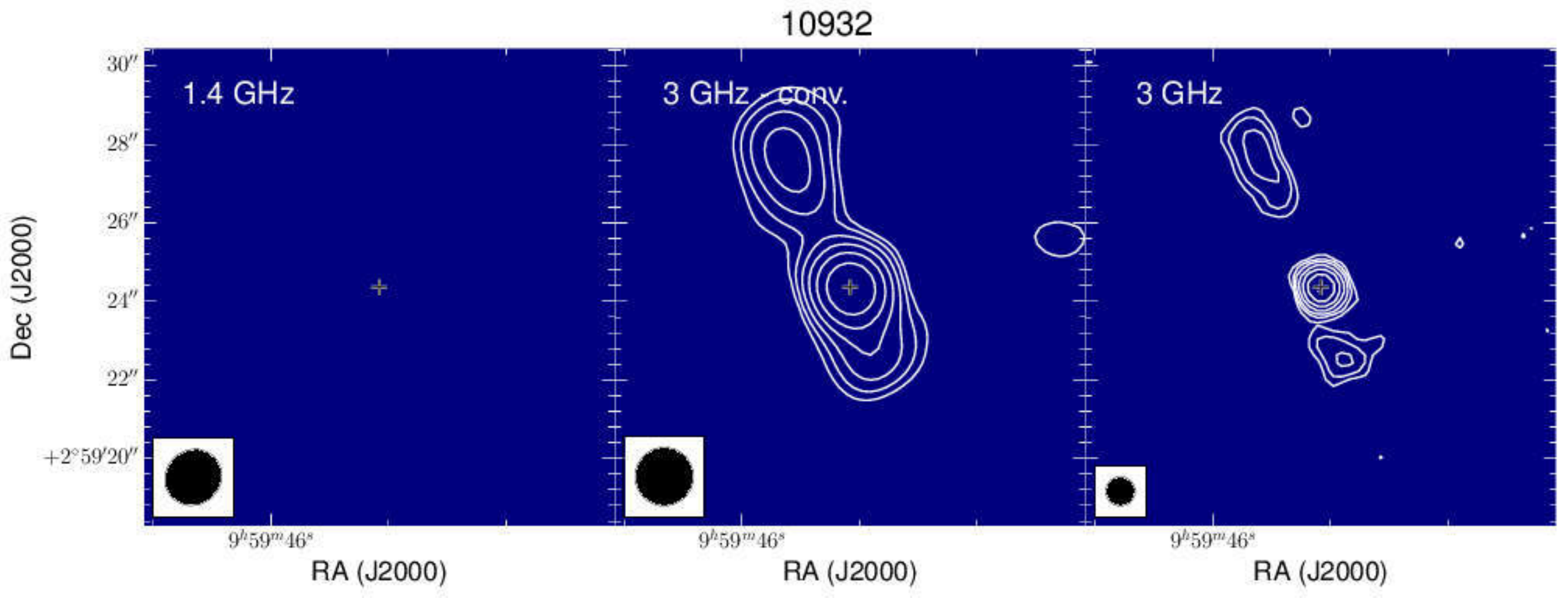}
 \includegraphics{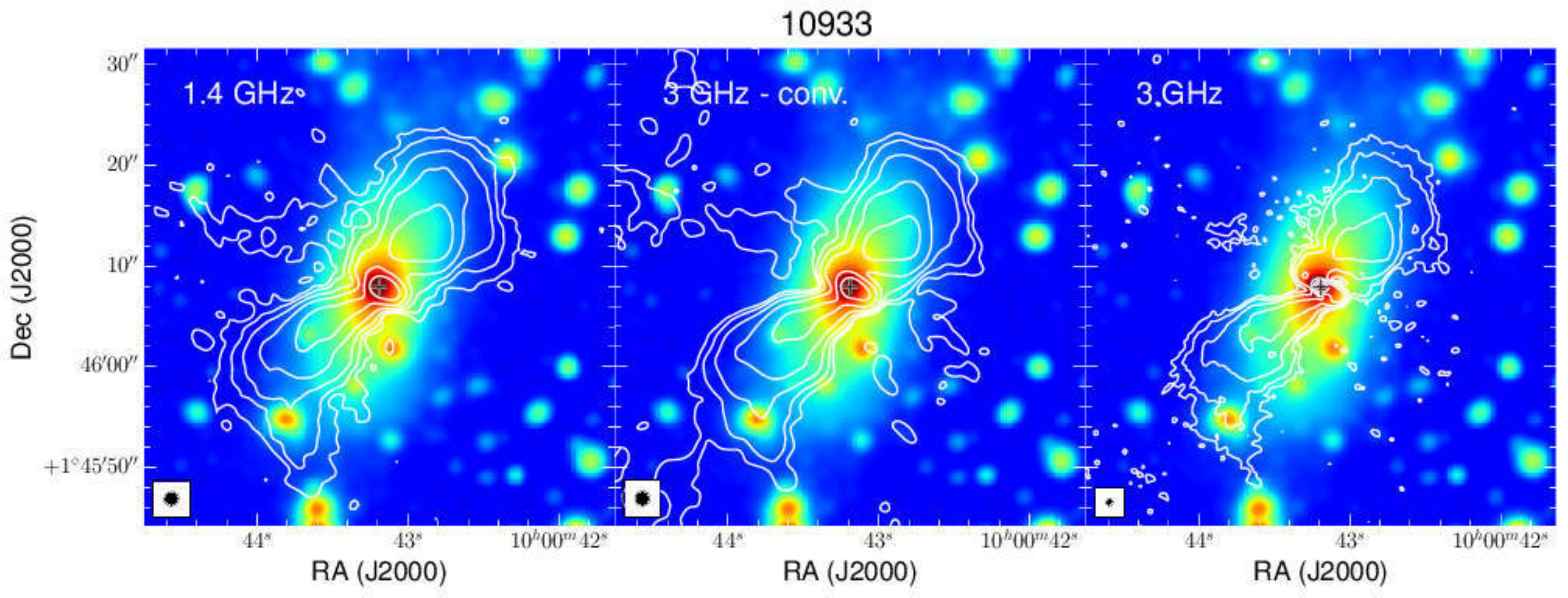}
            }
            \\ \\
  \resizebox{\hsize}{!}
 {\includegraphics{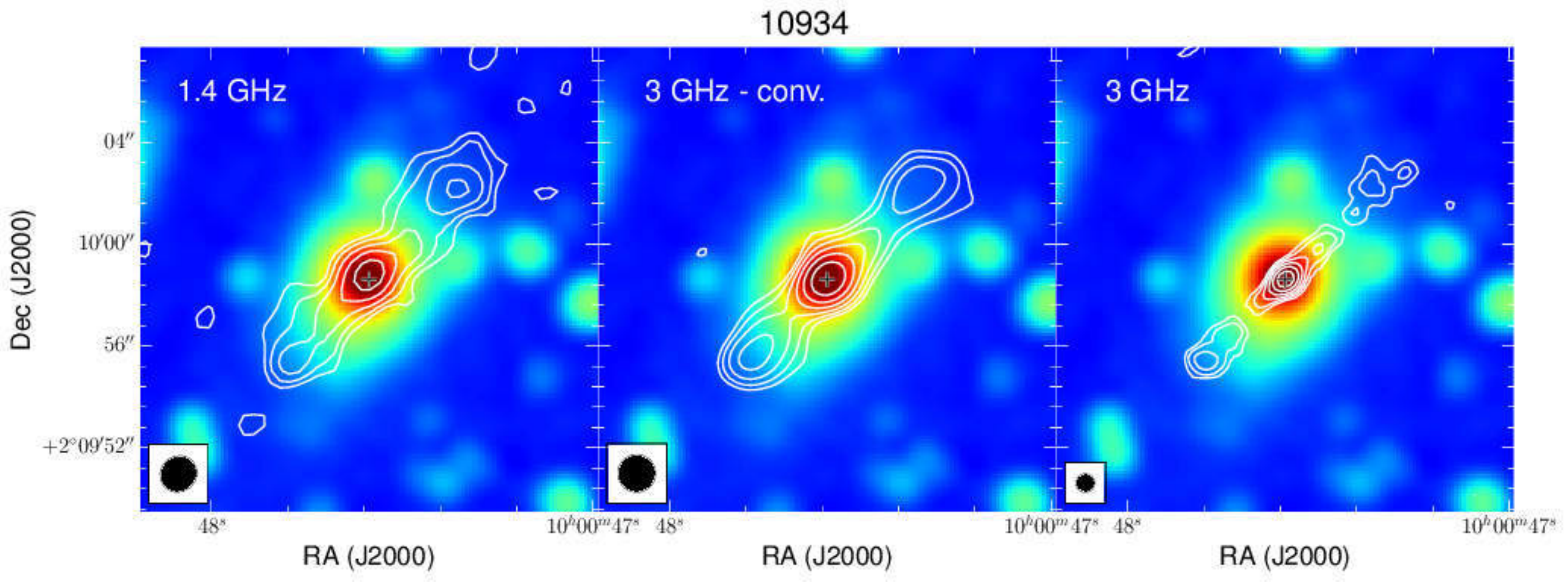}
    \includegraphics{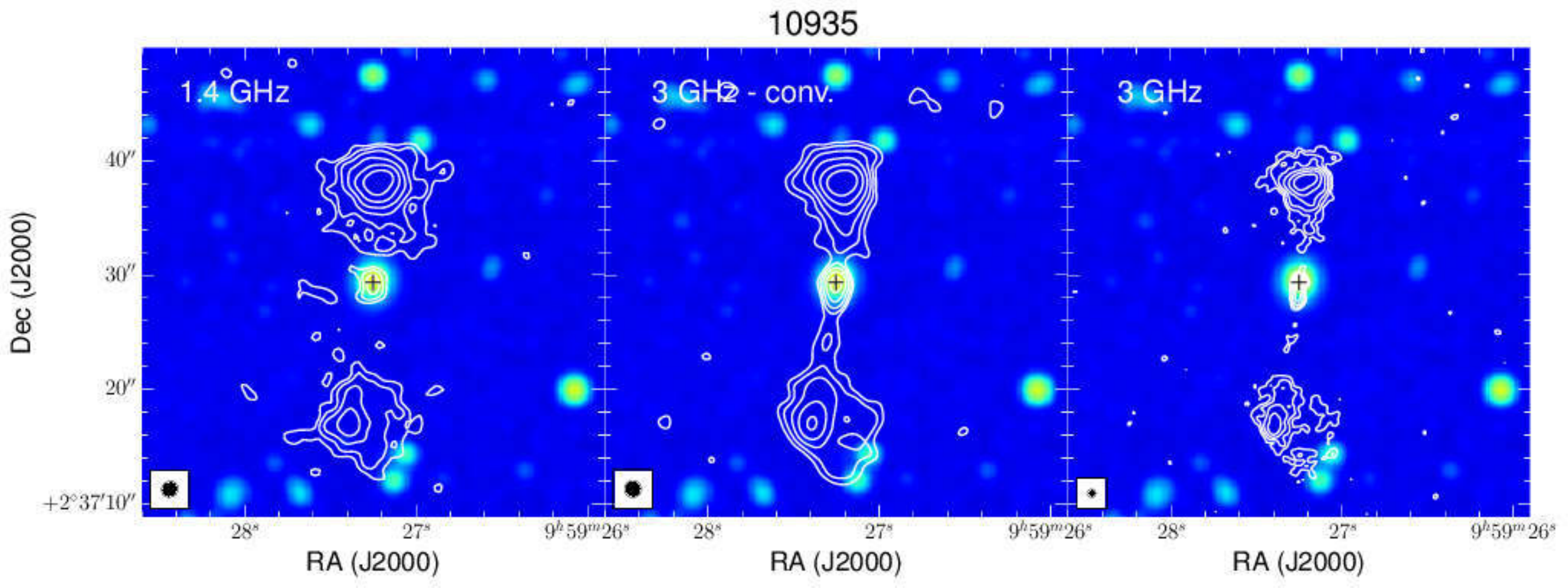}
            }
             \\ \\ 
      \resizebox{\hsize}{!}
       {\includegraphics{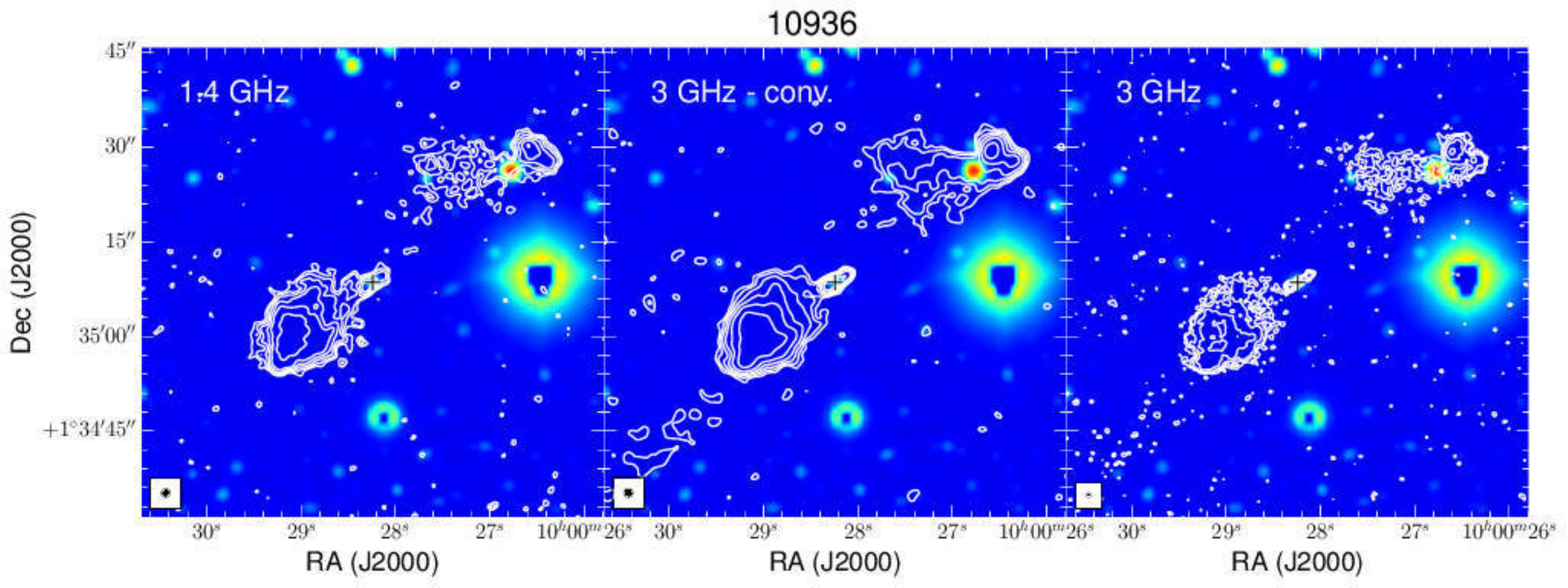}
        \includegraphics{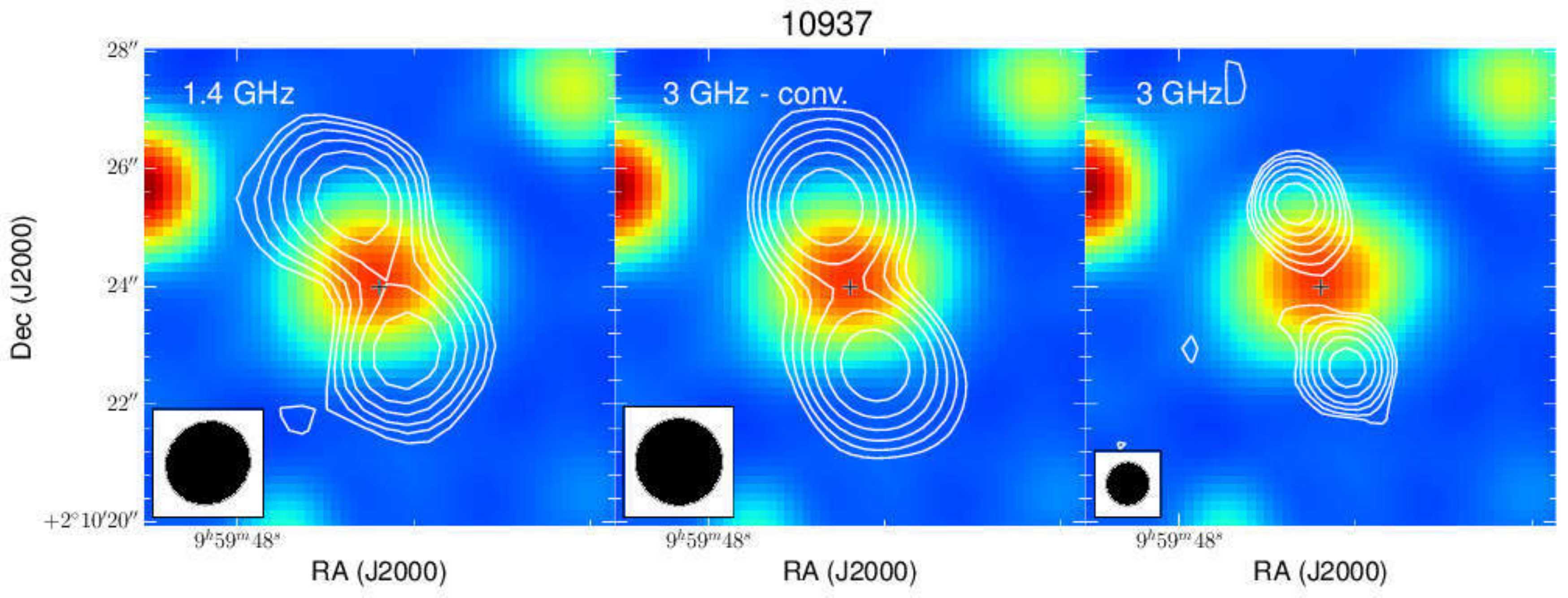}
            }
\\ \\
 \resizebox{\hsize}{!}
{\includegraphics{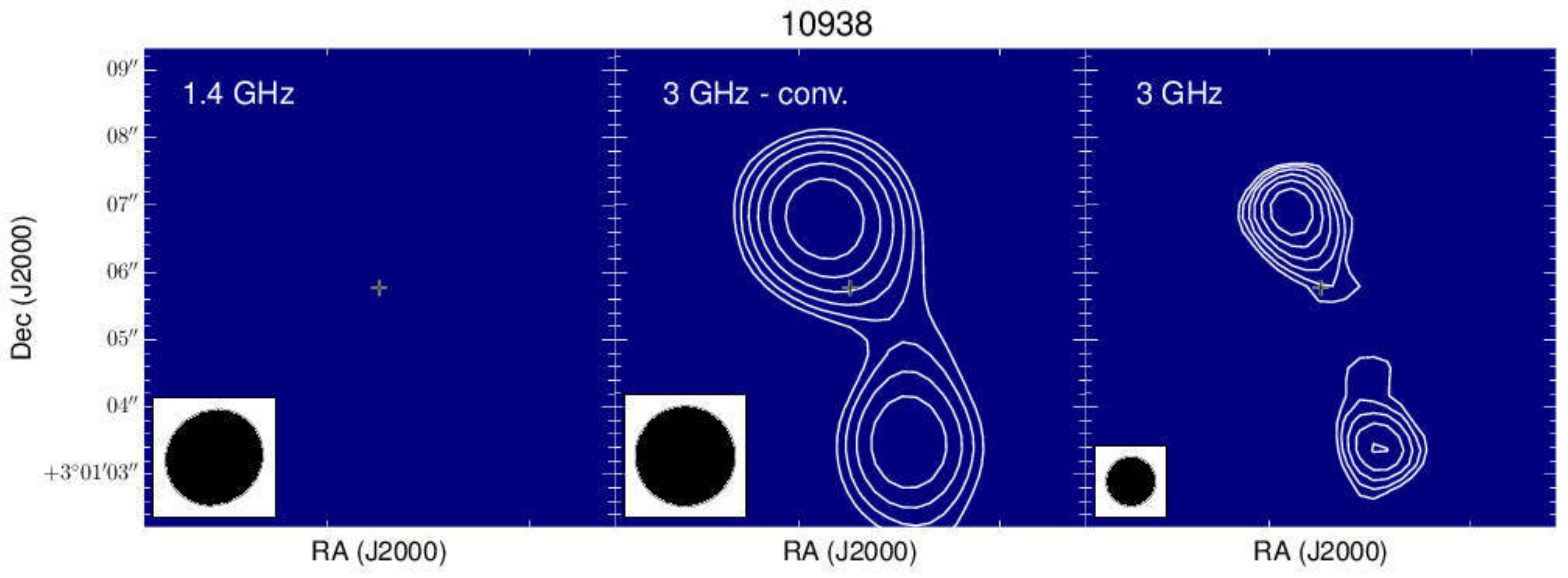}
 \includegraphics{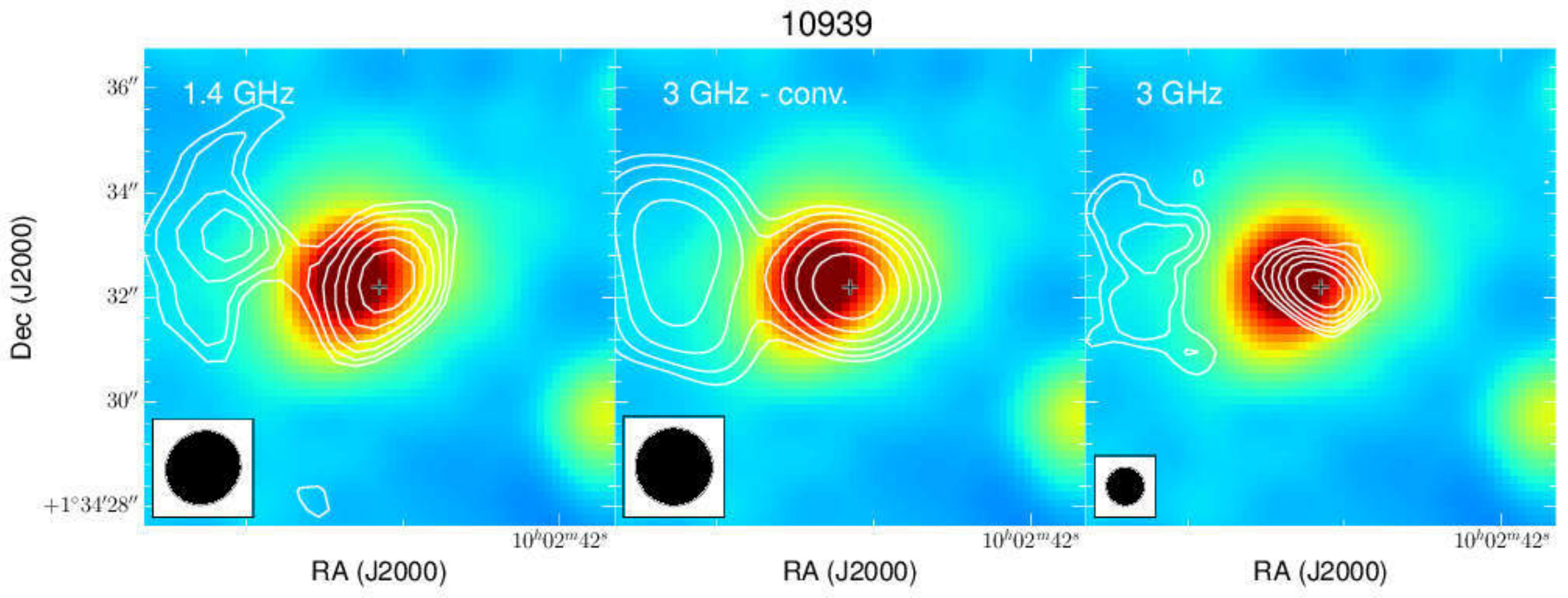}
            }
   \caption{(continued)
   }
              \label{fig:maps2}%
    \end{figure*}
\addtocounter{figure}{-1}
\begin{figure*}[!ht]
    \resizebox{\hsize}{!}
       {\includegraphics{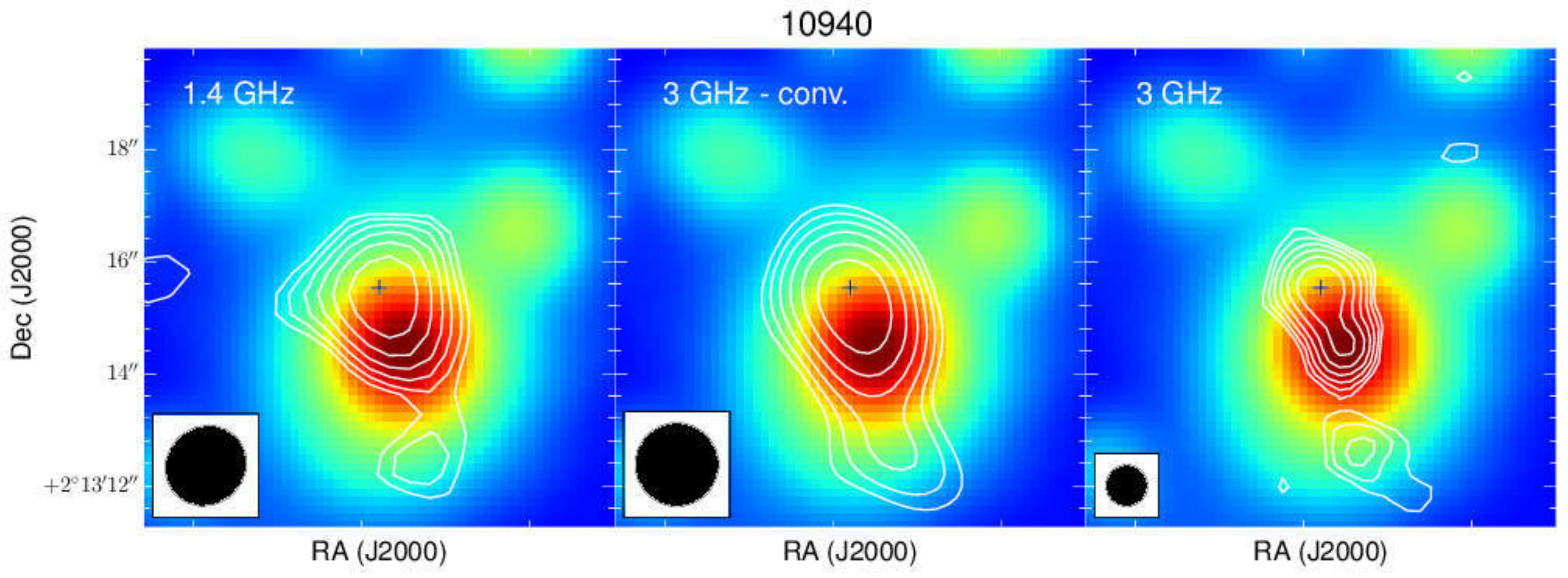}
        \includegraphics{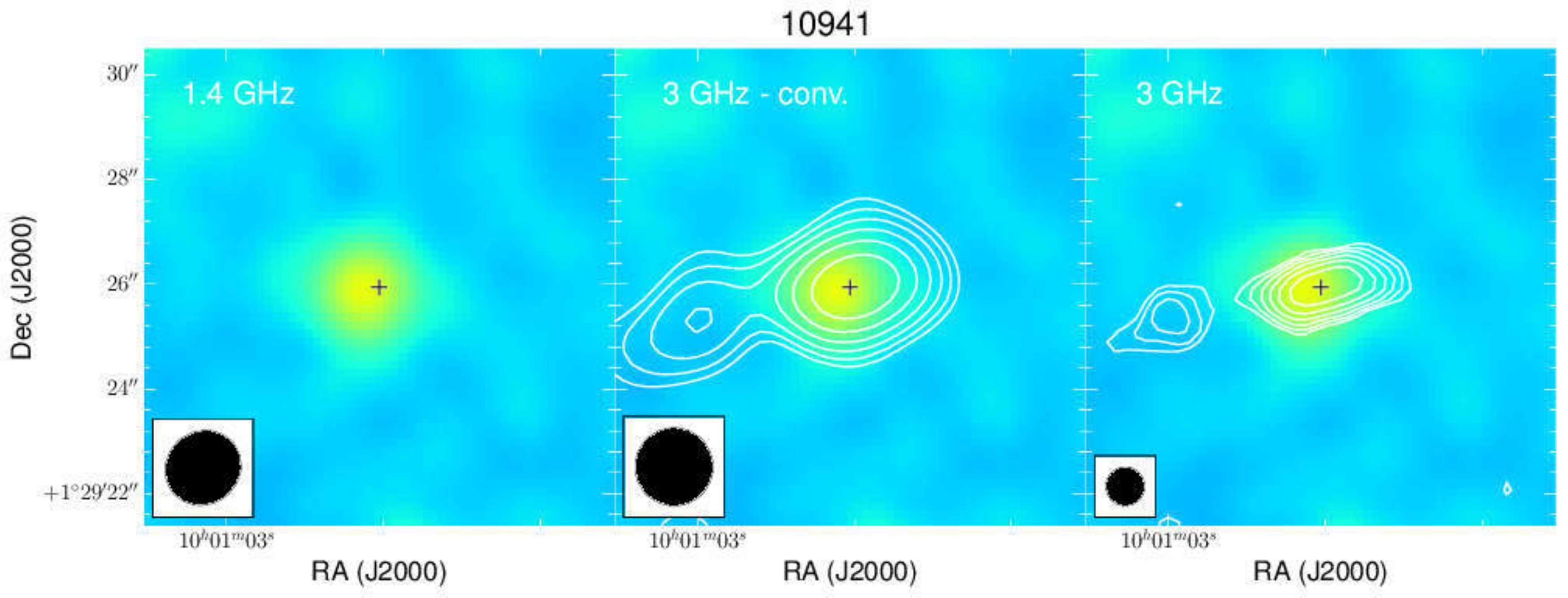}
            }
            \\ \\
 \resizebox{\hsize}{!}
{\includegraphics{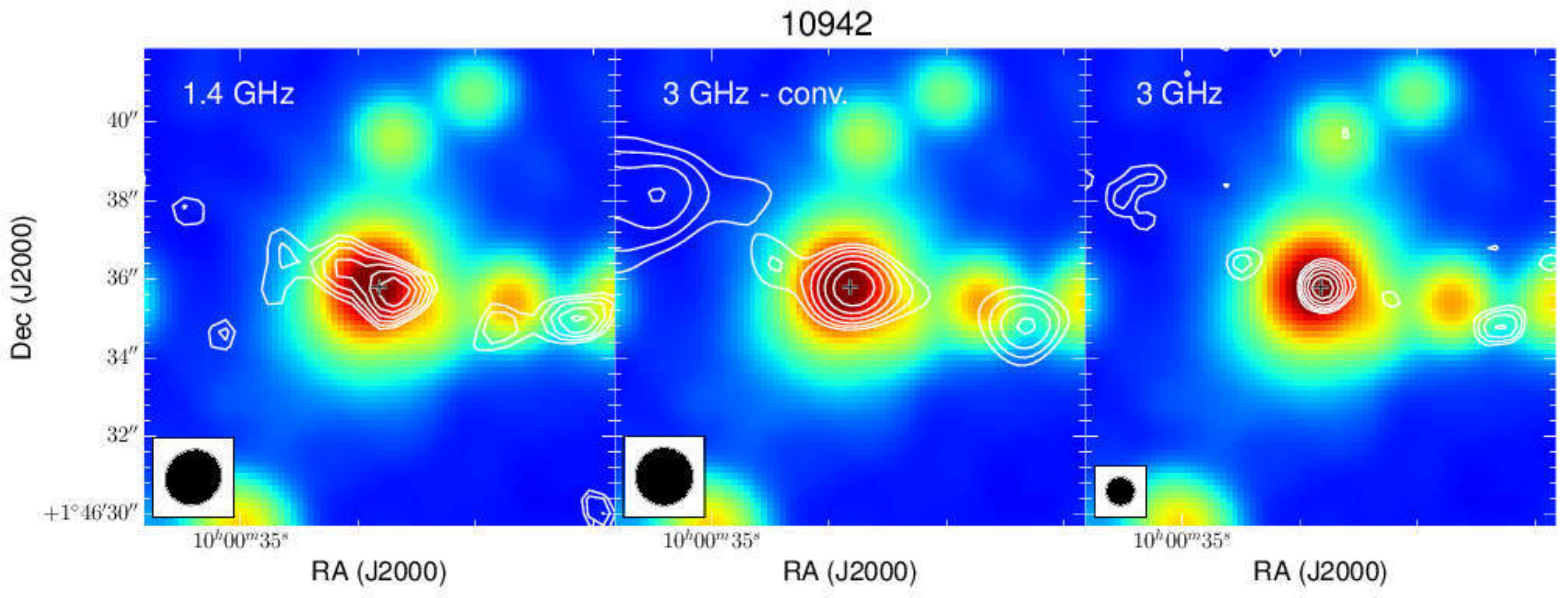}
 \includegraphics{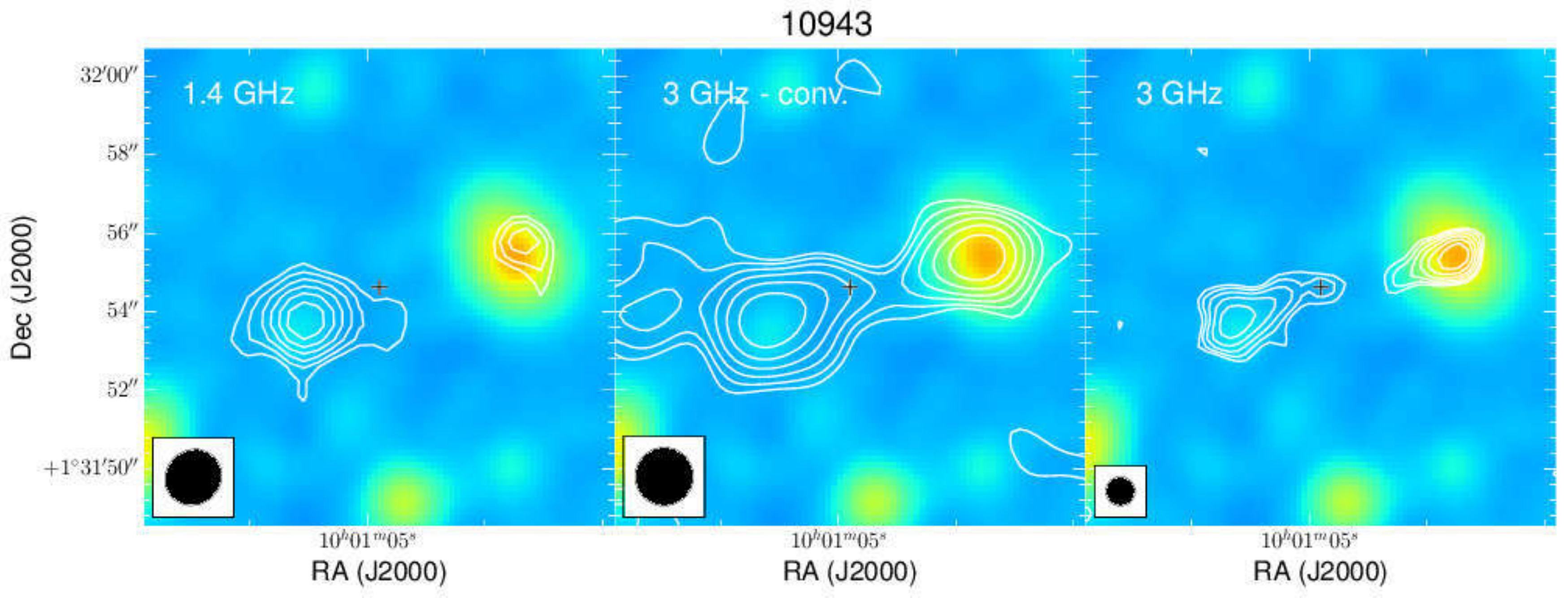}
            }
            \\ \\
  \resizebox{\hsize}{!}
 {\includegraphics{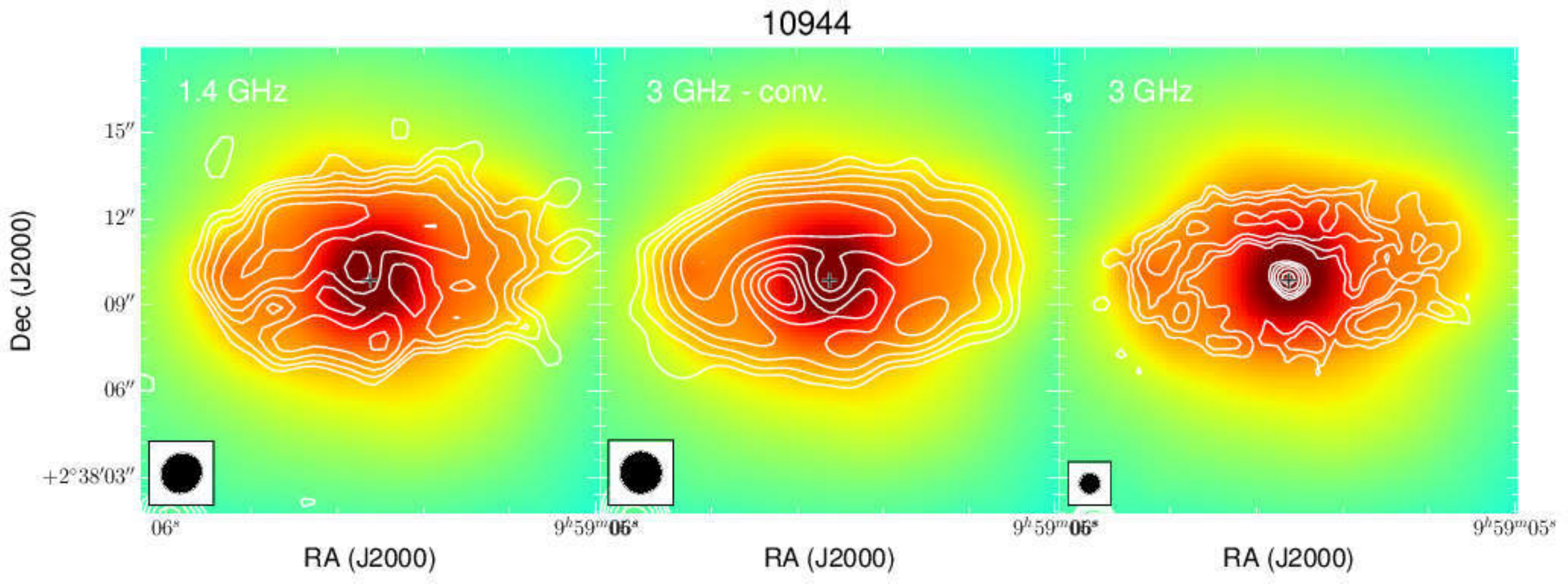}
    \includegraphics{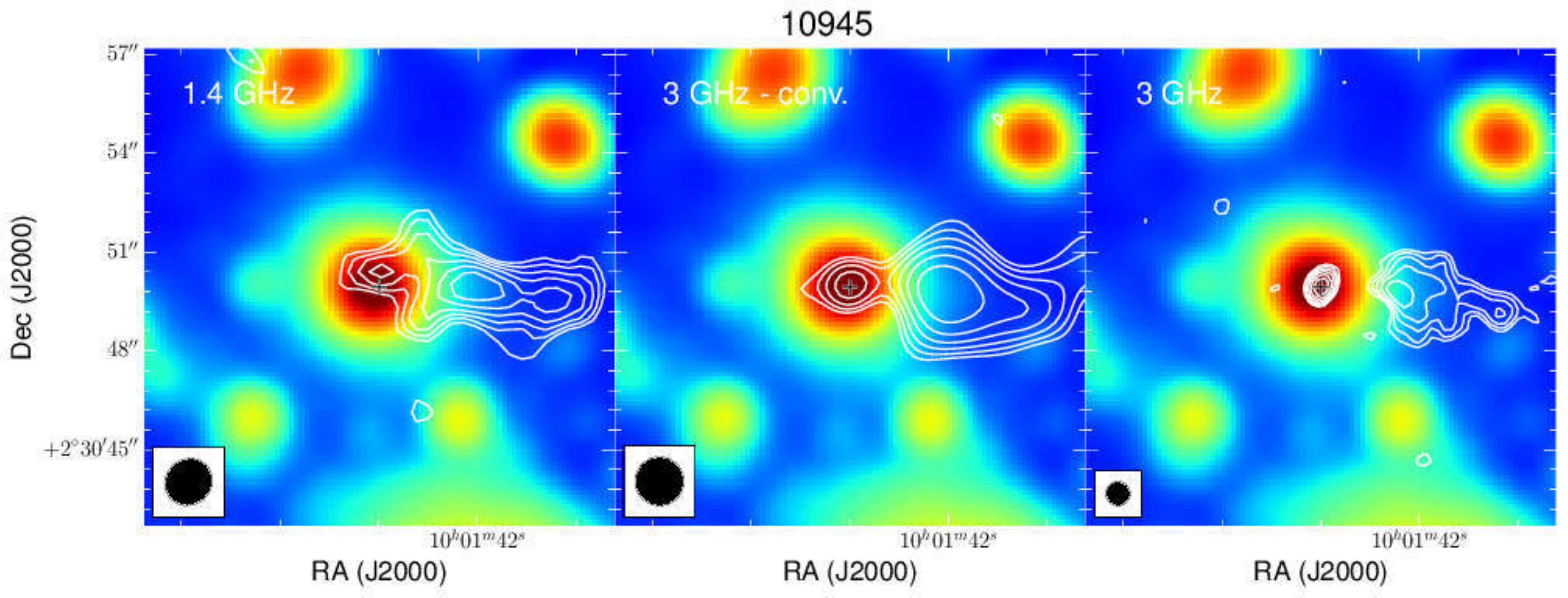}
            }
             \\ \\ 
      \resizebox{\hsize}{!}
       {\includegraphics{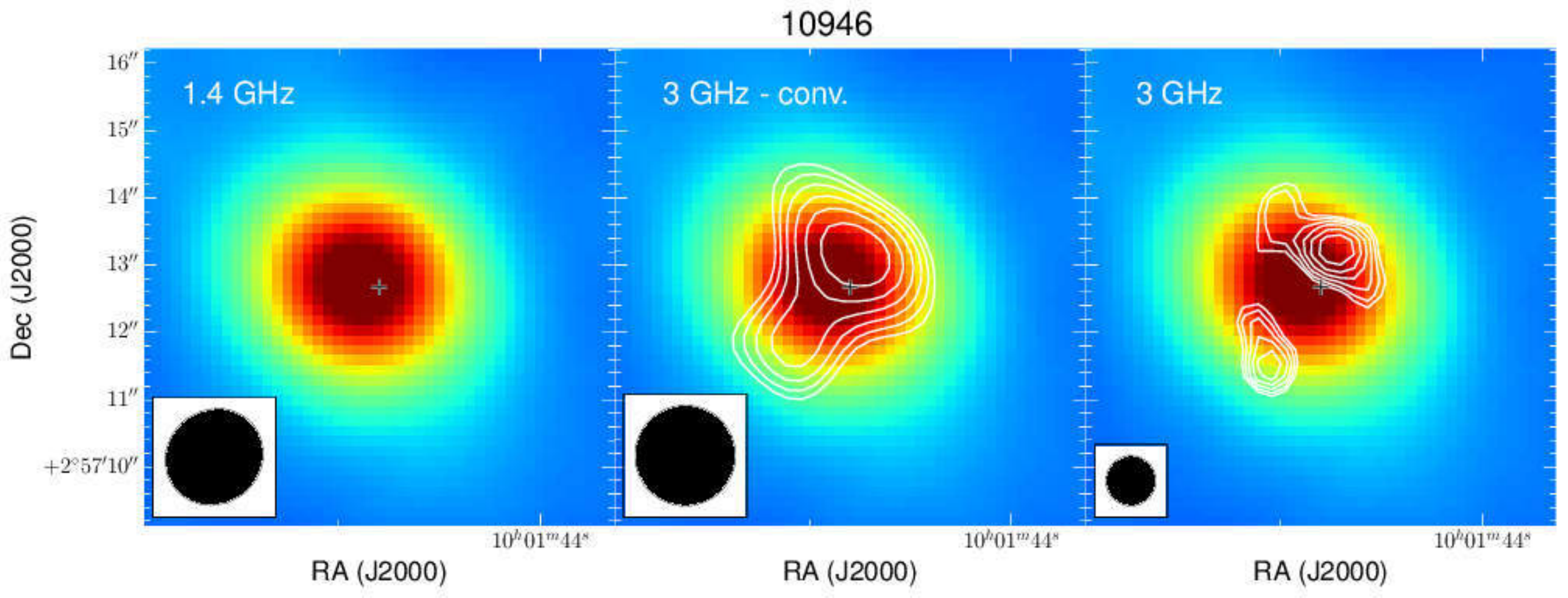}
        \includegraphics{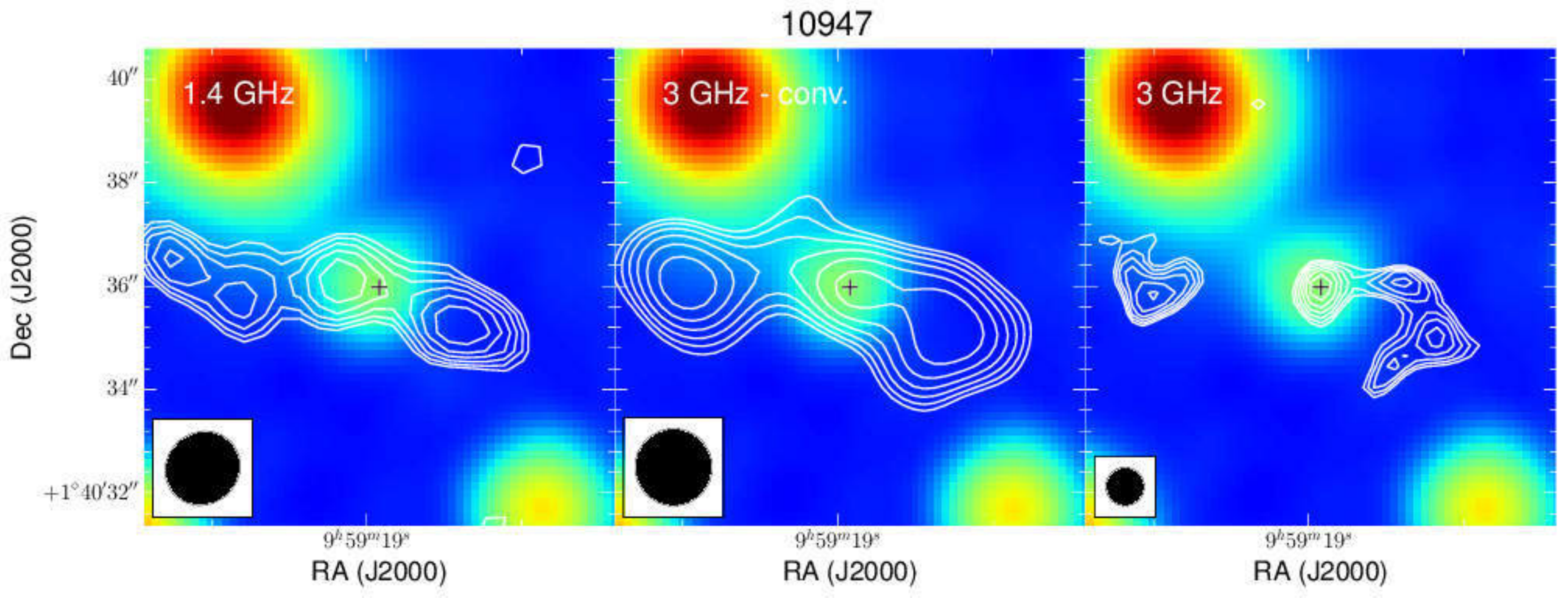}
            }
\\ \\
 \resizebox{\hsize}{!}
{\includegraphics{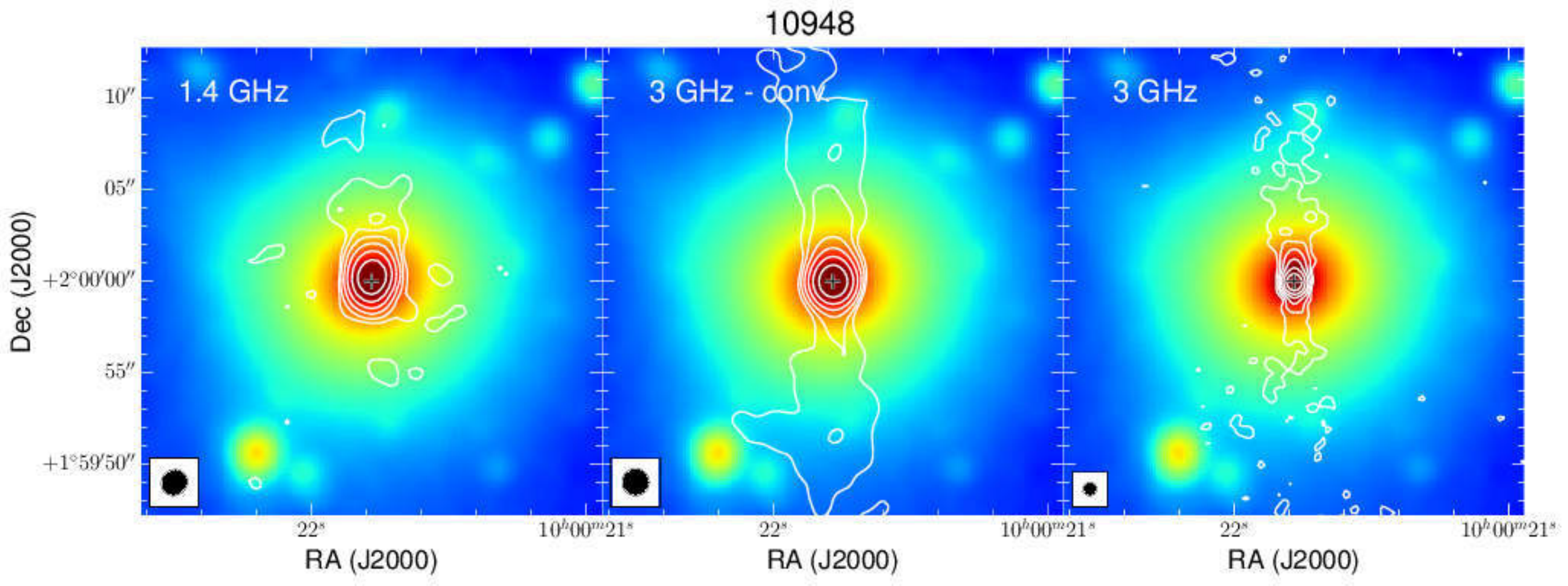}
 \includegraphics{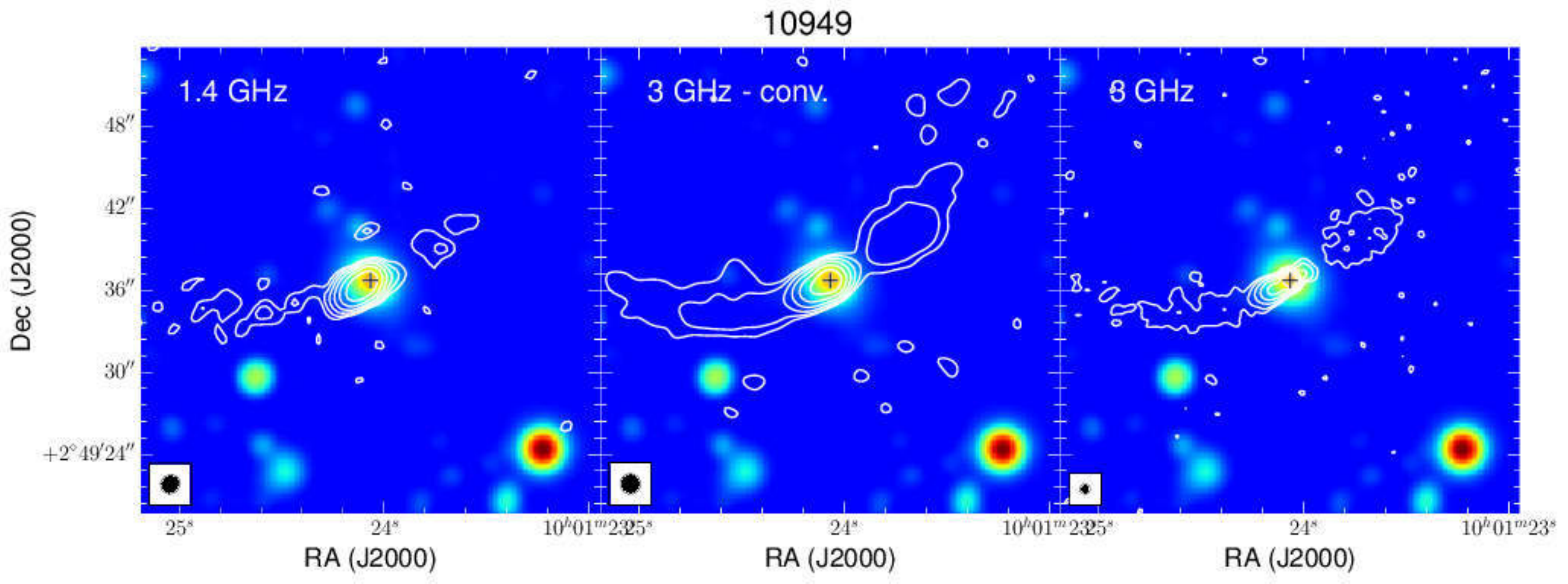}
            }
   \caption{(continued) 
      }
              \label{fig:maps2}%
    \end{figure*}
\addtocounter{figure}{-1}
\begin{figure*}[!ht]
    \resizebox{\hsize}{!}
       {\includegraphics{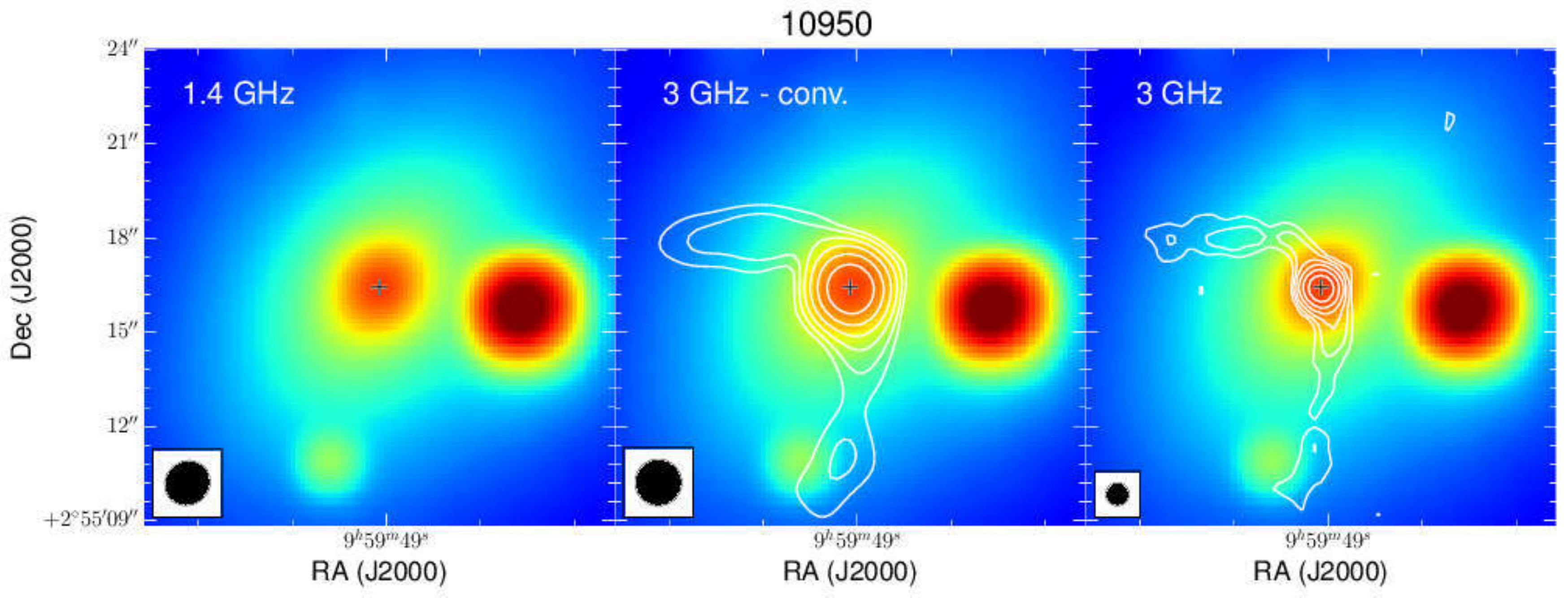}
        \includegraphics{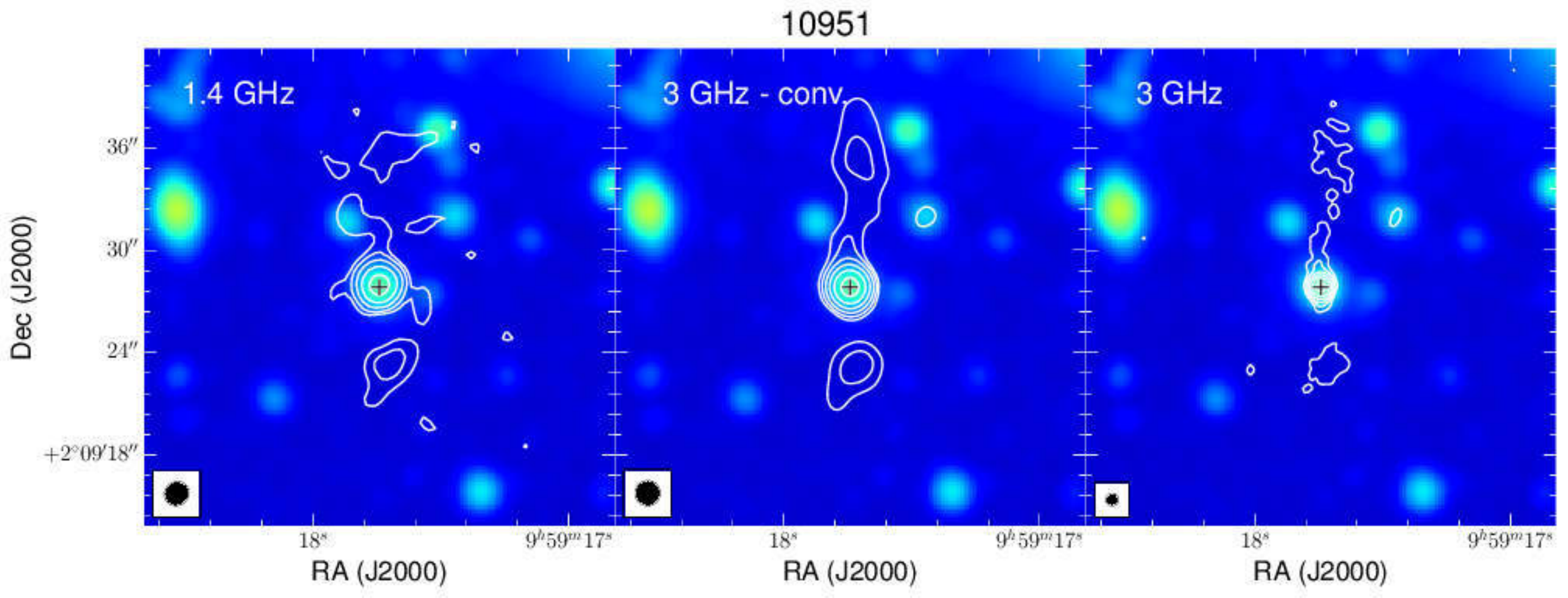}
            }
            \\ \\
 \resizebox{\hsize}{!}
{\includegraphics{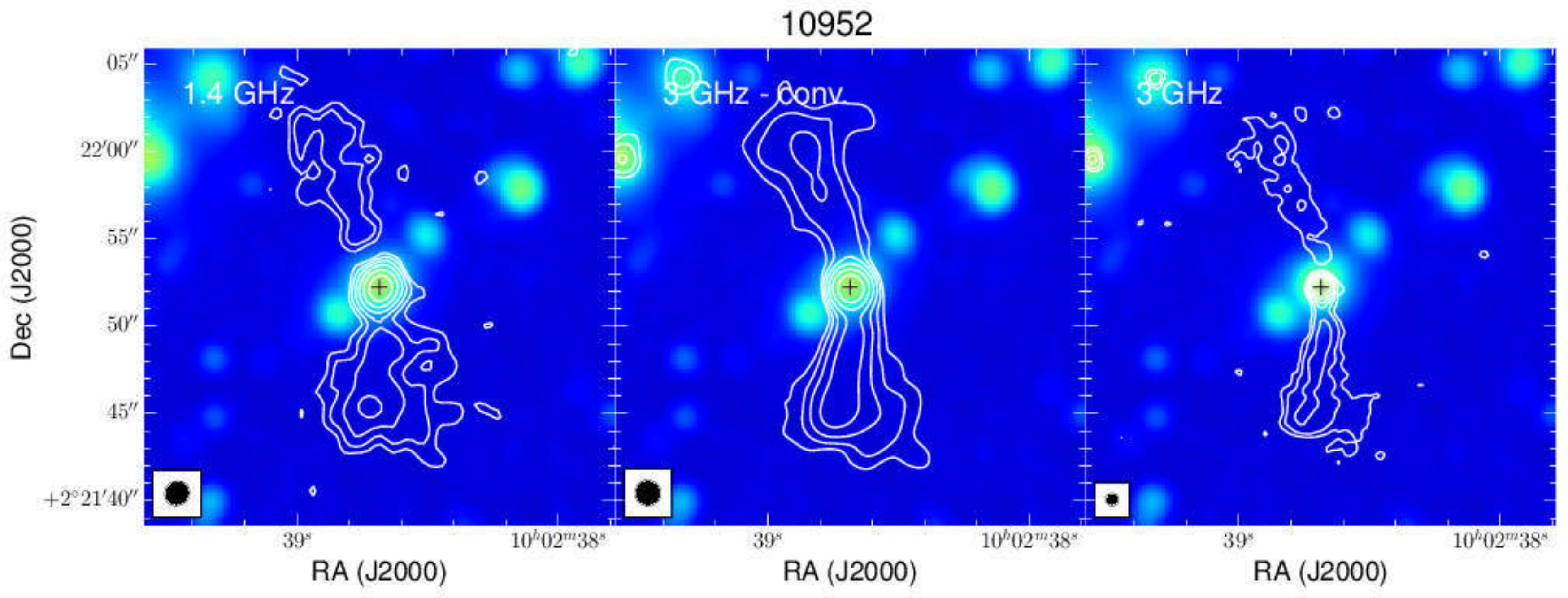}
 \includegraphics{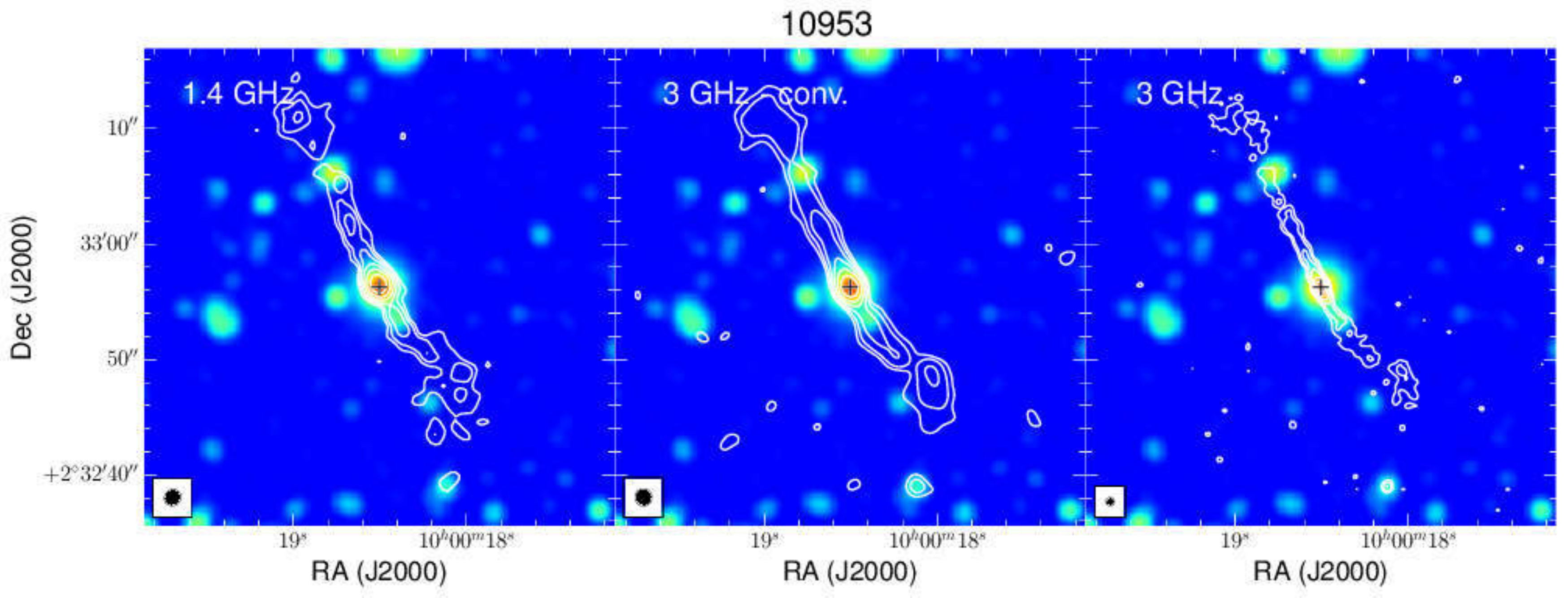}
            }
            \\ \\
  \resizebox{\hsize}{!}
 {\includegraphics{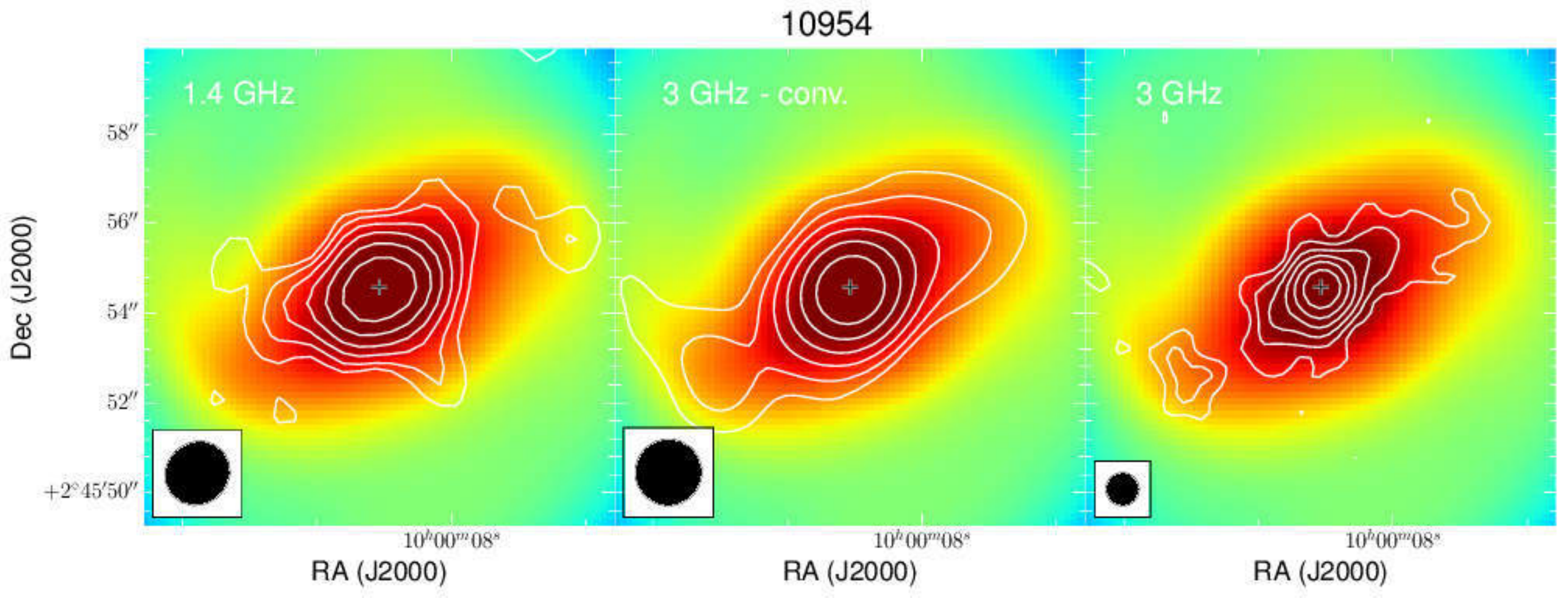}
    \includegraphics{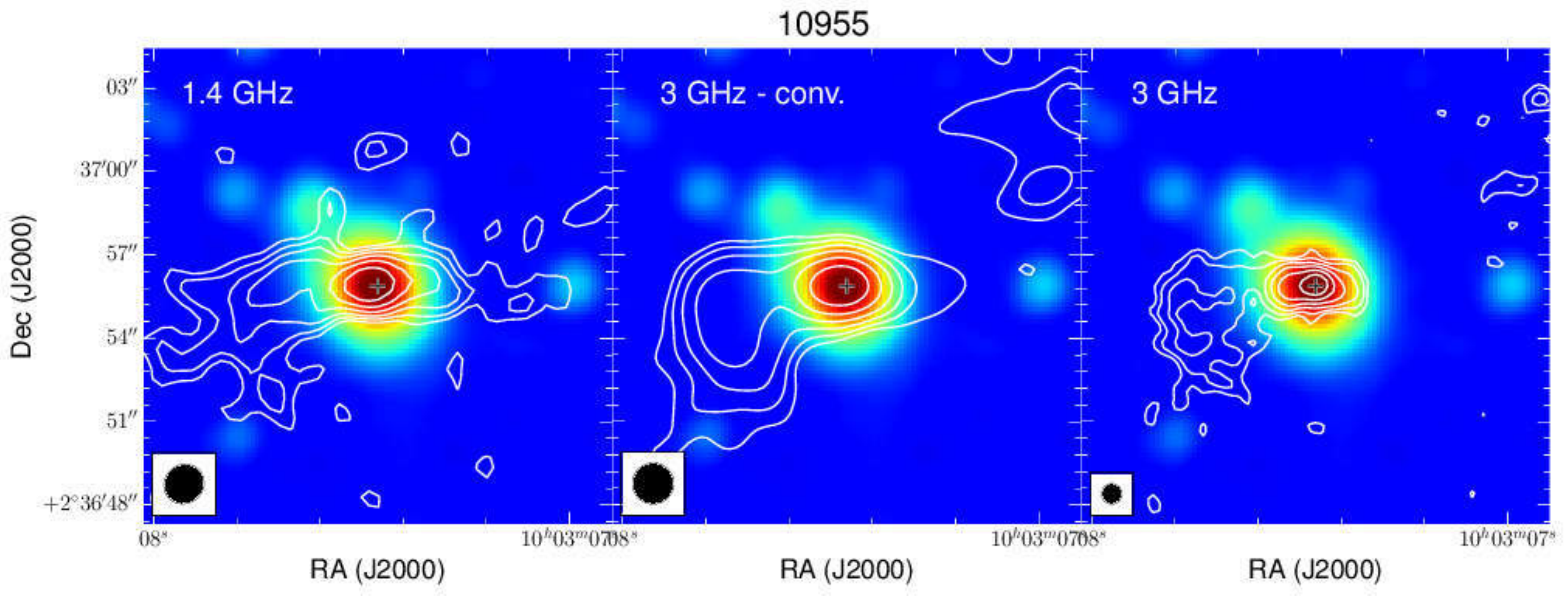}
            }
             \\ \\ 
      \resizebox{\hsize}{!}
       {\includegraphics{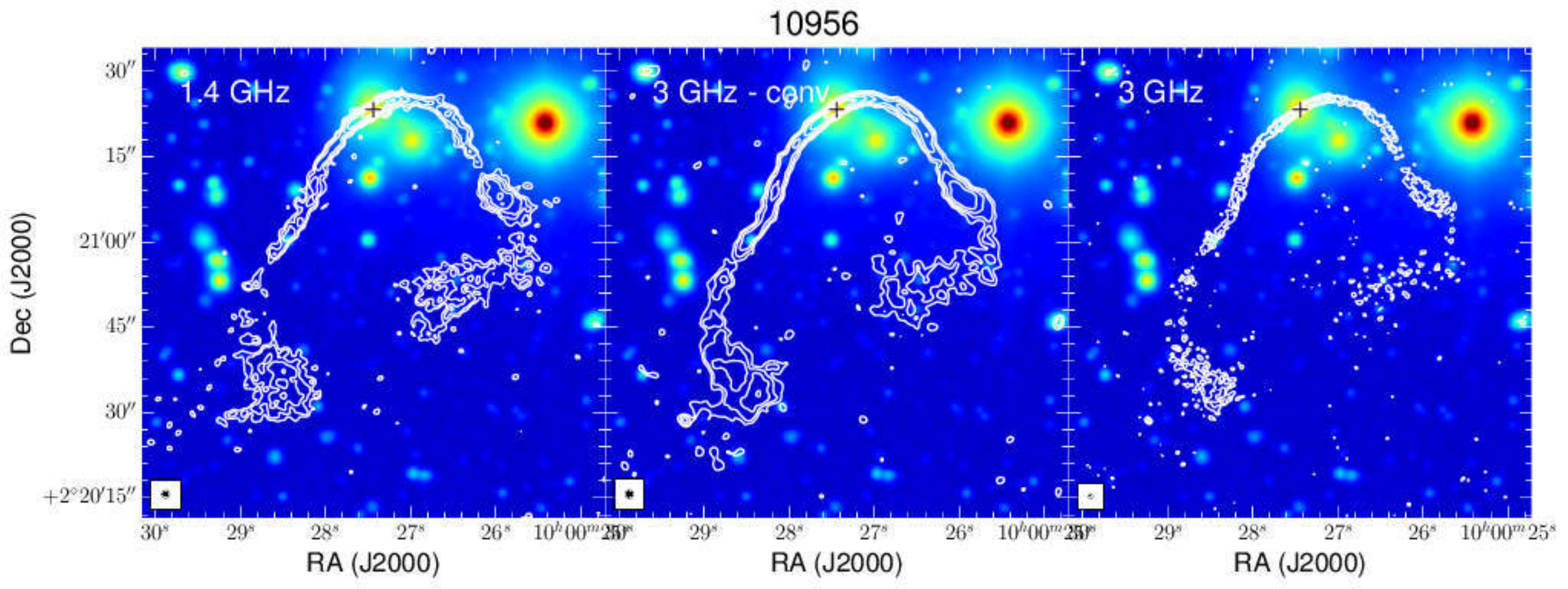}
        \includegraphics{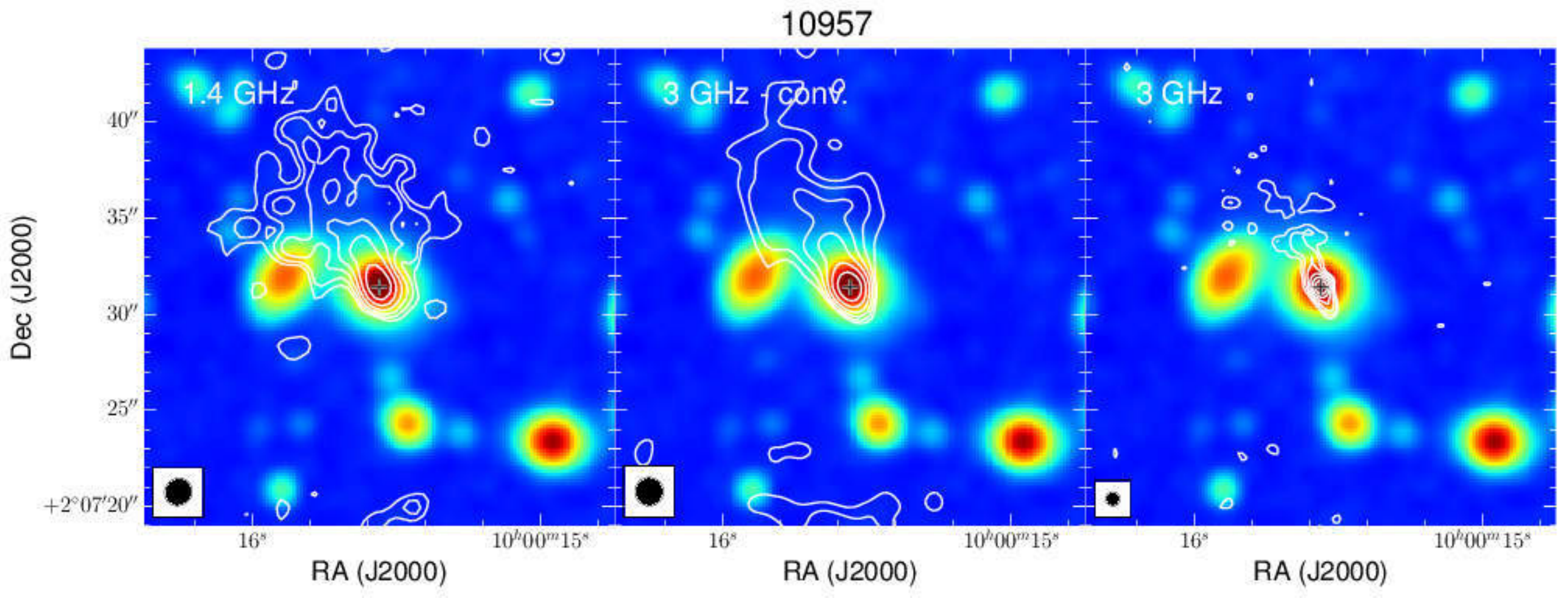}
            }
\\ \\
 \resizebox{\hsize}{!}
{\includegraphics{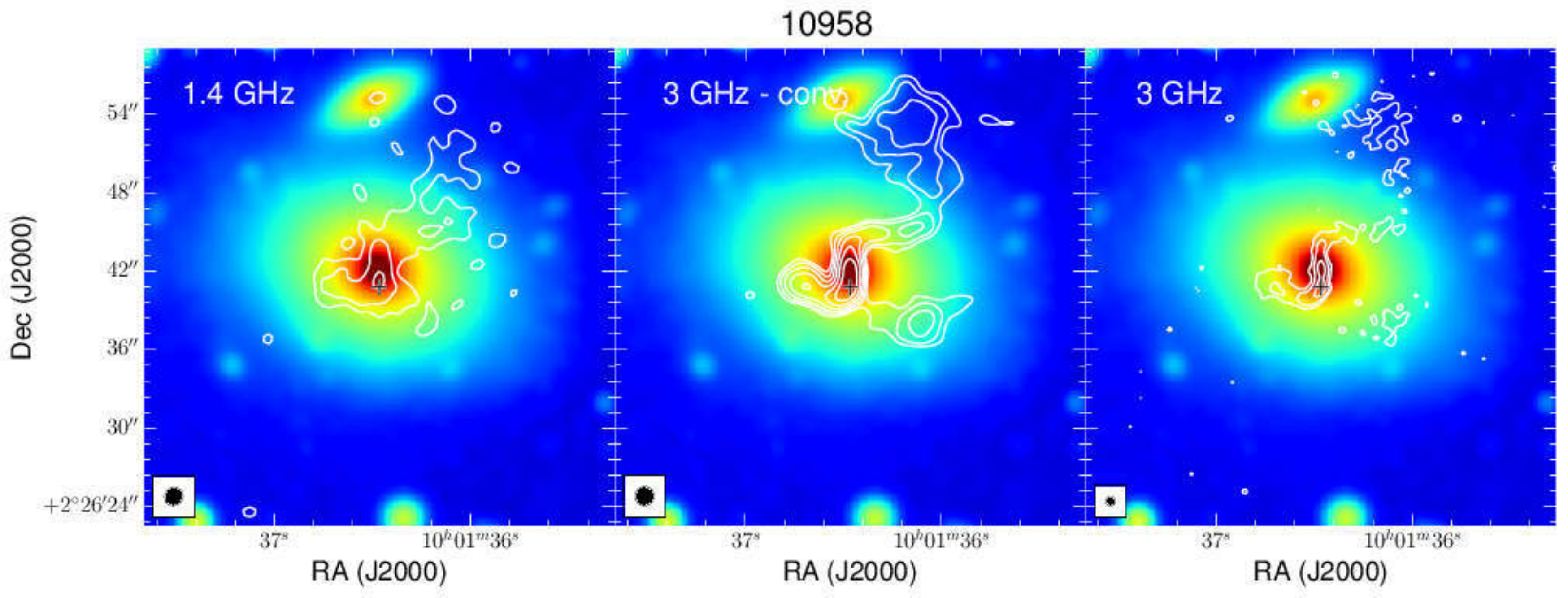}
 \includegraphics{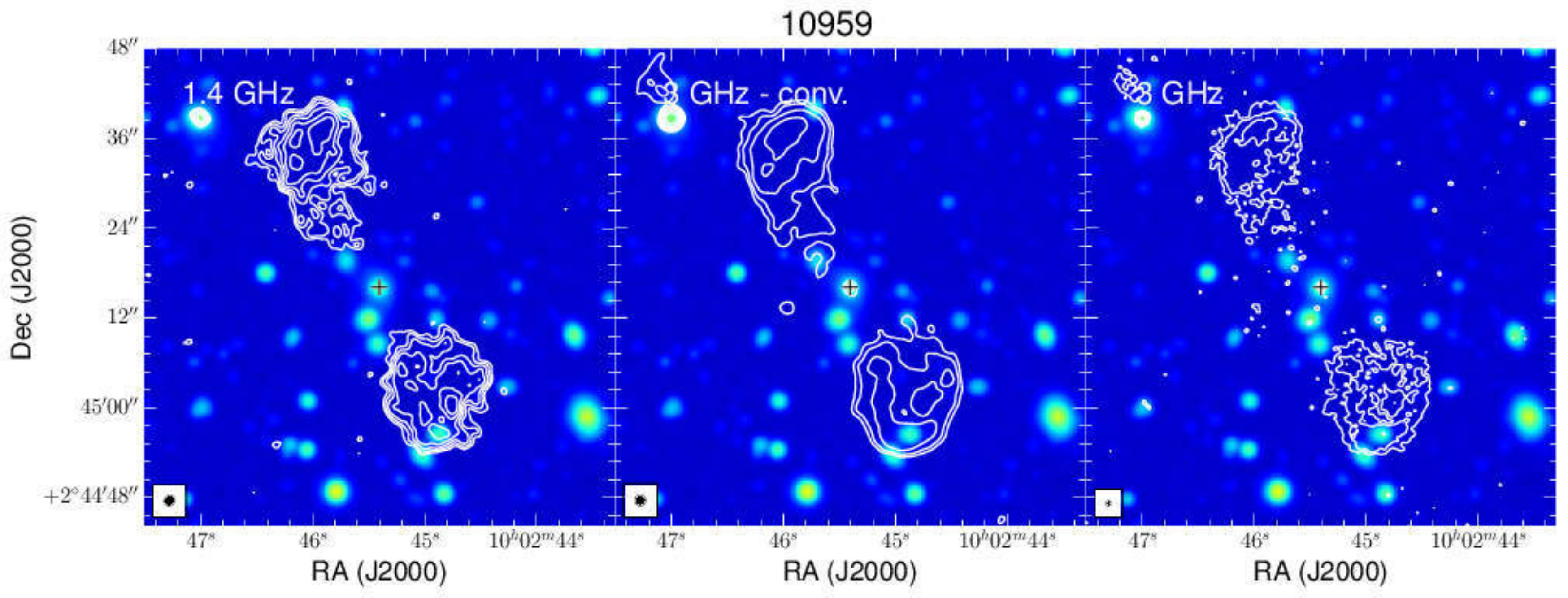}
            }
   \caption{(continued)
   }
              \label{fig:maps2}%
    \end{figure*}
\addtocounter{figure}{-1}
\begin{figure*}[!ht]
    \resizebox{\hsize}{!}
       {\includegraphics{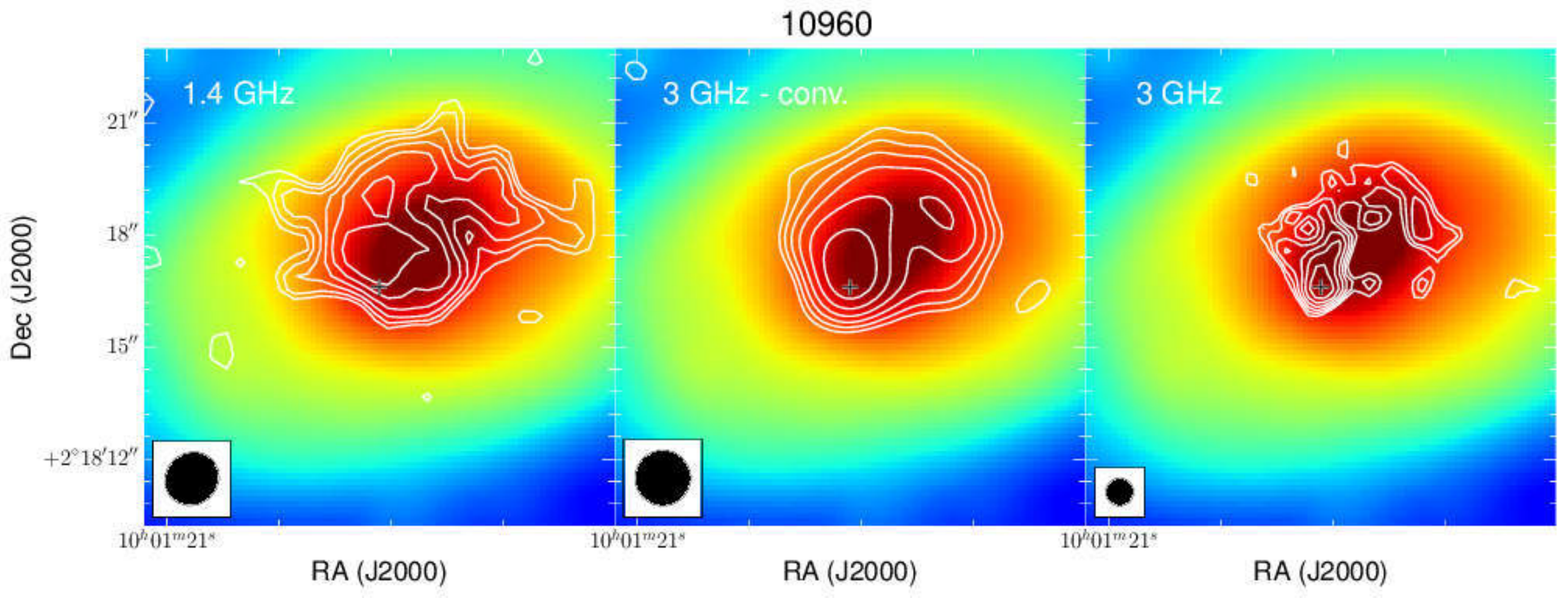}
        \includegraphics{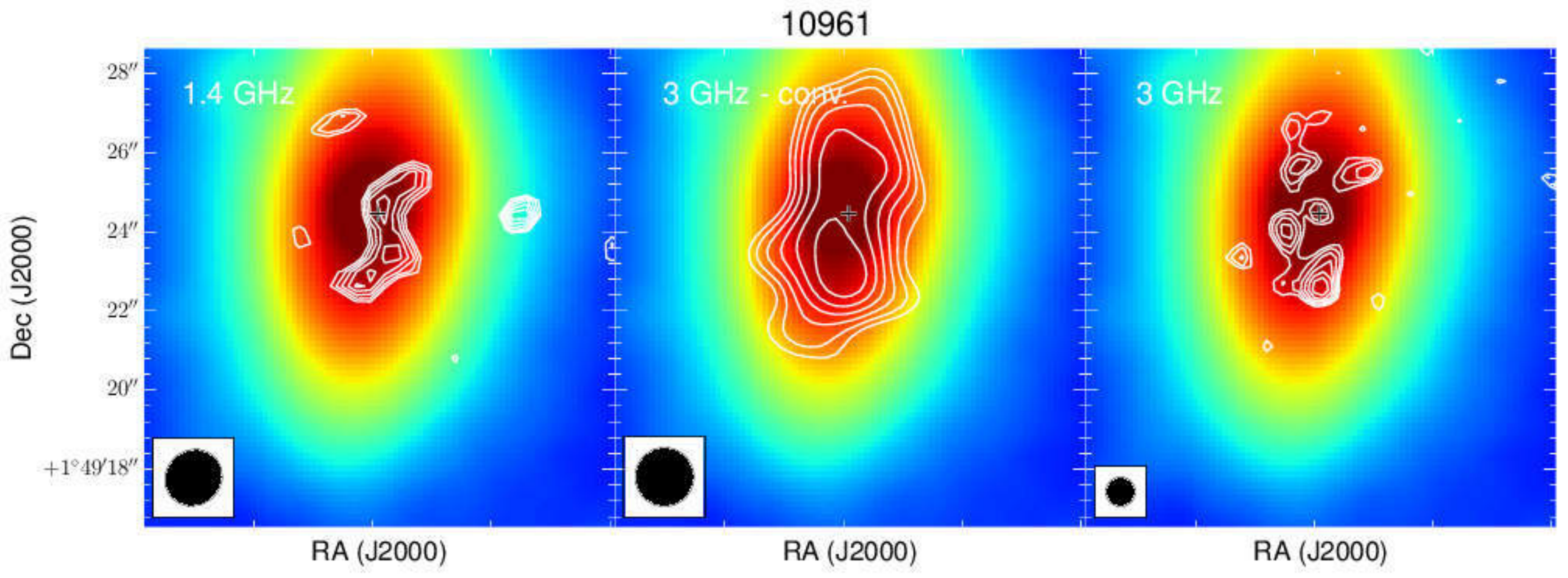}
            }
            \\ \\
 \resizebox{\hsize}{!}
{\includegraphics{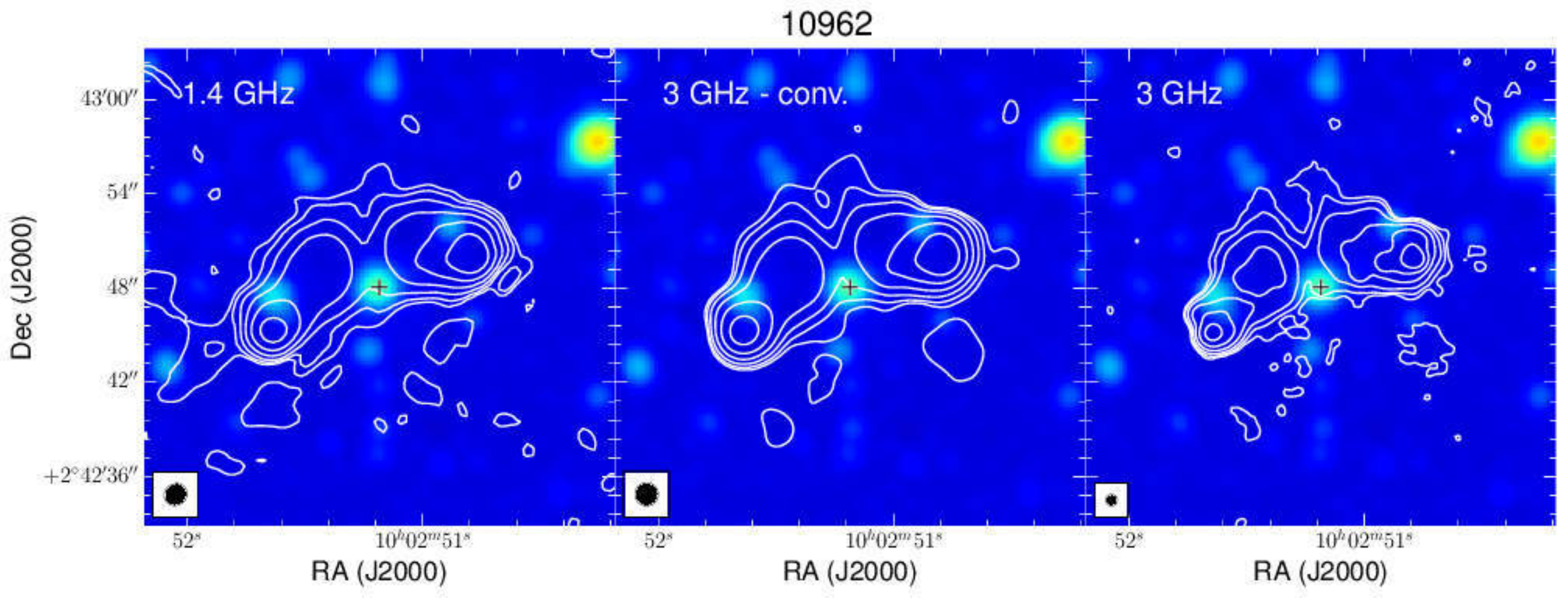}
 \includegraphics{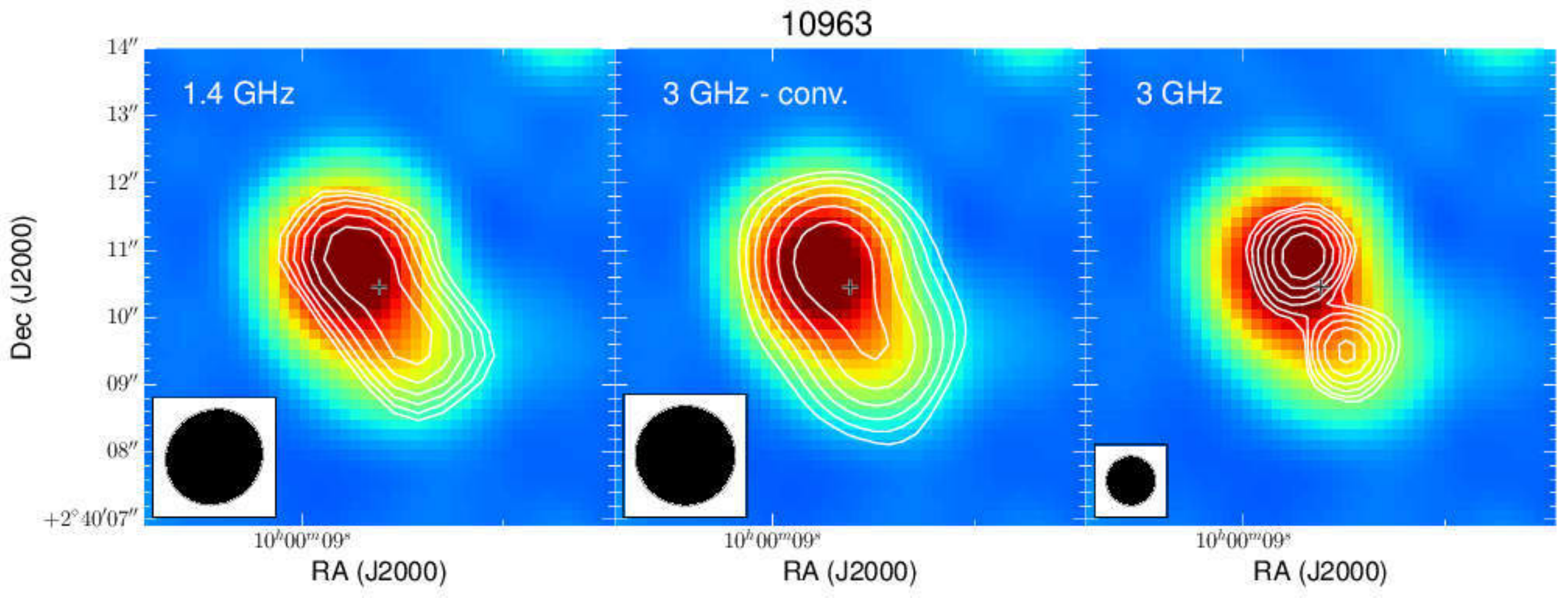}
            }
            \\ \\
  \resizebox{\hsize}{!}
 {\includegraphics{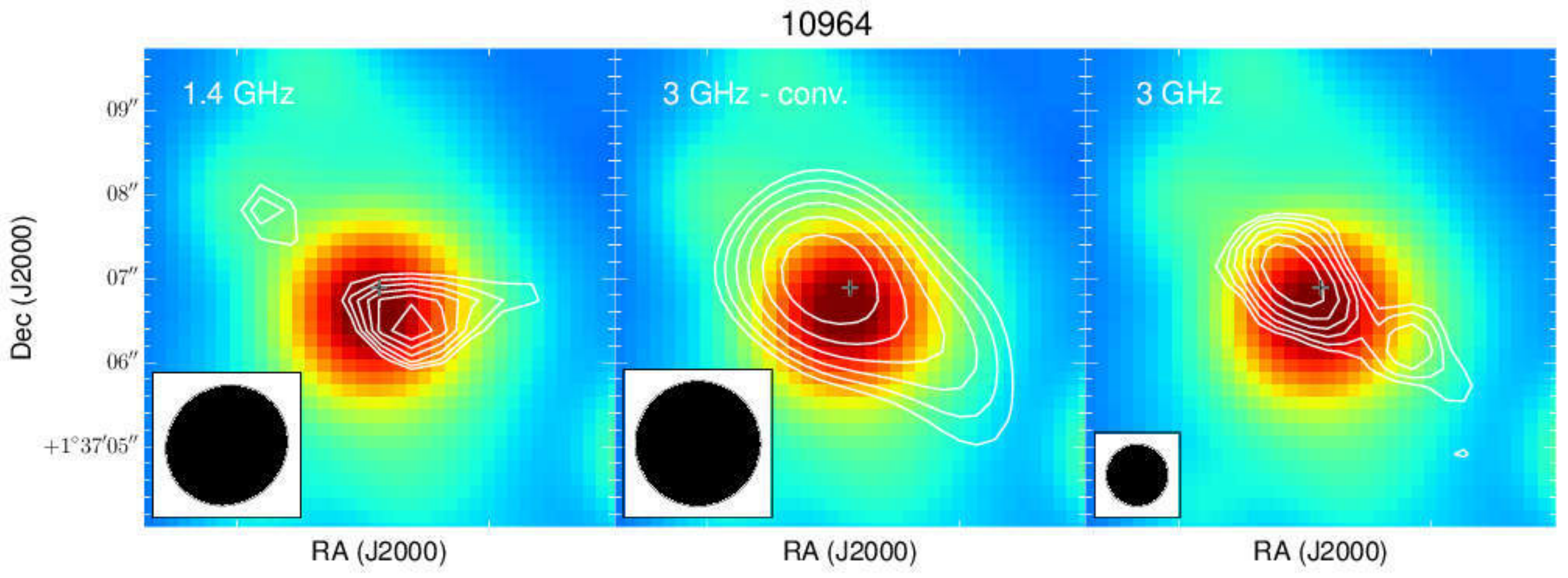}
    \includegraphics{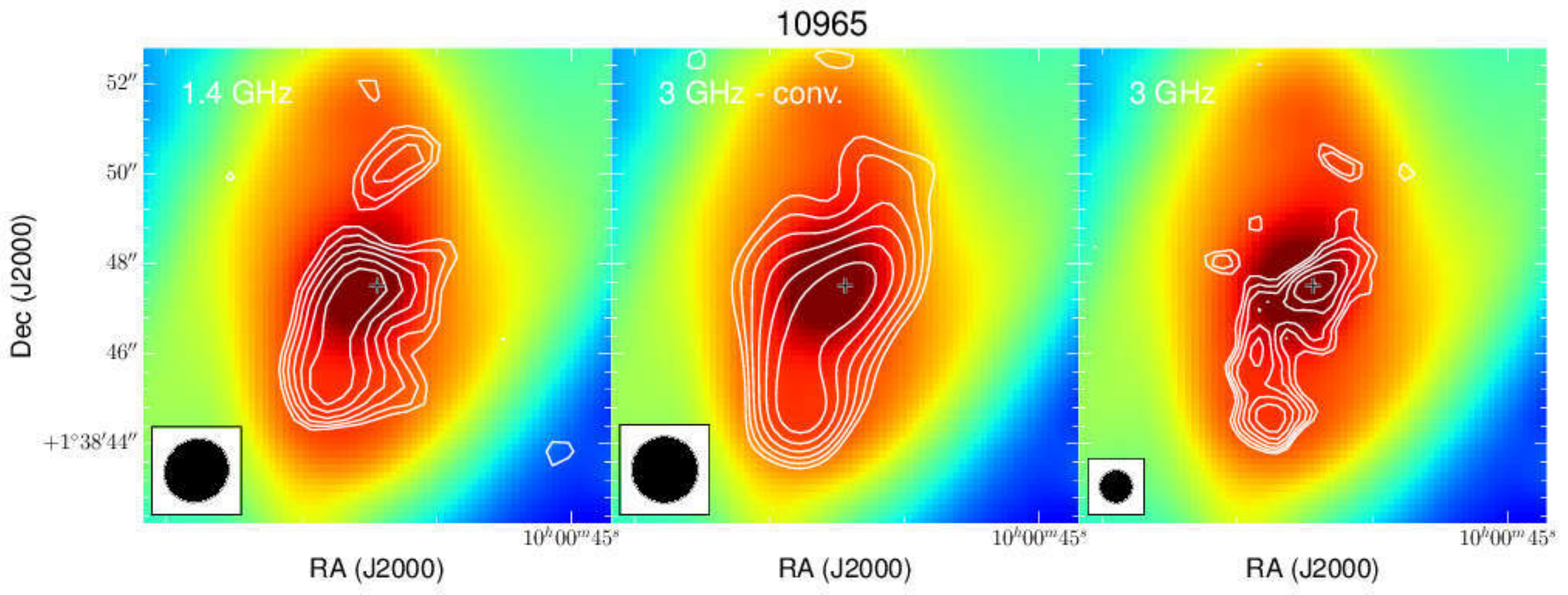}
            }
             \\ \\ 
      \resizebox{\hsize}{!}
       {\includegraphics[width=0.088cm]{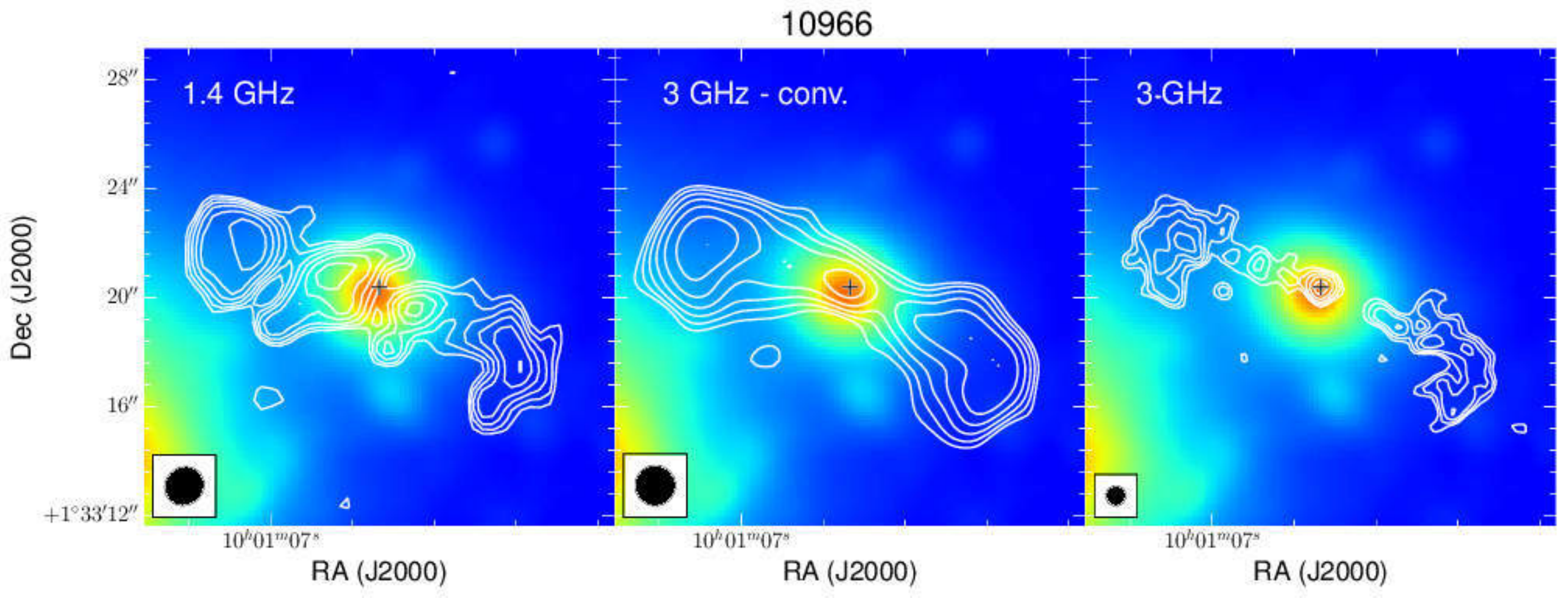}
       
            }
   \caption{(continued)
   }
              \label{fig:maps2}%
    \end{figure*}

\newpage

\end{document}